\documentclass[11pt]{article}

\usepackage[a4paper]{geometry}
\usepackage{authblk}
\usepackage{arydshln}
\usepackage{psfrag}
\usepackage{multirow}
\usepackage{changepage}
\usepackage{lscape}
\usepackage{url}
\usepackage{tikz}
\usepackage[symbol]{footmisc}
\usepackage{geometry}
\usepackage{setspace}
\usepackage{graphicx}
\usepackage{scalerel}

\usepackage[ruled,vlined]{algorithm2e}

\usepackage{changepage}
\usepackage[british]{babel}
\usepackage{float}
\usepackage{mathrsfs,amsmath}
\usepackage{wrapfig}
\usepackage{cases, mathtools}
\usepackage{amssymb, amsbsy} 
\usepackage[labelfont=bf]{caption}
\usepackage{subcaption}
\usepackage{mathrsfs,amsmath}
\usepackage{listings}
\usepackage{bm}

\usepackage[style=ieee, citestyle=numeric-comp, backend=biber,maxbibnames=99]{biblatex}
\definecolor{lightblue}{RGB}{6,69,173}
\definecolor{myRED}{rgb}{1,0,0.0745}
\definecolor{myBLUE}{rgb}{0.0039,0.3961,1}

\usepackage[colorlinks]{hyperref}
\usepackage{xcolor}
\hypersetup{
    colorlinks,
    linkcolor={lightblue},
    citecolor={lightblue},
    urlcolor={lightblue}
}

\usetikzlibrary{shapes.geometric}
\usepackage{rotating}
\usetikzlibrary{quotes,angles}
\usetikzlibrary{patterns}
\usetikzlibrary{decorations.pathmorphing}
\usetikzlibrary{decorations.markings}
\usetikzlibrary{matrix}
\usetikzlibrary{angles, quotes}
\usetikzlibrary{calc}

\def\XXint#1#2#3{{\setbox0=\hbox{$#1{#2#3}{\int}$}
     \vcenter{\hbox{$#2#3$}}\kern-.5\wd0}}

\usepackage{listings}
\usepackage{color}
 
\definecolor{codegreen}{rgb}{0,0.6,0}
\definecolor{codegray}{rgb}{0.5,0.5,0.5}
\definecolor{codepurple}{rgb}{0.58,0,0.82}
\definecolor{backcolour}{rgb}{0.95,0.95,0.92}
 
\lstdefinestyle{mystyle}{
    backgroundcolor=\color{backcolour},   
    commentstyle=\color{codegreen},
    keywordstyle=\color{magenta},
    numberstyle=\tiny\color{codegray},
    stringstyle=\color{codepurple},
    basicstyle=\footnotesize,
    breakatwhitespace=false,         
    breaklines=true,                 
    captionpos=b,                    
    keepspaces=true,                 
    numbers=left,                    
    numbersep=5pt,                  
    showspaces=false,                
    showstringspaces=false,
    showtabs=false,                  
    tabsize=2
}
 
\DeclareMathOperator{\sech}{sech}

\definecolor{myGreen}{RGB}{0,128,0}
\definecolor{myPurple}{RGB}{139,0,139}
\definecolor{myRed}{RGB}{162, 20, 47}
\definecolor{myYellow}{RGB}{255, 215, 0}
\definecolor{myGrey}{RGB}{189, 201, 199}

\newcommand*\circledP{\hspace*{-0.0ex}\raisebox{-0.9ex}{\tikz[]{
		\node[regular polygon, circle, draw = myRED, inner sep=2.5,rotate=0,line width=0.5mm, fill = myRED] at (0,0)  {};  
		\node[text = white] at (0,0) {$\displaystyle  \textbf{+}$}; 
		}}}

\newcommand*\circledM{\hspace*{-0.0ex}\raisebox{-0.9ex}{\tikz[]{
		\node[regular polygon, circle, draw = myBLUE, inner sep=2.5,rotate=0,line width=0.5mm, fill = myBLUE] at (0,0)  {};  
		\node[text = white] at (0,0) {\large $\displaystyle  -$}; 
		}}}

\newcommand*\circledPM{\hspace*{-0.0ex}\raisebox{-0.9ex}{\tikz[]{
		\node[regular polygon, circle, draw = myPurple, inner sep=2.5,rotate=0,line width=0.5mm, fill = myPurple] at (0,0)  {};  
		\node[text = white] at (0,0) {$\displaystyle  \boldsymbol{\pm}$}; 
		}}}		
		
\newcommand*\circledMP{\hspace*{-0.0ex}\raisebox{-0.9ex}{\tikz[]{
		\node[regular polygon, circle, draw = myPurple, inner sep=2.5,rotate=0,line width=0.5mm, fill = myPurple] at (0,0)  {};  
		\node[text = white] at (0,0) {$\displaystyle  \boldsymbol{\mp}$}; 
		}}}					

\usepackage{float,yfonts}

\title{Analytical solutions for Bloch waves in resonant phononic crystals: \\ Deep subwavelength energy splitting and mode steering between topologically protected interfacial and edge states}

\author[1]{R. Wiltshaw}
\author[2]{J. M. De Ponti}
\author[1]{R. V. Craster}
\affil[1]{Department of Mathematics,
Imperial College London,
London, SW7 2AZ, United Kingdom}
\affil[2]{Department of Civil and Environmental Engineering, Politecnico di Milano, Piazza Leonardo da Vinci, 32, 20133 Milano, Italy}

\date{ }

\begin{document}

\setstretch{1.2}
\maketitle
\setstretch{1.0}

\begin{abstract}
We derive analytical solutions based on singular Green's functions, which enable efficient computations of scattering simulations or Floquet-Bloch dispersion relations for waves propagating through an elastic plate, whose surface is patterned by periodic arrays of elastic beams. Our methodology is versatile and allows us to solve a range of problems regarding arrangements of multiple beams per primitive cell, over Bragg to deep-subwavelength scales; we cross-verify against finite element numerical simulations to gain further confidence in our approach, which relies upon the hypothesis of Euler-Bernoulli beam theory 
 considerably simplifying continuity conditions such that each beam can be replaced by point forces and moments applied to the neutral plane of the plate. The representations of Green's functions by Fourier series or Fourier transforms readily follows, yielding rapid and accurate analytical schemes.
The accuracy and flexibility of our solutions are demonstrated by engineering topologically non-trivial states, from primitive cells with broken spatial symmetries, following the phononic analogue of the Quantum Valley Hall Effect (QVHE). Topologically protected states are produced and coexist along: interfaces between adjoining chiral-mirrored bulk media and edges between one such chiral bulk and the surrounding bare elastic plate, allowing topological circuits to be designed with robust waveguiding; these topologically non-trivial states exist within near flexural resonances of the constituent beams of the phononic crystal, and hence can be tuned into a deep-subwavelength regime.



\end{abstract}

\section{Introduction}
Identifying degeneracies, such as locally dispersionless band structures meeting at a point, allows periodic structured media to be constructed with remarkable propagation properties, e.g. extraordinary transmission \cite{sepkhanov2007extremal,zhang2008extremal,bittner2012extremal}, cloaking \cite{huang2011dirac,guo2013dirac}, interfacial states \cite{makwana2020hybrid, lu2016topological, makwana2018designing, gao2017valley, dong2017valley, yang2018topological, kang2018pseudo, chen_tunable_2018, he2019silicon, lu_observation_2016, ye2017observation, zhang_topological_2018,  wu_direct_2017, jung2018midinfrared, gao2018topologically, zhang2019valley,  xia2018observation, shalaev2018experimental, liu2018tunable}, edge states \cite{wang2015topological,torrent2013elastic, skirlo2014multimode}, one-way propagation \cite{wang2015topological,wang2008reflection,raghu2008analogs,haldane2008possible,poo2011experimental,ao2009one}, or a near-zero refractive index \cite{huang2011dirac} - all of which can be engineered in a variety of physical settings. This has led to extremely active communities, predominantly in photonics \cite{zolla_foundations_2005,joannopoulos_photonic_2008} and phononics \cite{laude2015phononic,Craster2013},  extensively researching designs and devices that take advantage of this power to manipulate waves. 

The characteristic length scale of fundamental units forming a crystalline media can have a variety of scales depending on the frequencies of propagating waves to be controlled. Optical properties of nanomaterials/nanocomposites, for instance the lustre attained by potteries and glasses can be described through a modern photonic crystal framework and are indeed ancient nanoscale technologies \cite{colomban2009use}. Nature provides examples ranging from microphotonic crystals forming the iridescent colours of opals and certain butterfly species \cite{gralak2003structural}, to geophysical scales where forests have been shown to act as phononic crystals capable of controlling seismic waves \cite{colombi2016forests,lott2020evidence} 
 by manipulating surface Rayleigh waves via subwavelength band gaps. The focus of \cite{colombi2016forests} was on compressional resonances of $\mathcal{O}(10)$ Hz, providing strong attenuation of waves with wavelengths $5-10$ times that of the lattice spacing. Herein we design phononic crystals containing symmetry protected Dirac cones coinciding with the flexural resonances of beams atop a thin elastic plate which, once gapped, produce very low $\mathcal{O}(0.1)$ Hz frequency band gaps; allowing the manipulation of waves whose wavelengths are $35-40$ times that of the lattice spacing. Our designs produce topologically non-trivial states which are capable of confining modes along various interfaces and edges, following the elastic analogue of the QVHE; the topological protection produced by our designs show a variety of conventional and new properties at the deep subwavelength scale.

The valleytronics community \cite{makwana2020hybrid, lu2016topological, makwana2018designing, gao2017valley, dong2017valley, yang2018topological, kang2018pseudo, chen_tunable_2018, he2019silicon, lu_observation_2016, ye2017observation, zhang_topological_2018,  wu_direct_2017, jung2018midinfrared, gao2018topologically, zhang2019valley,  xia2018observation, shalaev2018experimental, liu2018tunable} often use the terms interfacial and edge states interchangeably. For clarity we will refer to states existing along the interface between two chiral-mirrored bulk media as interfacial states. The states which live on the edge of the crystal, between the bulk derived from one such chiral pair and the free space, will be referred to as edge states; we demonstrate how these topologically protected interfacial and edge states co-exist within the subwavelength band gap of the underlying bulk media, and how modal conversion or preservation allows modes to efficiently navigate corners or split energy between the interfaces and edges of our phononic crystal designs, as shown later in figs. \ref{fig:ScattSquareTopoArrange}, \ref{fig:ScattPertHexTopoArrange} \& \ref{fig:ScattPertHexTopoRobustArrange}. 
The near-resonant behaviour of our topologically distinct valleys allows us to tune our topologically protected states into a deep-subwavelength regime, where we demonstrate the interfacial and edge modes are sufficiently robust to navigate gentle and sharp \cite{makwana2018geometrically} corners, with minimal backscatter, due to the combination of low-frequency resonances  \cite{torrent2013elastic,wang2019topological} and topological protection \cite{wang2015topological,wiltshaw2020asymptotic}; our edgemodes either propagate around the entire perimeter of the crystal, or hybridise into corner states \cite{palmer2021berry, palmer2021revealing,chen2021corner} which readily shed energy into the surrounding free-space. 

Recent years have witnessed an increasing popularity of metamaterial concepts, based on  local resonance phenomenon, to control the propagation of electromagnetic  \cite{li2006locally}, acoustic \cite{Craster2013,porter2018plate} and elastic waves \cite{Craster2017} in artificially engineered media. In elasticity, resonant structures formed from arrays of beams yield superior characteristics for broadband wave focusing \cite{Aguzzi2022}, mode conversion \cite{colombi2016seismic, colombi2017enhanced,colombi2017elastic} and energy harvesting \cite{Deponti2020, DePonti2021Harvesting,Chaplain2020}. Novel works combining resonance and topological protection \cite{zhang2019subwavelength,chaunsali2018subwavelength,chaunsali2018experimental,wang2019topological,qi2022valley,fang2022valley,yves2017topological,yves2017crystalline,zhang2020dirac} achieve confinement of modes at the subwavelength scale; for instance \cite{zhang2019subwavelength,chaunsali2018subwavelength,chaunsali2018experimental,wang2019topological,qi2022valley,fang2022valley,yves2017topological,yves2017crystalline,zhang2020dirac} consider zero-line modes, the robustness of which is proven by introducing bends \cite{makwana2018designing,tang2020observations,lu_observation_2016,zhang_topological_2018,zhang_manipulation_2018,shalaev2018experimental,liu2019experimental,jung_active_2018,gao2017valley,chen_tunable_2018,wiltshaw2020asymptotic} or splitting energy \cite{makwana19a,proctor2020manipulating,makwana2019topological}. 

In a phononic setting, long-waves can have negative impacts; for example, those who live in flats or in close proximity to large scale transport infrastructure will often complain of long-wave disturbances, generated from low-frequency changes to pressure (e.g. music with deep bass) or vibrations from passing vehicles. 
Long-wave disturbances also have the potential to be highly damaging. Naturally occurring seismic events are well known to cause large scale, irreparable damage to sensitive infrastructure; for instance, earthquakes whose epicentre could be considered far field \cite{colombi2016seismic,aki2002quantitative} are of particular concern to densely populated urban environments. 
 There is considerable research around mechanisms and devices for low-frequency long waves to be manipulated and steered, with resonant phononic crystals showing great promise in areas of seismic and phononic passive protection. 


The extraordinary properties exhibited by our phononic crystal designs require symmetry-based arguments. We design our media through Floquet-Bloch \cite{brillouin1953wave,kittel1996introduction} analysis, allowing the dispersive nature of doubly periodic structured medium to be deduced from eigensolutions through primitive cells with Floquet-Bloch boundary conditions. Once the underlying structure has been suitably engineered, we test the localisation of our topologically non-trivial states and how they behave when excited; eigenproblems formed from supercells \cite{makwana2018designing} of these fundamental units, or scattering simulations of large finite collections of these cells are considered to test our topologically protected designs. These problems can be solved by general finite element method (FEM) analysis, however considering arrays of beams attached to plates, the number of nodes required to discretise the problem leads to computationally expensive results with long computation times. Nevertheless, FEM analysis leads to accurate results and is by far the most popular method considered by the metamaterial community. 

We aim to develop much faster and efficient methods based on singular Green's functions. Our solutions are cross-verified against FEM computations and are shown to be highly accurate, efficient and flexible enough to consider a variety of Floquet-Bloch eigensolutions and scattering simulations. Codes based on these analytical solutions are fast to run, thereby expediting the process of modifying the underlying media to produce the required dispersion. Moreover, once the crystal has been designed, we need to test its effectiveness in achieving the desired effect through direct scattering simulations of finite arrangements of scatterers under incidence. Analytically considering scatterers placed in fields with infinite domains allows for efficient computations. For instance, the scattering simulations in figs. \ref{fig:ScattSquareTopoArrange}, \ref{fig:ScattPertHexTopoArrange} \& \ref{fig:ScattPertHexTopoRobustArrange} are merely plots of our analytical solutions; not only do they contain $\mathcal{O}(10^4)$ beams judiciously placed atop the elastic plate, each beam scatters the field such that the Sommerfeld radiation condition is satisfied at infinity. Moreover, each computation took $\mathcal{O}(10)$ minutes for a single frequency even near resonance. Perfectly matched layers can be used within FEM computations to replicate these infinite domains and perform similar scattering simulations, however the computational time and power required for such results would be considerably larger than our analysis. 

The aims of this study are to outline all necessary analytical nuances to deal with singular Green's functions, allowing simple codes to be constructed such that Floquet-Bloch eigensolutions or scattering simulations can be computed for phononic crystals built from resonant objects. In section \ref{sec:SGFIE}, we derive the singular Green's function for one beam atop an elastic plate of infinite expanse; we extend this solution to consider dispersion relations and eigenmodes for Floquet-Bloch wave propagation through periodic infinite arrays of beams atop a plate in section \ref{sec:EigProb}, and scattering simulations for finite collections of beams atop an elastic plate under incident forcing in section \ref{sec:Foldy}. These problems are shown to simplify down to an algebraic eigenvalue problem or determining the inverse of a matrix. We test our analytical solutions against FEM computations in section \ref{sec::Testing}. Finally in section \ref{Sec:Topo}, we demonstrate the simplicity, accuracy and efficiency of our analytically constructed codes by designing a variety of topologically protected states, existing within deep-subwavelength topological band gaps of the underlying bulk medium; we also test these designs by performing large scale scattering simulations in which our phononic crystals, formed by finite arrays of these primitive cells, are subject to scattering from some incident source of energy.


\section{Singular Green's functions in elasticity} \label{sec:SGFIE}
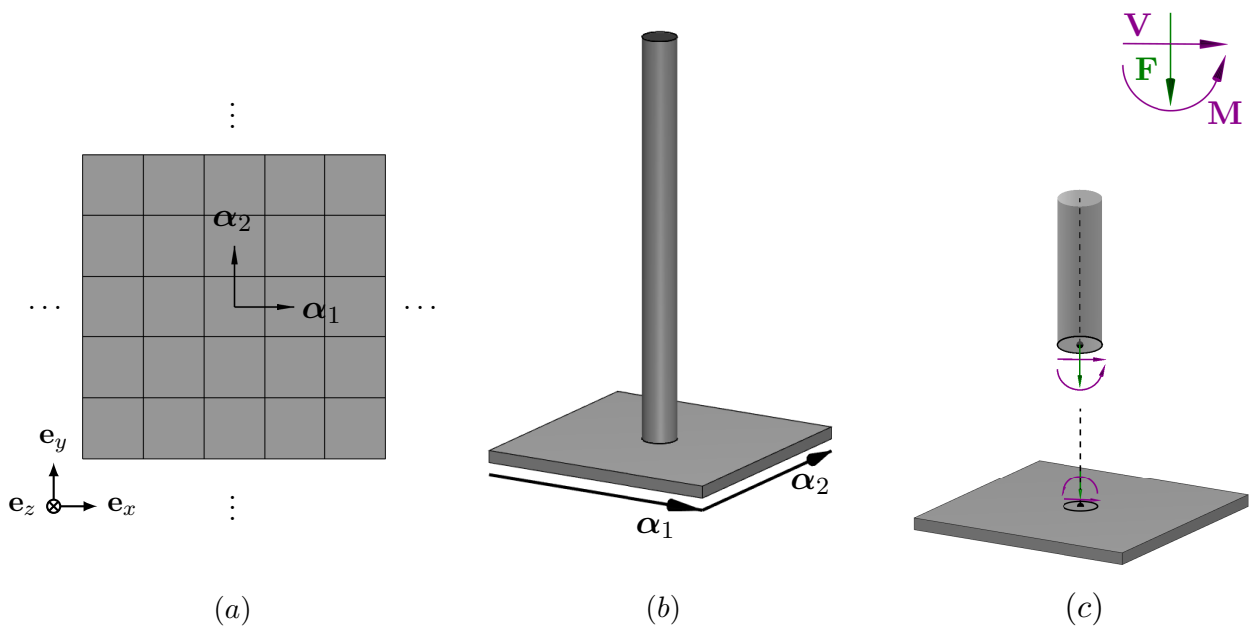
\begin{figure}[h]
\hspace*{-0.25 cm}
\begin{tikzpicture}[scale=0.58, transform shape,cross/.style={path picture={ 
  \draw[black]
(path picture bounding box.south east) -- (path picture bounding box.north west) (path picture bounding box.south west) -- (path picture bounding box.north east);
}}]

\begin{scope}[shift = {(-7,-1)}]
\draw (-4.625, -3.5) node[inner sep=0] {\includegraphics[scale=0.5]{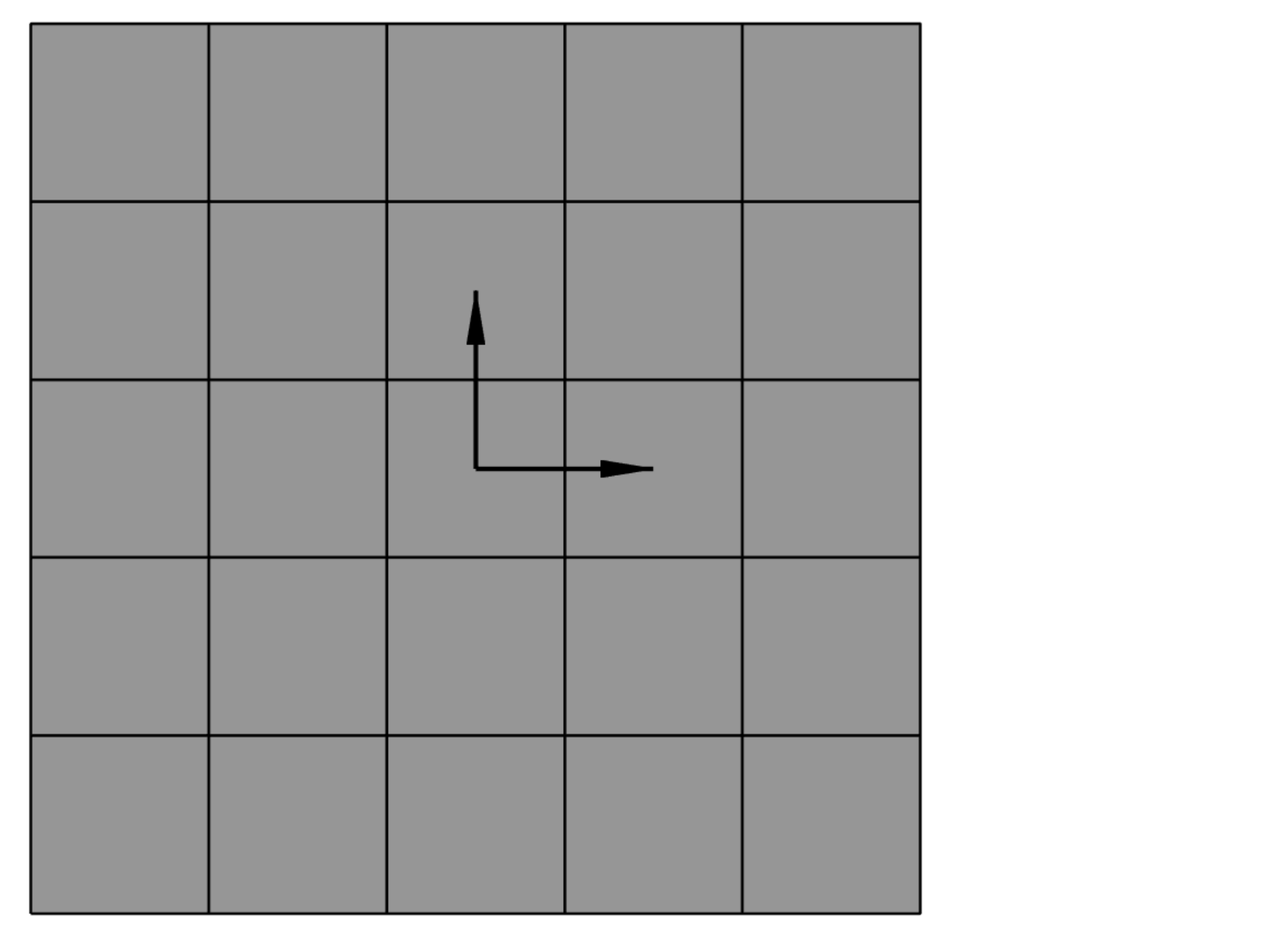}};
\node[below, scale=2,black] at (-3.85,-3.1) {$\displaystyle  \boldsymbol{\alpha}_{1}$}; 
\node[below, scale=2,black] at (-5.9,-3.1+2.1) {$\displaystyle  \boldsymbol{\alpha}_{2}$}; 
\node[below, scale=1.75,] at (-5.9+4.3,-3.2) {$\displaystyle  \ldots$}; 
\node[below, scale=1.75,] at (-5.9-4.3,-3.2) {$\displaystyle  \ldots$}; 
\node[below, scale=1.75,] at (-5.9,-3.2+5.0) {$\displaystyle  \vdots$}; 
\node[below, scale=1.75,] at (-5.9,-3.2-4.3+0.35) {$\displaystyle  \vdots$}; 
\node[below, scale=1.75,] at (-5.9,-3.2-4.3-2.3) {$\displaystyle  (a)$}; 
\end{scope}

\begin{scope}[shift={(-10,-9)}]
\draw[thick,-latex] (-7 , 0 ) -- (-6 , 0);
\draw[thick,-latex] (-7 , 0 ) -- (-7 , 1);
\draw[thick,cross,fill=white] (-7,0) circle (0.15);
\node[right, scale=1.75] at (-6,0) {$\displaystyle \textbf{e}_{x}$};
\node[above, scale=1.75] at (-7,1) {$\displaystyle \textbf{e}_{y}$};
\node[left, scale=1.75]at (-7.15,0) {$\displaystyle \textbf{e}_{z}$};
\end{scope}

 \begin{scope}[shift = {(3, 0)}]  
		\node[rectangle,draw, inner sep=6 cm,rotate=0,line width=0.0mm, white,
           path picture={
               \node[rotate=0] at (-2,0){
                   \includegraphics[scale=1]{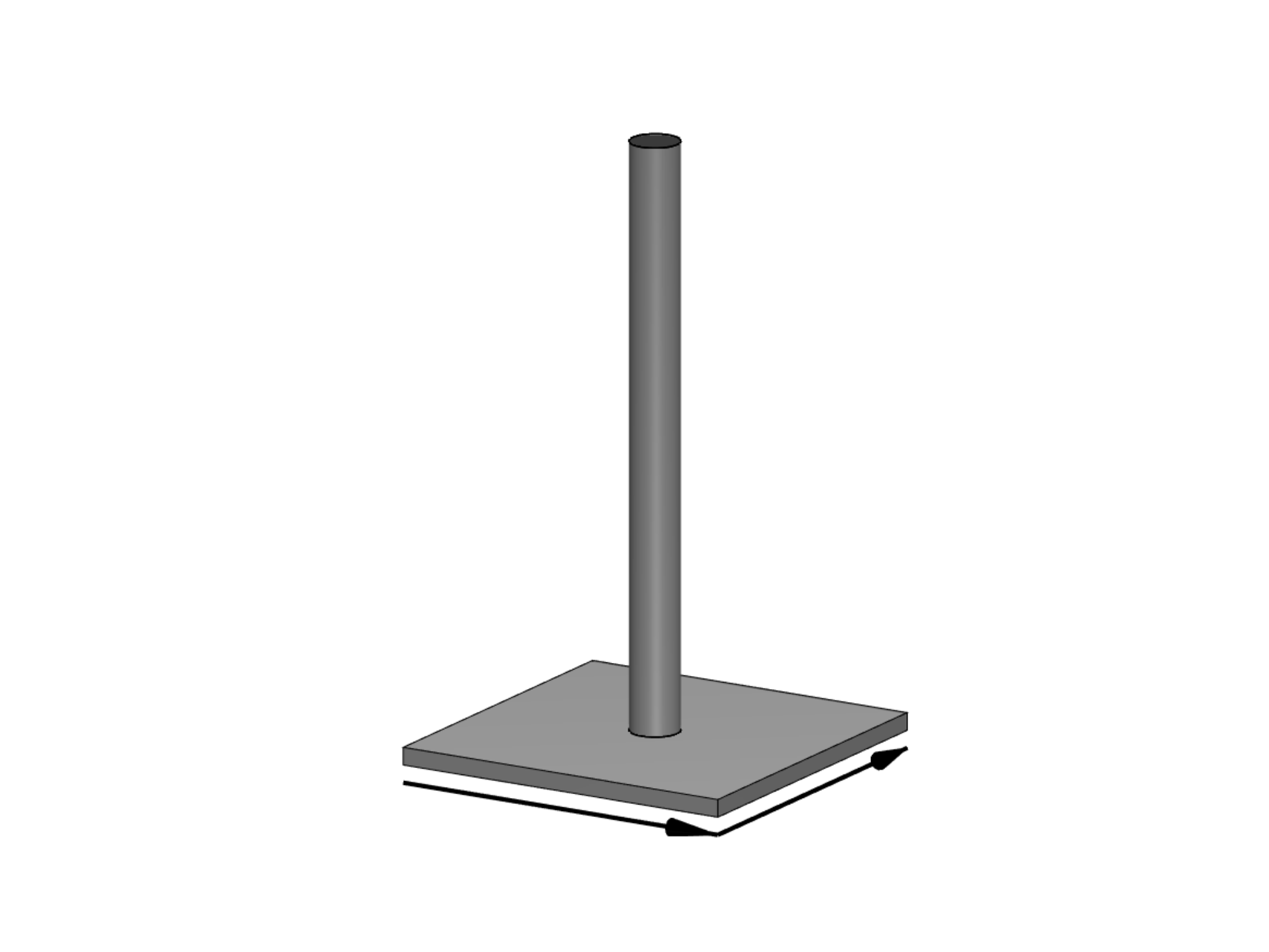}
               };
           }] at (-4.625+0.2, -3.5) {};
        \node[below, scale=1.75,] at (-4.625 + 0.25-1.65,-2.5 -8.275) {$\displaystyle  (b)$}; 
        \node[below, scale=2,black] at (-2.65,-8.0) {$\displaystyle  \boldsymbol{\alpha}_{2}$}; 
		\node[below, scale=2,black] at (-6.2,-3.1-5.9) {$\displaystyle  \boldsymbol{\alpha}_{1}$}; 
\end{scope}

 \begin{scope}[shift = {(10+1.0,-1.5-3.2)}]  
		\node[rectangle,draw, inner sep=4 cm,rotate=0,line width=0.0mm, white,
           path picture={
               \node[rotate=0] at (-0.2, 0){
                   \includegraphics[scale=0.5]{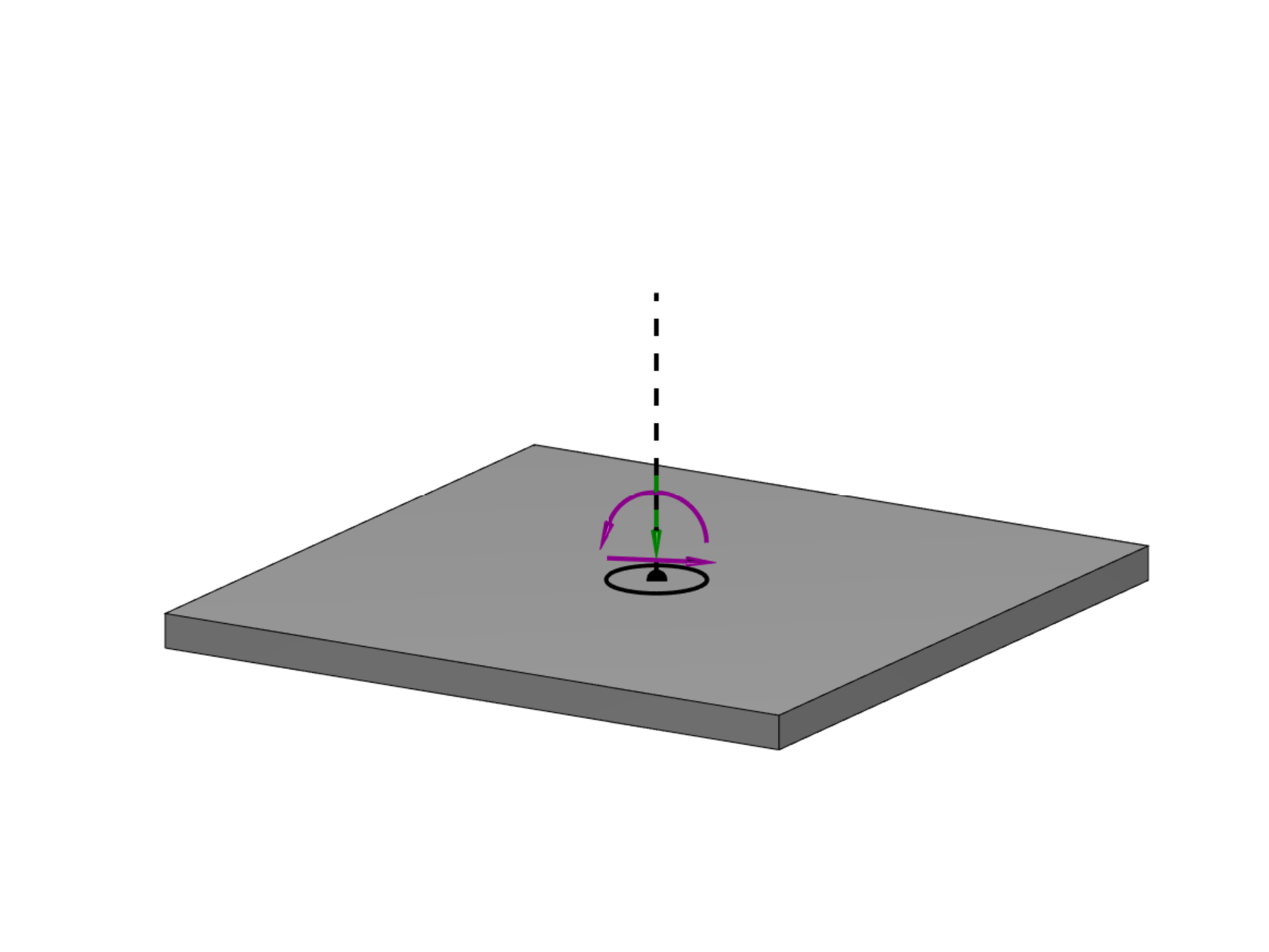}
               };
           }] at (-4.625+0.2, -3.5) {};
        \node[below, scale=1.95] at (-4.625 + 0.25,-2.5 -3.5) {$\displaystyle  (c)$};   
\end{scope}

 \begin{scope}[shift = {(10.15+0.6,1.5-3.15 + 1)}]  
		\node[rectangle,draw, inner sep=4 cm,rotate=0,line width=0.0mm, white,
           path picture={
               \node[rotate=0] at (0, -2.2){
                   \includegraphics[scale=0.5]{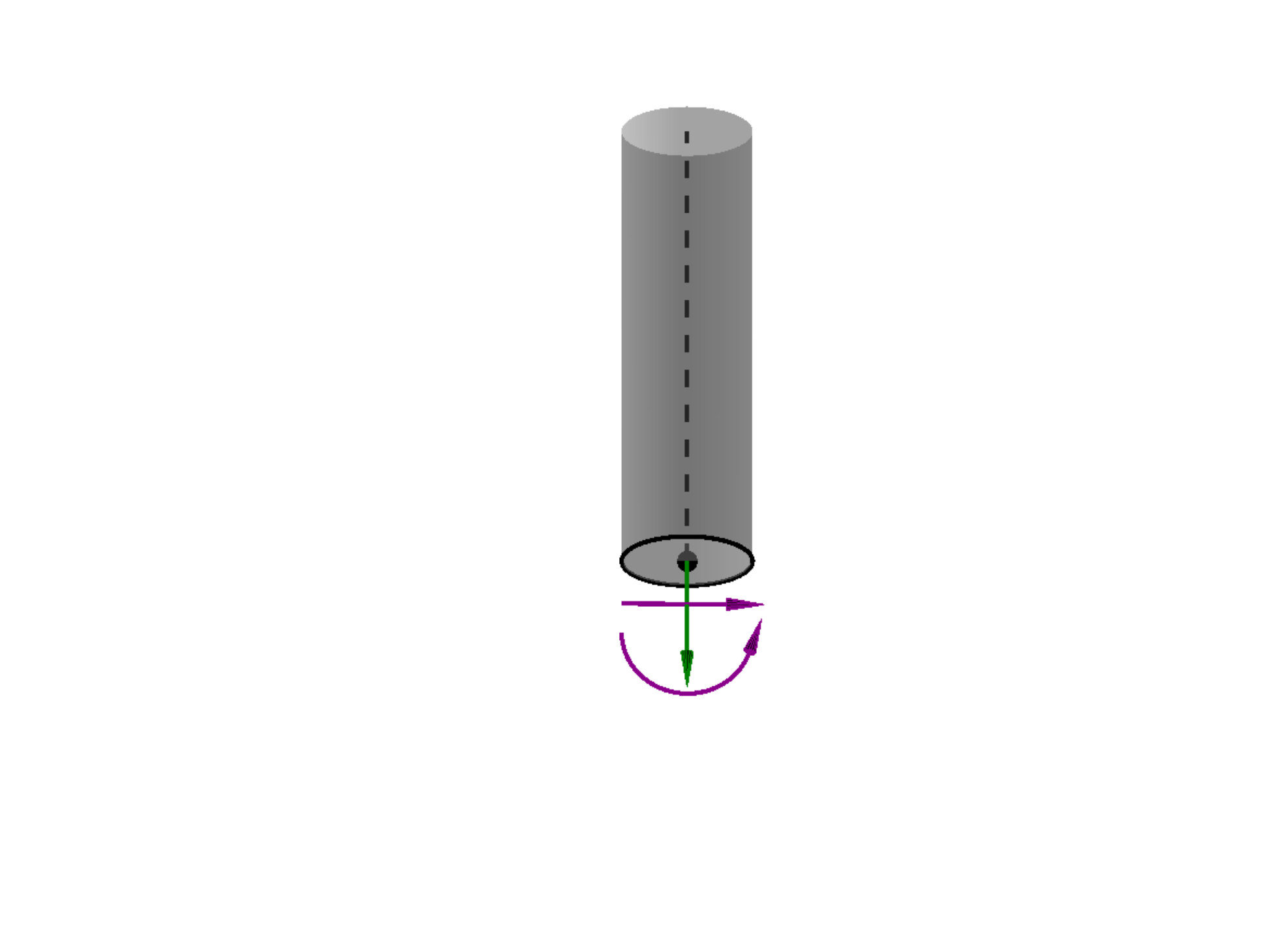}
               };
           }] at (-4.625, -3.5+1.7) {};
\end{scope}  
 
\begin{scope}[shift = {(13.15,5.5)}]
	\node[rectangle,draw, inner sep=2.5 cm,rotate=0,line width=0.0mm, white,
           path picture={
               \node[rotate=0] at (0, 0){
                   \includegraphics[scale=0.5]{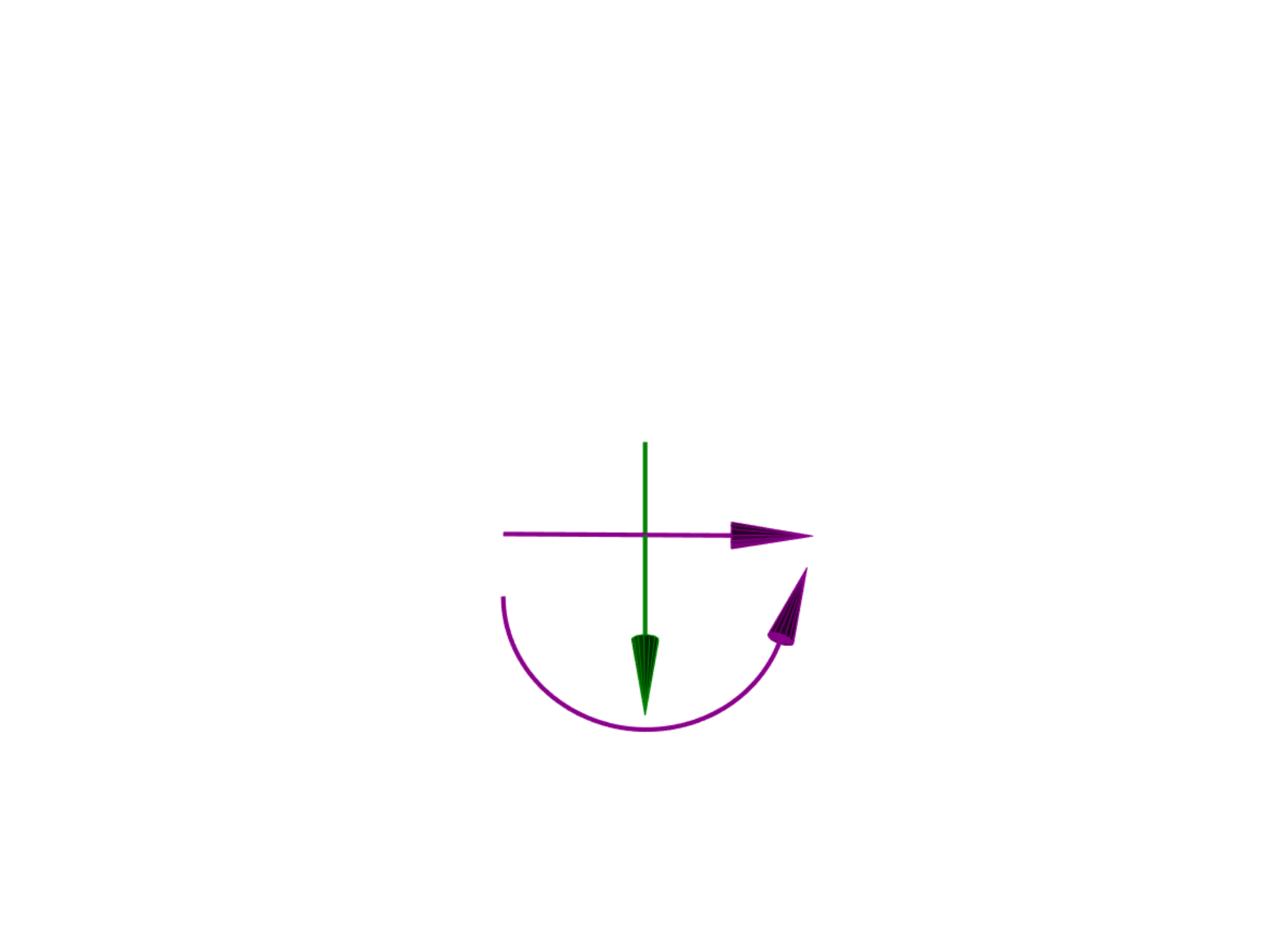}
               };
           }] at (-4.625, -3.5) {};
	\node[below, scale=1.95, myPurple] at (-3.8 + 0.5,-3.5-1.5) {$\displaystyle  \textbf{M}$};
	\node[below, scale=1.95, myPurple] at (-3.8 + -1.5,-3.5+0.5) {$\displaystyle  \textbf{V}$};   
	\node[below, scale=1.95, myGreen] at (-4.6 + 0.5-1,-3.5-2+1.5) {$\displaystyle  \textbf{F}$};   
\end{scope}
\end{tikzpicture}
\caption{A doubly periodic arrangement of beams patterning the surface of an elastic plate. Here $(a)$ shows the media in physical space, where the coordinate system is chosen such that the neutral plane of the elastic plate is spanned by $\textbf{e}_{x}$ and $\textbf{e}_{y}$ Cartesian basis vectors with $\textbf{e}_{z}$ out of the page. The media is formed by a tessellation of a primitive cell, defined by the primitive lattice vectors $\boldsymbol{\alpha}_{1}$ and $\boldsymbol{\alpha}_{2}$. In $(b)$ the simplest arrangement of one beam per primitive cell is shown. Panel $(c)$ shows the point forces and moments arising from the compressional ($\textbf{F}$) and shear ($\textbf{V}$) forces, and moments ($\textbf{M}$) within a beam interacting with the plate.
} 
\label{fig:PeriodicBeamsSkem}
\end{figure}

Consider a thin elastic plate of infinite extent, whose surface is patterned by some doubly periodic arrangement of beams as shown in fig.\ref{fig:PeriodicBeamsSkem}. The beams considered are of circular cross-section, centred in the $x-y$ plane on $\textbf{x} = \textbf{X}$ and of radius $r = \epsilon$, where we define the radial position $r = |\textbf{x}-\textbf{X}|$.

The Green's functions, whilst providing flexible, efficient and accurate results are not simple to utilise - they contain trigonometric and hyperbolic functions \cite{colquitt2017seismic} of frequency and singularities \cite{carta2020chiral,carta2020one,melnikov2001green,timoshenko1959theory,wiltshaw2020asymptotic} as one approaches the centre of a scatterer. Through careful consideration, the singularities present within Green's functions can be removed allowing semi-analytical computations to follow; direct scattering simulations utilising generalised Foldy's method \cite{martin2006multiple,martin2015scattering,schnitzer2017bloch,wiltshaw2020asymptotic} remove such singularities by examining inner limits of the external field for every scatterer. Moreover, conditionally convergent Fourier series can be rearranged in a novel way such that singularities present in the field naturally cancel from the problem \cite{schnitzer2017bloch}. 

Our Green's functions are related to those studied in \cite{carta2020one,carta2020chiral}, where gyro-topped massless beams without resonance were considered atop plates. Their Green's functions introduce similar logarithmic singularities to ours; however, their analysis was not leveraged to consider scattering simulations nor satisfies the Sommerfeld radiation condition in the far field. Similarly to Colquitt \textit{et al}.  \cite{colquitt2017seismic} we consider beams with mass, whose resonances manipulate waves at the subwavelength scale. In \cite{colquitt2017seismic}, only 1D arrays of beams were considered for quasi-2D media, here we consider 2D arrays of beams for quasi-3D media. The Green's functions of \cite{colquitt2017seismic} were constructed from the Poisson summation formula \cite{lighthill1958introduction} of the Fourier transformed medium, the infinite sums were analytically evaluated and the problem was shown to simplify down to finding the zeros of a $3 \times 3$ determinant. The solutions to the quasi-2D problem remain well behaved, even near resonance, and elegantly determine the dispersive properties of the media.

Unfortunately, the quasi-3D problem is not so well behaved. To analytically model the problem, we decompose the displacement fields following classical Euler-Bernoulli beam theory and Kirchhoff–Love thin plate theory \cite{graff1975wave,d1989theory}, in which respective solutions vary only about the neutral axis or planes of the beams and plates. The Green's functions satisfying the time-harmonic problem contain logarithmic \cite{schnitzer2017bloch, carta2020one, carta2020chiral} and algebraic \cite{wiltshaw2020asymptotic} singularities in the field as one approaches the centre of a beam along the neutral plane. 

The singular Green's functions in Carta \textit{et al}. \cite{carta2020one,carta2020chiral} are regularised in a manner as proposed by \cite{reissner1929unsymmetrische,timoshenko1959theory} (and later in a periodic setting by \cite{cai2016multiple,cai2016movable}), where one may assume sections of the plate beneath the beam are rigid and hence described by some unknown (finite) translation and rotation. The lattice sums present in \cite{carta2020one,carta2020chiral} ignore the singular behaviour of the Green's functions by regularising the solutions on the boundary of the scatterer at $r = \epsilon$. We consider the same regularisation, but include in-plane motion, and derive schemes in which the Green's functions behave as the singular expressions they are in the near field, and scatter energy such that the Sommerfeld radiation condition is satisfied in the far field.

We apply the approach developed by Schnitzer and Craster \cite{schnitzer2017bloch} to solve the Floquet-Bloch dispersion problem, or direct scattering problems, for arrangements of scatterers which singularly perturb a wavefield. The point forcing approximations for the isotropic case in \cite{schnitzer2017bloch} was extended to consider anisotropic forcing in \cite{wiltshaw2020asymptotic}. We replace the singular asymptotic matching framework, which considered Dirichlet \cite{schnitzer2017bloch} or Neumann \cite{wiltshaw2020asymptotic} inclusions within a Helmholtz wavefield, by assuming the displacement field at the junction \cite{carta2020one,carta2020chiral} between a beam and plate follows the hypothesis of Euler-Bernoulli beam theory \cite{graff1975wave}, and hence acts like a rigid disk. We opt to simplify the analysis via the aforementioned regularisation procedure instead of considering a formal matching procedure since a variety of asymptotic regimes exist \cite{kaplunov1998dynamics}.

The static problem considered in \cite{timoshenko1959theory} is appropriate when considering the simplified junction conditions for compressional and flexural beam motion coupling into flexural plate motion\footnote{Refer to §63 and $w = w_{0} + w_{1}$  in \cite{timoshenko1959theory} for assumptions and conditions on the boundary of rigid scatterers under forcing and §75 for derivation of the asymptotics of the Green's functions for point forces ($w_{0}$) and moments ($w_{1}$) applied to the surface of an elastic plate}. Examining the singular asymptotics of our dynamic Green's function (see equation \eqref{OutOfPlaneGreens}), we find agreement with \cite{timoshenko1959theory}, our dipole term $w_{\mathrm{dipole}}$ arising from the moment behaves like
\begin{equation}
w_{\mathrm{dipole}} = \frac{\textbf{e}_{\theta} \cdot \textbf{M}(\Omega)}{8D} \left[ \frac{2}{\pi} \left( \log \frac{r \sqrt{\Omega}}{2} + \gamma_{E} - \frac{1}{2} \right) - \frac{i}{2} \right] r + \mathcal{O}(r^{3}), \quad \quad \mbox{as $r \to 0$}, \label{MomentDiPoleSingularity}
\end{equation}
where $\gamma_{E}$ is the Euler-Mascheroni constant. Therefore, the moment term produces a logarithmic singularity in the gradient of the flexural displacement field of the plate as $r \to 0$,  the components of which are required for continuity of rotation. The related works of Mead \cite{mead1990plates} and Mace \cite{mace1996vibration} also consider the quasi-3D analogue of \cite{colquitt2017seismic} by employing integral transform techniques and applying the Poisson summation formula to derive the solution. The treatment of the summations within \cite{mead1990plates} and \cite{mace1996vibration} simply truncate the conditionally convergent series and hence ignore the singular asymptotics of $\nabla w_{\mathrm{dipole}}$; as a result, their solutions are only appropriate away from flexural dominated regimes of motion within the beams.

\subsection{Doubly periodic arrangements of beams atop an elastic plate} \label{sec:modelling}

We model the beams using Euler-Bernoulli beam theory, see appendix section \ref{AppModelling} equations \eqref{BeamComp} and \eqref{BeamFlex} and the boundary conditions \eqref{NotFreeEnd} for details. The forces and moments arising from the beams, applied to the surface of the plate may be analytically expressed in terms of unknown plate quantities; henceforth, we approximate each beam by point monopole and dipole source terms, whose coefficients are carefully determined to satisfy conservation of displacements, rotations, forces and moments where each beam meets the plate. For any arbitrary doubly periodic arrangement of beams we consider $N$ primitive cells, each of which containing $P$ beams. We introduce $I = 1, \ldots, N$ and $J = 1, \ldots, P$ to enumerate quantities belonging to the $IJ$th beam by subscript $IJ$, and follow \cite{carta2018elastic,carta2020one,carta2020chiral} for the convention of moments arising from the beams. 

We consider the full Kirchhoff–Love plate system, whose flexural motion is due to pure bending and may be superimposed with purely longitudinal deformations \cite{d1989theory, graff1975wave}. Refer to the appendix \ref{AppModelling} equation \eqref{PlateDispFe} for the assumed form of the displacement field, and equations \eqref{KL} and \eqref{2DE} for the expressions governing the displacement field in the plate; which is assumed to act under plane stress, as appropriate for thin plates \cite{d1989theory,achenbach2012wave}. We consider the problem in the time-harmonic regime, using the point force and moment approximation of the beams, we consider the following dimensionless system governing the displacement field of the plate 
\begin{align}
\left[ \nabla^{4} - \Omega^{2} \right] \textbf{w} =  \frac{1}{D} \sum_{I=1}^{N} \sum_{J=1}^{P} L \textbf{F}_{IJ} \delta(\textbf{x} - \textbf{X}_{IJ}) - \textbf{M}_{IJ} \times \nabla \delta(\textbf{x} - \textbf{X}_{IJ}), \label{ForcedKL} \\
\left[ \nabla^{2} + \frac{\Omega^{2} h^{2}}{12 L^{2}} \right] \phi = \frac{12 L}{\rho \Omega^{2} h^{3}} \frac{1}{\alpha^{2}} \sum_{I=1}^{N} \sum_{J=1}^{P} \nabla \cdot \textbf{V}_{IJ} \delta(\textbf{x} - \textbf{X}_{IJ}) , \label{ForcedDilation} \\
\left[ \nabla^{2} + \frac{\Omega^{2} h^{2}}{12 L^{2}} \frac{\alpha^{2}}{\beta^{2}} \right] \boldsymbol{\psi} = - \frac{12 L}{\rho \Omega^{2} h^{3}} \frac{1}{\alpha^{2}} \sum_{I=1}^{N} \sum_{J=1}^{P} \nabla \times  \textbf{V}_{IJ}  \delta(\textbf{x} - \textbf{X}_{IJ}). \label{ForcedShear}
\end{align}
Here, $\textbf{w} = w \textbf{e}_{z}$,  $\phi$, $\boldsymbol{\psi} = \psi \textbf{e}_{z}$ respectively denote dimensionless out-of-plane displacement and the dilational and shear potentials in the $x-y$ plane. $L$ denotes some characteristic length scale, $\Omega^{2} = \omega^{2} \frac{\rho h L^{4}}{D}$ the dimensionless frequency parameter, $\alpha^{2} = \frac{4 \mu( \lambda + \mu)}{\rho (\lambda + 2 \mu)} = \frac{E}{\rho (1-\nu^{2})}$ and $\beta^{2} = \frac{\mu}{\rho} = \frac{1}{2} \frac{E}{\rho (1 + \nu)}$ denote the dilational (P) and shear (S) wave speeds (squared) for elastic waves propagating in thin elastic plates acting under plane stress \cite{achenbach2012wave,d1989theory}. We define dimensionless quantities $\hat{\alpha}^{2} = \frac{\hat{\rho}}{\hat{E}} \frac{D}{\rho h L^{2}}$ and $\hat{\beta}^{4} = \frac{\hat{\rho} \hat{S}}{\hat{E} \hat{I}} \frac{D}{\rho h}$ within the beams. $\lambda$ and $\mu$ denote the Lam\`{e} constants of the plate of thickness $h$, flexural rigidity $D = \frac{E h^{3}}{12 (1 - \nu^{2})}$ and Poisson's ratio $\nu$. Further, we denote $\rho$ as the density, $E$ Young's Modulus, $S$ cross-sectional area and $I$ the second moment of area of either the plate (regular quantities) or the beams (hatted quantities). We denote $\epsilon_{IJ}$ as the dimensionless cross-sectional radius of the beams. Throughout, $\nabla = \textbf{e}_{x} \frac{\partial}{\partial x} + \textbf{e}_{y} \frac{\partial}{\partial y}$ denotes the in-plane two dimensional gradient. Lastly, $\delta(\textbf{x})$ denotes the Dirac delta function. 

The source terms $\textbf{F}_{IJ}$, $\textbf{M}_{IJ}$ and $\textbf{V}_{IJ}$ are deduced by solving the compressional and flexural equations governing displacement in the beams, satisfying free-end conditions at the free end and continuity of displacements and rotations at the other. The dilational and shear components of $\textbf{V}_{IJ}$ coupling into $\phi$ and $\boldsymbol{\psi}$ have been determined, as in Hudson \cite{hudson1980excitation}, via inspection of equation \eqref{2DE}.  For brevity, we shall assume $P=1$ knowing that generalisations to consider arbitrary $P$ are trivial for the techniques we use. Subsequently, it is sufficient to consider the contribution from the $\textbf{x} = \textbf{X}$th beam and drop any subscript $IJ$ and any summations over $IJ$; these subscripts and sums will be reinstated as required to consider the contributions of more beams. It can be shown the forces and moments at the base of the beams applied to the surface of the plate are \cite{graff1975wave, d1989theory} (see appendix section \ref{AppModelling})

\begin{equation} 
\textbf{F}( \Omega) = - \hat{S} \hat{E} \frac{\partial \hat{\textbf{w}}}{\partial z} \Big|_{z = \frac{h}{2L}} = \hat{S} \hat{E} \hat{\alpha} \Omega \tan (\hat{\alpha} \Omega \hat{\ell} ) \textbf{w}( \textbf{X}), \label{ComFor}
\end{equation}
\begin{equation}
\begin{split}
\textbf{M}( \Omega) = \frac{ \hat{E} \, \hat{I}}{L}  \frac{\partial }{\partial z} \Big[ \nabla \times \hat{\textbf{u}} \Big] \Big|_{z = \frac{h}{2L}} = \quad \quad \quad \quad \quad \quad \quad \quad \quad \quad \quad \quad \quad \quad \quad \quad \quad \quad \quad \quad \quad \quad \quad \quad \quad \quad \quad \quad \\ 
= \frac{ \hat{E} \, \hat{I}}{L} \frac{ \hat{\beta} \sqrt{\Omega} \Big[ \nabla \times \textbf{w} \Big] \Big|_{\textbf{x} = \textbf{X}} \Big\lbrace \frac{h}{2L} \hat{\beta} \sqrt{\Omega}\sin  \hat{\beta} \sqrt{\Omega} \hat{\ell}  \tanh \hat{\beta} \sqrt{\Omega} \hat{\ell}   +   2 \Big( \sin  \hat{\beta} \sqrt{\Omega} \hat{\ell}  - \cos  \hat{\beta} \sqrt{\Omega} \hat{\ell}  \tanh  \hat{\beta} \sqrt{\Omega} \hat{\ell} \Big)   \Big\rbrace }{\cos \hat{\beta} \sqrt{\Omega} \hat{\ell} + \sech \hat{\beta} \sqrt{\Omega} \hat{\ell} }, \label{FlexMom}
\end{split} 
\end{equation}
\begin{equation}
\begin{split}
\textbf{V}( \Omega) = - \frac{ \hat{E} \, \hat{I}}{L^{2}} \frac{\partial^{3} \hat{\textbf{u}}}{\partial z^{3}} \Big|_{z = \frac{h}{2L}}  =  \quad \quad \quad \quad \quad \quad \quad \quad \quad \quad \quad \quad \quad \quad \quad \quad \quad \quad \quad \quad \quad \quad \quad \quad \quad \quad \quad \quad \quad \quad \\ 
= - \frac{ \hat{E} \, \hat{I}}{L^{2}} \frac{\hat{\beta}^{2} \Omega \Big[ \nabla \textbf{w} \Big] \Big|_{\textbf{x} = \textbf{X}} \Big\lbrace    2 \sin  \hat{\beta} \sqrt{\Omega} \hat{\ell}  \tanh  \hat{\beta} \sqrt{\Omega} \hat{\ell}  +  \frac{h}{2L} \hat{\beta} \sqrt{\Omega} \Big( \cos \hat{\beta} \sqrt{\Omega} \hat{\ell} \tanh \hat{\beta} \sqrt{\Omega} \hat{\ell} + \sin \hat{\beta} \sqrt{\Omega} \hat{\ell} \Big) \Big\rbrace  }{\cos \hat{\beta} \sqrt{\Omega} \hat{\ell} + \sech \hat{\beta} \sqrt{\Omega} \hat{\ell} }. \label{FlexForce}
\end{split} 
\end{equation} 
Here, we set $\textbf{u}(\textbf{X}) =  \textbf{0}$ to satisfy continuity of moments, otherwise the bending moment coupling into the plate is not due to pure bending and inconsistent with the  full Kirchhoff–Love plate model \cite{d1989theory}. The system \eqref{ForcedKL} - \eqref{ForcedShear} can be solved for $N$ finite or infinite, depending on how one treats the unknowns. In section \ref{sec:EigProb}, we treat $N$ as infinite and, through the use of Floquet-Bloch analysis, express the solution through the fundamental cell (where $I=1$) as a polynomial eigenvalue problem. In section \ref{sec:Foldy}, we treat $N$ as finite and express the solution for direct scattering simulations in the unbounded $\textbf{x}$ space, utilising generalised Foldy's method.

The Green's functions satisfying equations \eqref{ForcedKL}-\eqref{ForcedShear} may be determined using the Fourier transform \cite{wiltshaw2020asymptotic,graff1975wave,gradshteyn2014table}, where we find 
\begin{equation}
w = \frac{L}{D} F( \Omega ) \Big[ H_{0} (r \sqrt{\Omega} ) - H_{0} ( i r \sqrt{\Omega} ) \Big] \frac{i}{8 \Omega}   - \frac{\textbf{e}_{\theta} \cdot \textbf{M}(\Omega)}{D}  \Big[ i H_{1} (r \sqrt{\Omega} ) + H_{1} ( i r \sqrt{\Omega} ) \Big] \frac{1}{8 \sqrt{\Omega}} , \label{OutOfPlaneGreens}  
\end{equation}
\begin{equation}
    \phi = \frac{i \sqrt{3}}{2} \frac{\textbf{e}_{r} \cdot \textbf{V}(\Omega)}{\rho \Omega h^{2} \alpha^{2}} H_{1} (\frac{\Omega h}{2 L \sqrt{3}} r ),  \label{DilationGreens}
\end{equation}
\begin{equation}
\psi= \frac{\sqrt{3}}{2 i} \frac{\textbf{e}_{\theta} \cdot \textbf{V}(\Omega)}{\rho \Omega h^{2} \alpha \beta} H_{1} (\frac{\alpha}{\beta} \frac{\Omega h}{2 L \sqrt{3}} r ) .  \label{ShearGreens}
\end{equation}
Here, we denote $H_{n}(x) = H^{(1)}_{n}(x)$ to be Hankel functions of the first kind and $n$th order - we suppress the superscript $(1)$ since we only consider Hankel functions of the first kind, those which correspond to scattered waves outgoing at infinity which satisfy the Sommerfeld radiation condition and correspond to physically meaningful solutions.   Note \eqref{OutOfPlaneGreens} contains logarithmically singular terms within its gradient, as in equation \eqref{MomentDiPoleSingularity}. Since the components of $\nabla w$  are required in the forcing terms \eqref{FlexMom} \& \eqref{FlexForce}, we must examine $\nabla w$ as $r \to 0$. For our inner problem, regarding a thin beam attached to a thin elastic plate, many parameters could be small ($\epsilon$, $h$, $\Omega$) or large (source terms \eqref{FlexMom} - \eqref{FlexForce} approaching resonance) meaning many different asymptotic regimes could exist within a formal matching procedure.

We work in a regime where $\epsilon \ll \mathrm{min}(|\boldsymbol{\alpha}_{1}|,|\boldsymbol{\alpha}_{2}|)$ and require frequencies that are not too high, such that the asymptotics in appendix \ref{ResidualTermsApp} hold, and any expansions of Bessel or Hankel functions remain accurate.  When deriving equations \eqref{BeamComp} and \eqref{BeamFlex}, two main assumptions have been made about beam motion. Firstly, we completely ignore rotatory inertia and hence neglect torsion about the axis of the beam. Secondly, shear deformations within beams are also neglected; cross-sections perpendicular to the neutral axis of the undeformed problem will always remain plane and perpendicular to the neutral axis during deformation  \cite{graff1975wave}. Under these assumptions, an asymptotically formal inner solution may be replaced by considering the beam-plate intersection as rigid \cite{reissner1929unsymmetrische,timoshenko1959theory,cai2016multiple,cai2016movable}. We subsequently alter our matching procedure in \cite{wiltshaw2020asymptotic} to regularise the problem in a similar fashion to \cite{carta2020chiral,carta2020one}, by introducing a rigid disk characterised by a finite, but unknown, displacement $w(\textbf{X})$ and rotation $\Big[ \nabla \times \textbf{w} \Big] \Big|_{\textbf{x} = \textbf{X}}$
\begin{align}
w = w(\textbf{X}) - r  \textbf{e}_{\theta } \cdot \Big( \Big[ \nabla \times \textbf{w} \Big] \Big|_{\textbf{x} = \textbf{X} } \Big) \quad \quad \mbox{for  $r  \le \epsilon$}, \label{RigidConditionsOutofPlane} \\
\textbf{u} =  \textbf{0}  \quad \quad \mbox{for  $r  \le \epsilon$}. \label{RigidConditionsInPlane}
\end{align}
Again, we set $\textbf{u} =  \textbf{0}$ on the boundary to satisfy continuity of moments.

\subsection{Green's functions incorporating boundary conditions} \label{sec:modellingBCs}

The Green's functions \eqref{OutOfPlaneGreens} - \eqref{ShearGreens} satisfy the forced equations \eqref{ForcedKL} - \eqref{ForcedShear}, their current form ignores the physical behaviour of the actual scatterer about $r = \epsilon$. Hence, we require from \eqref{RigidConditionsOutofPlane} 
\begin{align}
w \Big|_{r = \epsilon} = w(\textbf{X}) - \epsilon \textbf{e}_{\theta} \cdot \Big( \Big[ \nabla \times \textbf{w} \Big] \Big|_{\textbf{x} = \textbf{X}} \Big),  \label{DispCond1} \\
\frac{\partial w}{\partial r} \Big|_{r = \epsilon}  =  -  \textbf{e}_{\theta} \cdot \Big( \Big[ \nabla \times \textbf{w} \Big] \Big|_{\textbf{x} = \textbf{X}} \Big). \label{DispCond2}
\end{align}
Further our potentials need to be consistent with equations \eqref{RigidConditionsInPlane} and \eqref{plateCondRot}, and hence 
\begin{equation}
    \phi \Big|_{r = \epsilon} = \Phi(\textbf{X}), \quad \quad \quad \quad \quad \quad \psi \Big|_{r = \epsilon} = \Psi(\textbf{X}). \label{PotentialConds}
\end{equation}
Currently, \eqref{OutOfPlaneGreens} - \eqref{ShearGreens} solve the inhomogeneous equations \eqref{ForcedKL} - \eqref{ForcedShear}. We consider the complementary solutions to the homogeneous part of these equations where, for a certain choice of constants, the above conditions are satisfied and a solution with the correct behaviour as $r \to \infty$ constructed. We make the assumption that the complementary solution quickly decays to zero, therefore in the far field only \eqref{OutOfPlaneGreens} - \eqref{ShearGreens} remain and hence the Sommerfeld radiation condition is satisfied. Consider
\begin{equation}
w = w_{P} + w_{C}, \quad \phi = \phi_{P} + \phi_{C}, \quad \psi = \psi_{P} + \psi_{C}, \label{TotalSolns}
\end{equation}
where subscript $P$ and $C$ terms respectively denote the particular and complementary solutions of \eqref{OutOfPlaneGreens} - \eqref{ShearGreens}. We assume the complementary solutions  are finite and will behave as follows, provided $r$ is small
\begin{equation}
w_{C} = A_{0} J_{0} (\sqrt{\Omega} r) + B_{0} J_{0}(i \sqrt{\Omega} r) + \textbf{e}_{\theta} \cdot \textbf{A} J_{1}(\sqrt{\Omega} r) + \textbf{e}_{\theta} \cdot \textbf{B} J_{1}(i\sqrt{\Omega} r), \label{Wc} 
\end{equation}
\begin{equation}
\phi_{C} =  C J_{0}(\frac{\Omega h}{2 L \sqrt{3}} r ) + \textbf{e}_{r} \cdot \textbf{C} J_{1}(\frac{\Omega h}{2 L \sqrt{3}} r ),  \label{phiC} 
\end{equation}
\begin{equation}
\psi_{C} =  D J_{0}(\frac{\alpha}{\beta} \frac{\Omega h}{2 L \sqrt{3}} r ) + \textbf{e}_{\theta} \cdot \textbf{D}  J_{1}(\frac{\alpha}{\beta} \frac{\Omega h}{2 L \sqrt{3}} r ).  \label{psiC}
\end{equation}
The coefficients of the above are chosen such that \eqref{DispCond1} - \eqref{PotentialConds} are satisfied and are given in appendix \ref{AppCompCoeffs}. To be clear, for $r>\epsilon$ these complimentary solutions are ignored - they should be thought of as standing waves existing only in the vicinity of the scatterer, in order to apply conditions \eqref{DispCond1}-\eqref{PotentialConds} and not invalidate the Sommerfeld radiation condition at infinity \cite{wiltshaw2020asymptotic}.

\section{An eigenvalue problem for periodic infinite arrays} \label{sec:EigProb}

In this section we assume $N$ from equations \eqref{ForcedKL}-\eqref{ForcedShear} is infinite. We consider the centroid of each primitive cell in fig. \ref{fig:PeriodicBeamsSkem} to coincide with the vertices of a 2D  Bravais lattice
\begin{equation}
\textbf{R} = n \boldsymbol{\alpha}_{1} + m \boldsymbol{\alpha}_{2}, \label{BravLat}
\end{equation}
for arbitrary integers $n$ and $m$. We also define $\boldsymbol{\beta}_{1}$ and $\boldsymbol{\beta}_{2}$ as a basis in reciprocal space, satisfying the following orthogonality condition
\begin{equation}
\boldsymbol{\alpha}_{i} \cdot \boldsymbol{\beta}_{j} = 2 \pi \delta_{ij} \quad \mbox{for $i,j = 1,2$} .
\end{equation}
Here, $\delta_{ij}$ denotes the Kronecker delta function. Similarly we can define the reciprocal lattice vector
\begin{equation}
\textbf{G} = n \boldsymbol{\beta}_{1} + m \boldsymbol{\beta}_{2},
\end{equation}
again for arbitrary integers $n$ and $m$. By Bloch's theorem \cite{kittel1996introduction,brillouin1953wave}, waves propagating through the periodic medium satisfy the Floquet-Bloch conditions
\begin{align}
w = W(\textbf{x}) \exp( i \boldsymbol{\kappa} \cdot \textbf{x}), \quad \phi = \Phi(\textbf{x}) \exp( i \boldsymbol{\kappa} \cdot \textbf{x}), \quad  \psi = \Psi(\textbf{x}) \exp( i \boldsymbol{\kappa} \cdot \textbf{x}), \label{BlochPerForm} \\
\mbox{where} \quad W(\textbf{x}) = W(\textbf{x} + \textbf{R}), \quad \quad  \Phi(\textbf{x}) = \Phi(\textbf{x} + \textbf{R}), \quad \quad  \Psi(\textbf{x}) = \Psi(\textbf{x} + \textbf{R}).
\end{align}
Here, $\boldsymbol{\kappa}$ is the Bloch-wavevector. Furthermore $W$, $\Phi$ and $\Psi$ are periodic functions, with period derived from the lattice \eqref{BravLat}, whose solutions are naturally expressed as a Fourier series
\begin{equation}
W(\textbf{x}) = \sum_{\textbf{G}} W_{\textbf{G}} \exp(i \textbf{G} \cdot \textbf{x}), \quad \Phi(\textbf{x}) = \sum_{\textbf{G}} \Phi_{\textbf{G}} \exp(i \textbf{G} \cdot \textbf{x}), \quad \Psi(\textbf{x}) = \sum_{\textbf{G}} \Psi_{\textbf{G}} \exp(i \textbf{G} \cdot \textbf{x}). \quad \quad \label{FourierSeries}
\end{equation}
Here, $W_{\textbf{G}}$, $\Phi_{\textbf{G}}$ and $\Psi_{\textbf{G}}$ denote Fourier coefficients of the expanded variables. We seek solutions to $w$, $\phi$ and $\psi$ in the form of \eqref{BlochPerForm}. The forces and moments in \eqref{ComFor} - \eqref{FlexForce} can also be expressed using equation \eqref{BlochPerForm}, as follows
\begin{equation}
\textbf{F}( \Omega, \boldsymbol{\kappa}) = \textbf{F}_{p} ( \Omega ) \exp(i \boldsymbol{ \kappa} \cdot \textbf{X}), \quad \textbf{M}( \Omega, \boldsymbol{\kappa}) = \textbf{M}_{p} ( \Omega, \boldsymbol{\kappa}) \exp(i \boldsymbol{ \kappa} \cdot \textbf{X}), \quad \textbf{V}( \Omega , \boldsymbol{\kappa}) = \textbf{V}_{p} ( \Omega , \boldsymbol{\kappa} ) \exp(i \boldsymbol{ \kappa} \cdot \textbf{X}). \label{BlochFormForcing}
\end{equation}
Also, $A_{0}$, $B_{0}$, $\textbf{A}$, $\textbf{B}$, $\textbf{C}$ and $\textbf{D}$ (see appendix \ref{AppCompCoeffs}) can be expressed by equation \eqref{BlochPerForm} as
\begin{equation}
A_{0} = A_{0 p}  \exp(i \boldsymbol{ \kappa} \cdot \textbf{X}),
\end{equation}
similarly for the other coefficients in equations \eqref{Wc} - \eqref{psiC}. Utilizing \eqref{BlochFormForcing} and \eqref{FourierSeries}  within \eqref{ForcedKL} - \eqref{ForcedShear}, multiplying by $\exp( - i \textbf{K}_{\textbf{G}'} \cdot \textbf{x})$ and integrating over the fundamental cell, one finds by orthogonality
\begin{align}
\Big[ (\textbf{K}_{\textbf{G}} \cdot \textbf{K}_{\textbf{G}})^{2} - \Omega^{2} \Big] \textbf{W}_{\textbf{G}} = \frac{1}{\mathscr{A}  D} \Big[ L \textbf{F}_{p} - i \textbf{M}_{p} \times \textbf{K}_{\textbf{G}} \Big] \exp( -i \textbf{G} \cdot \textbf{X}) \label{WFourCoeffs}, \\
i \mathscr{A} \alpha^{2} \frac{\rho \Omega^{2} h^{3}}{12 L} \Big[ \textbf{K}_{\textbf{G}} \cdot \textbf{K}_{\textbf{G}} - \frac{\Omega^{2} h^{2}}{12 L^{2}} \Big] \Phi_{\textbf{G}} =  \textbf{K}_{\textbf{G}} \cdot \textbf{V}_{p} \exp( -i \textbf{G} \cdot \textbf{X}) \label{PhiFourCoeffs}, \\
i \mathscr{A} \alpha^{2} \frac{\rho \Omega^{2} h^{3}}{12 L} \Big[ \textbf{K}_{\textbf{G}} \cdot \textbf{K}_{\textbf{G}} - \frac{\Omega^{2} h^{2}}{12 L^{2}} \frac{\alpha^{2}}{\beta^{2}} \Big] \boldsymbol{\Psi}_{\textbf{G}} = -   \textbf{K}_{\textbf{G}} \times \textbf{V}_{p} \exp( -i \textbf{G} \cdot \textbf{X}) \label{PsiFourCoeffs}.
\end{align}
Here, $\mathscr{A}$ is the physical area of the fundamental cell and $\textbf{K}_{\textbf{G}} = \boldsymbol{\kappa} + \textbf{G}$. Notably, the $\exp(i \boldsymbol{\kappa} \cdot \textbf{X})$ terms cancel from these equations governing the Fourier coefficients. Say we wish to include $M$ modes within \eqref{FourierSeries}, we have expressed our linear system of differential equations as a system of equations and unknowns. The unknowns in our system  \eqref{WFourCoeffs} - \eqref{PsiFourCoeffs} from $w$ are $W(\textbf{X})$, $\partial_{x} W(\textbf{X})$, $\partial_{y} W(\textbf{X})$ and $W_{\textbf{G}_{i}}$ for $i = 1, \ldots M$; similar unknowns exist for $\phi$ and $\psi$ (without gradient terms) and we currently  have a total of $3M$ equations for $3M + 5P$ unknowns. The other $5P$ equations come from inserting \eqref{WFourCoeffs} - \eqref{PsiFourCoeffs} into the Fourier expanded versions of \eqref{BlochPerForm} and examining $w$, $\nabla w$, $\phi$ and $\psi$ all in the limit as $\textbf{x} \to \textbf{X}$, or equivalently $r \to 0$, for each beam in the fundamental cell. 

\subsection{The limit of Green's functions as $r \to 0$}
The Green's functions \eqref{OutOfPlaneGreens}-\eqref{ShearGreens} and \eqref{BlochPerForm} are two different representations of our solution. The singular asymptotics of expressions \eqref{DilationGreens}-\eqref{ShearGreens} and $\nabla$\eqref{OutOfPlaneGreens} have at best $\log(r)$, and at worst $\frac{1}{r}$ behaviour as $r \to 0$. Similarly, the series representations (inserting  \eqref{FourierSeries} in \eqref{BlochPerForm}) are conditionally convergent, where at $\textbf{x} = \textbf{X}$ the exponential terms stop oscillating and the series diverge. However, as $r \to 0$, both the Fourier series and Hankel function representations of our solutions will diverge to the same value, and singularities in $r$ can be carefully arranged such that they cancel.

We introduce some $R \gg 1$ denoting a truncation radius in Fourier space containing $M$ modes and split the Fourier series expansions as follows:
\begin{equation}
w =  w_{\mathrm{tr}} + w_{\mathrm{res}}, \quad \quad 
\phi = \phi_{\mathrm{tr}} + \phi_{\mathrm{res}}, \quad \quad
\psi = \psi_{\mathrm{tr}} + \psi_{\mathrm{res}} \label{wPhiPsiTRres},
\end{equation}
\begin{equation}
\mbox{here} \quad w_{\mathrm{tr}} = \sum_{ |\textbf{G}| < R} W_{\textbf{G}} \exp(i \textbf{K}_{\textbf{G}} \cdot \textbf{x}), \quad w_{\mathrm{res}} = \sum_{ |\textbf{G}| > R} W_{\textbf{G}} \exp(i \textbf{K}_{\textbf{G}} \cdot \textbf{x}).  \label{TruncAndRes}
\end{equation}
The series for $\phi$ and $\psi$ are split up in an identical fashion to expression \eqref{TruncAndRes}, where subscript $\mathrm{tr}$ and $\mathrm{res}$  terms respectively denote the truncated and residual portion of the series before and after the truncation position $R$. Examining $\lim_{r \to 0} \exp(-i \boldsymbol{\kappa} \cdot \textbf{x}) \cdot$\eqref{wPhiPsiTRres} we find
\begin{equation}
\lim_{r \to 0} \sum_{|\textbf{G}| < R} W_{\textbf{G}} \exp (i \textbf{G} \cdot \textbf{x}) = W(\textbf{X}),\label{Disp_rTo0}
\end{equation}
indeed since the $\lim_{r \to 0}  w_{\mathrm{res}} = 0$. Additionally we consider the following, all in the limit as $r \to 0$
\begin{align}
\exp(- i \boldsymbol{\kappa} \cdot \textbf{x}) \nabla \Big[ w -  w_{\mathrm{res}} \Big] & = \exp(- i \boldsymbol{\kappa} \cdot \textbf{x}) \nabla w_{\mathrm{tr}} , \\
\exp(- i \boldsymbol{\kappa} \cdot \textbf{x}) \Big[ \phi -  \phi_{\mathrm{res}} \Big] & =  \exp(- i \boldsymbol{\kappa} \cdot \textbf{x}) \phi_{\mathrm{tr}} , \\
\exp(- i \boldsymbol{\kappa} \cdot \textbf{x}) \Big[ \psi -  \psi_{\mathrm{res}} \Big] & =  \exp(- i \boldsymbol{\kappa} \cdot \textbf{x}) \psi_{\mathrm{tr}}.
\end{align}
Substituting the required expressions from \eqref{wPhiPsiTRres}, \eqref{TotalSolns} and appendix section \ref{ResidualTermsApp} into the above, expanding the Bessel (and integral Bessel \eqref{BesselIntBehave}) functions and Hankel functions as $r \to 0$, and assuming $rR \ll 1$, the following expressions can be found
\begin{align}
\lim_{r \to 0}  \textbf{e}_{x} \cdot \sum_{|\textbf{G}| < R} i \textbf{K}_{\textbf{G}} W_{\textbf{G}} \exp(i \textbf{G} \cdot \textbf{x}) = -\frac{\textbf{e}_{y} \cdot \textbf{M}_{p}}{4 \pi D} \Big[ \log \frac{\epsilon R}{2} + \gamma_{E} - 1 \Big] + W_{x} (\textbf{X}) + i \kappa_{1} W(\textbf{X}) , \label{GradDisp_rTo0} \\
\lim_{r \to 0}  \textbf{e}_{y} \cdot \sum_{|\textbf{G}| < R} i \textbf{K}_{\textbf{G}} W_{\textbf{G}} \exp(i \textbf{G} \cdot \textbf{x}) = \frac{\textbf{e}_{x} \cdot \textbf{M}_{p}}{4 \pi D} \Big[ \log \frac{\epsilon R}{2} + \gamma_{E} - 1 \Big] + W_{y} (\textbf{X}) + i \kappa_{2} W(\textbf{X}),
\end{align}

\begin{equation}
\lim_{r \to 0} \frac{\rho \Omega^{2} h^{3}}{12 L}  \sum_{|\textbf{G}| < R} \Phi_{\textbf{G}} \exp(i \textbf{G} \cdot \textbf{x})    =  \frac{1}{\alpha^{2}} \Big[ \frac{\textbf{V}_{p} \cdot \boldsymbol{\kappa}}{4 i \pi } \Big] + \frac{\rho \Omega^{2} h^{3}}{12 L} \Phi(\textbf{X}),
\end{equation}

\begin{equation}
\lim_{r \to 0}  \frac{\rho \Omega^{2} h^{3}}{12 L} \sum_{|\textbf{G}| < R} \Psi_{\textbf{G}} \exp(i \textbf{G} \cdot \textbf{x})   = \frac{1}{\alpha^{2}} \Big[ \frac{i (\boldsymbol{\kappa} \times \textbf{V}_{p} ) \cdot \textbf{e}_{z}}{4 \pi } \Big] + \frac{\rho \Omega^{2} h^{3}}{12 L} \Psi(\textbf{X}) \label{Psi_rTo0}.
\end{equation}
Both components of $\boldsymbol{\kappa} = \boldsymbol{\kappa}(\Omega)$ are required, hence we introduce the following change of variables
\begin{equation}
\boldsymbol{\kappa} = \eta \Big[ \cos \theta_{\eta} \textbf{e}_{x} + \sin \theta_{\eta} \textbf{e}_{y} \Big] + \zeta. \label{WaveVectOriginShift}
\end{equation}
Here, $\zeta$ can be set to shift the origin of the wavevector to any convenient (high-symmetry) point within the Brillouin zone, $\theta_{\eta}$ set to consider any direction along said point, and $\eta$ subsequently determined as the eigenvalue of such a problem for any frequency considered. Rewriting equations \eqref{WFourCoeffs} - \eqref{PsiFourCoeffs}, \eqref{Disp_rTo0}, \eqref{GradDisp_rTo0} - \eqref{Psi_rTo0} in terms of $\eta$, one finds the following polynomial eigenvalue problem whose solution yields all unknowns in the system
\begin{equation}
\Big[ \eta^{4} \mathcal{A}(\Omega) + \eta^{3} \mathcal{B}(\Omega) + \eta^{2} \mathcal{C}(\Omega) + \eta \mathcal{D}(\Omega) +  \mathcal{E}(\Omega) \Big] \boldsymbol{ \Theta } = \textbf{0}. \label{EVPdispBloch}
\end{equation}
Here, $\mathcal{A}$, $\mathcal{B}$, $\ldots$, $\mathcal{E}$ denote matrices of dimension $(3 M + 5P) \times (3 M + 5P)$ containing various trigonometric and hyperbolic functions of $\Omega$. The eigenvalue $\eta$ is inserted into \eqref{WaveVectOriginShift}, hence one finds $\boldsymbol{\kappa} = \boldsymbol{\kappa}(\Omega)$ by looping through all required frequencies. The eigenvector $\boldsymbol{ \Theta }$ contains the $(3 M + 5P)$ unknowns $W_{\textbf{G}}$'s, $\Phi_{\textbf{G}}$'s, $\Psi_{\textbf{G}}$'s, $W(\textbf{X})$'s, $W_{x}(\textbf{X})$'s, $W_{y}(\textbf{X})$'s, $\Phi(\textbf{X})$'s, $\Psi(\textbf{X})$'s.   

For the asymptotic analysis within appendix \ref{ResidualTermsApp} to remain accurate, one needs to ensure $R$ and hence $M$ is sufficiently large. Appropriate magnitudes of $M$ depend on the problem one wishes to consider. Practically speaking, $M$ being in the tens or hundreds gives sufficiently small errors when considering Floquet-Bloch dispersion through the fundamental cell; however, for ribbon or interfacial problems containing many cells (as in figs \ref{fig:PertSquareTopoArrange} $(e)$ or \ref{fig:PertHexArrange} $(e)$) $\mathcal{O}(M) = 10^3$ is required. In such cases, the polynomial eigenvalue problem of dimension $(3 M + 5P) \times (3 M + 5P)$ will be computationally expensive to solve.  However, utilising \cite{bridges1984differential} the eigenvalue problem \eqref{EVPdispBloch} can be simplified to an algebraic eigenvalue problem, whose eigenvalues and eigenvectors can be computed quickly and cheaply, as shown in the appendix section \ref{CompLE}.


\section{The generalised Foldy problem: A finite arrangement of scatterers under incident forcing}   \label{sec:Foldy}
Consider a phononic crystal formed from a finite portion of the infinitely repeating periodic media examined in section \ref{sec:EigProb}; provided this finite section is large enough the Floquet-Bloch waves, whose dispersion properties are a priori known by the analysis of section \ref{sec:EigProb}, still persist throughout the medium. Subsequently, we can  tailor the constituent primitive cells to control how energy propagates from some incident source within the crystal. Propagation about the edge of such media can also be determined \cite{joannopoulos_photonic_2008} by considering conserved quantities of $\boldsymbol{\kappa}$. 

In this section, we assume $N$ from equations \eqref{ForcedKL}-\eqref{ForcedShear} is finite.  We now consider the crystal under incidence, and derive the solutions describing the displacement fields over the entire space. In this section, we re-write the summations present in \eqref{ForcedKL}-\eqref{ForcedShear} as follows

\begin{align}
\left[ \nabla^{4} - \Omega^{2} \right] \textbf{w} = \tilde{\textbf{w}}_{\mathrm{inc}}(\textbf{x}) +  \frac{1}{D} \sum_{j=1}^{m}  L \textbf{F}_{j} \delta(\textbf{x} - \textbf{X}_{j}) - \textbf{M}_{j} \times \nabla \delta(\textbf{x} - \textbf{X}_{j}), \label{ForcedKLFoldy} \\
\left[ \nabla^{2} + \frac{\Omega^{2} h^{2}}{12 L^{2}} \right] \phi = \tilde{\phi}_{\mathrm{inc}} (\textbf{x}) + \frac{12 L}{\rho \Omega^{2} h^{3}} \frac{1}{\alpha^{2}} \sum_{j=1}^{m} \nabla \cdot \textbf{V}_{j} \delta(\textbf{x} - \textbf{X}_{j}) , \label{ForcedDilationFoldy} \\
\left[ \nabla^{2} + \frac{\Omega^{2} h^{2}}{12 L^{2}} \frac{\alpha^{2}}{\beta^{2}} \right] \boldsymbol{\psi} = \tilde{\boldsymbol{\psi}}_{\mathrm{inc}}(\textbf{x}) - \frac{12 L}{\rho \Omega^{2} h^{3}} \frac{1}{\alpha^{2}} \sum_{j=1}^{m} \nabla \times  \textbf{V}_{j}  \delta(\textbf{x} - \textbf{X}_{j}),  \label{ForcedShearFoldy}
\end{align}
Here, index $j$ enumerates the $m = NP$ total number of beams forming the phononic crystal. Additionally, $\tilde{\textbf{w}}_{\mathrm{inc}}$, $\tilde{\phi}_{\mathrm{inc}}$ and $\tilde{\boldsymbol{\psi}}_{\mathrm{inc}}$ denotes some forcing incident upon the overall structure, the Green's functions of which we respectively denote $\textbf{w}_{\mathrm{inc}}$, $\phi_{\mathrm{inc}}$ and $\boldsymbol{\psi}_{\mathrm{inc}}$.  Herein, we consider a purely out-of-plane monopole point source of strength $\varpi_{\mathrm{inc}}$, centred on $\textbf{x} = \textbf{X}_{\mathrm{inc}}$, by setting
\begin{equation}
\tilde{\textbf{w}}_{\mathrm{inc}} = \varpi_{\mathrm{inc}} \delta(\textbf{x} - \textbf{X}_{\mathrm{inc}}) \textbf{e}_{z}, \quad \mbox{and} \quad \tilde{\phi}_{\mathrm{inc}}(\textbf{x}) = \tilde{\psi}_{\mathrm{inc}}(\textbf{x}) = 0.  \label{ForcePOPinc}
\end{equation}
The Green's functions for the above source are given by 
\begin{equation}
w_{\mathrm{inc}} (\textbf{x}) = \frac{i}{8 \Omega} \varpi_{\mathrm{inc}} \Big[ H_{0} (|\textbf{x} - \textbf{X}_{\mathrm{inc}}| \sqrt{\Omega} ) - H_{0} ( i |\textbf{x} - \textbf{X}_{\mathrm{inc}}|  \sqrt{\Omega} ) \Big], \quad \quad \phi_{\mathrm{inc}} = \psi_{\mathrm{inc}} = 0. \label{ForcePOPincGREENS}
\end{equation}
 We proceed by re-writing the forcing terms from \eqref{ComFor} - \eqref{FlexForce}, as follows
\begin{equation}
\textbf{F}_{j}(\Omega) = F_{j} \Omega w( \textbf{X}_{j} ) \textbf{e}_{z}, \quad \quad \textbf{M}_{j} = M_{j} \sqrt{\Omega} (\nabla \times \textbf{w})\Big|_{\textbf{x} = \textbf{X}_{j}}, \quad \quad \textbf{V}_{j} = V_{j} \Omega \nabla w\Big|_{\textbf{x} = \textbf{X}_{j}},
\end{equation} 
where
\begin{align}
F_{j} = \hat{S}_{j} \hat{E}_{j} \hat{\alpha}_{j} \tan (\hat{\alpha}_{j} \Omega \hat{\ell}_{j} ),  \\
M_{j} = \frac{ \hat{E}_{j} \, \hat{I}_{j}}{L} \frac{ \hat{\beta}_{j} \Big\lbrace \frac{h}{2L} \hat{\beta}_{j} \sqrt{\Omega}\sin  \hat{\beta}_{j} \sqrt{\Omega} \hat{\ell}_{j}  \tanh \hat{\beta}_{j} \sqrt{\Omega} \hat{\ell}_{j}   +   2 \Big( \sin  \hat{\beta}_{j} \sqrt{\Omega} \hat{\ell}_{j}  - \cos  \hat{\beta}_{j} \sqrt{\Omega} \hat{\ell}_{j}  \tanh  \hat{\beta}_{j} \sqrt{\Omega} \hat{\ell}_{j} \Big)   \Big\rbrace }{\cos \hat{\beta}_{j} \sqrt{\Omega} \hat{\ell}_{j} + \sech \hat{\beta}_{j} \sqrt{\Omega} \hat{\ell}_{j} }, \\
V_{j} = - \frac{ \hat{E}_{j} \, \hat{I}_{j}}{L^{2}} \frac{\hat{\beta}_{j}^{2}  \Big\lbrace    2 \sin  \hat{\beta}_{j} \sqrt{\Omega} \hat{\ell}_{j}  \tanh  \hat{\beta}_{j} \sqrt{\Omega} \hat{\ell}_{j}  +  \frac{h}{2L} \hat{\beta}_{j} \sqrt{\Omega} \Big( \cos \hat{\beta}_{j} \sqrt{\Omega} \hat{\ell}_{j} \tanh \hat{\beta}_{j} \sqrt{\Omega} \hat{\ell}_{j} + \sin \hat{\beta}_{j} \sqrt{\Omega} \hat{\ell}_{j} \Big) \Big\rbrace  }{\cos \hat{\beta}_{j} \sqrt{\Omega} \hat{\ell}_{j} + \sech \hat{\beta}_{j} \sqrt{\Omega} \hat{\ell}_{j} }.
\end{align}
The Green's functions governing the total displacement fields throughout the structure are as follows
\begin{equation}
\begin{split}
w(\textbf{x}) = w_{\mathrm{inc}} (\textbf{x}) + \quad \quad  \quad \quad \quad  \quad \quad  \quad \quad \quad \quad \quad  \quad \quad \quad  \quad \quad  \quad \quad \quad \quad \quad  \quad \quad \quad  \quad \quad  \quad \quad \quad  \\ + \sum_{j=1}^{m} \left\lbrace \frac{i L}{8D} F_{j} w(\textbf{X}_{j}) \Big[ H_{0} (r_{j} \sqrt{\Omega} ) - H_{0} ( i r_{j} \sqrt{\Omega} ) \Big]  - \textbf{e}_{\theta \, j} \cdot (\nabla \times \textbf{w})\Big|_{\textbf{x} = \textbf{X}_{j}} \frac{M_{j}}{8D}  \Big[ i H_{1} (r_{j} \sqrt{\Omega} ) + H_{1} ( i r_{j} \sqrt{\Omega} ) \Big]  \right\rbrace
\end{split}, \label{GreensCrystalW}
\end{equation}

\begin{equation}
\phi(\textbf{x}) = \phi_{\mathrm{inc}} (\textbf{x}) + \sum_{j=1}^{m} \frac{i V_{j} \sqrt{3}}{2 \rho  h^{2} \alpha^{2}} \textbf{e}_{r \, j}  \cdot \nabla w\Big|_{\textbf{x} = \textbf{X}_{j}}  H_{1} (\frac{\Omega h}{2 L \sqrt{3}} r_{j} ), \label{GreensCrystalphi}
\end{equation}

\begin{equation}
\psi(\textbf{x}) = \psi_{\mathrm{inc}} (\textbf{x}) - \sum_{j=1}^{m}  \frac{i V_{j} \sqrt{3}}{2 \rho h^{2} \alpha \beta} \textbf{e}_{\theta \, j} \cdot \nabla w\Big|_{\textbf{x} = \textbf{X}_{j}}  H_{1} (\frac{\alpha}{\beta} \frac{\Omega h}{2 L \sqrt{3}} r_{j} ) . \label{GreensCrystalpsi}  
\end{equation}

Here, $\textbf{e}_{r \, j}$ and $\textbf{e}_{\theta \, j}$ are the local radial polar coordinate basis vectors centred on $\textbf{x} = \textbf{X}_{j}$ and  $r_{j} = |\textbf{x} - \textbf{X}_{j}|$. We apply Foldy's method \cite{foldy1945multiple} in which singularities, as $r_{j} \to 0$, are naturally removed by examining the external field. Foldy's hypothesis \cite{foldy1945multiple} states that the ``strength" of the scattered field is proportional to the external field; the constant of proportionality, known as the scattering coefficient, is typically leveraged to determine all unknowns in the system - as in \cite{martin2015scattering,wiltshaw2020asymptotic}. The approach taken by Schnitzer and Craster \cite{schnitzer2017bloch} is much more straightforward and requires examination of the external field via matched asymptotic analysis, revealing a system of linear equations readily solved by inverting a matrix. 

Similarly to \cite{schnitzer2017bloch}, we opt not to consider scattering coefficients but the external field as defined in the appendix \ref{AppExtFie} equations \eqref{wExternal} - \eqref{psiExternal}, in the limit as $\textbf{x}_{n} \to 0$. On the one hand, inserting \eqref{GreensCrystalW} - \eqref{GreensCrystalpsi}  into \eqref{wExternal} - \eqref{psiExternal} we find the contribution of the $n$th scatter removed and the summations, as $\textbf{x}_{n} \to 0$, are now finite everywhere. On the other hand, we note the singular asymptotics of the field as $r_{n} \to 0$ will be dominated by the $n$th scatterer, whose presence causes the singularities to occur - in this limit, we may make use of equation \eqref{TotalSolns} where,  as $r_{n} \to 0$, the various finite Bessel function terms are retained in order to apply the conditions \eqref{DispCond1}-\eqref{PotentialConds}. Equating these two ways of examining the the external fields, as $r_{n} \to 0$, one finds from equations \eqref{wExternal}, $\textbf{e}_{x} \cdot \nabla$\eqref{wExternal}, $\textbf{e}_{y} \cdot \nabla$\eqref{wExternal},  \eqref{phiExternal} and \eqref{psiExternal}:
\begin{equation}
\begin{split}
\left[ 1 - \frac{i L}{8 D} F_{n} \right] w(\textbf{X}_{n}) - \sum_{\substack{j=1 \\ j \neq n}}^{m} \frac{i L}{8D} F_{j} w(\textbf{X}_{j}) \Big[ H_{0} (r_{nj} \sqrt{\Omega} ) - H_{0} ( i r_{nj} \sqrt{\Omega} ) \Big]  - \quad \quad \quad  \quad \\ 
- \sum_{\substack{j=1 \\ j \neq n}}^{m} \frac{M_{j}}{8D} \Big\lbrace \cos \theta_{nj} w_{x} (\textbf{X}_{j}) + \sin \theta_{nj} w_{y} (\textbf{X}_{j}) \Big\rbrace  \Big[ i H_{1} (r_{nj} \sqrt{\Omega} ) + H_{1} ( i r_{nj} \sqrt{\Omega} ) \Big]  = w_{\mathrm{inc}}(\textbf{X}_{n}), \label{Foldy_w}
\end{split} 
\end{equation}
\begin{equation}
\begin{split}
w_{x}(\textbf{X}_{n}) + \frac{M_{n} \sqrt{\Omega}}{8 D} \left[ \frac{2}{\pi} \Big( \log\frac{\epsilon_{n} \sqrt{\Omega}}{2} + \gamma_{E} - 1 \Big) - \frac{i}{2} \right] w_{x}(\textbf{X}_{n}) - \quad \quad  \quad  \quad  \quad  \quad \quad  \quad  \quad  \quad \\
- \sum_{\substack{j=1 \\ j \neq n}}^{m} \cos \theta_{nj} \frac{i L}{8D} F_{j} w(\textbf{X}_{j}) i \sqrt{\Omega} \Big[ H_{1} (i r_{nj} \sqrt{\Omega} ) + i H_{1} ( r_{nj} \sqrt{\Omega} ) \Big]  -  \quad \quad  \quad  \quad  \quad  \quad \quad  \quad  \quad  \quad \\ 
- \sum_{\substack{j=1 \\ j \neq n}}^{m} \frac{i \cos \theta_{nj} M_{j} \sqrt{\Omega}}{8D}  \Big\lbrace \cos \theta_{nj} w_{x} (\textbf{X}_{j}) + \sin \theta_{nj}  w_{y} (\textbf{X}_{j}) \Big\rbrace 
 \Big[ \frac{ \scaleto{H_{0} (i r_{nj} \sqrt{\Omega} ) + H_{0} ( r_{nj} \sqrt{\Omega} ) - H_{2} (i r_{nj} \sqrt{\Omega} ) -  H_{2} (r_{nj} \sqrt{\Omega})}{10 pt} }{ \scaleto{2}{7.5 pt}} \Big] + \quad \\
 + \sum_{\substack{j=1 \\ j \neq n}}^{m} \frac{\sin \theta_{nj}}{r_{nj}} M_{j}  \Big\lbrace \cos \theta_{nj} w_{y} (\textbf{X}_{j}) - \sin \theta_{nj}  w_{x} (\textbf{X}_{j}) \Big\rbrace  \frac{1}{8D} \Big[ H_{1} (i r_{nj} \sqrt{\Omega} ) + i H_{1} ( r_{nj} \sqrt{\Omega} ) \Big]  = \quad \quad \quad \\ = (\textbf{e}_{x} \cdot \nabla w_{\mathrm{inc}})|_{\textbf{x} = \textbf{X}_{n}}, \quad \quad  \quad  \quad  \quad  \quad \quad  \quad  \quad  \quad  \quad \quad  \quad   \quad  \quad   \quad  \quad  
\end{split} \label{Foldy_w_x}
\end{equation}

\begin{equation}
\begin{split}
w_{y}(\textbf{X}_{n}) + \frac{M_{n} \sqrt{\Omega}}{8 D} \left[ \frac{2}{\pi} \Big( \log\frac{\epsilon_{n} \sqrt{\Omega}}{2} + \gamma_{E} - 1 \Big) - \frac{i}{2} \right]  w_{y}(\textbf{X}_{n}) - \quad \quad  \quad  \quad  \quad  \quad \quad  \quad  \quad  \quad  \\
- \sum_{\substack{j=1 \\ j \neq n}}^{m} \sin \theta_{nj} \frac{i L}{8D} F_{j} w(\textbf{X}_{j}) i \sqrt{\Omega} \Big[ H_{1} (i r_{nj} \sqrt{\Omega} ) + i H_{1} ( r_{nj} \sqrt{\Omega} ) \Big]  - \quad \quad  \quad  \quad  \quad  \quad \quad  \quad  \quad  \quad \\ 
- \sum_{\substack{j=1 \\ j \neq n}}^{m} \frac{i\sin \theta_{nj} M_{j} \sqrt{\Omega}}{8D} \Big\lbrace \cos \theta_{nj} w_{x} (\textbf{X}_{j}) + \sin \theta_{nj}  w_{y} (\textbf{X}_{j}) \Big\rbrace 
 \Big[ \frac{ \scaleto{H_{0} (i r_{nj} \sqrt{\Omega} ) + H_{0} ( r_{nj} \sqrt{\Omega} ) - H_{2} (i r_{nj} \sqrt{\Omega} ) -  H_{2} (r_{nj} \sqrt{\Omega})}{10 pt} }{ \scaleto{2}{7.5 pt}} \Big] -  \quad \\
 - \sum_{\substack{j=1 \\ j \neq n}}^{m} \frac{\cos \theta_{nj}}{r_{nj}} M_{j}  \Big\lbrace \cos \theta_{nj} w_{y} (\textbf{X}_{j}) - \sin \theta_{nj}  w_{x} (\textbf{X}_{j}) \Big\rbrace  \frac{1}{8D} \Big[ H_{1} (i r_{nj} \sqrt{\Omega} ) + i H_{1} ( r_{nj} \sqrt{\Omega} ) \Big]   = \quad \quad \quad \\ = (\textbf{e}_{y} \cdot \nabla w_{\mathrm{inc}})|_{\textbf{x} = \textbf{X}_{n}}, \quad \quad  \quad  \quad  \quad  \quad \quad  \quad  \quad  \quad  \quad \quad  \quad   \quad  \quad   \quad  \quad  
\end{split}  \label{Foldy_w_y}
\end{equation}

\begin{equation}
 \phi(\textbf{X}_{n}) - \sum_{\substack{j=1 \\ j \neq n}}^{m} \frac{i V_{j} \sqrt{3}}{2 \rho  h^{2} \alpha^{2}}  \Big\lbrace \cos \theta_{nj} w_{x} (\textbf{X}_{j}) + \sin \theta_{nj} w_{y} (\textbf{X}_{j}) \Big\rbrace  H_{1} (\frac{\Omega h}{2 L \sqrt{3}} r_{nj} ) = \phi_{\mathrm{inc}}(\textbf{X}_{n}), \label{Foldy_phi}
\end{equation}

\begin{equation}
 \psi(\textbf{X}_{n}) - \sum_{\substack{j=1 \\ j \neq n}}^{m} \frac{V_{j} \sqrt{3}}{2 i \rho h^{2} \alpha \beta}  \Big\lbrace \cos \theta_{nj} w_{y} (\textbf{X}_{j}) - \sin \theta_{nj} w_{x} (\textbf{X}_{j}) \Big\rbrace  H_{1} (\frac{\alpha}{\beta} \frac{\Omega h}{2 L \sqrt{3}} r_{nj} )  = \psi_{\mathrm{inc}}(\textbf{X}_{n}).  \label{Foldy_psi}
\end{equation}
Here $\theta_{nj}$ denotes the angle between the $n$th and $j$th scatterer, centred on $\textbf{X}_{j}$ and $r_{nj} = |\textbf{X}_{n}-\textbf{X}_{j}|$. To find the above, Bessel functions of the first kind were asymptotically expanded as $r_{nn}  \to 0$. Similarly, complementary coefficients \eqref{EqnAB0comp}-\eqref{EqnAB12comp} are expanded assuming $\epsilon_{n}$ is a small parameter. Successively considering  $n$ to take the values from $1$ to $m$, equations \eqref{Foldy_w} - \eqref{Foldy_psi} represent a system of $5m$ equations for $5m$ unknowns and are factorised as follows
\begin{equation}
\mathcal{G} \boldsymbol{\Lambda} = \boldsymbol{\Lambda}_{\mathrm{inc}}. \label{GeneralisedFoldySystem}
\end{equation} 
Here, $\mathcal{G}$ represents the $5m \times 5m$ matrix formed from equations \eqref{Foldy_w}-\eqref{Foldy_psi}, $\boldsymbol{\Lambda}$ a vector containing our unknowns $w(\textbf{X}_{n})$, $\partial_{x} w(\textbf{X}_{n})$, $\partial_{y} w(\textbf{X}_{n})$, $\phi(\textbf{X}_{n})$ and $\psi(\textbf{X}_{n})$ and finally $\boldsymbol{\Lambda}_{\mathrm{inc}}$ a vector containing our known incident source $w_{\mathrm{inc}}(\textbf{X}_{n})$, $\partial_{x} w_{\mathrm{inc}}(\textbf{X}_{n})$, $\partial_{y} w_{\mathrm{inc}}(\textbf{X}_{n})$, $\phi_{\mathrm{inc}}(\textbf{X}_{n})$ and $\psi_{\mathrm{inc}}(\textbf{X}_{n})$. Once $\boldsymbol{\Lambda} $ is determined, the total displacement field is calculated everywhere in the phononic crystal by use of \eqref{GreensCrystalW}-\eqref{GreensCrystalpsi}. The inversion of $\mathcal{G}$ is simple and accurate unless near resonance, where matrices have poor condition numbers.  As in appendix \ref{CompLE}, we opt to use the minimum norm least-squares solution to compute $\boldsymbol{\Lambda}$ near resonance.

\section{FEM verification of the analytical solutions} \label{sec::Testing}
In this section, we demonstrate the reliability of our analytical solutions by comparing our eigenvalues from \eqref{EVPactualalgebraicDispBloch}, to computations using FEM simulations for the actual beam plate structure from ABAQUS \textsuperscript{\textregistered} software. The FEM model is defined using 4-node doubly curved general-purpose shell elements for the plate, and 2-node cubic beam elements for the beams. Provided the dispersion relation computed from the eigenvalue problem is accurate, our generalized Foldy solution \eqref{GeneralisedFoldySystem} must also be accurate, as these two solutions are nothing more than the same Green's function written in a slightly different fashion; one utilizing a Fourier series and the other a Fourier transform. Additionally, one can observe whether the direct scattering solution has the same dispersive properties as predicted by the eigenvalue problem, further demonstrating the relationship of the two solutions. 

 Moreover, we test the model by removing various terms from the system \eqref{ForcedKL} - \eqref{ForcedShear}, and hence various rows and columns from the matrices forming \eqref{GeneralisedFoldySystem} or \eqref{EVPactualalgebraicDispBloch}, to consider the following cases:

\begin{itemize}
\item Case 1 - purely compressional forcing from the beam coupling into purely flexural motion from the plate, only equation \eqref{ForcedKL} with the moment term $\textbf{M}_{IJ}$ removed governs this case.
\item Case 2 - compressional and flexural forcing from the beam coupling into purely flexural motion from the plate, hence only equation \eqref{ForcedKL} is included.
\item Case 3 - the full system \eqref{ForcedKL} - \eqref{ForcedShear}.
\end{itemize}

Here, cases 1, 2 and 3 contain compressional resonances, arising through $\textbf{F}$ and given by
\begin{equation}
\hat{\alpha} \Omega \hat{\ell}  =  \left( q - \frac{1}{2} \right) \pi \quad \mbox{for $q \in \mathbb{Z}$. }  \label{CompressionalRes}
\end{equation}
Cases 2 and 3 also contain flexural resonances arising from $\textbf{M}$ and $\textbf{V}$. Figs \ref{fig:ThickerPlateFEMcomp} \& \ref{fig:ThinnerPlateFEMcomp}  reveal two different regimes exist for the flexural resonances within the system, depending on the ratio of $\epsilon$ to $h$. As in Colquitt \textit{et al}. \cite{colquitt2017seismic}, when the plate is thicker than the beam, the resonances coincide with the natural frequencies of the beam satisfying clamped-free boundary conditions applied at either end - the clamped-free natural frequencies satisfy \cite{colquitt2017seismic,graff1975wave}
\begin{equation}
\cos \hat{\beta} \sqrt{\Omega} \hat{\ell} + \sech \hat{\beta} \sqrt{\Omega} \hat{\ell}  = 0. \label{ClampedFreeRes}
\end{equation}
When the plate is thinner than the beam, we observe that the resonances shift to coincide with the natural frequencies of the pinned-free beam, where $\Omega$ satisfies \cite{graff1975wave}
\begin{equation}
\sin \hat{\beta} \sqrt{\Omega} \hat{\ell} - \cos \hat{\beta} \sqrt{\Omega} \hat{\ell} \tanh \hat{\beta} \sqrt{\Omega} \hat{\ell}  = 0. \label{PinnedFreeRes}
\end{equation}
The roots arising from equation \eqref{ClampedFreeRes} coincide with the denominator terms of $\textbf{M}$ and $\textbf{V}$, hence approaching a resonance from \eqref{ClampedFreeRes}, terms within \eqref{EVPactualalgebraicDispBloch} or \eqref{GeneralisedFoldySystem}  will become large and cause poorly conditioned matrices. However, any matrix manipulations either involve computing the minimum norm least-squares solution or the solution to an algebraic eigenvalue problem, we find these methods to be sufficiently numerically stable for our needs near resonance - as corroborated by the observed shift in flexural resonances from \eqref{ClampedFreeRes} to \eqref{PinnedFreeRes}, as in figs \ref{fig:ThickerPlateFEMcomp} to \ref{fig:ThinnerPlateFEMcomp}, where our analytical solutions indeed match the FEM computations.

\begin{table}[h]
\begin{center}
\begin{tabular}{||c c||} 
 \hline
 Parameter & Value  \\ [0.5ex] 
 \hline\hline
 $L$ & $1$ $\mathrm{m}$  \\ 
 \hline
$\rho$  & $2710$ $\mathrm{kg \, m^{-3}}$ \\
 \hline
 $E$ & $69 \times 10^{9}$  $\mathrm{Pa}$ \\
 \hline
 $\nu$ & $0.33$   \\ [1ex] 
 \hline
\end{tabular}
\end{center}
\caption{The material parameters used within figs \ref{fig:ThickerPlateFEMcomp} - \ref{fig:ScattPertHexTopoArrange} modelling aluminium. We set $\hat{\rho}$, $\hat{E}$ and $\hat{\nu}$ to equal their unhatted counterparts within the proceeding computations. Any other required parameters will be given in the appropriate captions.}
\end{table}

\subsection{Verification for one beam per primitive cell}

\begin{figure}[h]
\centering
\hspace*{0cm}
\begin{tikzpicture}[scale=0.35, transform shape]

\begin{scope}[xshift=0cm, yshift=65cm]
\draw (0, 0) node[inner sep=0] {\includegraphics[scale=0.45]{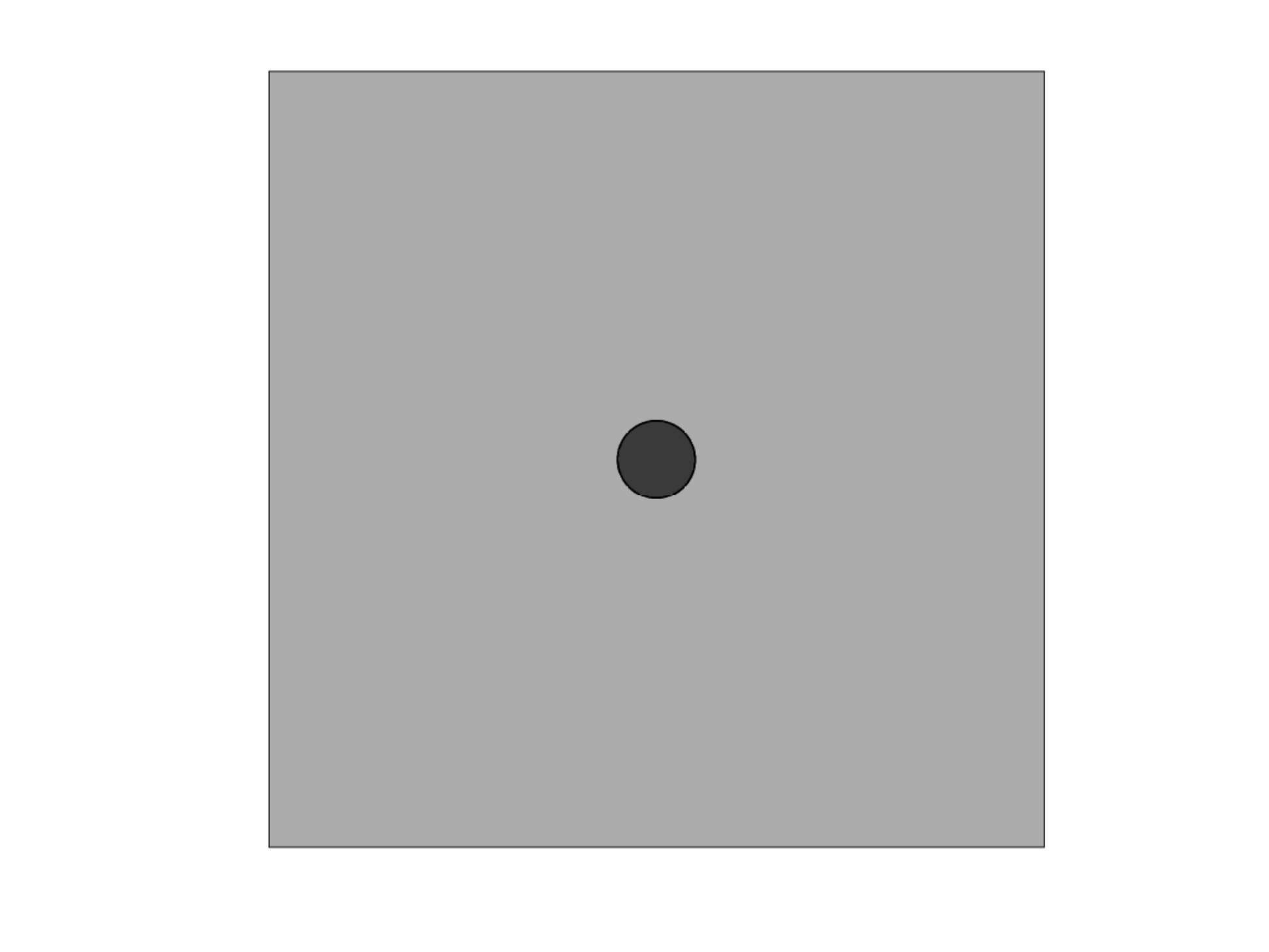}};
\node[below, scale=2, black] at (4,1) {$\displaystyle (f)$};
\end{scope}

\begin{scope}[xshift=0cm, yshift=52cm,scale=3, transform shape]
\draw (0, 0) node[inner sep=0] {\includegraphics[scale=0.45]{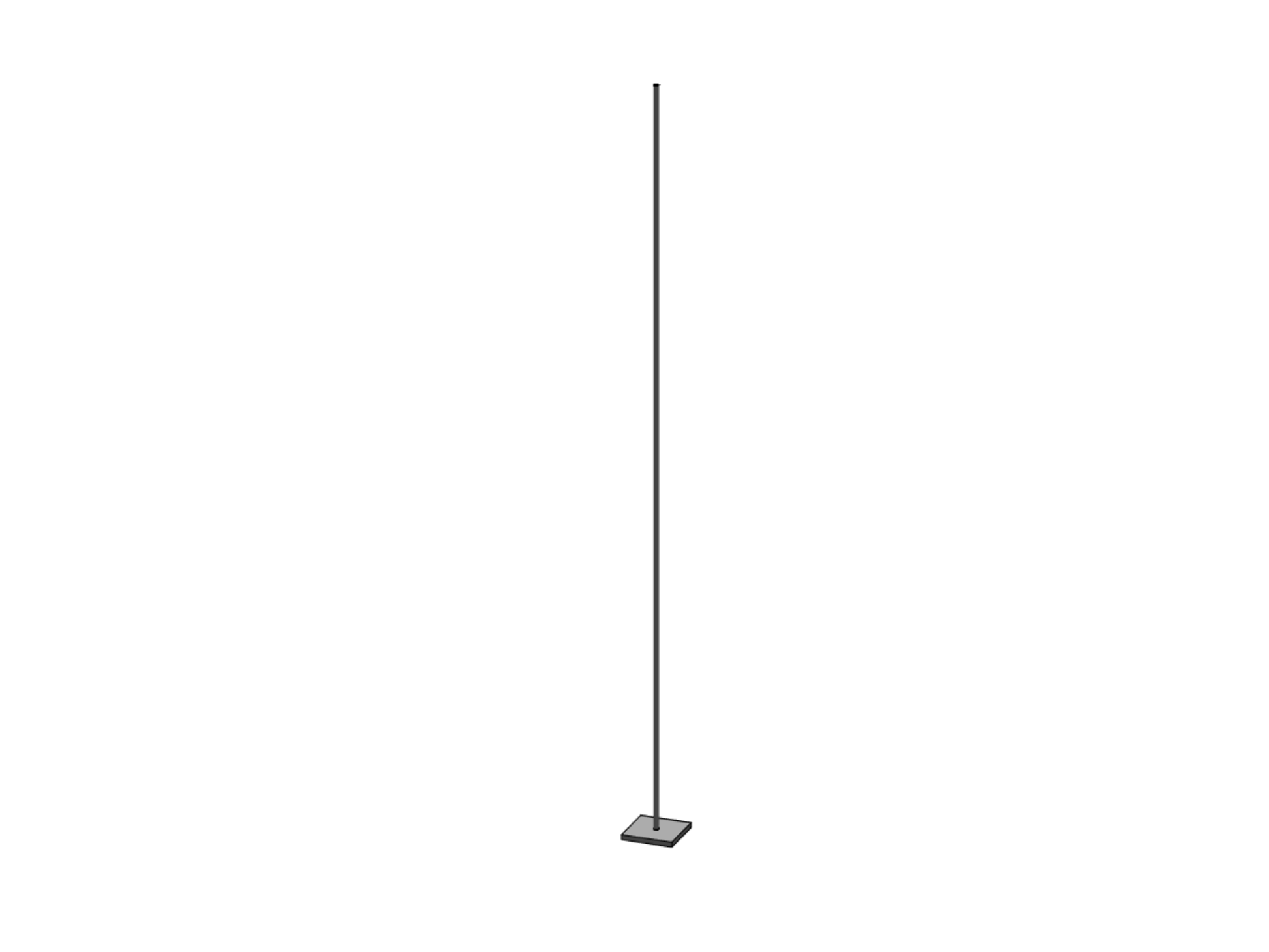}};
\node[below, scale=2*0.33, black] at (0.25,-3) {$\displaystyle (g)$};
\end{scope}

\begin{scope}[xshift=-30cm, yshift=69cm]
		\node[regular polygon, regular polygon sides=4,draw, inner sep=6.5cm,rotate=0,line width=0.0mm, white,
           path picture={
               \node[rotate=0] at (0.5,0){
                   \includegraphics[scale=1.25]{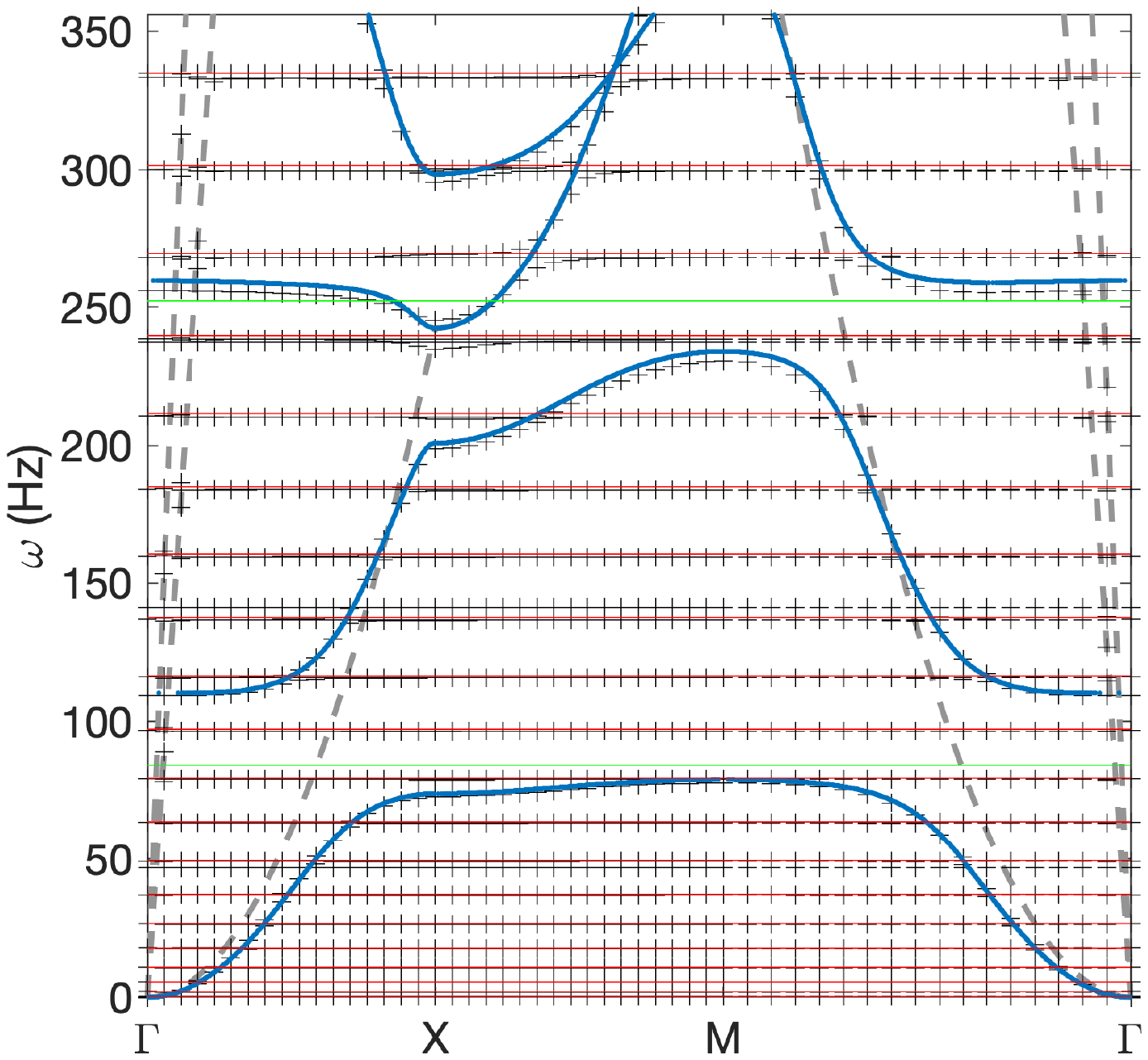}
               };
           }]{};
           		\node[below, scale=2, black] at (-8,9) {$\displaystyle (a)$};     
\end{scope}       

\begin{scope}[xshift=-12cm, yshift=69cm]
		\node[regular polygon, regular polygon sides=4,draw, inner sep=6.5cm,rotate=0,line width=0.0mm, white,
           path picture={
               \node[rotate=0] at (0.5,0){
                   \includegraphics[scale=1.25]{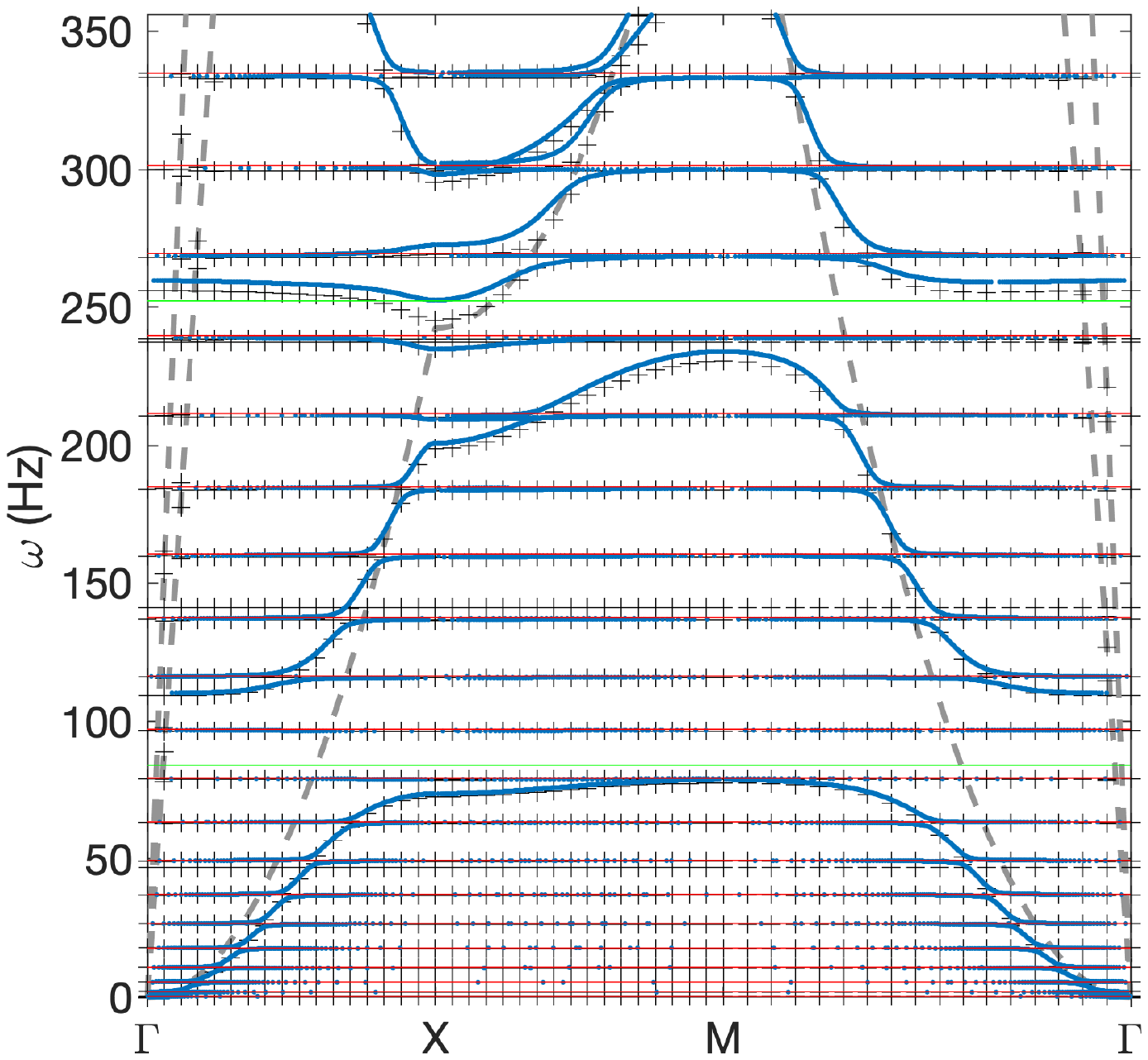}
               };
           }]{};
                      		\node[below, scale=2, black] at (-8,9)  {$\displaystyle (b)$};     
\end{scope}

\begin{scope}[xshift=-30cm, yshift=50cm]
		\node[regular polygon, regular polygon sides=4,draw, inner sep=6.5cm,rotate=0,line width=0.0mm, white,
           path picture={
               \node[rotate=0] at (0.5,1){
                   \includegraphics[scale=1.25]{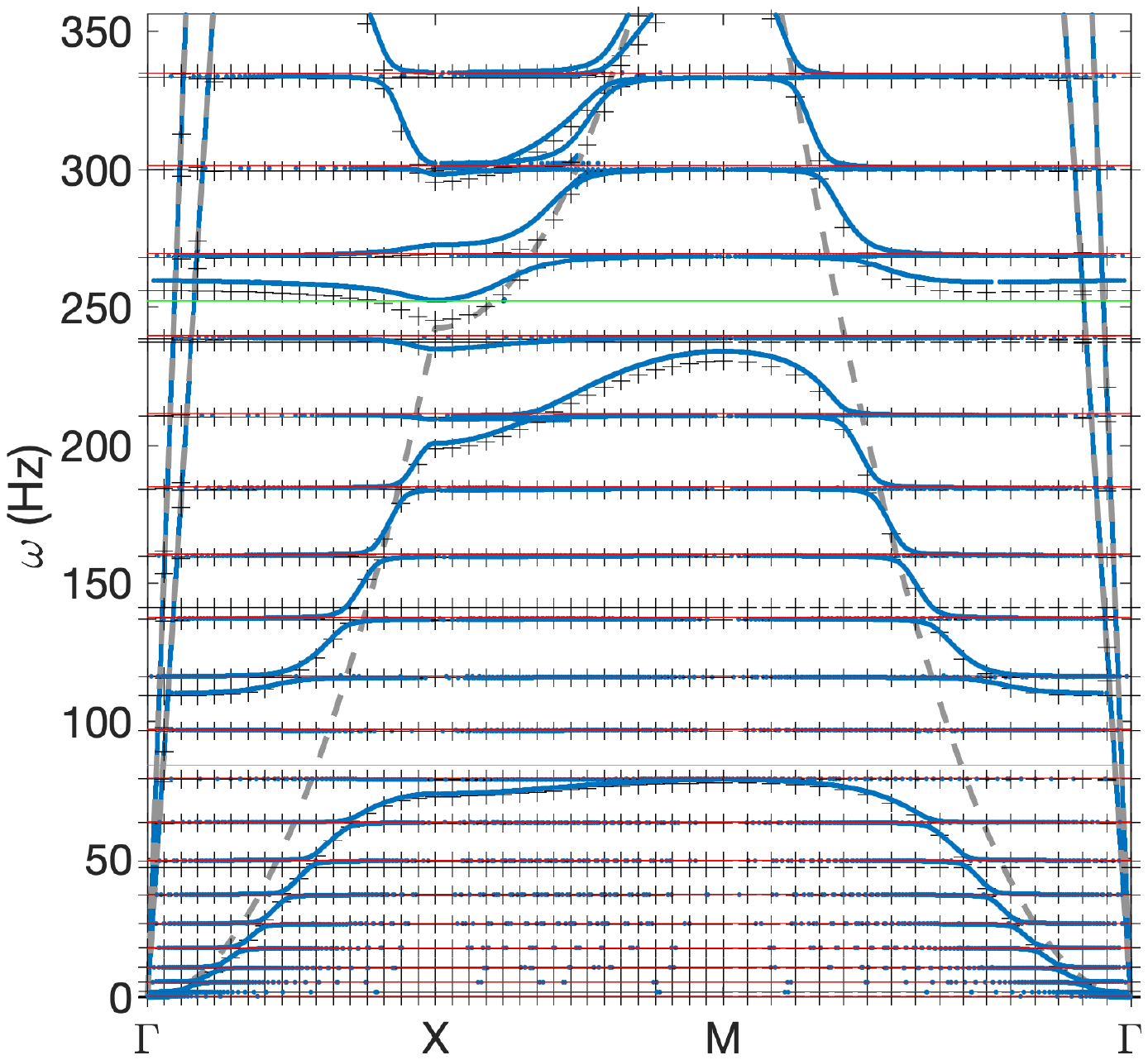}
               };
           }]{};
                      		\node[below, scale=2, black] at (-8,10)  {$\displaystyle (c)$};     
\end{scope}  

 \begin{scope}[xshift=-12cm, yshift=50cm]
		\node[regular polygon, regular polygon sides=4,draw, inner sep=6.5cm,rotate=0,line width=0.0mm, white,
           path picture={
               \node[rotate=0] at (0.5,1){
                   \includegraphics[scale=1.25]{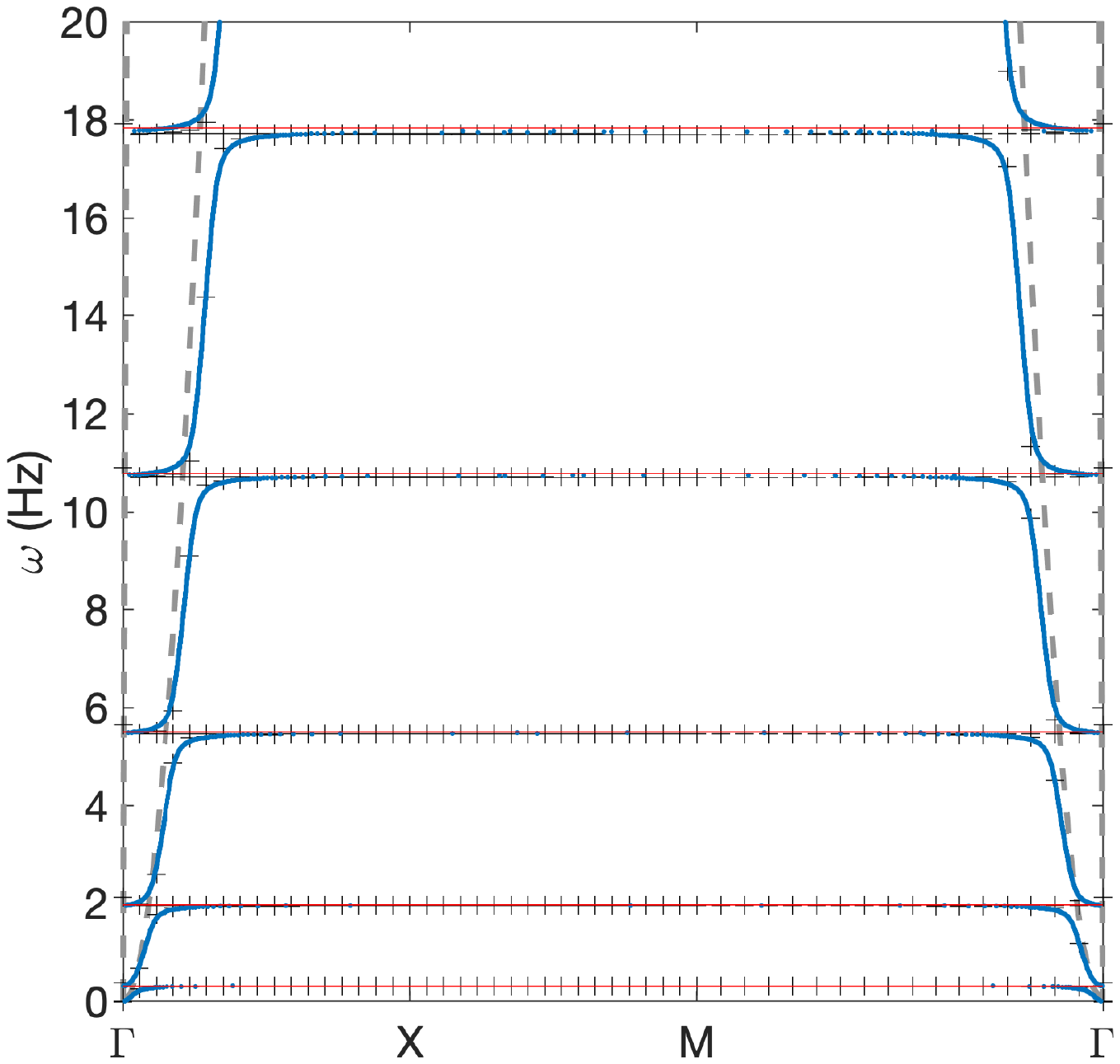}
               };
           }]{};
                      		\node[below, scale=2, black] at (-8,10)  {$\displaystyle (d)$};     
\end{scope} 

\begin{scope}[xshift=0cm, yshift=79cm,scale=1.5, transform shape]
\node[regular polygon, regular polygon sides=4, draw, inner sep=0.5*6.28*10.0 pt,rotate=0] at (0 pt,-6.28*10.0*1.5 pt) {};
\draw[line width=0.5mm,gray,-] (0pt,-6.28*10.0*1.5 pt) -- (0+ 0.5*6.28*10*1.41 pt,-6.28*10.0*1.5 pt);
\draw[line width=0.5mm,gray,-] (0+ 0.5*6.28*10*1.41 pt,-6.28*10.0*1.5 pt) -- (0+ 0.5*6.28*10*1.41  pt, 0.5*6.28*10*1.41 -6.28*10.0*1.5 pt);
\draw[line width=0.5mm,gray,-] (0+ 0.5*6.28*10*1.41  pt, 0.5*6.28*10*1.41 -6.28*10.0*1.5 pt) -- (0pt,-6.28*10.0*1.5 pt);
\node[below,left,scale=1.75] at (0pt,-6.28*10.0*1.5 pt) {$\displaystyle  \Gamma$}; 
\node[below,right,scale=1.75] at (0+ 0.5*6.28*10*1.41 pt,-6.28*10.0*1.5 pt) {$\displaystyle  X$};
\node[above,right,scale=1.75] at (0+ 0.5*6.28*10*1.41  pt, 0.5*6.28*10*1.41 -6.28*10.0*1.5 pt) {$\displaystyle M$};
\node[regular polygon, circle, draw, inner sep=1.25pt,rotate=0,line width=0.5mm,shading=fill,outer color=gray,gray] at (0pt,-6.28*10.0*1.5 pt)  {};
\node[regular polygon, circle, draw, inner sep=1.25pt,rotate=0,line width=0.5mm,shading=fill,outer color=gray,gray] at (0+ 0.5*6.28*10*1.41 pt,-6.28*10.0*1.5 pt)  {};
\node[regular polygon, circle, draw, inner sep=1.25pt,rotate=0,line width=0.5mm,shading=fill,outer color=gray,gray] at (0+ 0.5*6.28*10*1.41  pt, 0.5*6.28*10*1.41 -6.28*10.0*1.5 pt)  {};
\node[below, scale=1.333] at (0pt,-0.5*6.28*10*1.41 -6.28*10.0*1.5 pt) {$\displaystyle  (e)$};
\end{scope}

\end{tikzpicture}
\caption{Comparisons of the dispersive curves computed from the eigenvalue problem \eqref{EVPactualalgebraicDispBloch} (blue dots) and FEM computations (black crosses) for the primitive cell given in panels $(f)$ (top view)  and $(g)$ (side view). Here, $h=0.1$ m and the dimensionless parameters (scaled with $L$) are: $\boldsymbol{\alpha}_{1} = \textbf{e}_{x}$, $\boldsymbol{\alpha}_{2} = \textbf{e}_{y}$, $\epsilon = 0.05$, $\hat{\ell}=15$. Panels $(a)$, $(b)$ and $(c)$ consider cases 1, 2 and 3 respectively. Panel $(d)$ shows a zoomed in section of the flexural resonances from $(b)$. Here, $\omega = \omega(\boldsymbol{\kappa})$ is plotted throughout the irreducible Brillouin zone  $(e)$. The green and red lines correspond to the resonances from equations \eqref{CompressionalRes} and \eqref{ClampedFreeRes} respectively. The grey dashed lines represents the flexural (lowest), shear (middle) and dilational (highest) free space `light' lines.} 
\label{fig:ThickerPlateFEMcomp}
\end{figure}

\begin{figure}[h]
\centering
\begin{tikzpicture}[scale=0.4, transform shape]

\begin{scope}[xshift=-12cm, yshift=52cm]
		\node[regular polygon, regular polygon sides=4,draw, inner sep=6.5cm,rotate=0,line width=0.0mm, white,
           path picture={
               \node[rotate=0] at (0.5,-5){
                   \includegraphics[scale=0.95]{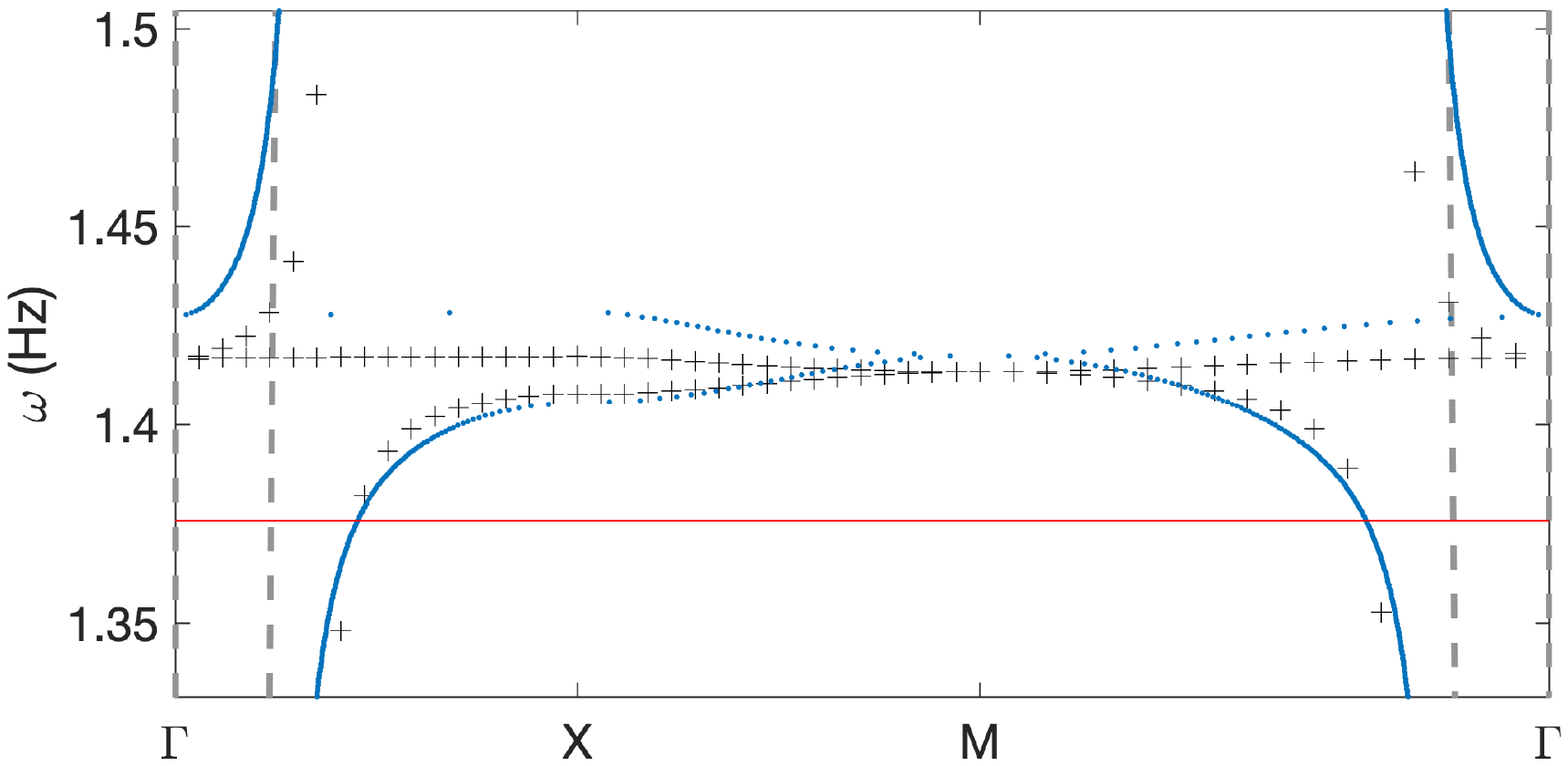}
               };
           }]{};
                               \node[below, scale=2, black] at (-8,-1.25) {$\displaystyle (e)$}; 
\end{scope}  

\begin{scope}[xshift=-12cm, yshift=60.5cm]
		\node[regular polygon, regular polygon sides=4,draw, inner sep=6.5cm,rotate=0,line width=0.0mm, white,
           path picture={
               \node[rotate=0] at (0.5,-5){
                   \includegraphics[scale=0.95]{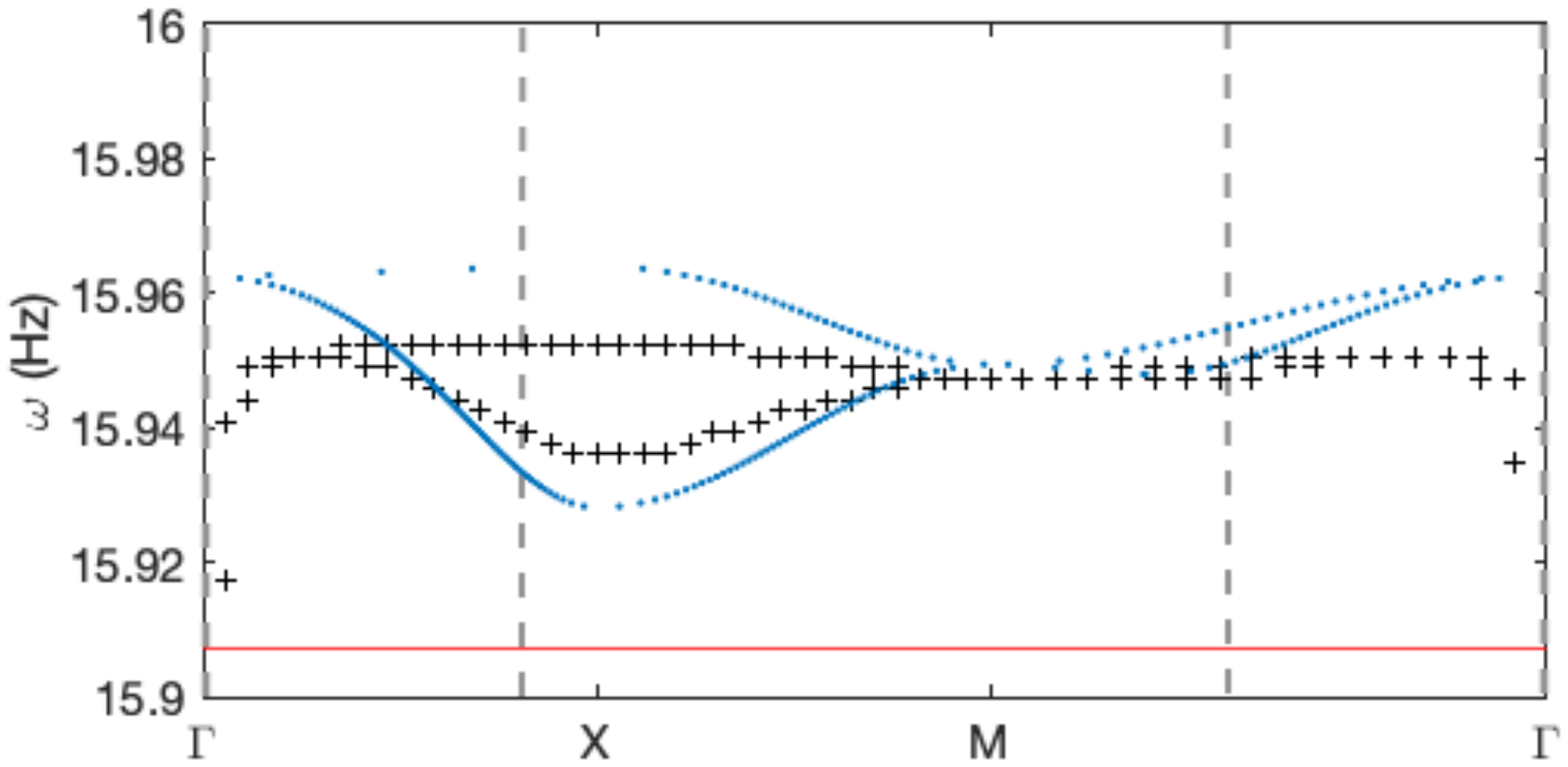}
               };
           }]{};
                               \node[below, scale=2, black] at (-8,-1.25) {$\displaystyle (d)$}; 
\end{scope}

\begin{scope}[xshift=-30cm, yshift=69cm]
		\node[regular polygon, regular polygon sides=4,draw, inner sep=6.5cm,rotate=0,line width=0.0mm, white,
           path picture={
               \node[rotate=0] at (0.5,0){
                   \includegraphics[scale=1.25]{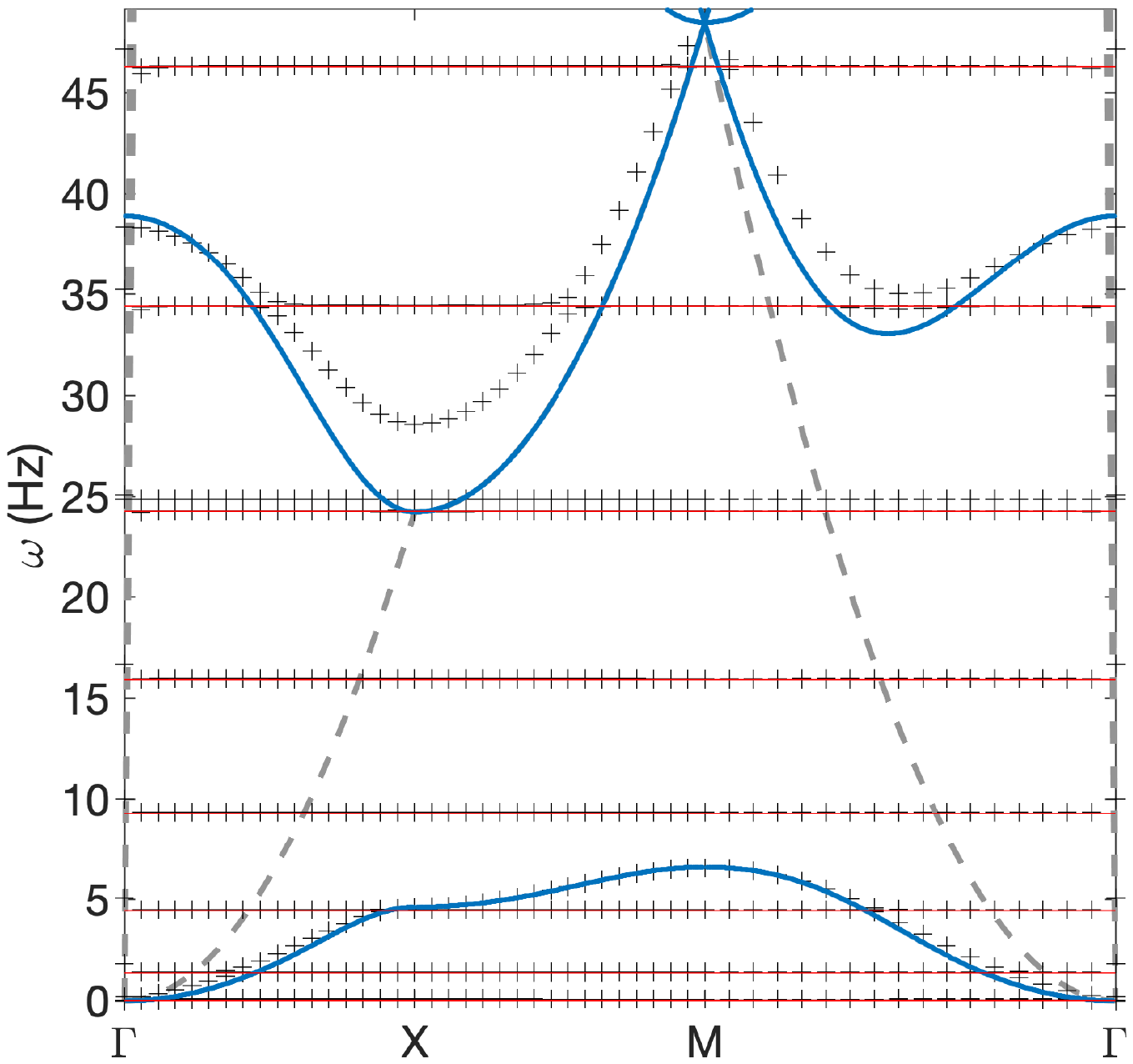}
               };
           }]{};
           		\node[below, scale=2, black] at (-7.5,8.75) {$\displaystyle (a)$};     
\end{scope}       

\begin{scope}[xshift=-12cm, yshift=69cm]
		\node[regular polygon, regular polygon sides=4,draw, inner sep=6.5cm,rotate=0,line width=0.0mm, white,
           path picture={
               \node[rotate=0] at (0.5,0){
                   \includegraphics[scale=1.25]{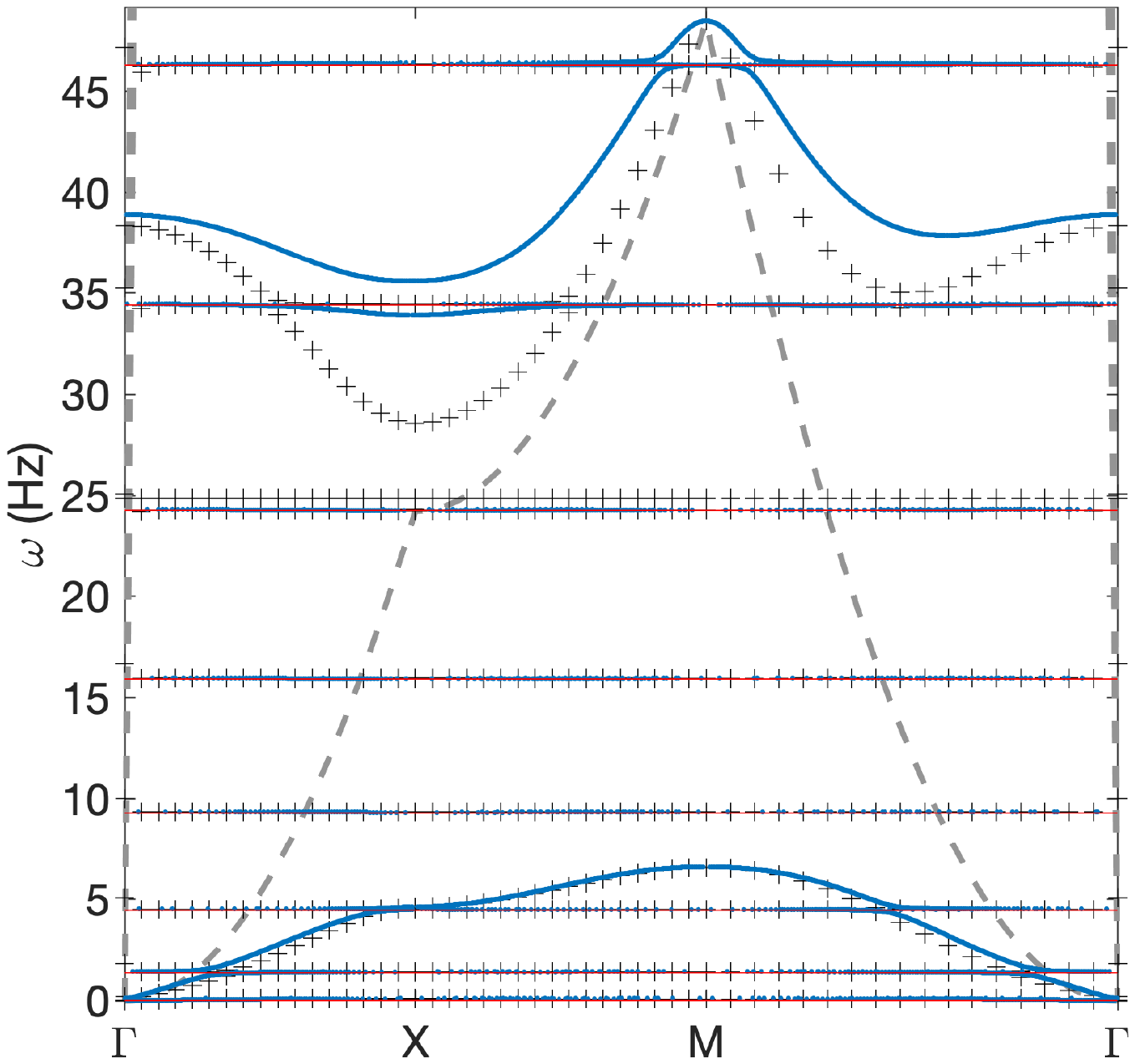}
               };
           }]{};
                      		\node[below, scale=2, black] at (-7.5,8.75) {$\displaystyle (b)$};     
\end{scope}

\begin{scope}[xshift=-30cm, yshift=50cm]
		\node[regular polygon, regular polygon sides=4,draw, inner sep=6.5cm,rotate=0,line width=0.0mm, white,
           path picture={
               \node[rotate=0] at (0.5,1){
                   \includegraphics[scale=1.25]{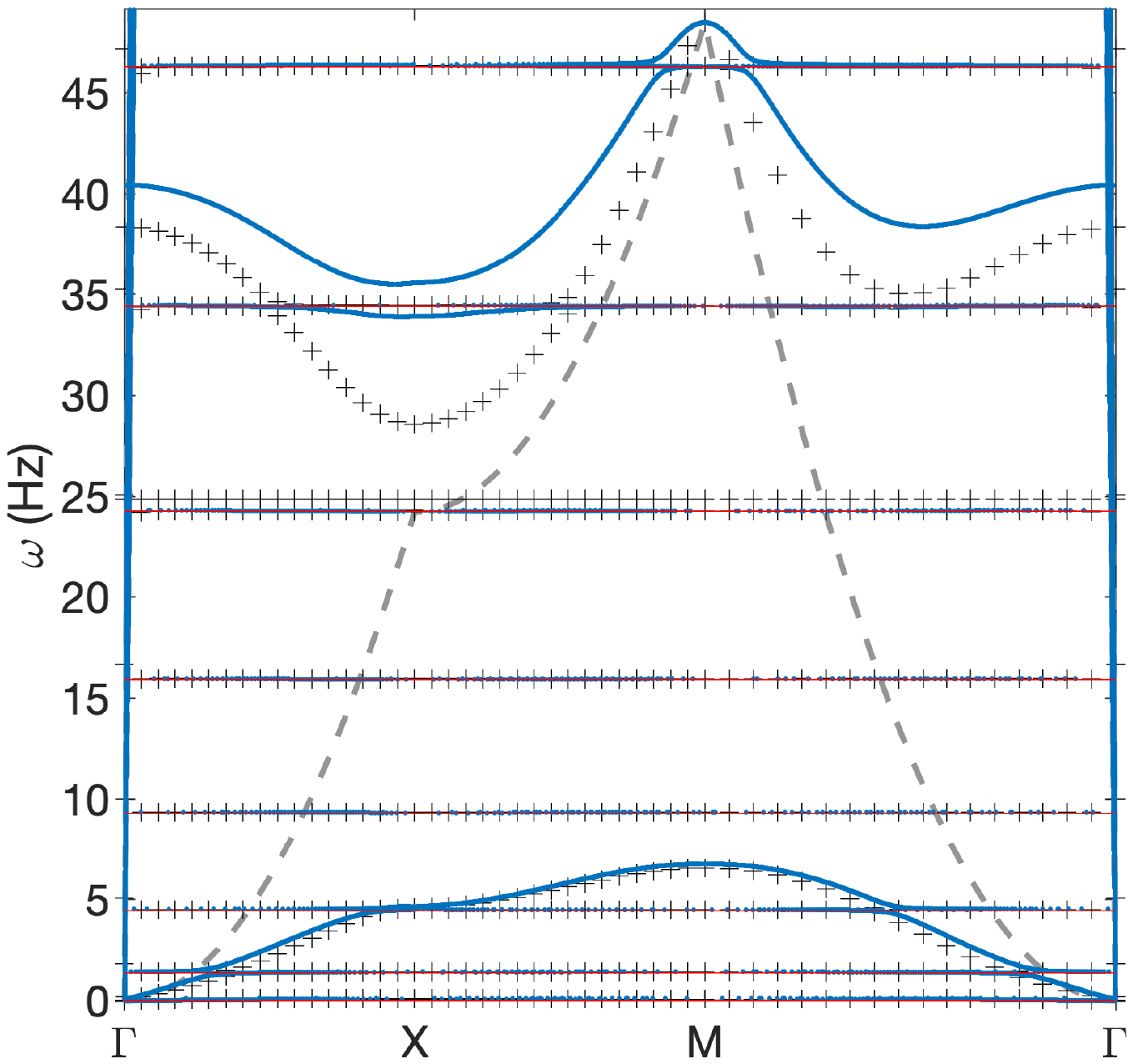}
               };
           }]{};
                      		\node[below, scale=2, black] at (-7.5,9.75) {$\displaystyle (c)$};     
\end{scope}  

\end{tikzpicture}
\caption{Same as fig. \ref{fig:ThickerPlateFEMcomp}, but with $h=0.01$ m and the red lines correspond to resonances from \eqref{PinnedFreeRes}. Here, panels $(d)$ and $(e)$ show zoomed in sections of the resonances from panel $(b)$.  } 
\label{fig:ThinnerPlateFEMcomp}
\end{figure}

Referring to figs \ref{fig:ThickerPlateFEMcomp} \& \ref{fig:ThinnerPlateFEMcomp}, we observe that the branches from our eigensolutions match those provided by the $3D$ FEM computations, but slightly differ from case to case. From fig. \ref{fig:ThickerPlateFEMcomp} $(a)$, we  see the eigenmodes from case 1 correspond to strongly perturbed flexural Kirchhoff–Love free-space modes, where the compressional resonances introduce large band gaps within the band structure. Comparing case 1 to 2 we see $\textbf{M}$ weakly perturbs the case 1 eigenmodes and introduces narrow band gaps at resonance \eqref{ClampedFreeRes}. Comparing case 2 to 3 we see the addition of weakly perturbed light lines corresponding to presence of dilational and shear waves now included in the system. 

The effect observed by the flexural beam motion is regime dependent. For instance, clamped-free resonances \eqref{ClampedFreeRes} introduce very flat resonant bands which cut through any case $1$ modes and create narrow bandgaps - as in fig. \ref{fig:ThickerPlateFEMcomp} $(d)$ and Colquitt \textit{et al}. \cite{colquitt2017seismic} - in such clamped-free regimes the dispersive nature of the system is dominated by compressional beam motion, and hence case $1$ provides sufficient accuracy provided one operates away from flexural resonances. As for pinned-free resonances \eqref{PinnedFreeRes}, their presence introduces nearly flat bands which are not associated with the generation of bandgaps - as in figs \ref{fig:ThinnerPlateFEMcomp} $(d)$ \& $(e)$. When the beams motion is in the pinned-free regime we observe only the first case $1$ branch is dominated by compressional beam motion, and for higher frequencies the flexural beam motion cannot be ignored; additionally, from the FEM analysis, above the Kirchhoff–Love `light' lines we observe beam motion with non-negligible torsional coupling, where our model does not fully capture the actual motion of the resonator. 


\begin{figure}[h]
\centering
\begin{tikzpicture}[scale=0.3, transform shape]

\begin{scope}[xshift=-27.5cm, yshift=56.0cm,scale=1.6]
		\node[regular polygon, regular polygon sides=4,draw, inner sep=7cm,rotate=0,line width=0.0mm, white,
           path picture={
               \node[rotate=0] at (-17.0,0){
                   \includegraphics[scale=1.25]{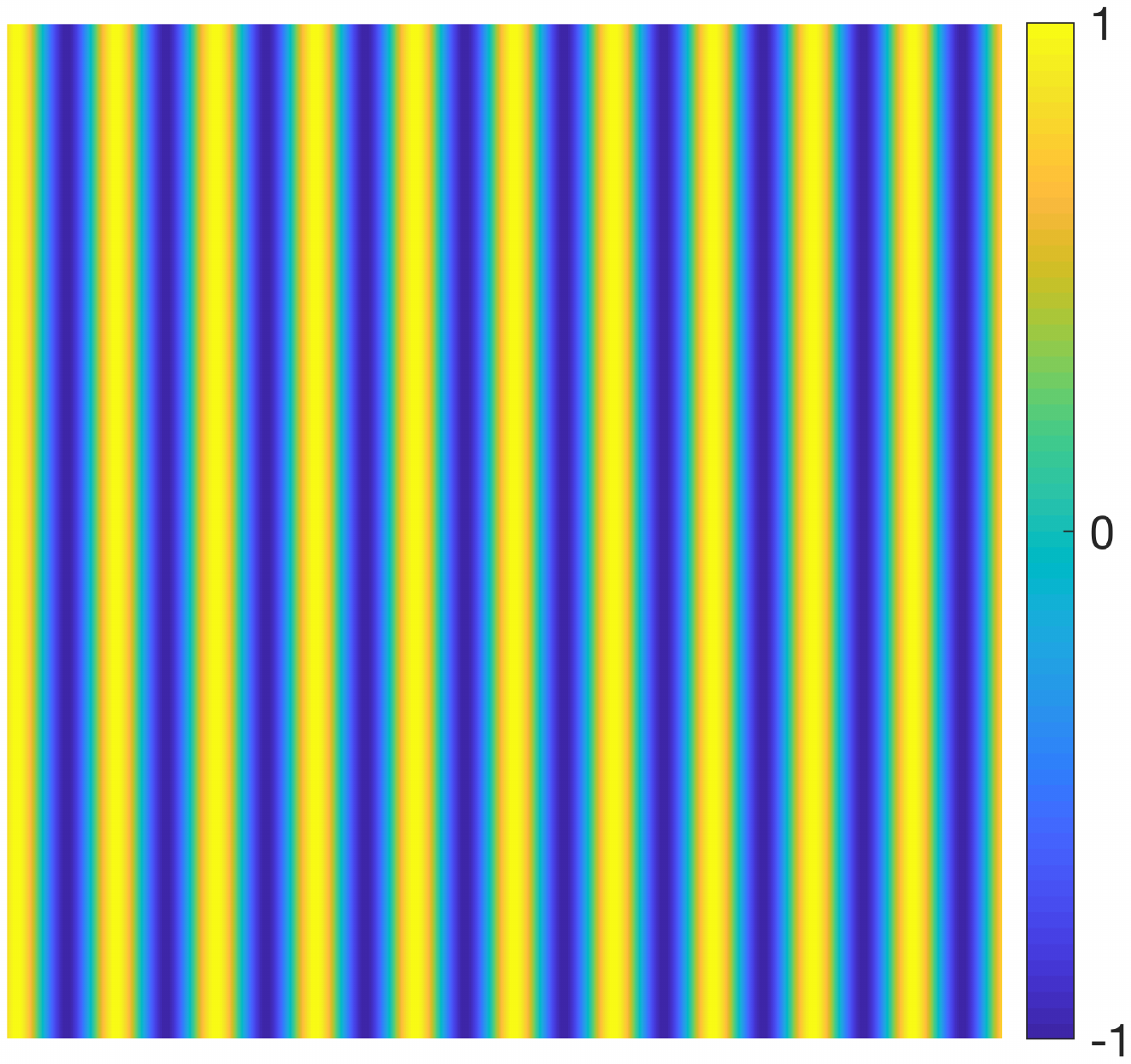}
               };
           }]{};
\end{scope}  

\begin{scope}[xshift=-30cm, yshift=67cm]
		\node[regular polygon, regular polygon sides=4,draw, inner sep=6.75cm,rotate=0,line width=0.0mm, white,
           path picture={
               \node[rotate=0] at (-0.5,-0.5){
                   \includegraphics[scale=1.5]{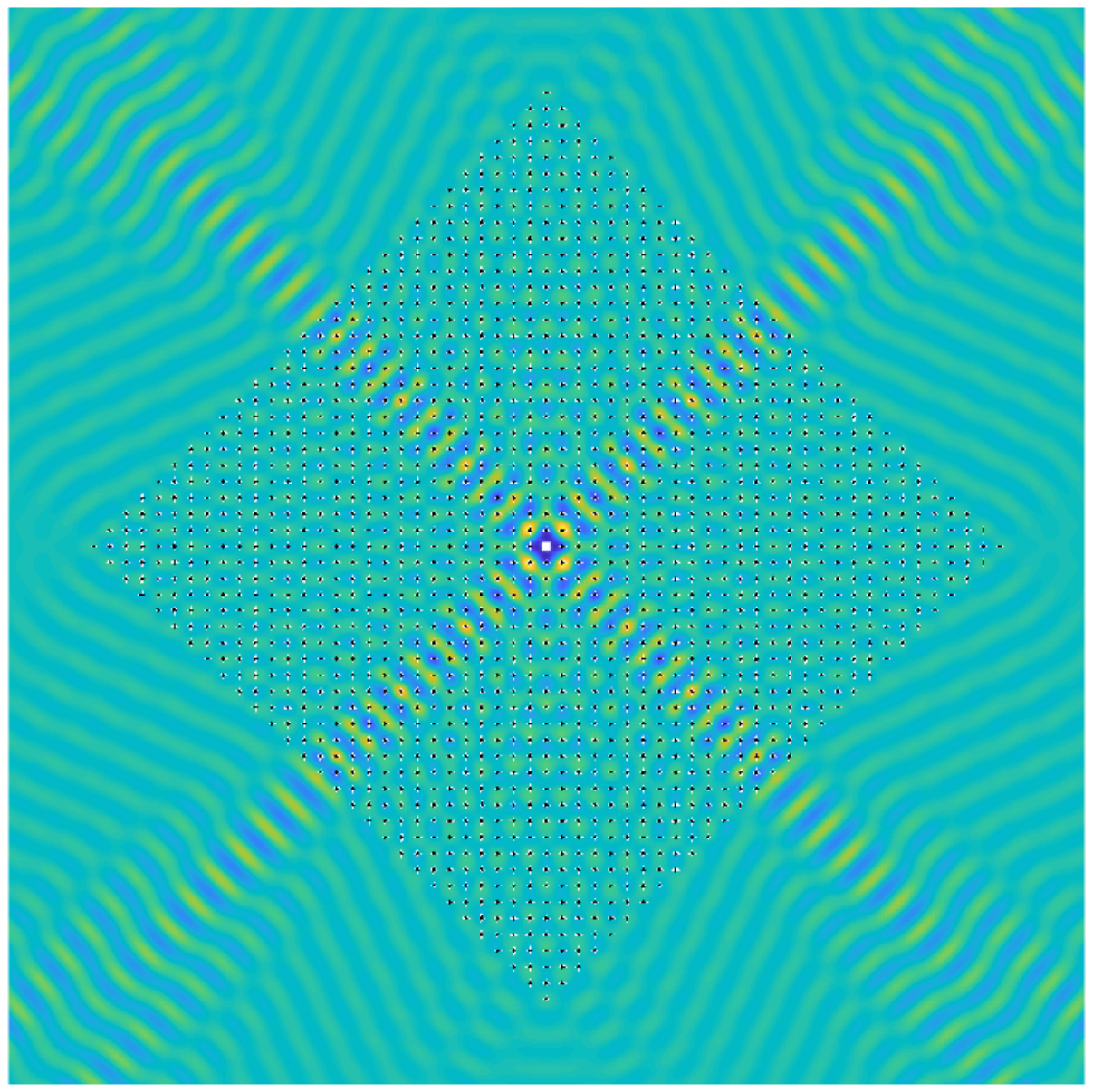}
               };
           }]{};
           \node[below, scale=4, black] at (-10.0,9.5) {$\displaystyle (a)$};
\end{scope}  

\begin{scope}[xshift=-8cm, yshift=67cm]
		\node[regular polygon, regular polygon sides=4,draw, inner sep=6.75cm,rotate=0,line width=0.0mm, white,
           path picture={
               \node[rotate=0] at (-0.5,-0.5){
                   \includegraphics[scale=1.5]{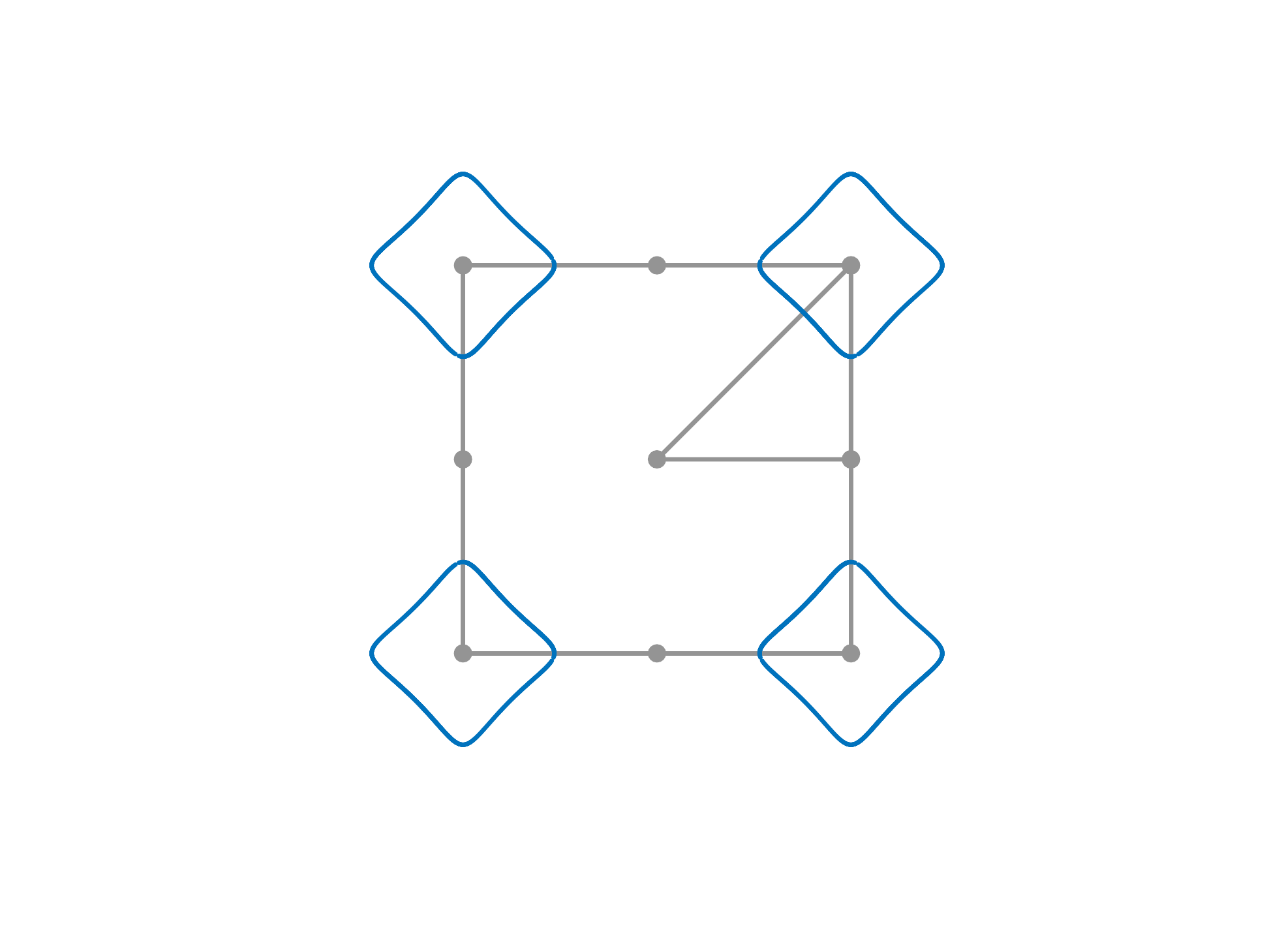}
               };
           }]{};
 			\node[below,left,scale=4] at (0.0,-0.5) {$\displaystyle  \Gamma$}; 
			\node[below,right,scale=4] at (4.25,-0.5) {$\displaystyle  X$};
			\node[above,right,scale=4] at (4.25, 4.5) {$\displaystyle M$};
          	\node[below, scale=4, black] at (-10.0,9.5) {$\displaystyle (b)$};
\end{scope}

\begin{scope}[xshift=-30cm, yshift=46cm]
		\node[regular polygon, regular polygon sides=4,draw, inner sep=6.75cm,rotate=0,line width=0.0mm, white,
           path picture={
               \node[rotate=0] at (-0.5,-0.5){
                   \includegraphics[scale=1.5]{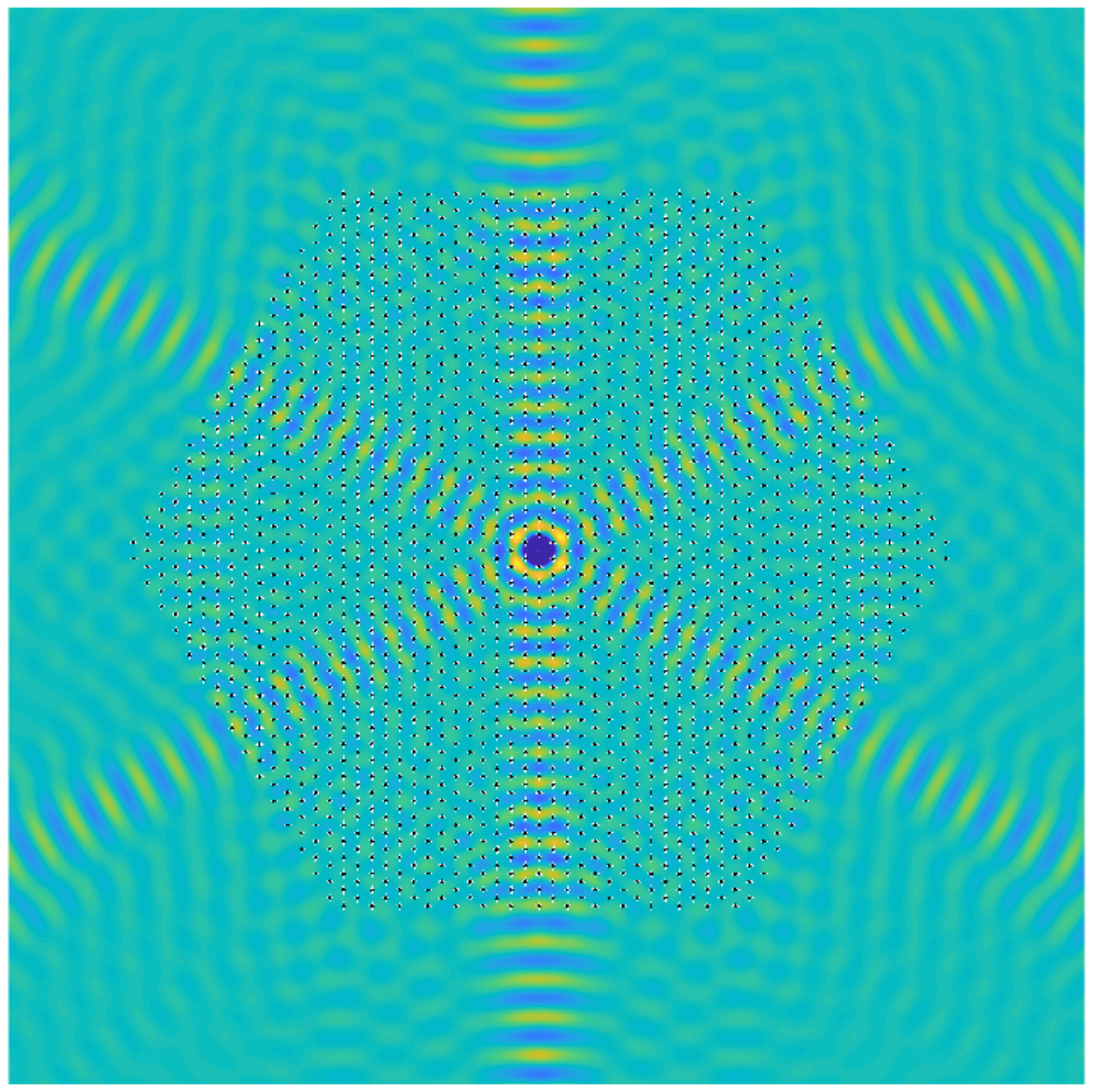}
               };
           }]{};
           \node[below, scale=4, black] at (-10.0,9.5) {$\displaystyle (c)$};
\end{scope}  

\begin{scope}[xshift=-8cm, yshift=46cm]
		\node[regular polygon, regular polygon sides=4,draw, inner sep=6.75cm,rotate=0,line width=0.0mm, white,
           path picture={
               \node[rotate=0] at (-0.5,-0.5){
                   \includegraphics[scale=1.5]{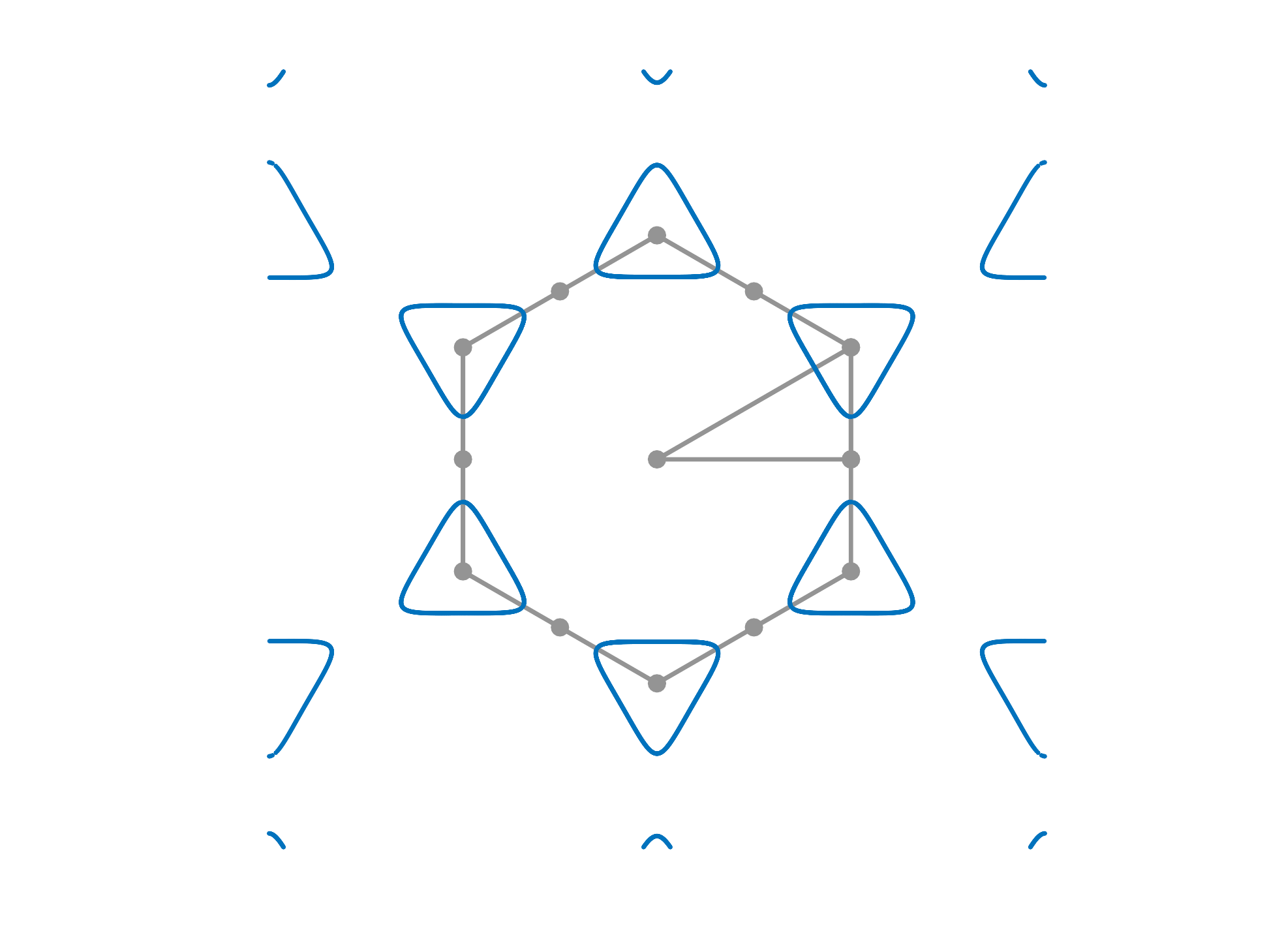}
               };
           }]{};
           \node[below,left,scale=4] at (0.0,-0.5) {$\displaystyle  \Gamma$}; 
			\node[below,right,scale=4] at (4.25,-0.5) {$\displaystyle  M$};
			\node[above,right,scale=4] at (4.25, 2.75) {$\displaystyle K$};
           \node[below, scale=4, black] at (-10.0,9.5) {$\displaystyle (d)$};
\end{scope}

\end{tikzpicture}
\caption{Comparisons between isofrequency contours computed from \eqref{EVPactualalgebraicDispBloch} case $2$ and our generalised Foldy solution \eqref{GeneralisedFoldySystem} case $3$. Panels $(a)$ and $(b)$ take the parameters from fig. \ref{fig:ThickerPlateFEMcomp},  the isofrequency contours and scattering simulations were calculated at $\Omega = 9$ ($\omega =  221.0287$ Hz) for a purely out-of-plane monopole incident source, setting  $\varpi_{\mathrm{inc}}=1$ within \eqref{ForcePOPinc}. Panels $(c)$ and $(d)$ are identical to  $(a)$ and $(b)$, other than setting $\boldsymbol{\alpha}_{1} = \cos(\frac{\pi}{6}) \textbf{e}_{x} + \sin(\frac{\pi}{6}) \textbf{e}_{y}$ and $\Omega = 7.5$ ($\omega =  184.1906$ Hz). The colour bar refers to the normalised real parts of the total out-of-plane displacement field of the plate within panels $(a)$ and $(c)$, where the location of beams is shown by the small black dots. The first and irreducible Brillouin zone are plotted in grey in $(b)$ and $(d)$.} 
\label{fig:ScatteringIsoFig}
\end{figure}

Figure \ref{fig:ScatteringIsoFig} compares the isofrequency contours (case $2$) to scattering simulations from the generalised Foldy solution (case $3$), we see that solutions from the closely related Green's functions \eqref{EVPactualalgebraicDispBloch} and \eqref{GeneralisedFoldySystem} show one and the same solution -  one over physical space and the other in reciprocal space. Additionally, when considering low frequency flexural modes propagating in the plate, case $2$ provides sufficient accuracy. In the following sections, we consider purely out-of-plane incident sources and operate at low frequencies such that case $2$ eigensolutions \eqref{EVPactualalgebraicDispBloch} or scattering solutions \eqref{GeneralisedFoldySystem} provide sufficient accuracy. 

\subsection{Multiple beams per primitive cell - symmetry  induced Dirac points?}
\begin{figure}[h]
\centering
\hspace*{2.5cm}
\begin{tikzpicture}[scale=0.4, transform shape]

\begin{scope}[xshift=10cm, yshift=48.0cm]
\node[regular polygon, regular polygon sides=4,draw, inner sep=9.0cm,rotate=0,line width=0.0mm, white,
           path picture={
               \node[rotate=0] at (0,0.5){
                   \includegraphics[scale=2.2]{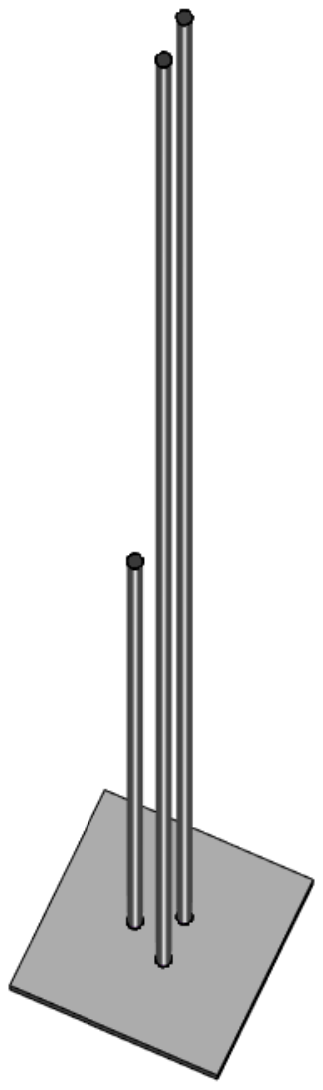}
               };
           }]{};
\node[below, scale=2, black] at (0.25,-12.0) {$\displaystyle (d)$};
\end{scope}

\begin{scope}[xshift=-4.5cm, yshift=50.5cm]
		\node[regular polygon, regular polygon sides=4,draw, inner sep=7.0cm,rotate=0,line width=0.0mm, white,
           path picture={
               \node[rotate=0] at (1,1){
                   \includegraphics[scale=1.25]{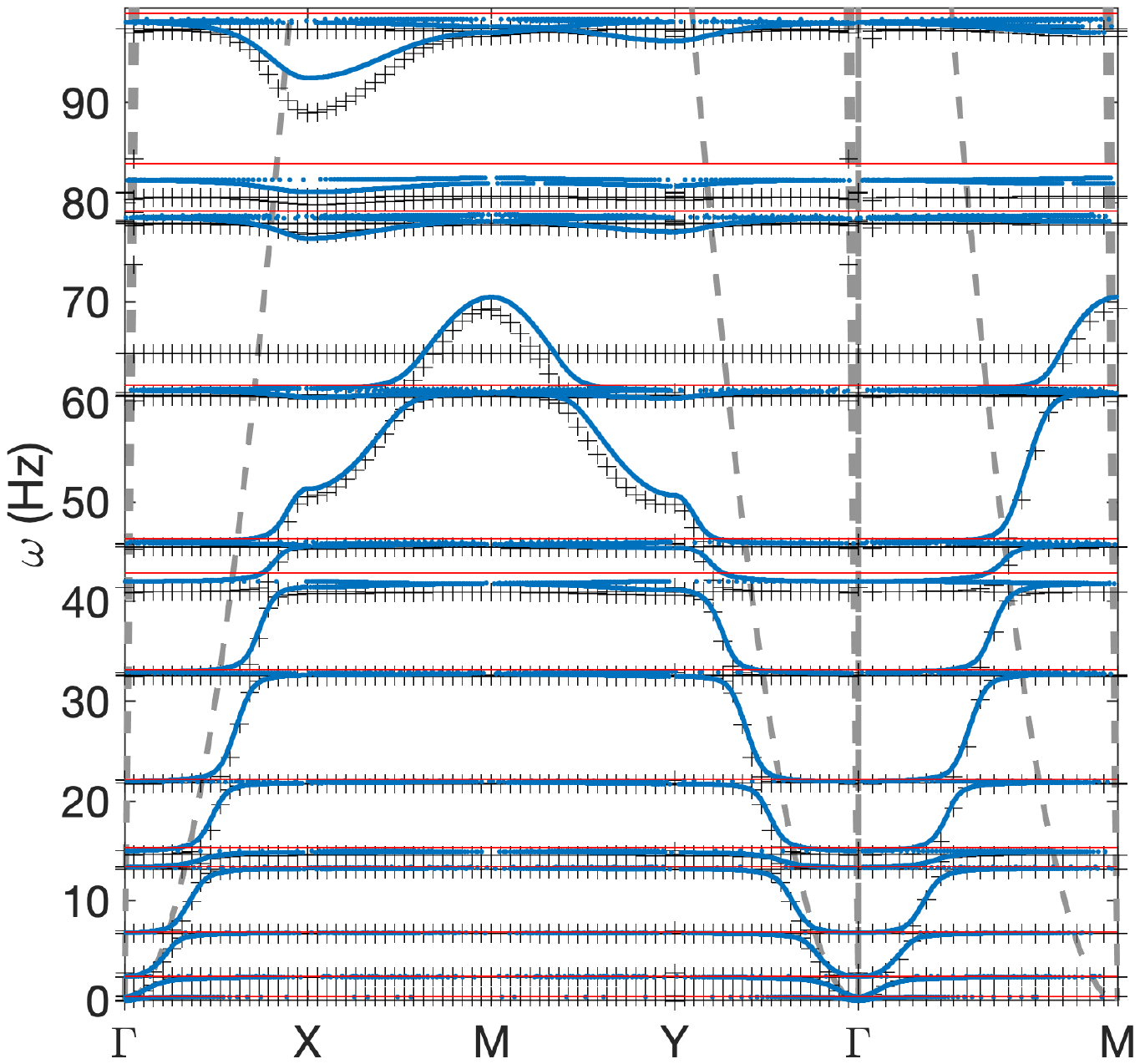}
               };
           }]{};
		\node[below, scale=2, black] at (-8,7.75) {$\displaystyle (a)$};         
\end{scope}  

\begin{scope}[xshift=-0.5cm, yshift=38cm,scale=0.5]
		\node[regular polygon, regular polygon sides=4,draw, inner sep=6cm,rotate=0,line width=0.0mm, white,
           path picture={
               \node[rotate=0] at (-0.5,0){
                   \includegraphics[scale=1.25]{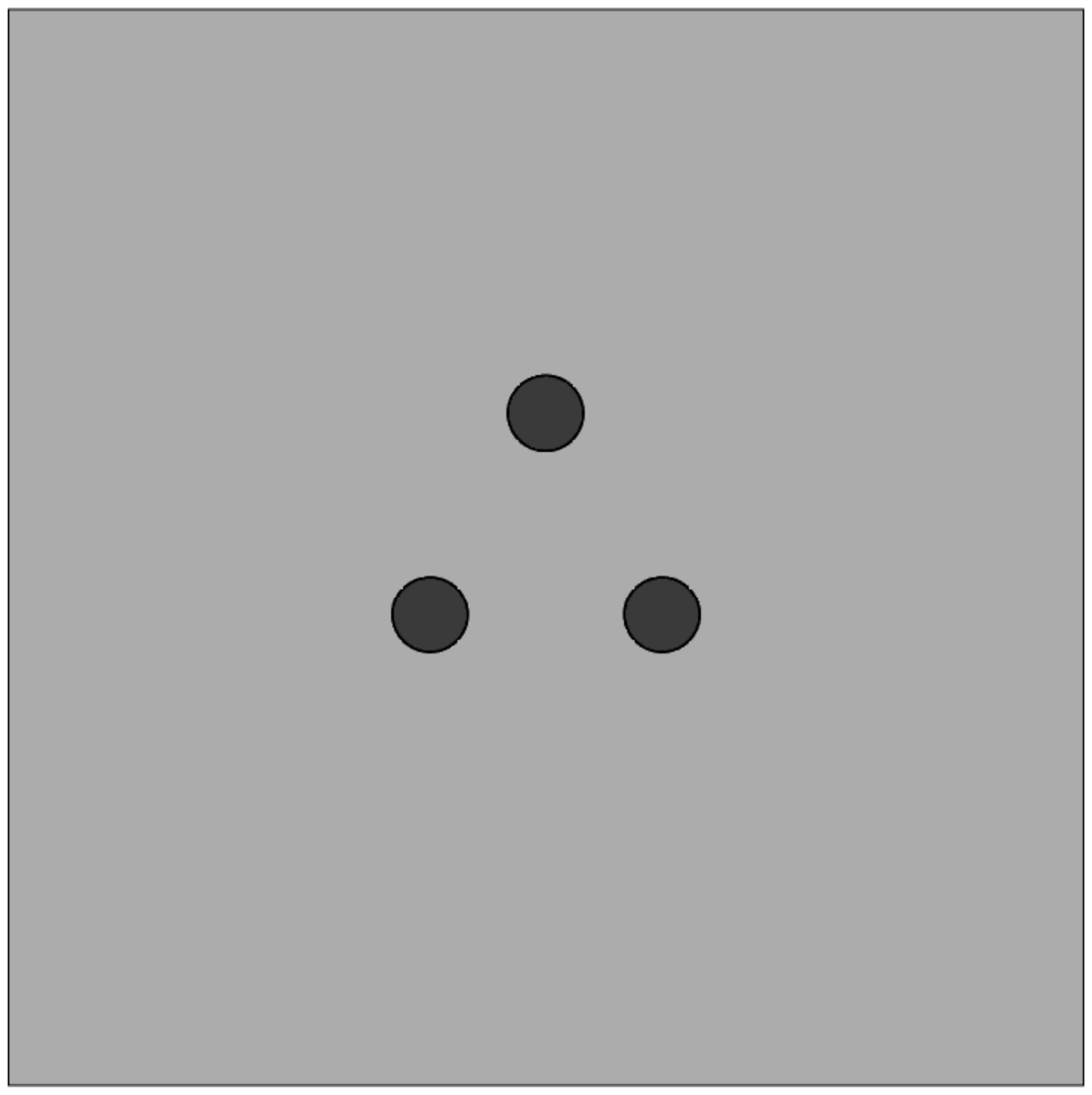}
               };
           }]{};
		\node[below, scale=4, black] at (-0.5,-6.95) {$\displaystyle (c)$};         
\end{scope}

\begin{scope}[xshift=-9.5cm, yshift=43.0cm,scale=1.5]
\node[regular polygon, regular polygon sides=4, draw, inner sep=0.5*6.28*10.0 pt,rotate=0] at (0 pt,-6.28*10.0*1.5 pt) {};
\draw[line width=0.5mm,gray,-] (0pt,-6.28*10.0*1.5 pt) -- (0+ 0.5*6.28*10*1.41 pt,-6.28*10.0*1.5 pt);
\draw[line width=0.5mm,gray,-] (0+ 0.5*6.28*10*1.41 pt,-6.28*10.0*1.5 pt) -- (0+ 0.5*6.28*10*1.41  pt, 0.5*6.28*10*1.41 -6.28*10.0*1.5 pt);
\draw[line width=0.5mm,gray,-] (0+ 0.5*6.28*10*1.41  pt, 0.5*6.28*10*1.41 -6.28*10.0*1.5 pt) -- (0pt,0.5*6.28*10*1.41 -6.28*10.0*1.5 pt);
\draw[line width=0.5mm,gray,-] (0pt,0.5*6.28*10*1.41 -6.28*10.0*1.5 pt) -- (0pt,-6.28*10.0*1.5 pt);
\draw[line width=0.5mm,gray,-] (0pt,-6.28*10.0*1.5 pt) -- (0+ 0.5*6.28*10*1.41  pt, 0.5*6.28*10*1.41 -6.28*10.0*1.5 pt);
\node[below,left,scale=1.75] at (0pt,-6.28*10.0*1.5 pt) {$\displaystyle  \Gamma$}; 
\node[below,right,scale=1.75] at (0+ 0.5*6.28*10*1.41 pt,-6.28*10.0*1.5 pt) {$\displaystyle  X$};
\node[above,right,scale=1.75] at (0+ 0.5*6.28*10*1.41  pt, 0.5*6.28*10*1.41 -6.28*10.0*1.5+10 pt) {$\displaystyle M$};
\node[above,left,scale=1.75] at (5pt,0.5*6.28*10*1.41 -6.28*10.0*1.5+10 pt) {$\displaystyle  Y$};
\node[regular polygon, circle, draw, inner sep=1.25pt,rotate=0,line width=0.5mm,shading=fill,outer color=gray,gray] at (0pt,-6.28*10.0*1.5 pt)  {};
\node[regular polygon, circle, draw, inner sep=1.25pt,rotate=0,line width=0.5mm,shading=fill,outer color=gray,gray] at (0+ 0.5*6.28*10*1.41 pt,-6.28*10.0*1.5 pt)  {};
\node[regular polygon, circle, draw, inner sep=1.25pt,rotate=0,line width=0.5mm,shading=fill,outer color=gray,gray] at (0+ 0.5*6.28*10*1.41  pt, 0.5*6.28*10*1.41 -6.28*10.0*1.5 pt)  {};
\node[regular polygon, circle, draw, inner sep=1.25pt,rotate=0,line width=0.5mm,shading=fill,outer color=gray,gray] at (0pt,0.5*6.28*10*1.41 -6.28*10.0*1.5 pt)  {};
\node[below, scale=1.33] at (0pt,-0.5*6.28*10*1.41 -6.28*10.0*1.5 pt) {$\displaystyle  (b)$};
\end{scope}

\end{tikzpicture}
\caption{Comparisons of the dispersion curves, in panel $(a)$, computed from the eigenvalue problem \eqref{EVPactualalgebraicDispBloch} case 2  (blue dots) and FEM computations (black crosses); here, the fundamental cell is shown in panels $(c)$ (top view) and $(d)$ (side view) where $h=0.05$ m and the dimensionless parameters (again scaled with $L$) are $\boldsymbol{\alpha}_{1} = \textbf{e}_{x}$, $\boldsymbol{\alpha}_{2} = \textbf{e}_{y}$,   $\textbf{X}_{1J} = 0.125\left[ \cos(\frac{2\pi (J-1)}{3} + \frac{\pi}{2}) \textbf{e}_{x} + \sin(\frac{2\pi (J-1)}{3} + \frac{\pi}{2})  \textbf{e}_{y}\right]$, $\hat{\ell}_{11} = 4.50$, $\hat{\ell}_{12} = \hat{\ell}_{13}  = 11.25$, $\epsilon_{1J} = 0.035$, for $J = 1,2,3$. Here, we plot $\omega = \omega(\boldsymbol{\kappa})$ throughout the irreducible Brillouin zone  $(b)$. The green and red lines correspond to the resonances from equations \eqref{CompressionalRes} (higher frequency) and \eqref{ClampedFreeRes} respectively.} 
\label{fig:SquareTopoArrange}
\end{figure}

Extending the eigenvalue problem \eqref{EVPactualalgebraicDispBloch} to consider any $P$ beams per primitive cell is straightforward, by modifying equations \eqref{WFourCoeffs}-\eqref{PsiFourCoeffs} to sum over the contribution from each beam in the cell. Additionally, one  considers \eqref{Disp_rTo0}, \eqref{GradDisp_rTo0} - \eqref{Psi_rTo0} in the limit as $r_{1J} \to 0$ for every $J=1, \ldots, P$; these asymptotic expressions remain unchanged, other than reinstating indices, since the $J$th object is assumed dominant as $r_{1J} \to 0$. Fig. \ref{fig:SquareTopoArrange} shows that our scheme, from \eqref{EVPactualalgebraicDispBloch} case $2$, is capable of determining the dispersion relations for systems containing multiple beams per cell - the comparison to the FEM dispersion curves is excellent beneath $50$Hz. Above this frequency observe the branches from \eqref{EVPactualalgebraicDispBloch} start to deviate from the correct values due to the asymptotics breaking down. 

\section{Deep subwavelength topologically protected states} \label{Sec:Topo}
Broken degeneracies produce states either classed as topologically trivial or non-trivial, depending on whether the initial degeneracy was induced trivially, solely because of the underlying lattice, or has symmetry protection due to non-trivial spacial symmetries within primitive cells. For instance, the topologically trivial edge states considered by Torrent \textit{et al}. \cite{torrent2013elastic} are induced from breaking a trivial degeneracy through the introduction of resonance. In this section, we design topologically non-trivial states arising from deliberate broken symmetries within primitive cells.

Chiral induced interfacial states can be created without the use of topology by breaking time-reversal-symmetry. Active methods are required in a phononic setting, for instance Carta \textit{et al}. \cite{carta2020chiral,carta2020one} consider gyro-topped beams affixed to elastic plates and demonstrate robust one-way interfacial states between media with counter-rotating chiral flux. Wang \textit{et al}. \cite{wang2015topological} melds the ideas surrounding topology and actively inducing chirality to consider a symmetry induced degeneracy actively broken, by gyricity, generating robust one-way edgestates. Within photonics, externally applied magnetic fields allow time-reversal-symmetry to be passively broken to generate states with a single chirality, as in \cite{wang2008reflection, wang2009observation,skirlo2014multimode,skirlo2015experimental}.   

Our topologically non-trivial states exist passively and are not restricted to one-way propagation. Our symmetry protected Dirac cones coalesce with the flexural resonances of the constituent beams, and hence can be tuned to a required frequency by simply changing the parameters of the beams; moreover, they can be tuned into a deep sub-wavelength regime, to frequencies far lower than traditional Bragg scattering \cite{zhang2020dirac}. However, the near resonant degeneracies lie on flat bands corresponding to slow sound within the wave spectrum; once these degeneracies are broken they result in narrow-band energy gaps which still generate topologically non-trivial states, additional graded rainbow \cite{Tsakmakidis2007} effects could be applied to increase the effective band gap of the media. 

\begin{figure}[h!]
\centering
\hspace*{0.0cm}
\begin{tikzpicture}[scale=0.25, transform shape]

\begin{scope}[xshift=-19cm]

\begin{scope}[xshift=14.5cm, yshift=22cm,scale=3]
	\node[draw, line width=0.5mm, minimum size=4cm, regular polygon,  regular polygon sides=4,fill = myGrey] (polygon) {};
	
	\foreach \x in {2,3}{
		\node[regular polygon, circle, draw = myRED, inner sep=2.5,rotate=0,line width=0.5mm, fill = myRED,xshift = 30] at (polygon.corner \x)  {};  
		\node[text = white,xshift = 30] at (polygon.corner \x) {$\displaystyle  \textbf{+}$}; 
		}

	\foreach \x in {1,4}{
		\node[regular polygon, circle, draw = myBLUE, inner sep=2.5pt,rotate=0,line width=0.5mm, fill = myBLUE,xshift = -30] at (polygon.corner \x)  {};  
		\node[text = white,xshift = -30] at (polygon.corner \x) {\large $\displaystyle  -$}; 
		}


  \foreach \x in {1,3}{
       \draw [myPurple,dashed, shorten <=-0.75cm,shorten >=-0.75cm](polygon.center) -- (polygon.side \x);}
        \node[above, right, text = myPurple,yshift=20] at (polygon.side 1) {\huge $\, \sigma_{v}$};


\end{scope}

\begin{scope}[xshift=14.5cm, yshift=3cm,scale=3]
	\node[draw, line width=0.5mm, minimum size=4cm, regular polygon,  regular polygon sides=4,fill = myYellow] (polygon) {};
	
	\foreach \x in {1,4}{
		\node[regular polygon, circle, draw = myRED, inner sep=2.5,rotate=0,line width=0.5mm, fill = myRED,xshift = -30] at (polygon.corner \x)  {};  
		\node[text = white,xshift = -30] at (polygon.corner \x) {$\displaystyle  \textbf{+}$}; 
		}

	\foreach \x in {2,3}{
		\node[regular polygon, circle, draw = myBLUE, inner sep=2.5pt,rotate=0,line width=0.5mm, fill = myBLUE,xshift = 30] at (polygon.corner \x)  {};  
		\node[text = white,xshift = 30] at (polygon.corner \x) {\large $\displaystyle  -$}; 
		}


  \foreach \x in {1,3}{
       \draw [myPurple,dashed, shorten <=-0.75cm,shorten >=-0.75cm](polygon.center) -- (polygon.side \x);}
        \node[above, right, text = myPurple,yshift=20] at (polygon.side 1) {\huge $\, \sigma_{v}$}; 
        \node[below, text = black,yshift=-35] at (polygon.side 3) {\Large $\displaystyle (a)$}; 
        \node[below, text = black,yshift=-35,xshift=5.5cm] at (polygon.side 3) {\Large $\displaystyle \, \, (b)$};


\end{scope}

\begin{scope}[xshift=31cm, yshift=16.75cm,scale=3]
	\node[draw, line width=0.5mm, minimum size=4cm, regular polygon,  regular polygon sides=4,fill = myGrey] (polygon) {};
	
	\foreach \x in {2}{
		\node[regular polygon, circle, draw = myRED, inner sep=2.5,rotate=0,line width=0.5mm, fill = myRED,xshift = 30] at (polygon.corner \x)  {};  
		\node[text = white,xshift = 30] at (polygon.corner \x) {$\displaystyle  \textbf{+}$}; 
		}

	\foreach \x in {1}{
		\node[regular polygon, circle, draw = myBLUE, inner sep=2.5pt,rotate=0,line width=0.5mm, fill = myBLUE,xshift = -30] at (polygon.corner \x)  {};  
		\node[text = white,xshift = -30] at (polygon.corner \x) {\large $\displaystyle  -$}; 
		}

\end{scope}

\begin{scope}[xshift=31cm, yshift=8.265cm,scale=3]
	\node[draw, line width=0.5mm, minimum size=4cm, regular polygon,  regular polygon sides=4,fill = myYellow] (polygon) {};
	
	\foreach \x in {4}{
		\node[regular polygon, circle, draw = myRED, inner sep=2.5,rotate=0,line width=0.5mm, fill = myRED,xshift = -30] at (polygon.corner \x)  {};  
		\node[text = white,xshift = -30] at (polygon.corner \x) {$\displaystyle  \textbf{+}$}; 
		}

	\foreach \x in {3}{
		\node[regular polygon, circle, draw = myBLUE, inner sep=2.5pt,rotate=0,line width=0.5mm, fill = myBLUE,xshift = 30] at (polygon.corner \x)  {};  
		\node[text = white,xshift = 30] at (polygon.corner \x) {\large $\displaystyle  -$}; 
		}

	\foreach \x in {1}{
		\node[regular polygon, circle, draw = myPurple, inner sep=2.5,rotate=0,line width=0.5mm, fill = myPurple,xshift = -30] at (polygon.corner \x)  {};  
		\node[text = white,xshift = -30] at (polygon.corner \x) {$\displaystyle  \boldsymbol{\mp}$}; 
		}

	\foreach \x in {2}{
		\node[regular polygon, circle, draw = myPurple, inner sep=2.5pt,rotate=0,line width=0.5mm, fill = myPurple,xshift = 30] at (polygon.corner \x)  {};  
		\node[text = white,xshift = 30] at (polygon.corner \x) {$\displaystyle  \boldsymbol{\pm}$}; 
		}

\end{scope}  

\end{scope}

\begin{scope}[xshift=19cm]
\begin{scope}[xshift=14.5cm, yshift=22cm,scale=3]
	\node[draw, line width=0.5mm, minimum size=4cm, regular polygon,  regular polygon sides=6, fill = myGrey] (polygon) {};
	
	\foreach \x in {1,3,...,5}{
		\node[regular polygon, circle, draw = myRED, inner sep=2.5pt,rotate=0,line width=0.5mm, fill = myRED] at (polygon.corner \x)  {};  
		\node[text = white] at (polygon.corner \x) {$\displaystyle  \textbf{+}$}; 
		}

	\foreach \x in {2,4,...,6}{
		\node[regular polygon, circle, draw = myBLUE, inner sep=2.5pt,rotate=0,line width=0.5mm, fill = myBLUE] at (polygon.corner \x)  {};  
		\node[text = white] at (polygon.corner \x) {\large $\displaystyle  -$}; 
		}


    \foreach \x in {1,2,...,6}{
        \draw [myPurple,dashed, shorten <=-0.75cm,shorten >=-0.75cm](polygon.center) -- (polygon.side \x);}
        \node[above, right, text = myPurple] at (polygon.side 6) {\huge $\quad  \sigma_{v}$};

\end{scope}

\begin{scope}[xshift=14.5cm, yshift=3cm,scale=3]
	\node[draw, line width=0.5mm, minimum size=4cm, regular polygon,  regular polygon sides=6, fill = myYellow] (polygon) {};
	
	\foreach \x in {2,4,...,6}{
		\node[regular polygon, circle, draw = myRED, inner sep=2.5pt,rotate=0,line width=0.5mm, fill = myRED] at (polygon.corner \x)  {};  
		\node[text = white] at (polygon.corner \x) {$\displaystyle  \textbf{+}$}; 
		}

	\foreach \x in {1,3,...,5}{
		\node[regular polygon, circle, draw = myBLUE, inner sep=2.5pt,rotate=0,line width=0.5mm, fill = myBLUE] at (polygon.corner \x)  {};  
		\node[text = white] at (polygon.corner \x) {\large $\displaystyle  -$}; 
		}


    \foreach \x in {1,2,...,6}{
        \draw [myPurple,dashed, shorten <=-0.75cm,shorten >=-0.75cm](polygon.center) -- (polygon.side \x);}
        \node[above, right, text = myPurple] at (polygon.side 6) {\huge $\quad  \sigma_{v}$}; 
        \node[below, text = black,yshift=-28] at (polygon.side 4) {\Large $\displaystyle (c)$}; 
        \node[below, text = black,yshift=-28,xshift=5.83cm] at (polygon.side 4) {\large $\displaystyle \, \, (d)$};


\end{scope}

\begin{scope}[xshift=32cm, yshift=17.75cm,scale=3]
	\node[draw, line width=0.5mm, minimum size=4cm, regular polygon,  regular polygon sides=6, fill = myGrey] (polygon) {};
	
	\foreach \x in {1,3}{
		\node[regular polygon, circle, draw = myRED, inner sep=2.5pt,rotate=0,line width=0.5mm, fill = myRED] at (polygon.corner \x)  {};  
		\node[text = white] at (polygon.corner \x) {$\displaystyle  \textbf{+}$}; 
		}

	\foreach \x in {2,6}{
		\node[regular polygon, circle, draw = myBLUE, inner sep=2.5pt,rotate=0,line width=0.5mm, fill = myBLUE] at (polygon.corner \x)  {};  
		\node[text = white] at (polygon.corner \x) {\large $\displaystyle  -$}; 
		}	
    
\end{scope}

\begin{scope}[xshift=32cm, yshift=7.35cm,scale=3]
	\node[draw, line width=0.5mm, minimum size=4cm, regular polygon,  regular polygon sides=6,fill = myYellow] (polygon) {};
	
	\foreach \x in {4,6}{
		\node[regular polygon, circle, draw = myRED, inner sep=2.5pt,rotate=0,line width=0.5mm, fill = myRED] at (polygon.corner \x)  {};  
		\node[text = white] at (polygon.corner \x) {$\displaystyle  \textbf{+}$}; 
		}
		
	\foreach \x in {1}{
		\node[regular polygon, circle, draw = myPurple, inner sep=2.5pt,rotate=0,line width=0.5mm, fill = myPurple] at (polygon.corner \x)  {};  
		\node[text = white] at (polygon.corner \x) {$\displaystyle  \boldsymbol{\pm}$}; 
		}

	\foreach \x in {2}{
		\node[regular polygon, circle, draw = myPurple, inner sep=2.5pt,rotate=0,line width=0.5mm, fill = myPurple] at (polygon.corner \x)  {};  
		\node[text = white] at (polygon.corner \x) {$\displaystyle  \boldsymbol{\mp}$}; 
		}

	\foreach \x in {3,5}{
		\node[regular polygon, circle, draw = myBLUE, inner sep=2.5pt,rotate=0,line width=0.5mm, fill = myBLUE] at (polygon.corner \x)  {};  
		\node[text = white] at (polygon.corner \x) {\large $\displaystyle  -$}; 
		}	
    
\end{scope}

\end{scope}

\end{tikzpicture}
\caption{The associated Berry curvature from $\sigma_{v}$ symmetry breaking for chiral-mirrored pairs of primitive cells \cite{makwana19a,makwana2019topological}. Here, grey and yellow cells represent any chiral-mirrored primitive cell pairs, derived from perturbations which break the $\sigma_{v}$ symmetries contained within the unperturbed cells; the vertical $\sigma_{v}$ symmetries, of which, are plotted by the purple dashed lines for square and hexagonal lattices in $(a)$ and $(c)$ respectively. Exemplar sign Berry curvatures, for the bands bounding any symmetry-protected gapped degeneracy (by $\sigma_{v}$ breaking), are denoted\protect\circledP  and\protect\circledM referring to locations with opposite (non-zero) sign in Berry curvature. Panels $(b)$ and $(d)$ show how the Berry curvature interacts when two such chiral-mirrored pairs are joined at an interface, where\protect\circledPM  and\protect\circledMP denote locations where\protect\circledP and\protect\circledM overlap, creating the topologically protected interfacial states known as zero line modes.} 
\label{fig:BerryCurvature}
\end{figure}
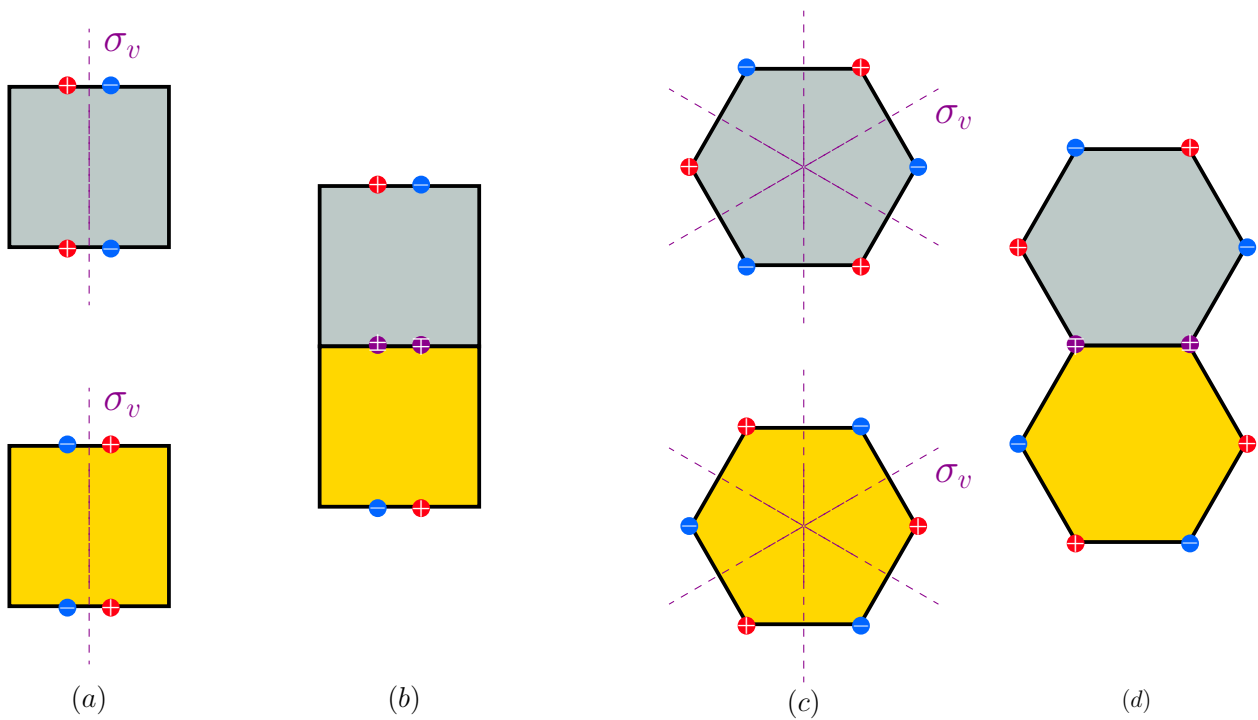

The polynomial eigenvalue problem \eqref{EVPdispBloch} is formed from Hermitian matrices, subsequently \cite{laforge2021acoustic} we expect bands to vary continuously unless approaching a degeneracy. Our focus is on vertical $\sigma_{v}$ symmetries for triangular arrangements of beams in both square  and hexagonal primitive cells, which induce symmetry protected degeneracies along the $XM$ path in square lattices \cite{makwana2019topological} and are guaranteed at the $KK'$ high symmetry points in hexagonal lattices \cite{makwana2018geometrically,makwana2020hybrid} - provided the structure in physical space has the $\sigma_{v}$ symmetries shown in fig. \ref{fig:BerryCurvature}. Breaking the $\sigma_{v}$ symmetries gaps such degeneracies and leads to the QVHE; where pairs of time-reversal-symmetry related valleys or extrema bounding band gaps have locally quadratic curvature, and correspond to topological vortex states with opposite chirality \cite{lu2016valley} in the bulk, which generate topologically non-trivial modes within the bulk band gap, as demonstrated pictorially in fig. 2 of \cite{makwana2020hybrid}. These modes correspond to interfacial \cite{wiltshaw2020asymptotic, makwana2018designing, he2015emergence, carta2020chiral} and edge \cite{ochiai2009photonic,wang2015topological,halperin1982quantized,wen1991gapless}  states with strong localisation and robustness due to the QVHE. 

Topological quantities such as the Chern number or Berry curvature \cite{makwana19a,palmer2021revealing}, characterise the direction of such vortex states and can be used to identify a topological phase transition occurring between the lower and upper valleys of a broken degeneracy \cite{palmer2021revealing}. Suitably engineered chiral-mirrored pairs, as shown in fig.\ref{fig:BerryCurvature}, also undergo a topological phase transition and have chiral-mirrored Berry curvatures and hence chiral mirrored vortex states. Therefore, placing two different bulk media - each of which derived from cells of one chiral pair - together results in interfaces in which Berry curvatures of opposite sign overlap; subsequently, topologically protected interfacial states emerge \cite{xiao_valley-contrasting_2007,ochiai2012photonic}, these valley Hall edge states are known as Zero Line Modes (ZLMs).  

These chiral states exist because of the topological properties derived from the bulk \cite{ochiai2014broken,yoshida2021bulk}. One could consider the bulk-boundary (or bulk-edge) correspondence \cite{hatsugai1993chern,qian2018theory} applicable to interfaces or edges, between two distinct bulk media, provided \cite{makwana19a} opposite sign topological quantities are not projected onto the same point (and hence the bulk-boundary correspondence cannot account for the creation of ZLMs as given in fig. \ref{fig:BerryCurvature}); in such cases, the bulk-boundary correspondence correlates non-zero Chern numbers (and hence Berry curvature) to the existence of topologically protected interfacial and edge states. 

The works of Ochiai  \cite{ochiai2010topological,ochiai2014broken,ochiai2021bulk} \textit{et al.} \cite{ochiai2009photonic} and Yao \textit{et al.} \cite{yao2009edge} remark that edges, between a media with topologically induced chiral states and the free space, will inherit the non-trivial topological properties of the topological bulk media. Provided modes cannot propagate into the surrounding free-space, they can be localised by one-sided chiral flux \cite{ochiai2014broken} and  give rise to topologically protected edgestates by the bulk-edge correspondence - when non-zero Berry curvature, within primitive cells derived from a sole chirality in fig. \ref{fig:BerryCurvature} $(a)$ or $(c)$, meets the free space. These edgestates are shown to be sufficiently robust to navigate corners, unless corner states exist which readily shed energy into the free space. 

Within a photonic setting \cite{ochiai2009photonic,ochiai2014broken}, physical intuition of the surrounding free space is important regarding the creation of edgemodes. For instance in \cite{ochiai2014broken}, the case of photons in air was considered and operating above the light line to induce leaky waves was necessary to confine energy propagation along edges, since energy of the photon propagating in the crystal was higher than the ``vacuum level'' in the free space (air). Otherwise, only the conservation \cite{joannopoulos_photonic_2008} of $\kappa_{\parallel}$, denoting the component of the wavevector parallel to the surface of the crystal, is crucial to determine the group velocity of waves conserved at such an edge, and whether propagation is allowed within the free space - as in fig. \ref{fig:ScatteringIsoFig} (or in a photonic setting \cite{yves2017crystalline, ochiai2009photonic}). Here, we operate at the subwavelength scale far beneath the analogous light lines and expect the near resonant behaviour responsible for these subwavelength states to confine energy to the edge of the crystal, as in \cite{wang2019topological}. We investigate these edgemodes in a similar fashion to Torrent \textit{et al}. \cite{torrent2013elastic}, where we consider ribbon strips containing edges of interest but confirm the confinement of states along the $\Omega = \Omega(\kappa_{\parallel})$ branches by visualising the edgemodes.

The creation of these topologically nontrivial states requires simple arguments based upon group theory \cite{dresselhaus_group_2008,atkins2011molecular} in a periodic setting \cite{heine_group_nodate,sakoda2004optical}, in which we can infer which bands are symmetry induced and hence which degeneracies have non-trivial symmetry protection. We denote $G_{\boldsymbol{\kappa}}$ to be the point group symmetry\footnote{For a formal definition of $G_{\boldsymbol{\kappa}}$ within hexagonal or square lattices refer to  \cite{makwana2018geometrically} or \cite{laforge2021acoustic} respectively.} of  a cell in reciprocal space at $\boldsymbol{\kappa}$, and remark that any eigenfunction can be used for an irreducible representation of $G_{\boldsymbol{\kappa}}$ \cite{sakoda2004optical}. A fitting choice of eigenfunction would be the eigenmode corresponding to $\boldsymbol{\kappa}$, i.e. the eigenstate \cite{makwana2019topological} corresponding to fixed Bloch momentum vector $\boldsymbol{\kappa}$.

As with dispersion, we expect our eigenmodes to continuously vary unless we approach dispersionless crossings; classifying eigenmodes by symmetry allows one to infer which bands are symmetry induced and which bands are allowed to cross, and hence form degeneracies. Typically this is done at the high symmetry points, after all, only  $G_{\Gamma}$ has the full point group symmetry of the lattice and $G_{\boldsymbol{\kappa} \ne \textbf{0}} \le G_{\Gamma}$ - that is to say $G_{\boldsymbol{\kappa} \ne \textbf{0}}$ is a subgroup of $G_{\Gamma}$. However, one can apply compatibility relations (see fig. $3.7$ and table $3.10$ in \cite{sakoda2004optical} for hexagonal lattices and fig. $34$ and table $26$ in \cite{heine_group_nodate} for square lattices) to relate irreducible relations for $\boldsymbol{\kappa}$ along $XM$ or $\Gamma K$, for instance, to those at the high symmetry points in $\boldsymbol{\kappa}$ space and vice versa.

Therefore, symmetry induced degeneracies can be identified by analysing the parity of eigenmodes immediately surrounding such a crossing along the irreducible Brillouin zone; since eigenvalues are not degenerate away from any crossing, one obtains information about the parity of branches and whether the dispersionless crossings are indeed symmetry protected and hence simple to gap. 
A telling sign that bands are symmetry induced is that the irreducible representations match the basis functions of the irreducible representations within the appropriate character tables; we proceed to design structured media in which we expect symmetry protected degeneracies to occur and confirm this behaviour by comparing the eigenmodes in the vicinity of such a degeneracy to the character tables. Subsequently, by simply breaking the symmetries within the cell, we gap such degeneracies and create topologically non-trivial states; these subwavelength states will be topologically protected, the robustness of which will be demonstrated in a number of scattering simulations. 

\subsection{Robust deep subwavelength topological waveguides within square arrays} \label{TopoSquareLattice}

\begin{figure}[h!]
\centering
\hspace*{-0.5cm}
\begin{tikzpicture}[scale=0.4, transform shape]

\begin{scope}[xshift=2cm, yshift=50.5cm]
		\node[regular polygon, regular polygon sides=4,draw, inner sep=7.0cm,rotate=0,line width=0.0mm, white,
           path picture={
               \node[rotate=0] at (1,1){
                   \includegraphics[scale=1.25]{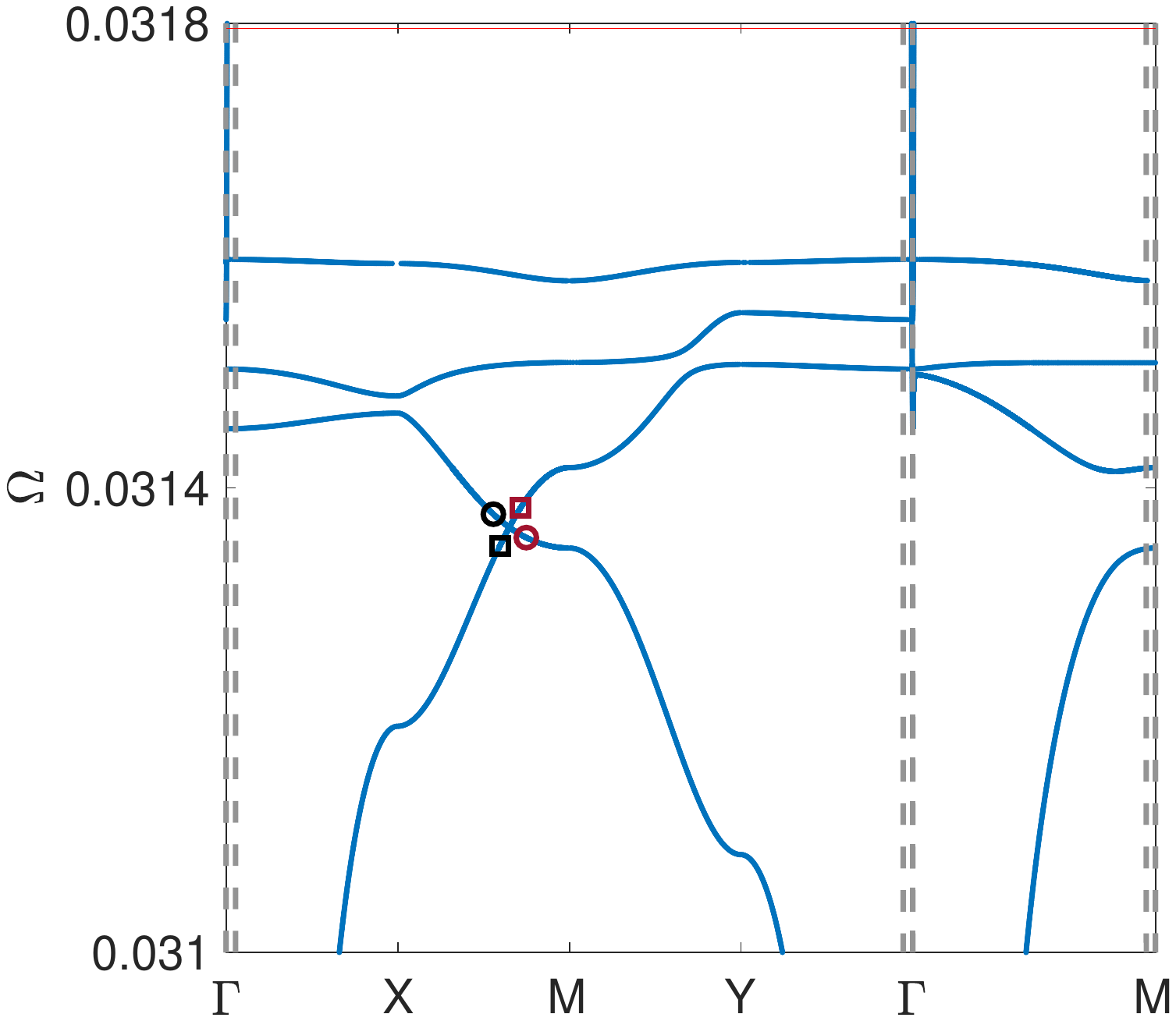}
               };
           }]{};
		\node[below, scale=2, black] at (1,-7.5) {$\displaystyle (a)$};           
\end{scope}

\begin{scope}[xshift=17cm, yshift=56cm,scale=0.45]
		\node[regular polygon, regular polygon sides=4,draw, inner sep=6.5cm,rotate=0,line width=0.0mm, white,
           path picture={
               \node[rotate=0] at (-0.35,-0.25){
                   \includegraphics[scale=1.4]{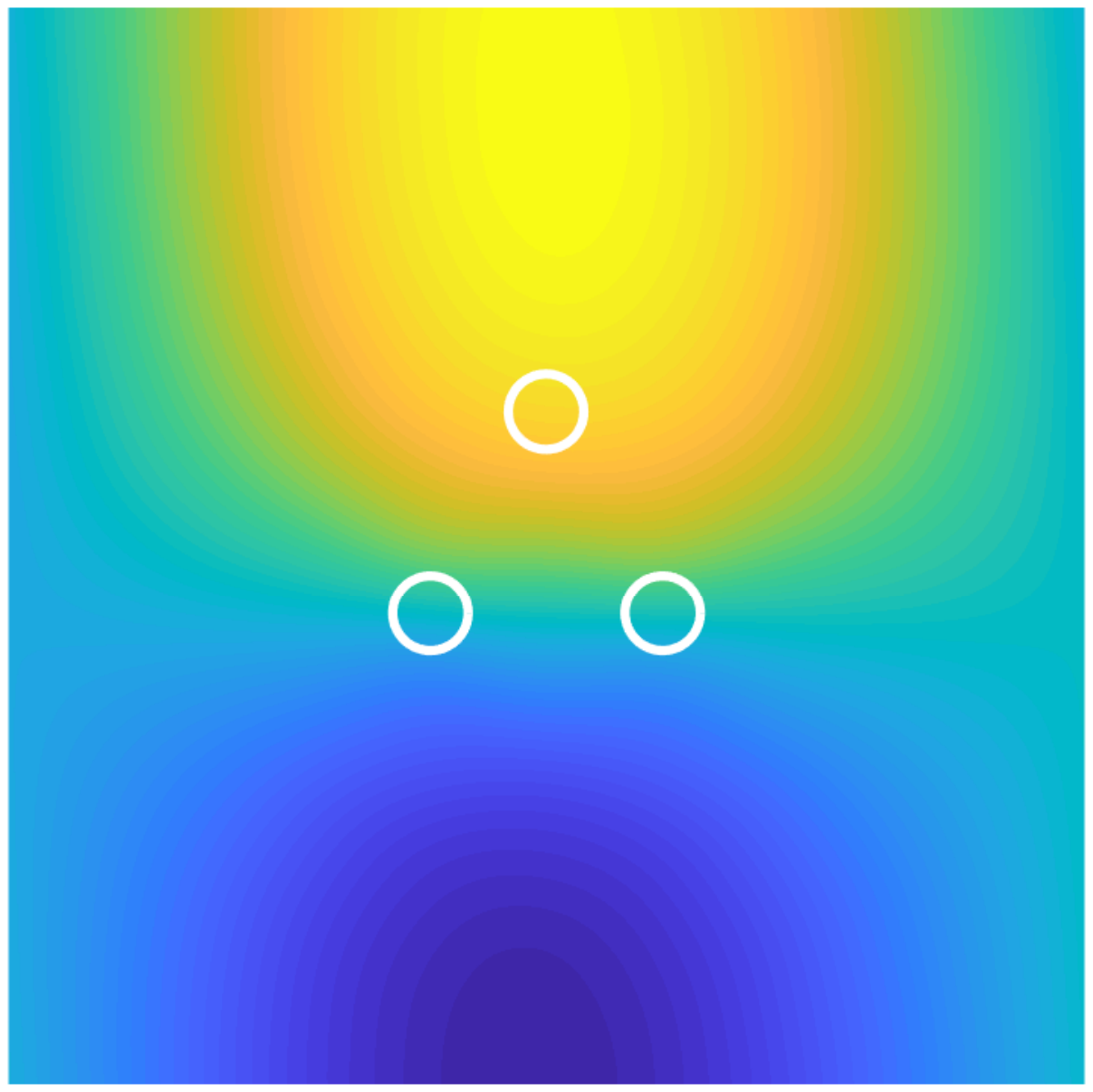}
               };
           }]{};
		\node[below, scale=4, black] at (-10,8.75) {$\displaystyle (b)$};           
\end{scope}

\begin{scope}[xshift=17cm, yshift=47.75cm,scale=0.45]
	\node[regular polygon, regular polygon sides=4,draw, inner sep=6.5cm,rotate=0,line width=0.0mm, white,
           path picture={
               \node[rotate=0] at (-0.35,-0.25){
                   \includegraphics[scale=1.4]{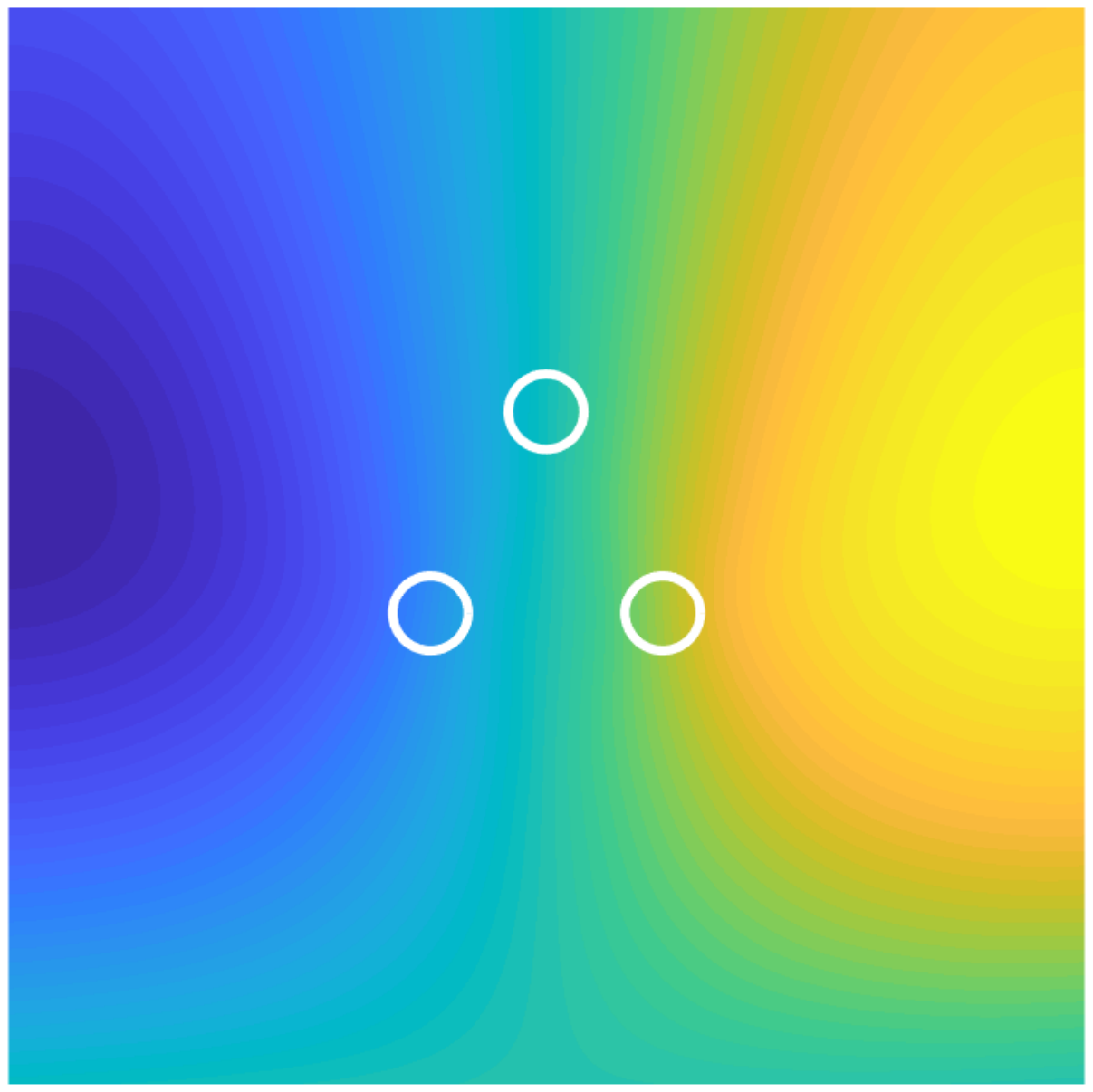}
               };
           }]{};
		\node[below, scale=4, black] at (-10,8.75) {$\displaystyle (c)$};          
\end{scope}  

\begin{scope}[xshift=25.25cm, yshift=56cm,scale=0.45]
	\node[regular polygon, regular polygon sides=4,draw, inner sep=6.5cm,rotate=0,line width=0.0mm, white,
           path picture={
               \node[rotate=0] at (-0.35,-0.25){
                   \includegraphics[scale=1.4]{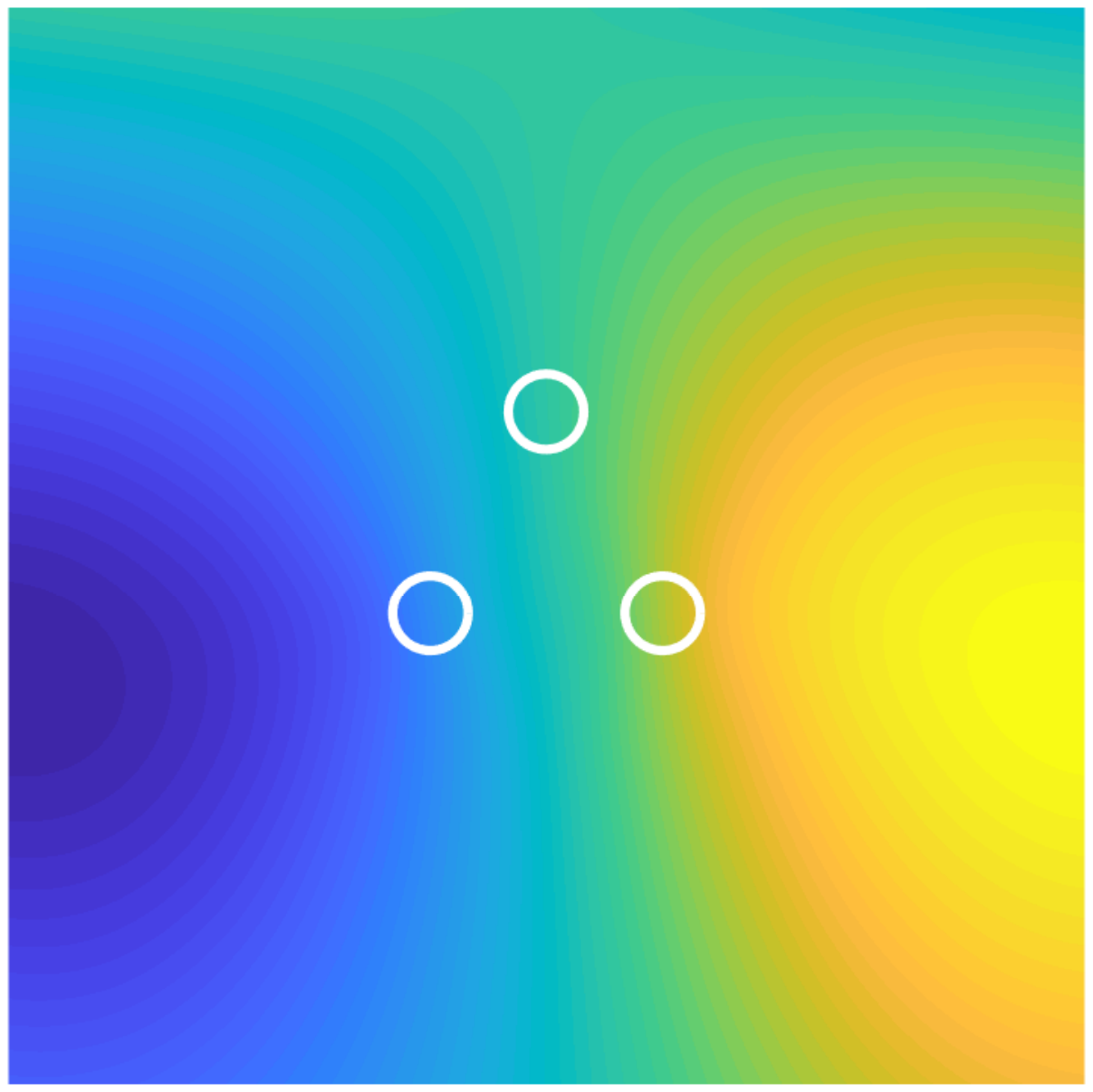}
               };
           }]{};
		\node[below, scale=4, black] at (10,8.75) {$\displaystyle (d)$};             
\end{scope}

\begin{scope}[xshift=25.25cm, yshift=47.75cm,scale=0.45]
		\node[regular polygon, regular polygon sides=4,draw, inner sep=6.5cm,rotate=0,line width=0.0mm, white,
           path picture={
               \node[rotate=0] at (-0.35,-0.25){
                   \includegraphics[scale=1.4]{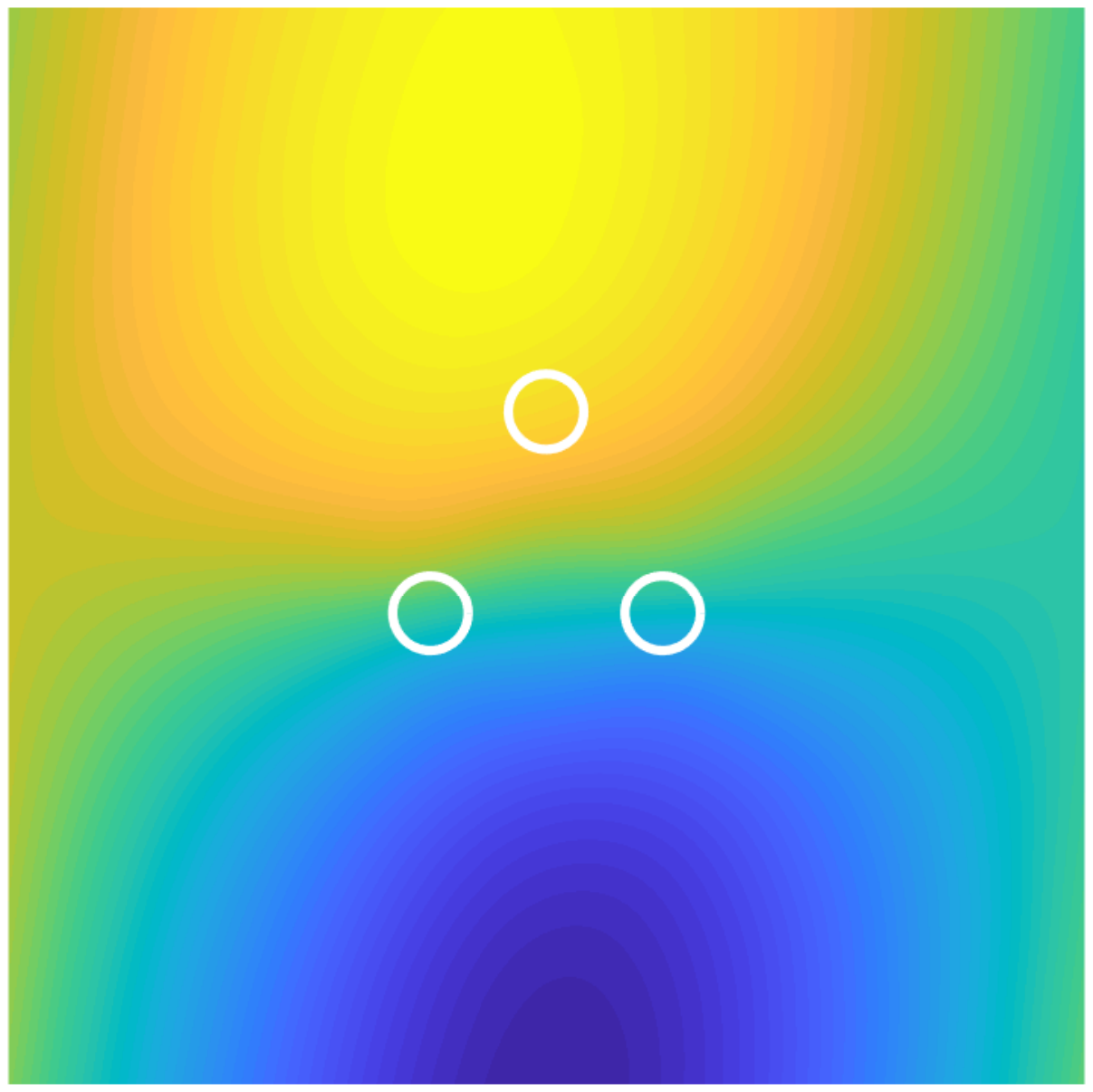}
               };
           }]{};
		\node[below, scale=4, black] at (10,8.75) {$\displaystyle (e)$};                  
\end{scope}  

\end{tikzpicture}
\caption{The Dirac point, in panel $(a)$, observed by zooming in about the first flexural resonance from figure \ref{fig:SquareTopoArrange} $(a)$. Here we plot the dimensionless dispersion curves $\Omega = \Omega( \boldsymbol{\kappa})$ from our eigenvalue problem \eqref{EVPactualalgebraicDispBloch}, the red line corresponds to the lowest flexural resonance satisfying equation \eqref{ClampedFreeRes}. The dimensional frequency of the Dirac point in $(a)$ is $\omega = 0.38516$ Hz, the FEM computed result is $\omega = 0.38127$ Hz hence our results agree with a relative error of $1.02 \% \,$. Panels $(b)$, $(c)$, $(d)$ and $(e)$ show the eigenmodes at the respective points $\boldsymbol{\bigcirc}$, $	\boldsymbol{\square}$, $\color{myRed} \boldsymbol{\square}$ and $\color{myRed} \boldsymbol{\bigcirc}$ from the bands in panel $(a)$, where the real part of the out-of-plane displacement fields are plotted in $(b)-(e)$.} 
\label{fig:SquareTopoArrangeDirac}
\end{figure}

Consider the square primitive cell in fig. \ref{fig:SquareTopoArrange} $(c)$ which has one vertical $\sigma_{v}$ reflectional symmetry and hence the required symmetries within fig. \ref{fig:BerryCurvature} (a). We denote $\boldsymbol{\kappa}_{XM}$ to be any $\boldsymbol{\kappa}$ along $XM$, and note $\sigma_{v}$ allows accidental degeneracies to occur for some $\boldsymbol{\kappa}_{XM}$  \cite{makwana19a} . Zooming in on the first resonance of figure \ref{fig:SquareTopoArrange}$(a)$, as shown in figure \ref{fig:SquareTopoArrangeDirac}$(a)$, we see such a dispersionless crossing. In \cite{makwana2019topological}, it was shown that the irreducible representations along $\boldsymbol{\kappa}_{XM}$ are compatible with those at $G_{X}$ and hence \cite{sakoda2004optical} the assignment of symmetry for $G_{X}$ is deduced by looking at any $G_{\boldsymbol{\kappa}_{XM}}$, and vice versa. 

Consider the eigenmodes in fig. \ref{fig:SquareTopoArrangeDirac} $(b)$, $(c)$, $(d)$ and $(e)$ corresponding to fig \ref{fig:SquareTopoArrangeDirac} $(a)$ points $\boldsymbol{\bigcirc}$, $\boldsymbol{\square}$, $\color{myRed} \boldsymbol{\square}$ and $\color{myRed} \boldsymbol{\bigcirc}$. The states here match the linear basis functions of the irreducible representations in table \ref{table:C2VCharacter}; the eigenmodes $\boldsymbol{\bigcirc}$ and $\color{myRed} \boldsymbol{\bigcirc}$ match the $y$ or linear $B_{2}$ basis, additionally $\boldsymbol{\square}$ and $\color{myRed} \boldsymbol{\square}$ match the $x$ or linear $B_{1}$ basis. Therefore, applying the above compatibility condition, we do indeed deduce $G_{X} = C_{2v}$. Moreover, since $G_{X} \le G_{\Gamma}$, then $G_{\Gamma} = C_{2v}$ is immediately confirmed. Further application of the compatibility relations along $\boldsymbol{\kappa}_{XM}$ ensures the band following $\boldsymbol{\square}$ and $\color{myRed} \boldsymbol{\square}$ will have irreducible representations corresponding to the $B_{1}$ basis, and hence is odd about the direction following the Bloch momentum vector $\boldsymbol{\kappa}_{XM}$ in which the states exists. Similarly, the band following $\boldsymbol{\bigcirc}$ and $\color{myRed} \boldsymbol{\bigcirc}$ will have irreducible representations corresponding to the $B_{2}$ basis, and be even about the $\boldsymbol{\kappa}_{XM}$ direction. Such a well ordered and opposite parity of branches, with irreducible representations of correct symmetry basis, allows the $\boldsymbol{\kappa}_{XM}$ crossing to exist and confirms its symmetry protected nature.

\begin{figure}[h!]
\centering
\hspace*{1.0cm} 
\begin{tikzpicture}[scale=0.3, transform shape]
\begin{scope}[xshift=14.5cm, yshift=22cm,scale=1.4]
		\node[regular polygon, regular polygon sides=4,draw, inner sep=7cm,rotate=0,line width=0.0mm, white,
           path picture={
               \node[rotate=0] at (1,0){
                   \includegraphics[scale=1.25]{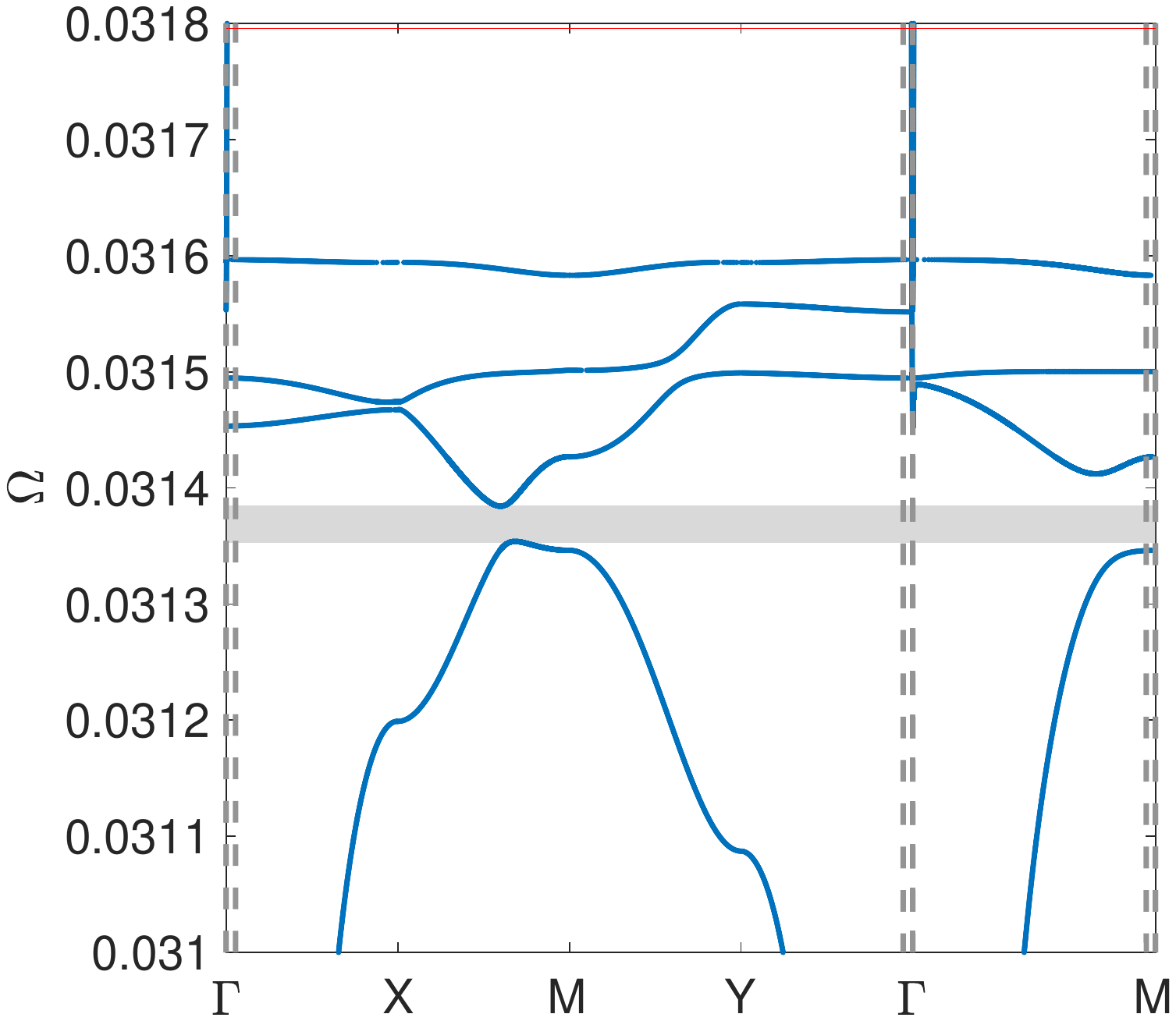}
               };
           }]{};
		\node[below, scale=2.86, black] at (-10,8.75) {$\displaystyle (a)$};           
\end{scope}  

\begin{scope}[xshift=35cm, yshift=27.5cm,scale=0.8]
		\node[regular polygon, regular polygon sides=4,draw, inner sep=6.5cm,rotate=0,line width=0.0mm, white,
           path picture={
               \node[rotate=0] at (2,0){
                   \includegraphics[scale=1.25]{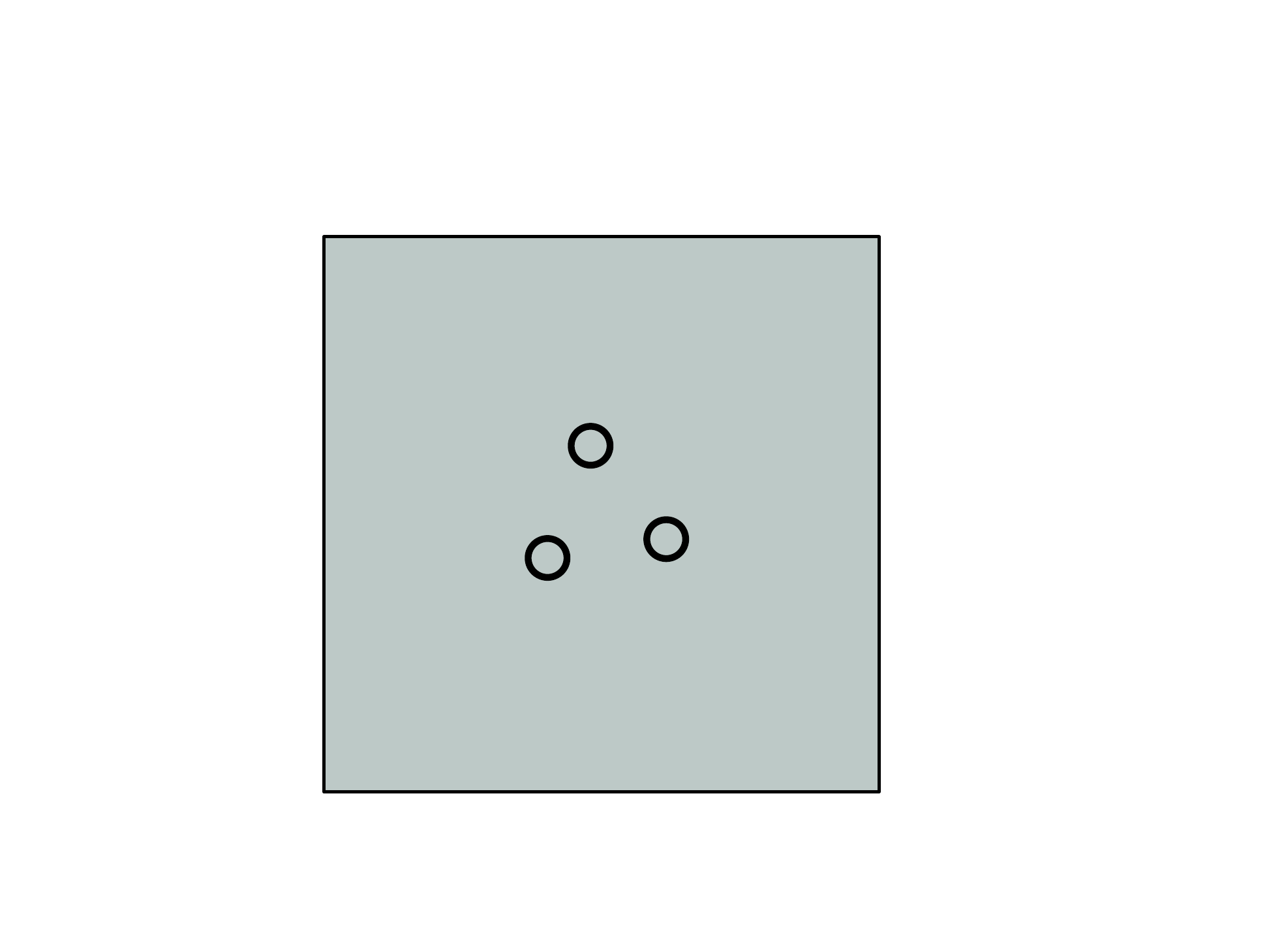}
               };
           }]{};
		\node[below, scale=5, black] at (-6,5.75) {$\displaystyle (b)$};           
\end{scope}

\begin{scope}[xshift=35cm, yshift=14.5cm,scale=0.8]
		\node[regular polygon, regular polygon sides=4,draw, inner sep=6.5cm,rotate=0,line width=0.0mm, white,
           path picture={
               \node[rotate=0] at (2,4){
                   \includegraphics[scale=1.25]{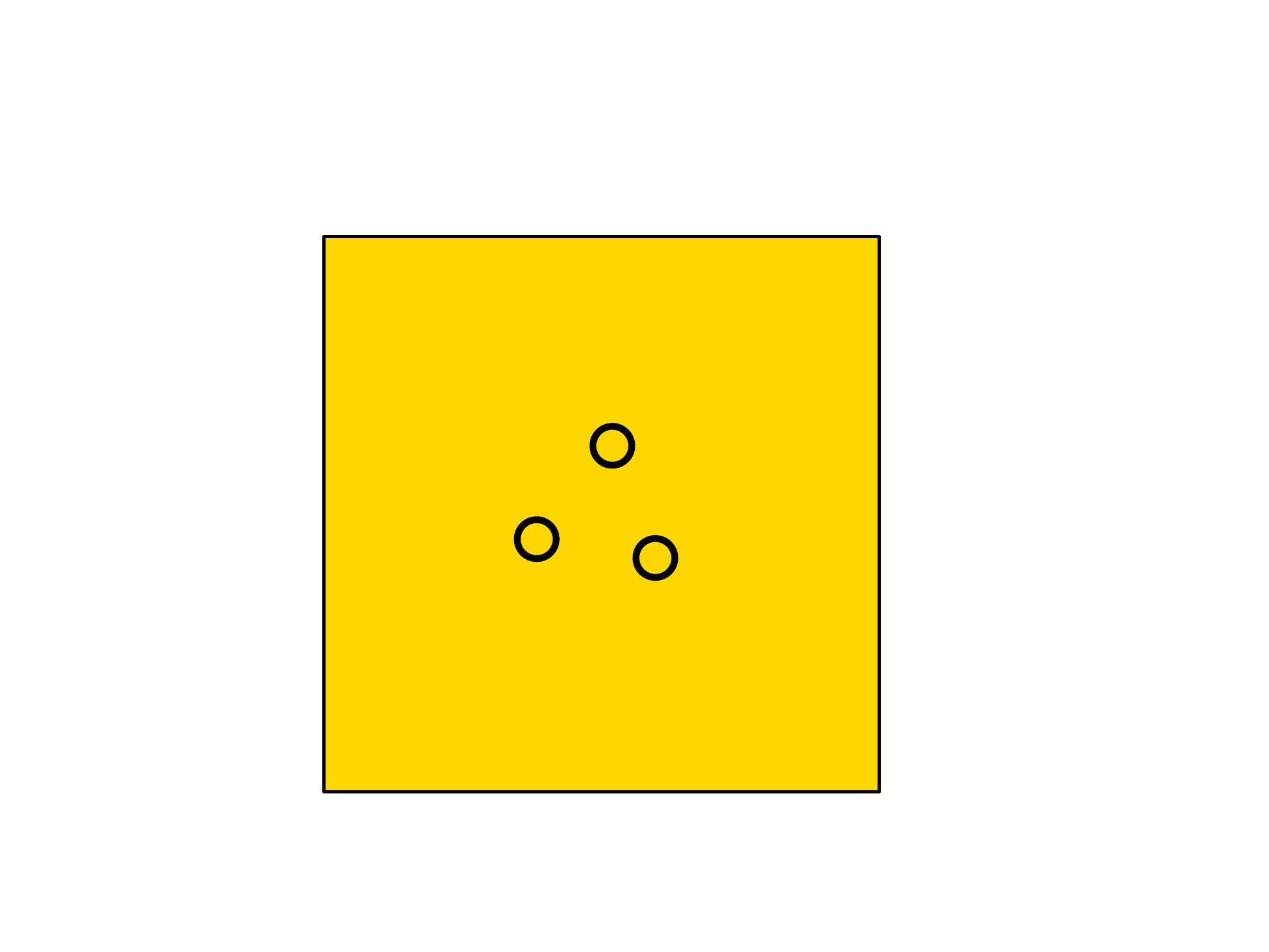}
               };
           }]{};
		\node[below, scale=5, black] at (-6,9.75) {$\displaystyle (c)$};           
\end{scope}  

\begin{scope}[xshift=14.5cm, yshift=-4cm,scale=1.4]
		\node[regular polygon, regular polygon sides=4,draw, inner sep=7cm,rotate=0,line width=0.0mm, white,
           path picture={
               \node[rotate=0] at (-9,0){
                   \includegraphics[scale=1.25]{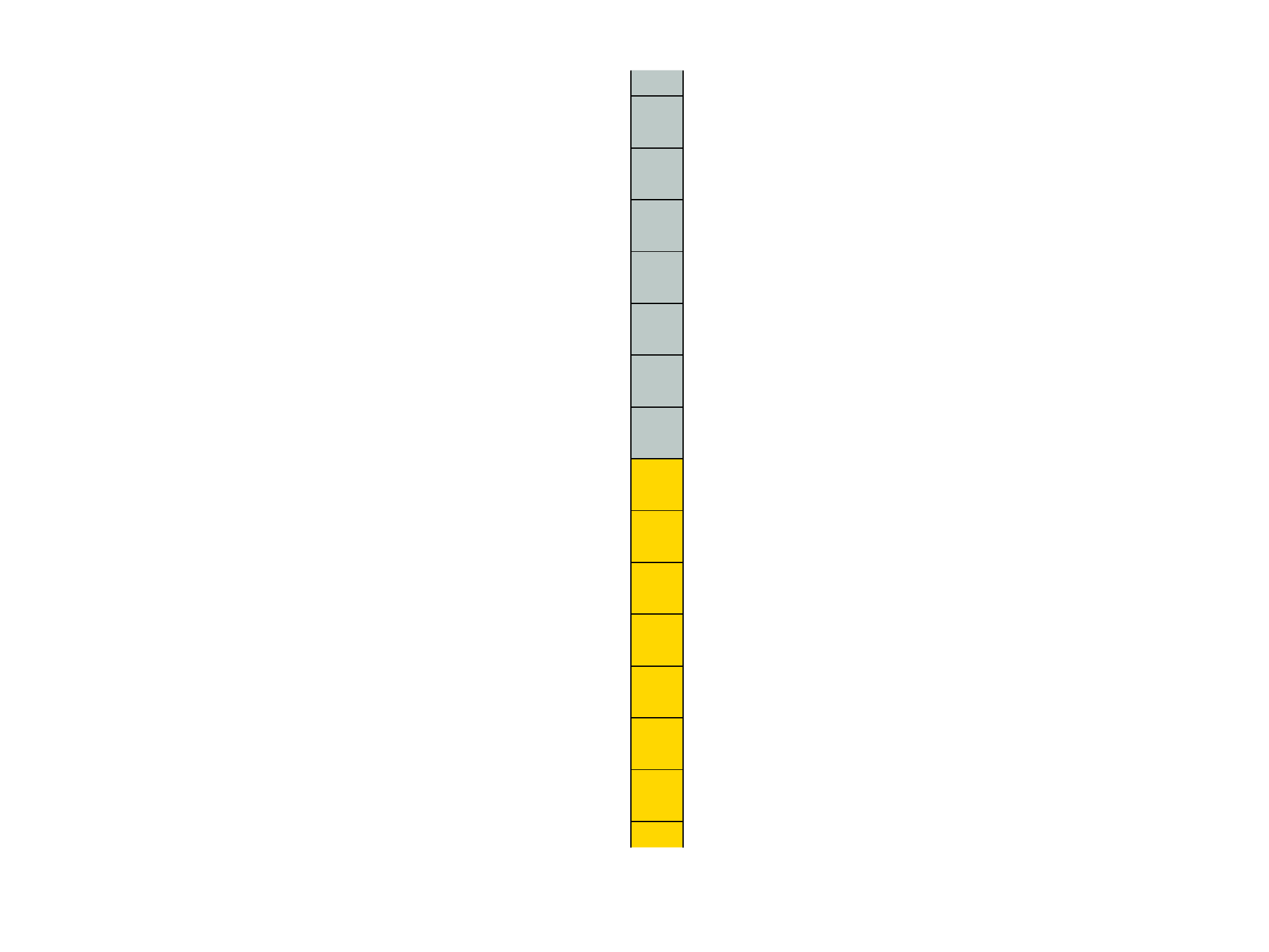}
               };
           }]{};
\end{scope}  

\begin{scope}[xshift=17.5cm, yshift=-4cm,scale=1.4]
		\node[regular polygon, regular polygon sides=4,draw, inner sep=7cm,rotate=0,line width=0.0mm, white,
           path picture={
               \node[rotate=0] at (-9,0){
                   \includegraphics[scale=1.25]{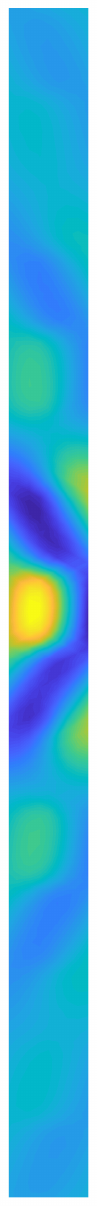}
               };
           }]{};
				\node[below, scale=2.86, black] at (-9.75,-7.5) {$\displaystyle (d)$};           
\end{scope}  

\begin{scope}[xshift=20.5cm, yshift=-4cm,scale=1.4]
		\node[regular polygon, regular polygon sides=4,draw, inner sep=7cm,rotate=0,line width=0.0mm, white,
           path picture={
               \node[rotate=0] at (1.5,0){
                   \includegraphics[scale=1.25]{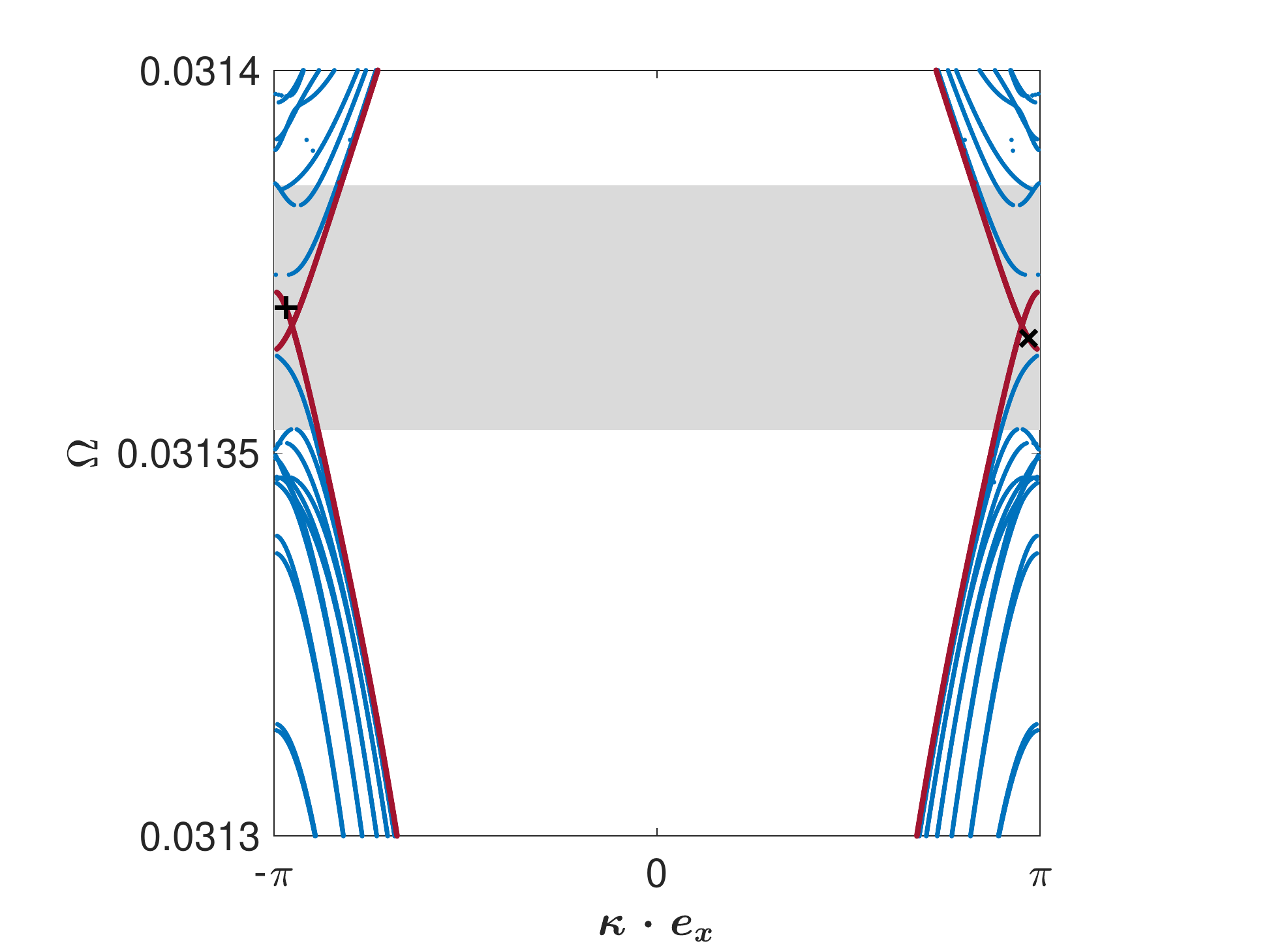}
               };
           }]{};
		\node[below, scale=2.86, black] at (-7,6.75) {$\displaystyle (e)$};           
\end{scope}  

\begin{scope}[xshift=48.5cm, yshift=-4cm,scale=1.4]
		\node[regular polygon, regular polygon sides=4,draw, inner sep=7cm,rotate=0,line width=0.0mm, white,
           path picture={
               \node[rotate=0] at (-9,0){
                   \includegraphics[scale=1.25]{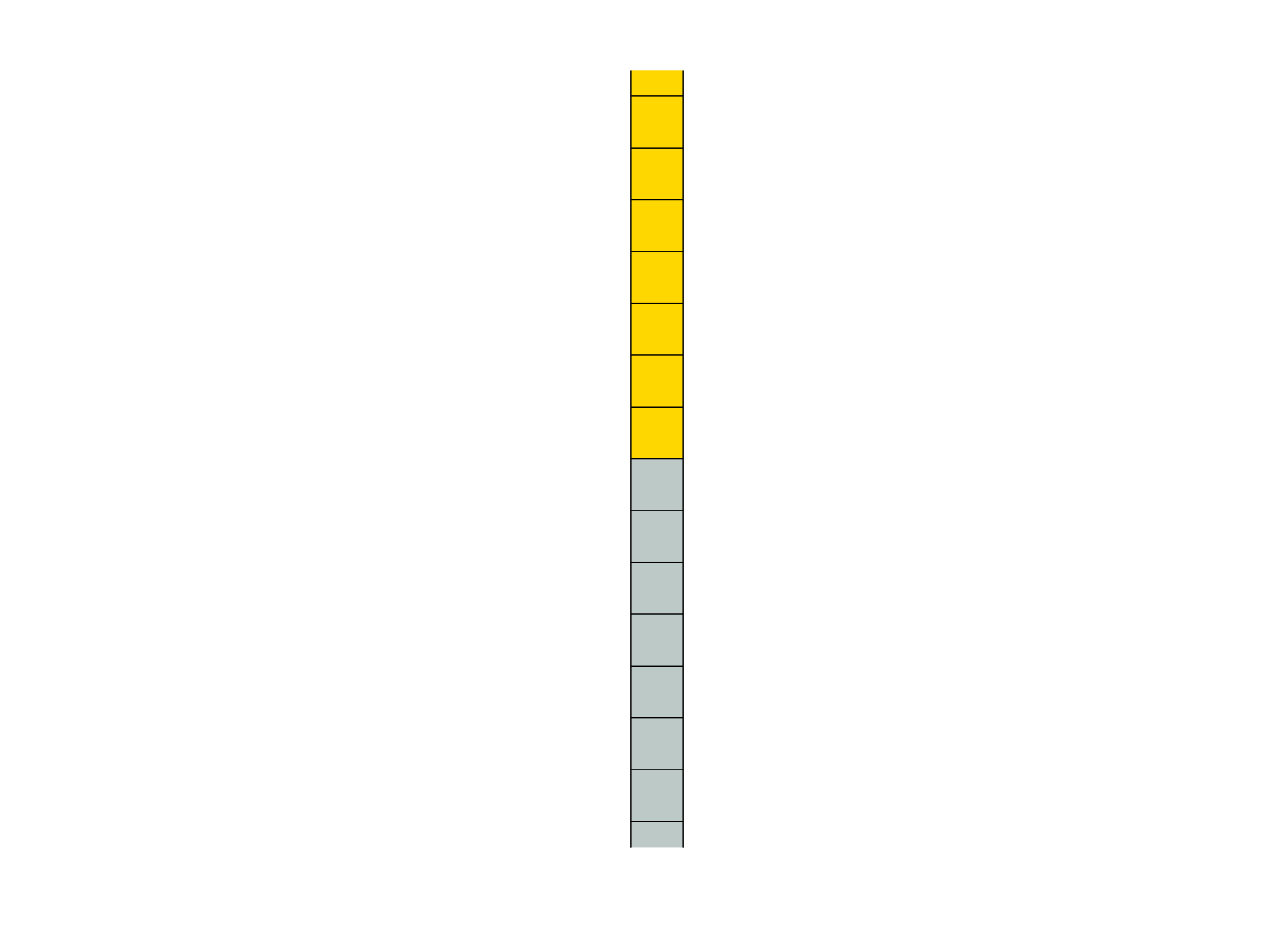}
               };
           }]{};
\end{scope}

\begin{scope}[xshift=51.5cm, yshift=-4cm,scale=1.4]
		\node[regular polygon, regular polygon sides=4,draw, inner sep=7cm,rotate=0,line width=0.0mm, white,
           path picture={
               \node[rotate=0] at (-9,0){
                   \includegraphics[scale=1.25]{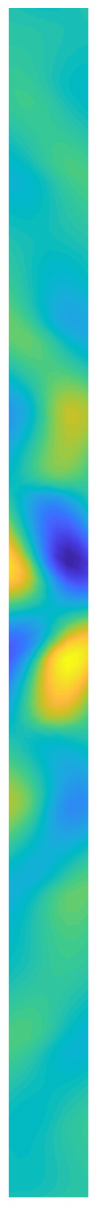}
               };
           }]{};
						\node[below, scale=2.86, black] at (-9.75,-7.5) {$\displaystyle (f)$}; 
\end{scope}  

\begin{scope}[xshift=54.5cm, yshift=-4cm,scale=1.4]
		\node[regular polygon, regular polygon sides=4,draw, inner sep=7cm,rotate=0,line width=0.0mm, white,
           path picture={
               \node[rotate=0] at (-10.25,0){
                   \includegraphics[scale=1.25]{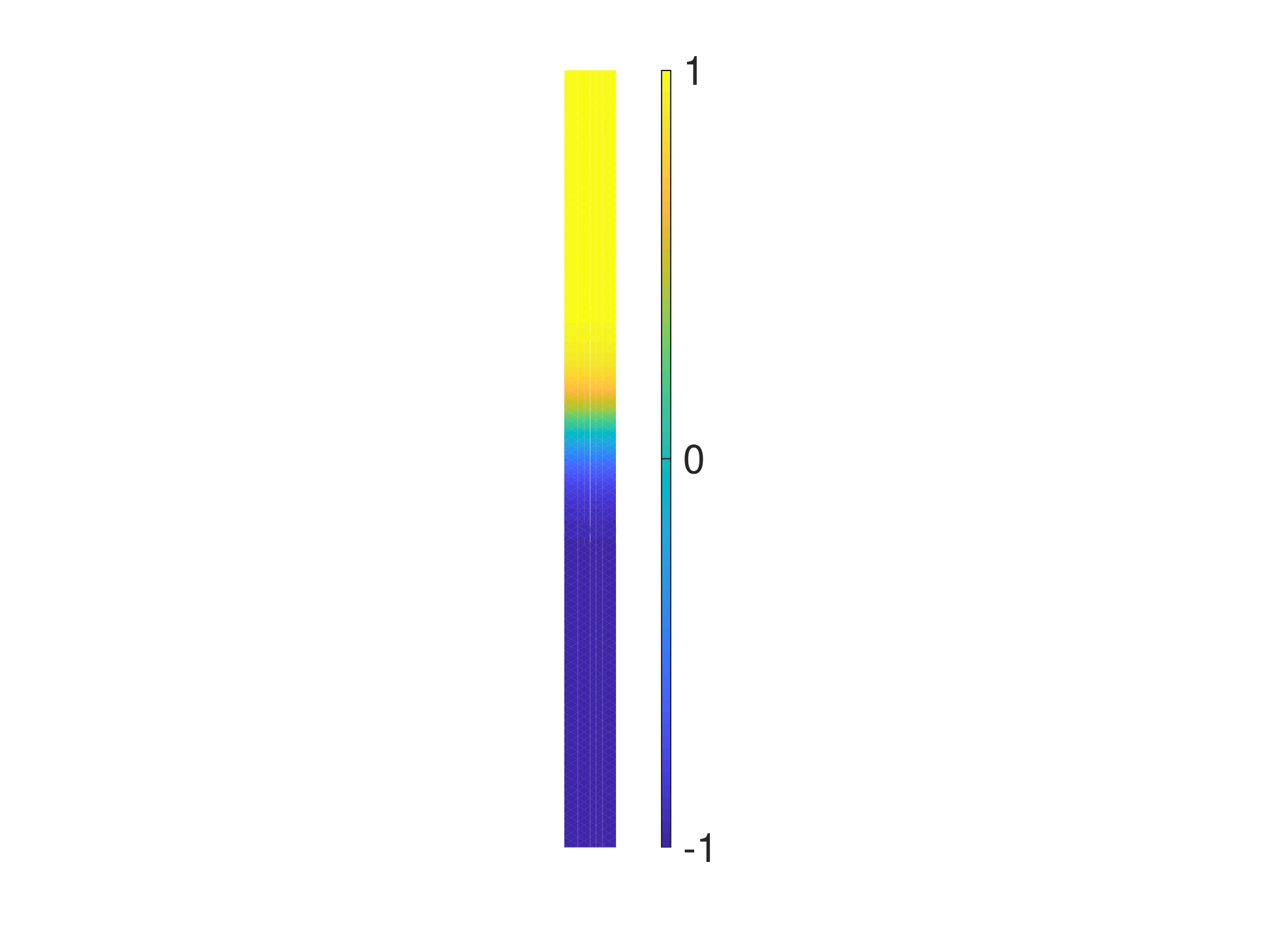}
               };
           }]{};
\end{scope}

\end{tikzpicture}
\caption{Panel $(a)$ shows the Floquet-Bloch dispersion branches of the primitive cell in $(b)$. Cells $(b)$ \& $(c)$ are obtained by perturbing the arrangement in fig.\ref{fig:SquareTopoArrange}$(c)$ via rotations, setting $\textbf{X}_{1J} = 0.125\left[ \cos(\frac{2\pi (J-1)}{3} + \frac{\pi}{2} + \theta') \textbf{e}_{x} + \sin(\frac{2\pi (J-1)}{3} + \frac{\pi}{2} + \theta')  \textbf{e}_{y}\right]$, the grey $(b)$ and yellow $(c)$ cells respectively correspond to positive and negative rotations in which $\theta' = \pm \frac{\pi}{20}$. The topological band gap is shaded in grey in $(a)$. Panels $(d)$ and $(f)$ show ribbon media, formed by stacking the chiral pairs in $(b)$ and $(c)$ as shown. The dispersion curves of these ribbon media, obtained from our Floquet-Bloch analysis \eqref{EVPactualalgebraicDispBloch}, are plotted in $(e)$ where the bulk band gap from panel $(a)$ is also highlighted in $(e)$. The interfacial modes of interest are plotted in red in $(e)$, the even $(d)$ and odd $(f)$ interfacial states are plotted next to the ribbon media in which they persist;  the eigenstates in $(d)$ and $(f)$ correspond to the  points $\boldsymbol{+}$ and  $\boldsymbol{\times}$ in panel $(e)$ respectively.} 
\label{fig:PertSquareTopoArrange}
\end{figure}

We perturb the primitive cell in fig. \ref{fig:SquareTopoArrange} $(c)$ by positively and negatively rotating the arrangement of beams to create the cells in fig. \ref{fig:PertSquareTopoArrange} $(b)$ \& $(c)$. The $\sigma_{v}$ symmetry breaking reduces the symmetry set from $\lbrace G_{\Gamma}, G_{X} \rbrace = \lbrace C_{2v} , C_{2v} \rbrace$ in cell \ref{fig:SquareTopoArrange} $(c)$, to $\lbrace G_{\Gamma}, G_{X} \rbrace = \lbrace C_{2} , C_{2} \rbrace$ in cells  \ref{fig:PertSquareTopoArrange} $(b)$ \& $(c)$; the perturbation affects the well ordered states along branches in \ref{fig:SquareTopoArrange} $(a)$ by mixing the parity of states along the branches, subsequently the first and second bands find themselves repulsed and the symmetry induced degeneracy is broken to form the subwavelength topologically non-trivial band gap in fig.  \ref{fig:PertSquareTopoArrange}. By stacking chiral-mirrored pairs of such structured media, as in fig. \ref{fig:PertSquareTopoArrange}, we observe the QVHE, in which we are guaranteed the creation of ZLMs by the mechanism presented in fig. \ref{fig:BerryCurvature} $(b)$. Indeed, figs \ref{fig:PertSquareTopoArrange}$(d)$ \& $(f)$ show the familiar even and odd parity ZLMs localised to two distinct interfaces, generated by the sole $\sigma_{v}$ symmetry breaking within the square cell - as in \cite{makwana2019topological,makwana19a}.

\begin{figure}[h!]
\centering
\hspace*{-0.75cm} 
\begin{tikzpicture}[scale=0.275, transform shape]

\begin{scope}[xshift=14.5cm, yshift=22cm,scale=1.4]
		\node[regular polygon, regular polygon sides=4,draw, inner sep=5.5cm,rotate=0,line width=0.0mm, white,
           path picture={
              \node[rotate=0] at (-0.5,-0.25){
                   \includegraphics[scale=1.25]{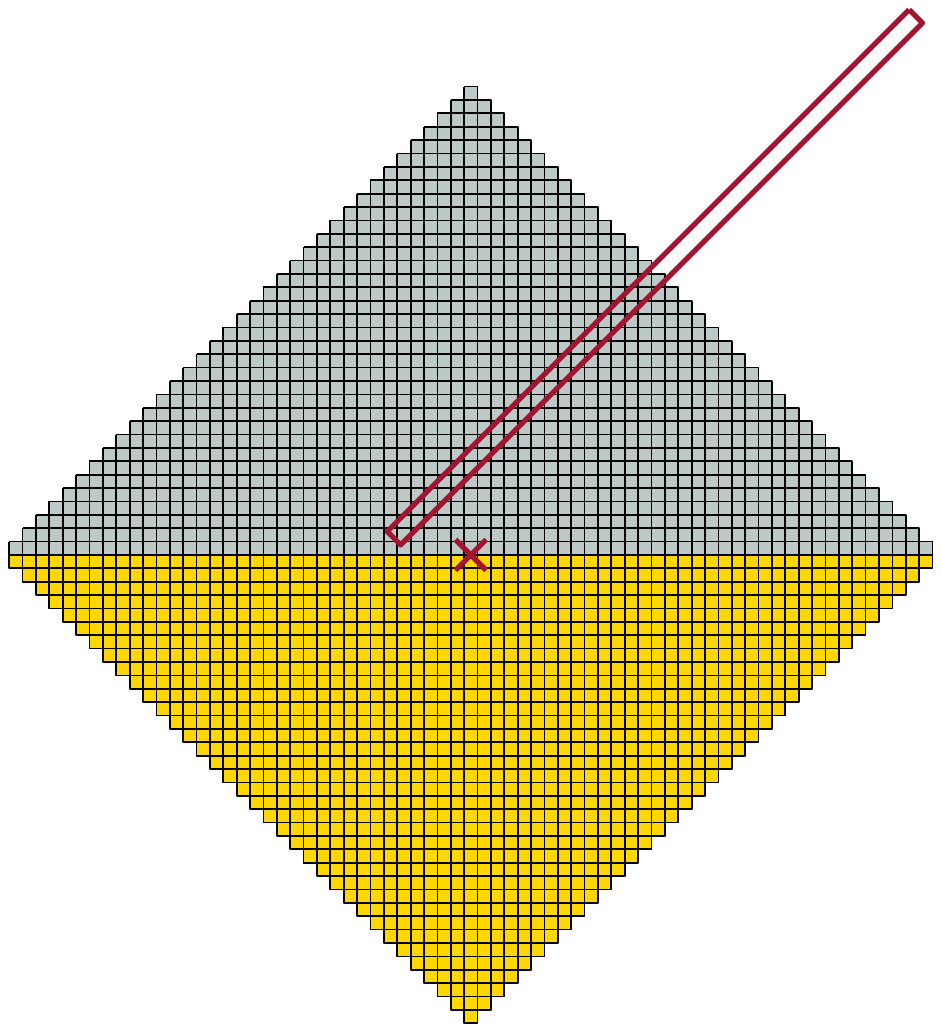}
               };    
           }]{};
		\node[below, scale=2.86, black] at (-8.5,8.75) {$\displaystyle (a)$};           
\end{scope}

\begin{scope}[xshift=40cm, yshift=22cm,scale=1.4]
		\node[regular polygon, regular polygon sides=4,draw, inner sep=5.5cm,rotate=0,line width=0.0mm, white,
           path picture={
               \node[rotate=0] at (-0.5,-0.25){
                   \includegraphics[scale=1.25]{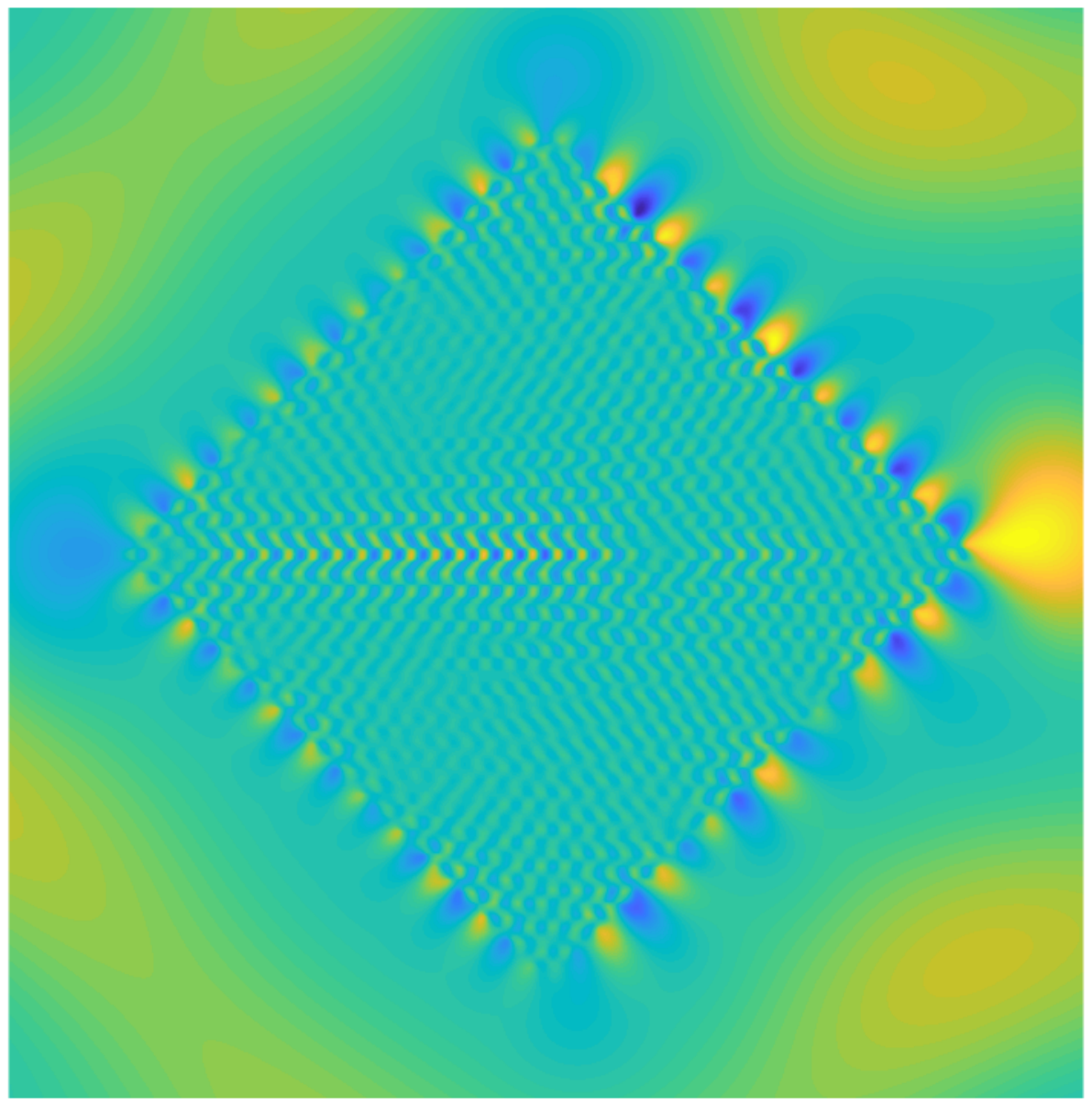}
               };
           }]{};
		\node[below, scale=2.86, black] at (-8.5,8.75) {$\displaystyle (b)$};           
\end{scope}

\begin{scope}[xshift=14.5cm, yshift=-4cm,scale=1.4]
               \node[rotate=0] at (-0.5,-0.25){
                   \includegraphics[scale=1.25]{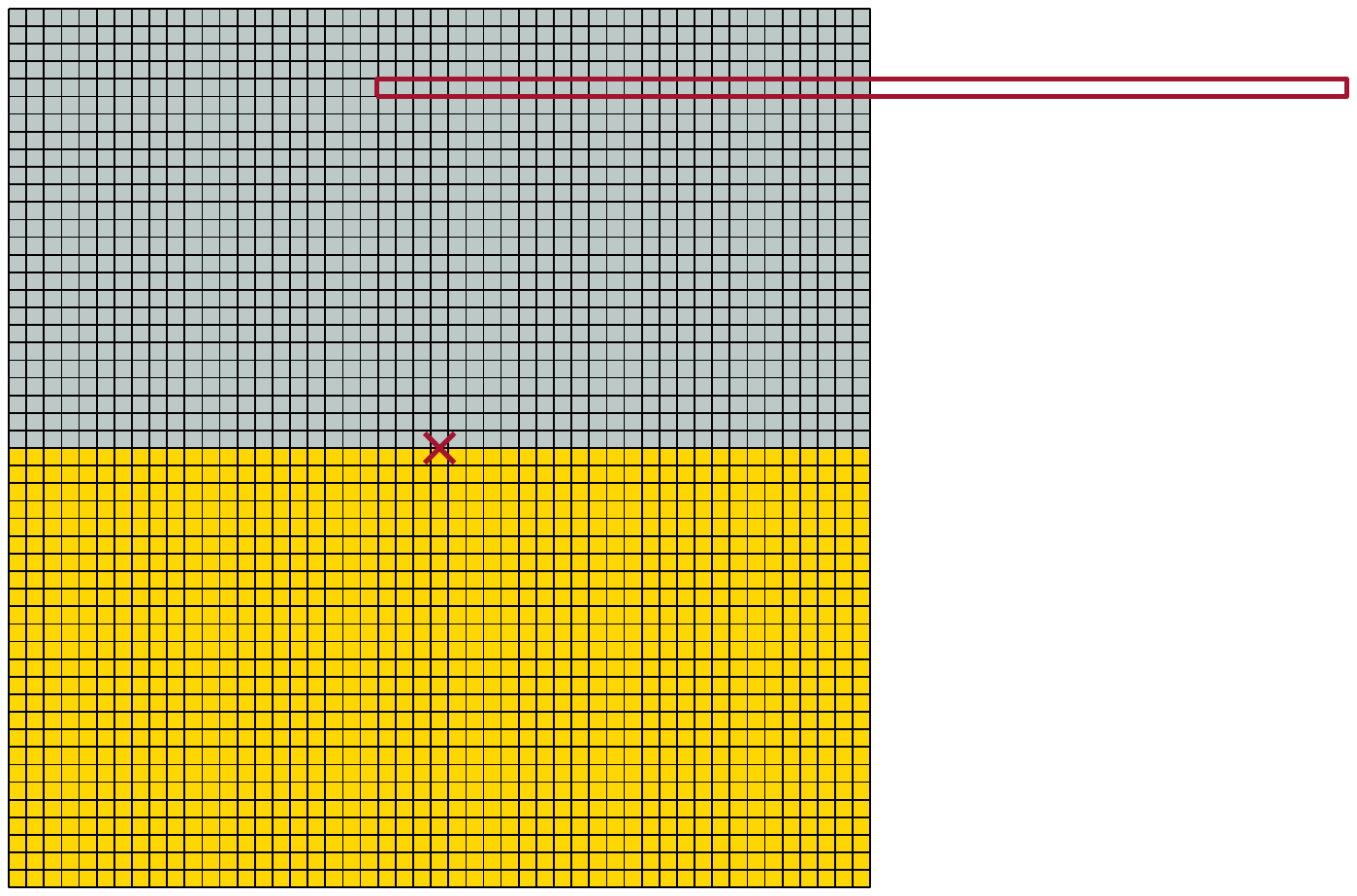}
               };
		\node[below, scale=2.86, black] at (-8.5,8.75) {$\displaystyle (c)$};           
\end{scope}

\begin{scope}[xshift=40cm, yshift=-4cm,scale=1.4]
		\node[regular polygon, regular polygon sides=4,draw, inner sep=5.5cm,rotate=0,line width=0.0mm, white,
           path picture={
               \node[rotate=0] at (-0.5,-0.25){
                   \includegraphics[scale=1.25]{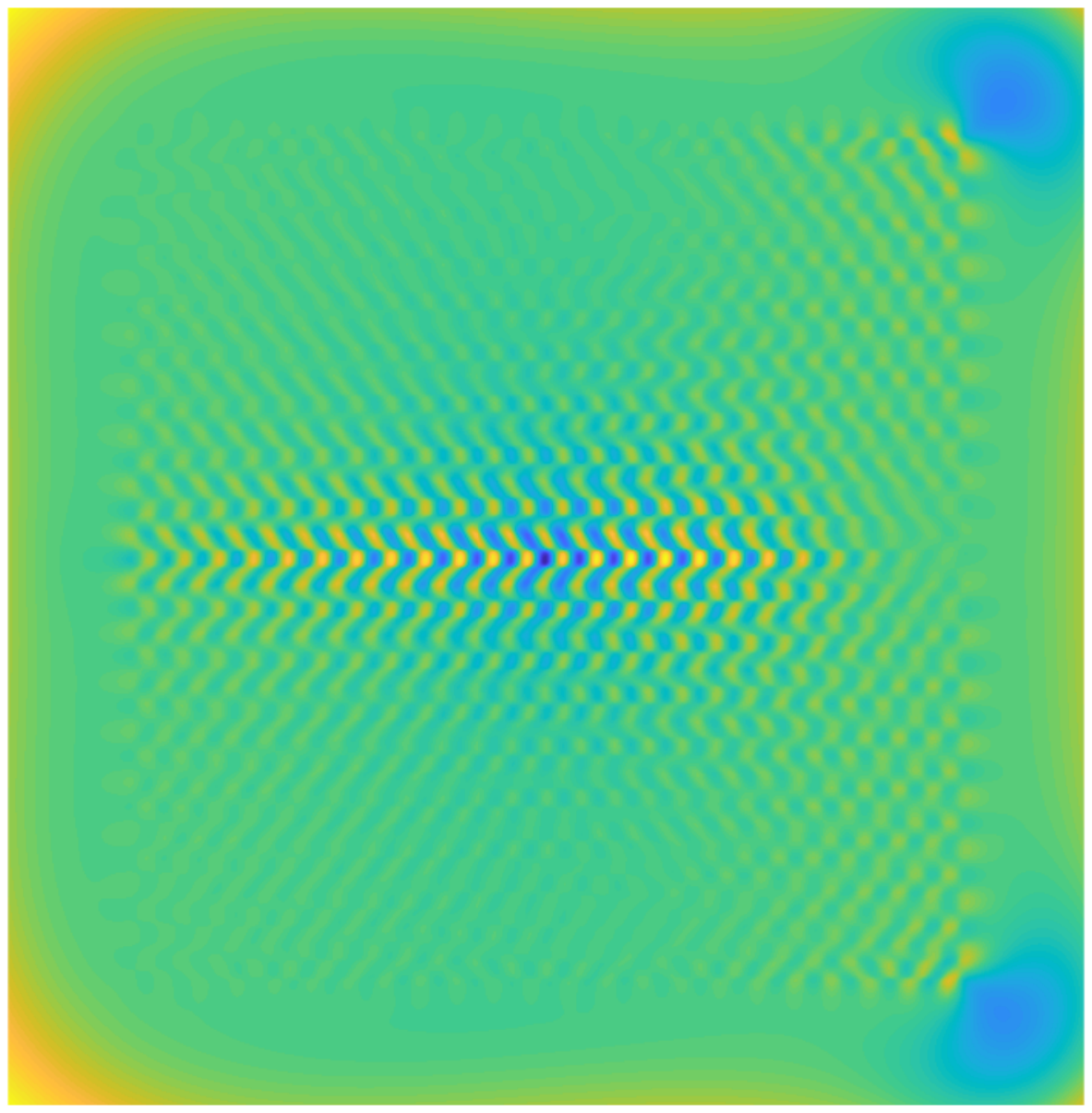}
               };
           }]{};
		\node[below, scale=2.86, black] at (-8.5,8.75) {$\displaystyle (d)$};           
\end{scope}  

\begin{scope}[xshift=12.0cm, yshift=-26cm,scale=1.2]
		\node[regular polygon, regular polygon sides=4,draw, inner sep=7cm,rotate=0,line width=0.0mm, white,
           path picture={
               \node[rotate=0] at (-9,0){
                   \includegraphics[scale=1.25]{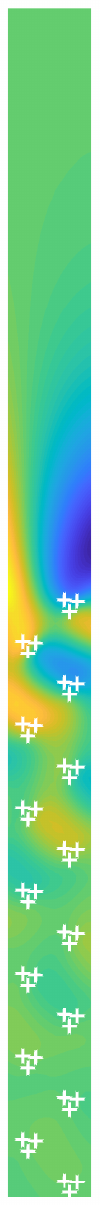}
               };
           }]{};
				\node[below, scale=2.45, black] at (-9.75,-7.5) {$\displaystyle (e)$};           
\end{scope}  

\begin{scope}[xshift=14.5cm, yshift=-26.5cm,scale=1.2]
		\node[regular polygon, regular polygon sides=4,draw, inner sep=7cm,rotate=0,line width=0.0mm, white,
           path picture={
               \node[rotate=0] at (1.5,0){
                   \includegraphics[scale=1.25]{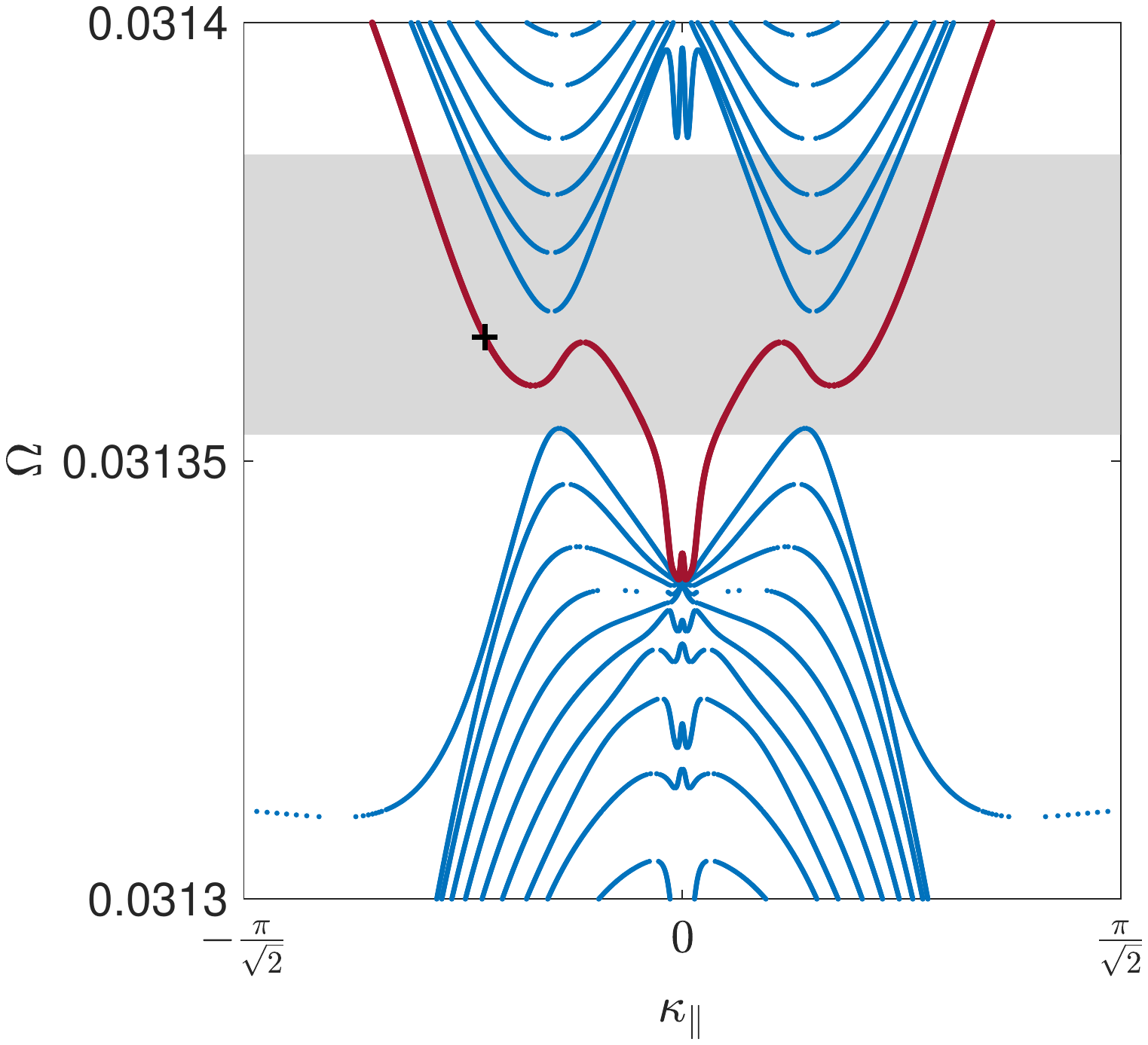}
               };
           }]{};
		\node[below, scale=2.45, black] at (-7,6.75) {$\displaystyle (f)$};           
\end{scope}  

\begin{scope}[xshift=39.5cm, yshift=-26.5cm,scale=1.2]
		\node[regular polygon, regular polygon sides=4,draw, inner sep=7cm,rotate=0,line width=0.0mm, white,
           path picture={
               \node[rotate=0] at (1.5,0){
                   \includegraphics[scale=1.25]{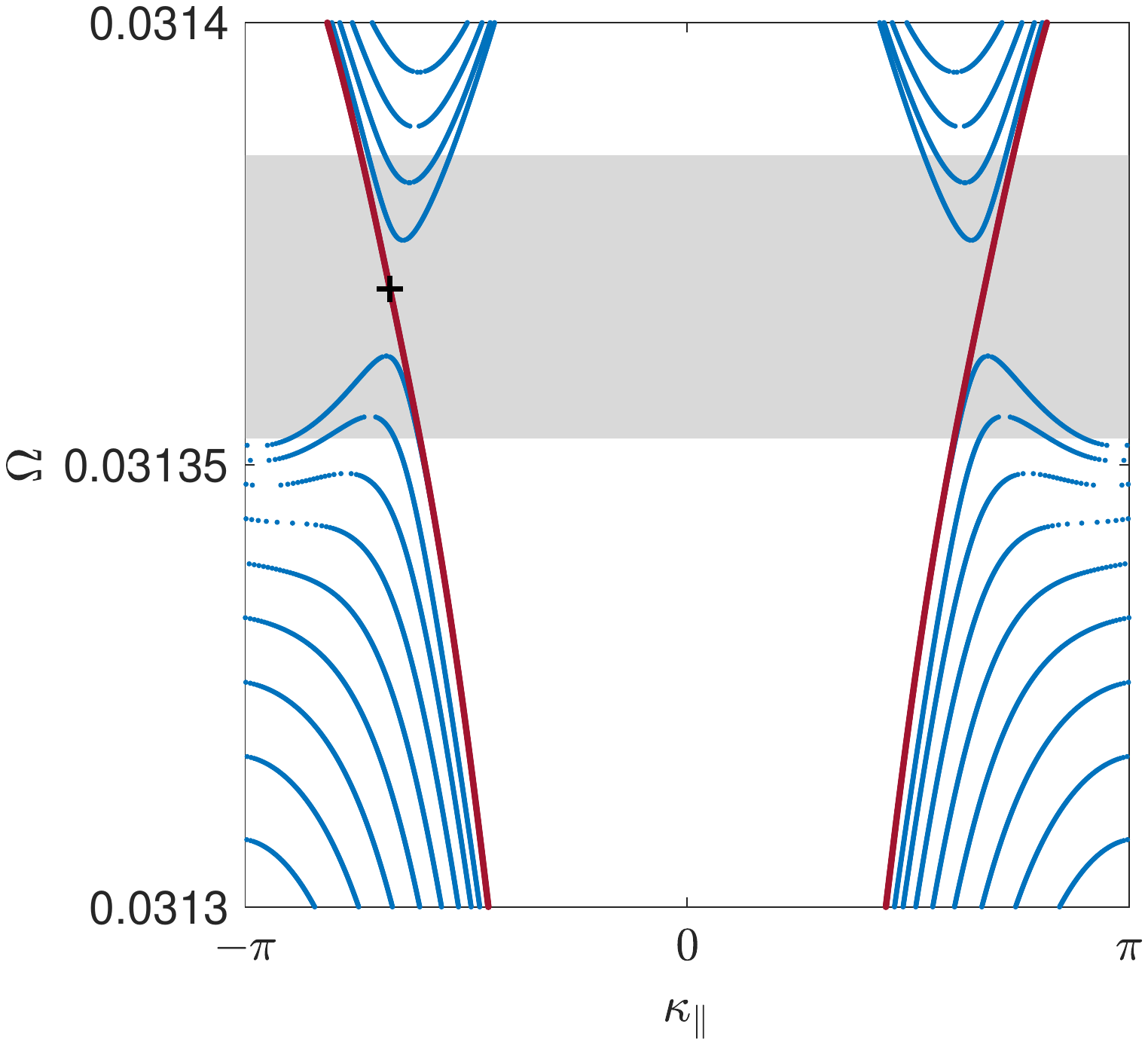}
               };
           }]{};
		\node[below, scale=2.45, black] at (-7,6.75) {$\displaystyle (g)$};           
\end{scope}  

\begin{scope}[xshift=62.5cm, yshift=-26cm,scale=1.2]
		\node[regular polygon, regular polygon sides=4,draw, inner sep=6.5cm,rotate=0,line width=0.0mm, white,
           path picture={
               \node[rotate=0] at (-9,0){
                   \includegraphics[scale=1.25]{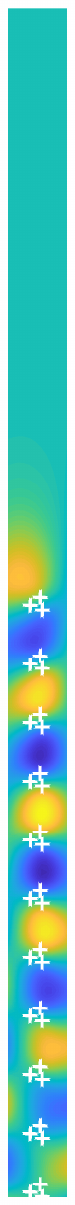}
               };
           }]{};
				\node[below, scale=2.45, black] at (-7.75,-7.5) {$\displaystyle (h)$};           
\end{scope}  

\begin{scope}[xshift=70cm, yshift=-4cm,scale=1.6]
		\node[regular polygon, regular polygon sides=4,draw, inner sep=7cm,rotate=0,line width=0.0mm, white,
           path picture={
               \node[rotate=0] at (-17.0,0){
                   \includegraphics[scale=1.25]{Figs/colourBarScattRe.pdf}
               };
           }]{};
\end{scope}  

\end{tikzpicture}
\caption{Generalised Foldy simulations $(b)$ and $(d)$ for the phononic crystal designs respectively shown in the schematics of panels $(a)$ and $(c)$; both $(a)$ and $(c)$ are formed from $2450$ cells, half grey and half yellow cells from fig. \ref{fig:PertSquareTopoArrange} $(b)$ \& $(c)$, for a total of $7350$ beams atop an elastic plate of infinite expanse. The incident sources, in $(b)$ and $(d)$, both consider monopoles with $\varpi_{\mathrm{inc}}=1$ from \eqref{ForcePOPinc} and $\textbf{X}_{\mathrm{inc}}$ is marked by $\color{myRed} \boldsymbol{\times}$ in $(a)$ and $(c)$. The frequencies of the point-sources both lie within the topological band gap of fig. \ref{fig:PertSquareTopoArrange} $(a)$,  and are set to $\Omega = 0.0313695$ for $(b)$ and $\Omega=0.0313698$ for $(d)$. The red rectangular strips in schematics $(a)$ and $(c)$ represent ribbon strips, half over the bare plate and half over the grey medium - panels $(f)$ and $(g)$ show the dispersion curves taken by considering Floquet-Bloch boundary conditions (from \eqref{EVPactualalgebraicDispBloch}) on the edges of these strips, the edge modes of interest are plotted in red and the bulk band gap is again highlighted in grey from fig. \ref{fig:PertSquareTopoArrange} (a). Here $\kappa_{\parallel}$ denotes the component of the wavevector  parallel to the surface of the crystal. Panels $(e)$ and $(h)$ corresponding to eigenmodes from $\boldsymbol{+}$ within panels $(f)$ and $(g)$ respectively, where the locations of the beams are shown by white crosses. The colour bar refers to the normalised displacement fields of panels $(b)$, $(d)$, $(e)$ and $(h)$, where the real part of the out-of-plane displacement field has been plotted. } 
\label{fig:ScattSquareTopoArrange}
\end{figure}

Consider fig. \ref{fig:ScattSquareTopoArrange} in which we test the designs of two phononic crystals $(a)$ \& $(c)$ with different edges, formed from tessellating the ribbon media in fig. \ref{fig:PertSquareTopoArrange}$(d)$. Note, from fig. \ref{fig:BerryCurvature} $(a)$, the edges of fig. \ref{fig:ScattSquareTopoArrange} $(a)$ have sections of non-zero Berry curvature meeting the surrounding free-space and hence, by the bulk-edge correspondence, we expect the existence of topologically protected edgestates. However the edges of fig. \ref{fig:ScattSquareTopoArrange} $(c)$ have negligible Berry curvature and hence cannot support chiral states. We consider a monopole point source, whose frequency lives within the bulk bandgap, placing this source as shown in figs. \ref{fig:ScattSquareTopoArrange} $(b)$ \& $(d)$ we excite the ZLM which resembles the eigenstates within figs \ref{fig:PertSquareTopoArrange}$(d)$ \& $(e)$; moreover, figs. \ref{fig:ScattSquareTopoArrange} $(b)$ \& $(d)$ indeed confirm our predictions regarding the existence of protected edgestates in designs figs \ref{fig:ScattSquareTopoArrange} $(a)$ \& $(c)$. 

The ribbon eigenmodal analysis in figs \ref{fig:ScattSquareTopoArrange} $(e)$-$(h)$ clearly shows the existence of edgemodes living within the topologically non-trivial band gap of fig.\ref{fig:PertSquareTopoArrange} $(a)$. Those in figs \ref{fig:ScattSquareTopoArrange} $(e)$ \& $(f)$ satisfy the bulk-edge correspondence and hence inherit topological protection from the structured bulk media and show similar decay as the ZLMs in fig. \ref{fig:PertSquareTopoArrange}$(d)$ - these modes are sufficiently robust to efficiently propagate around the edges of fig. \ref{fig:ScattSquareTopoArrange} $(a)$; they exist far beneath the analogous light line and are heavily localised to the edge.  The preferential direction observed in fig. \ref{fig:ScattSquareTopoArrange} $(b)$ is due to the anisotropic dipole nature of the scatterers, as expected at the near flexural resonances of the constituent beams. 

Fig. \ref{fig:ScattSquareTopoArrange} $(b)$ clearly show efficient mode conversion between the interfacial \ref{fig:PertSquareTopoArrange}$(d)$ states and edge \ref{fig:ScattSquareTopoArrange} $(e)$ states, where the absence of backscatter of energy at the corners of the designs is attributed to topological protection. The only energy losses observed here are due to the presence of sharp corners, in figs \ref{fig:ScattSquareTopoArrange}$(a)$, which readily produce corner  states \cite{palmer2021berry, palmer2021revealing,chen2021corner} and shed energy into the free space. We suspect the superior topological protection arising from the non-zero Berry curvature in the hexagonal case, fig. \ref{fig:BerryCurvature} $(c)$, will ensure a greater efficiency of the propagation between ZLMs and edgestates by forbidding corner states. The eigenmodes in figs \ref{fig:ScattSquareTopoArrange}$(g)$ \& $(h)$ do not benefit from topological protection. The lack of chiral-flux means they slowly decay within the non-trivial bulk and are only weakly excited in the scattering simulations of fig. \ref{fig:ScattSquareTopoArrange} $(d)$, where again we observe the presence of corner states. 

\begin{figure}[h!]
\centering
\hspace*{0.5cm} 
\begin{tikzpicture}[scale=0.3, transform shape]
\begin{scope}[xshift=14.5cm, yshift=22cm,scale=1.4]
		\node[regular polygon, regular polygon sides=4,draw, inner sep=5.5cm,rotate=0,line width=0.0mm, white,
           path picture={
               \node[rotate=0] at (-0.5,-0.25){
                   \includegraphics[scale=1.25]{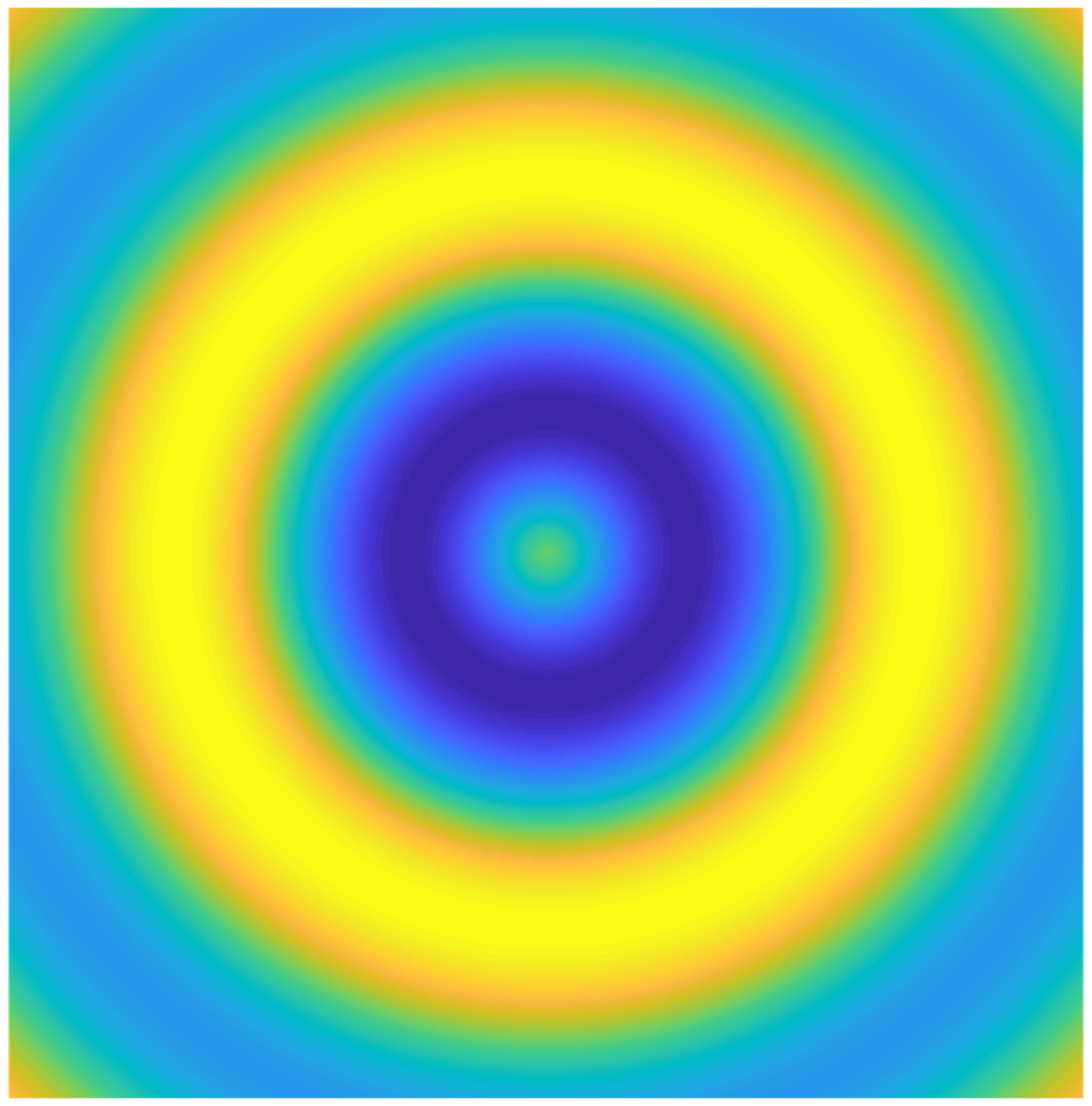}
               };
           }]{};
		\node[below, scale=2.86, black] at (-8.5,8.75) {$\displaystyle (a)$};           
\end{scope}

\begin{scope}[xshift=40cm, yshift=22cm,scale=1.4]
		\node[regular polygon, regular polygon sides=4,draw, inner sep=5.5cm,rotate=0,line width=0.0mm, white,
           path picture={
               \node[rotate=0] at (-0.5,-0.25){
                   \includegraphics[scale=1.25]{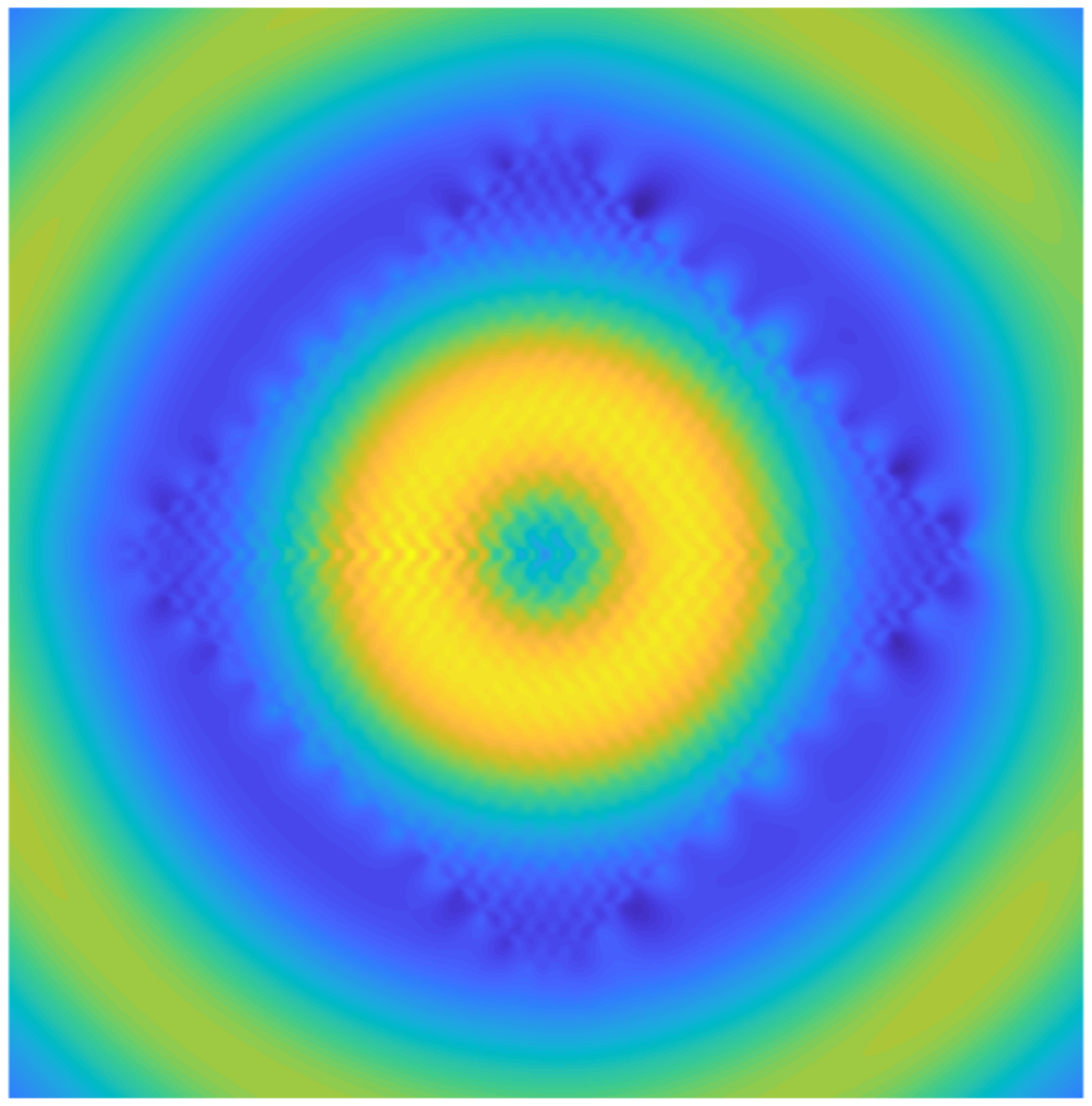}
               };
           }]{};
		\node[below, scale=2.86, black] at (-8.5,8.75) {$\displaystyle (b)$};           
\end{scope}  

\begin{scope}[xshift=65cm, yshift=22cm,scale=1.4]
		\node[regular polygon, regular polygon sides=4,draw, inner sep=7cm,rotate=0,line width=0.0mm, white,
           path picture={
               \node[rotate=0] at (-17.0,-0.25){
                   \includegraphics[scale=1.25]{Figs/colourBarScattRe.pdf}
               };
           }]{};
\end{scope}  

\end{tikzpicture}
\caption{Comparisons of the incident field $(a)$ and scattered field $(b)$, the addition of which yields the total field of fig. \ref{fig:ScattSquareTopoArrange} $(b)$. The incident source was set to $\varpi_{\mathrm{inc}}=1$ from \eqref{ForcePOPinc} and $\textbf{X}_{\mathrm{inc}}$ is marked by $\color{myRed} \boldsymbol{\times}$ in fig.\ref{fig:ScattSquareTopoArrange}  $(a)$. The colour bar shows the real part for the normalised wavefields.} 
\label{fig:IncScattPertSquareTopoArrange}
\end{figure}

The low-frequency, long-wavelength regime within which our designs manipulate the propagation of waves should not be understated. Refer to fig. \ref{fig:IncScattPertSquareTopoArrange}  comparing the incident \eqref{ForcePOPincGREENS} and scattered field; here, the incident source is of wavelength $30$-$40$ times that of the lattice spacing, observe how the respective peaks and troughs from the incident field align with the troughs and peaks in the scattered field, the addition of which yields the total field from fig. \ref{fig:ScattSquareTopoArrange} $(b)$.

\begin{table}[H]
\centering
\begin{tabular}{c|c c c c|c}
\cline{2-5} & \multicolumn{4}{ c| }{Classes} \\ 
	\hline
\multicolumn{1}{ ||c|  }{IRs} & $E$ & $C_{2}$ & $\sigma_{v}$ & $\sigma_{h}$ & \multicolumn{1}{ |c||  }{Basis} \\
	\hline\hline
	\multicolumn{1}{ ||c|  }{$B_{1}$} & $+1$	 & $-1$ & $+1$	 & $-1$ &	\multicolumn{1}{ |c||  }{$x$} \\
	\multicolumn{1}{ ||c|  }{$B_{2}$} & $+1$	 & $-1$ & $-1$	 & $+1$ &	\multicolumn{1}{ |c||  }{$y$} \\
	\hline	 
\end{tabular}
\caption{Excerpt of the $C_{2v}$ character table - we require only the linear basis functions spanning the irreducible representations (IRs). } 
\label{table:C2VCharacter}
\end{table}

\subsection{Superior topological protection: subwavelength states in hexagonal arrays} \label{TopoHexLattice}

The symmetry arguments in section \ref{TopoSquareLattice}, underpinning the creation of topologically protected states in square lattices also apply to hexagonal lattices, where the degeneracies now occur at the $KK'$ high symmetry points. In the hexagonal case larger band gaps can be produced as larger rotations, through $\theta '$, allow valleys to retain their locally quadratic behaviour in the vicinity of $KK'$. The pros and cons of different geometrically designed interfacial states for square and hexagonal lattices are summarised in fig. 12 of Makwana and Chaplain \cite{makwana19a}. 

\begin{figure}[h!]
\centering
\hspace*{-0.25cm}
\begin{tikzpicture}[scale=0.4, transform shape]

\begin{scope}[xshift=8cm, yshift=50.5cm]
\node[regular polygon, regular polygon sides=4,draw, inner sep=9.0cm,rotate=0,line width=0.0mm, white,
           path picture={
               \node[rotate=0] at (0,0){
                   \includegraphics[scale=2]{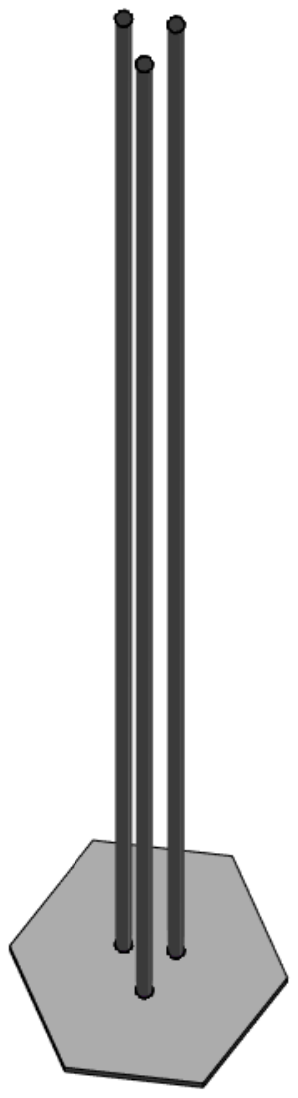}
               };
           }]{};
\node[below, scale=2, black] at (0.25,-11.5) {$\displaystyle (d)$};
\end{scope}

\begin{scope}[xshift=-4.5cm, yshift=50.5cm]
		\node[regular polygon, regular polygon sides=4,draw, inner sep=7.0cm,rotate=0,line width=0.0mm, white,
           path picture={
               \node[rotate=0] at (1,1){
                   \includegraphics[scale=1.25]{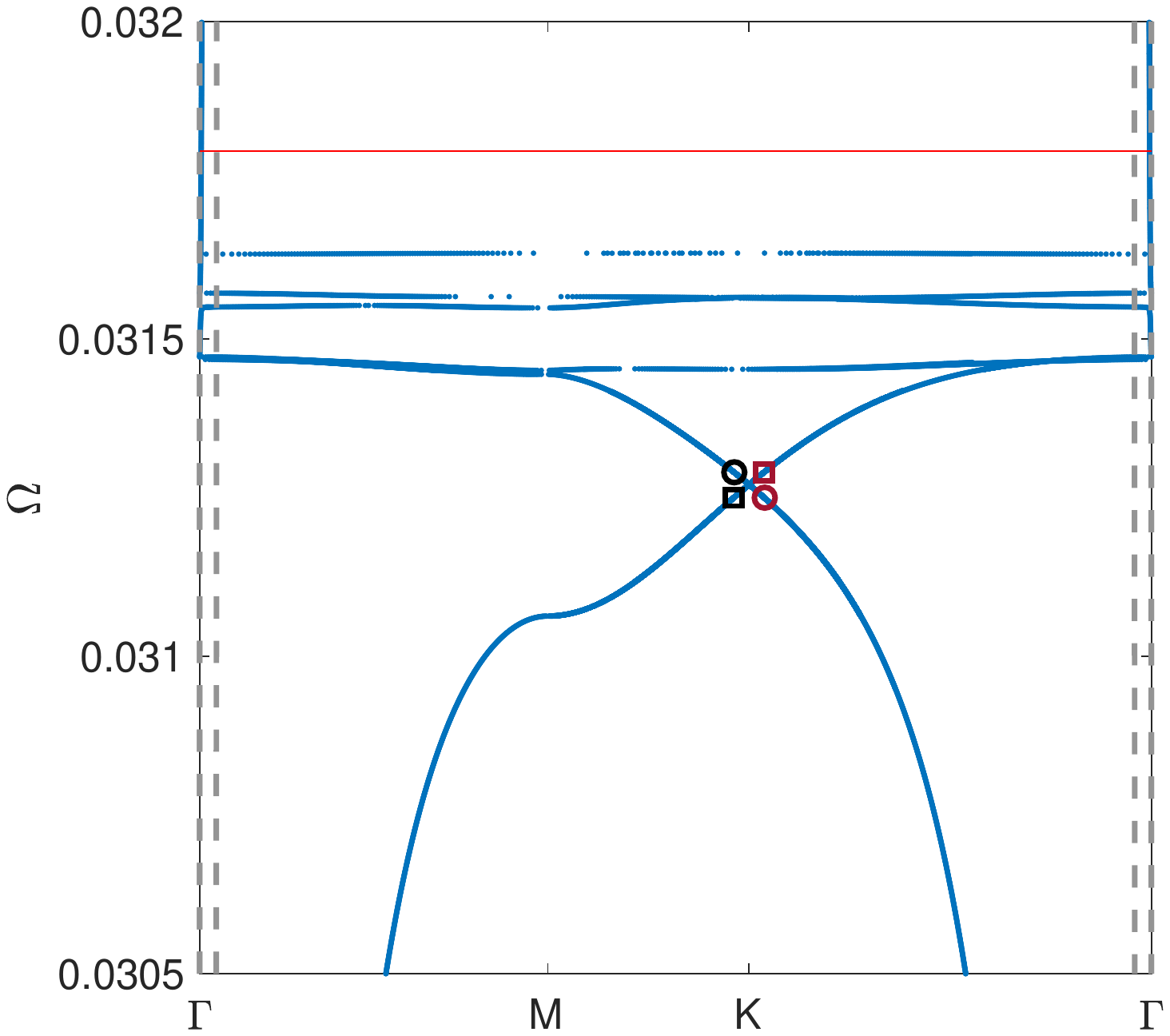}
               };
           }]{};
		\node[below, scale=2, black] at (-8,7.75) {$\displaystyle (a)$};         
\end{scope}  

\begin{scope}[xshift=17cm, yshift=56cm,scale=0.45]
		\node[regular polygon, regular polygon sides=6,draw, inner sep=6.25cm,rotate=0,line width=0.0mm, white,
           path picture={
               \node[rotate=0] at (-0.45,-0.35){
                   \includegraphics[scale=1.4]{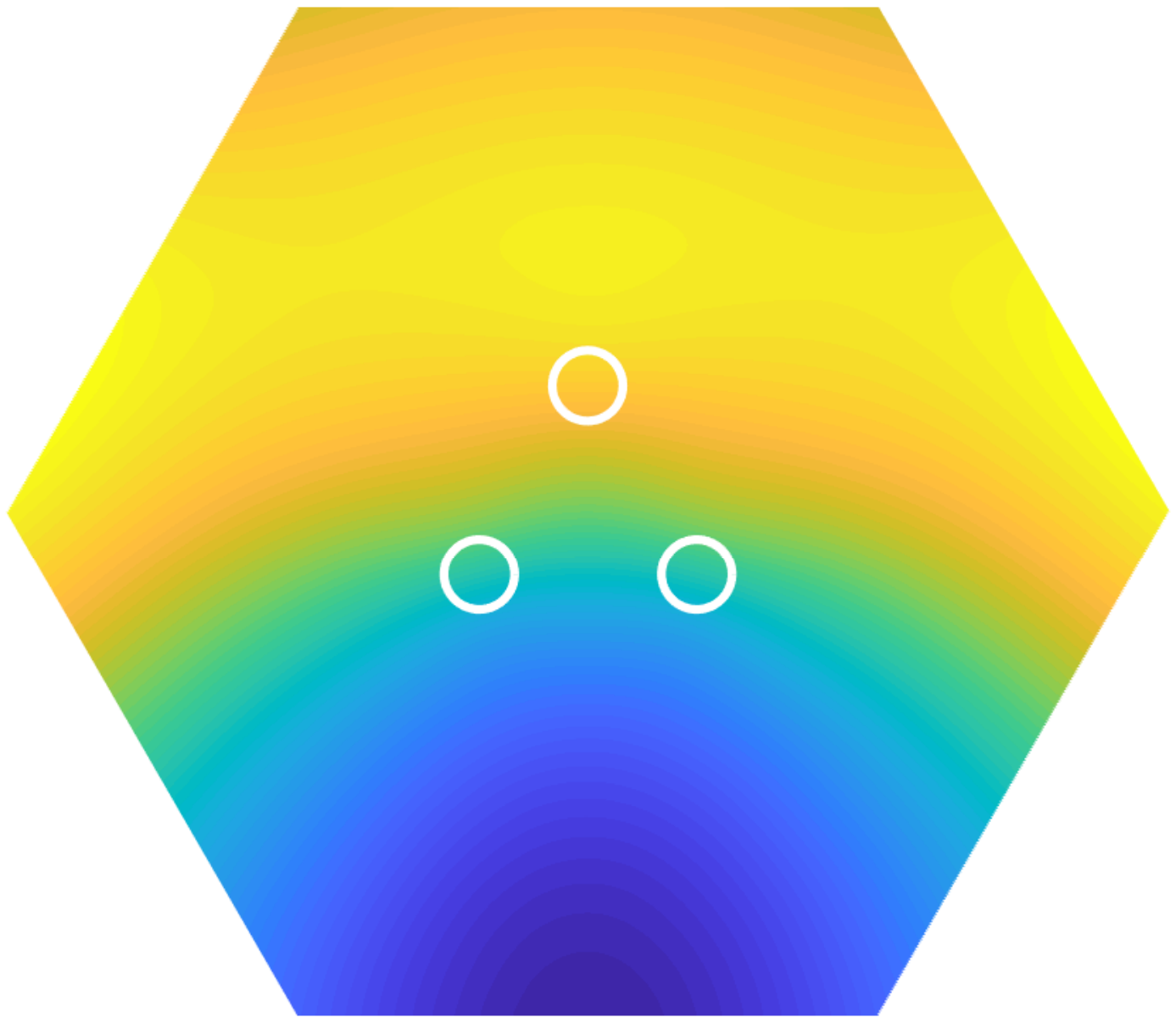}
               };
           }]{};
		\node[below, scale=4.4, black] at (-9.0,6.75) {$\displaystyle (e)$};           
\end{scope}  
\begin{scope}[xshift=17cm, yshift=48.05cm,scale=0.45]
		\node[regular polygon, regular polygon sides=6,draw, inner sep=6.25cm,rotate=0,line width=0.0mm, white,
           path picture={
               \node[rotate=0] at (-0.45,-0.35){
                   \includegraphics[scale=1.4]{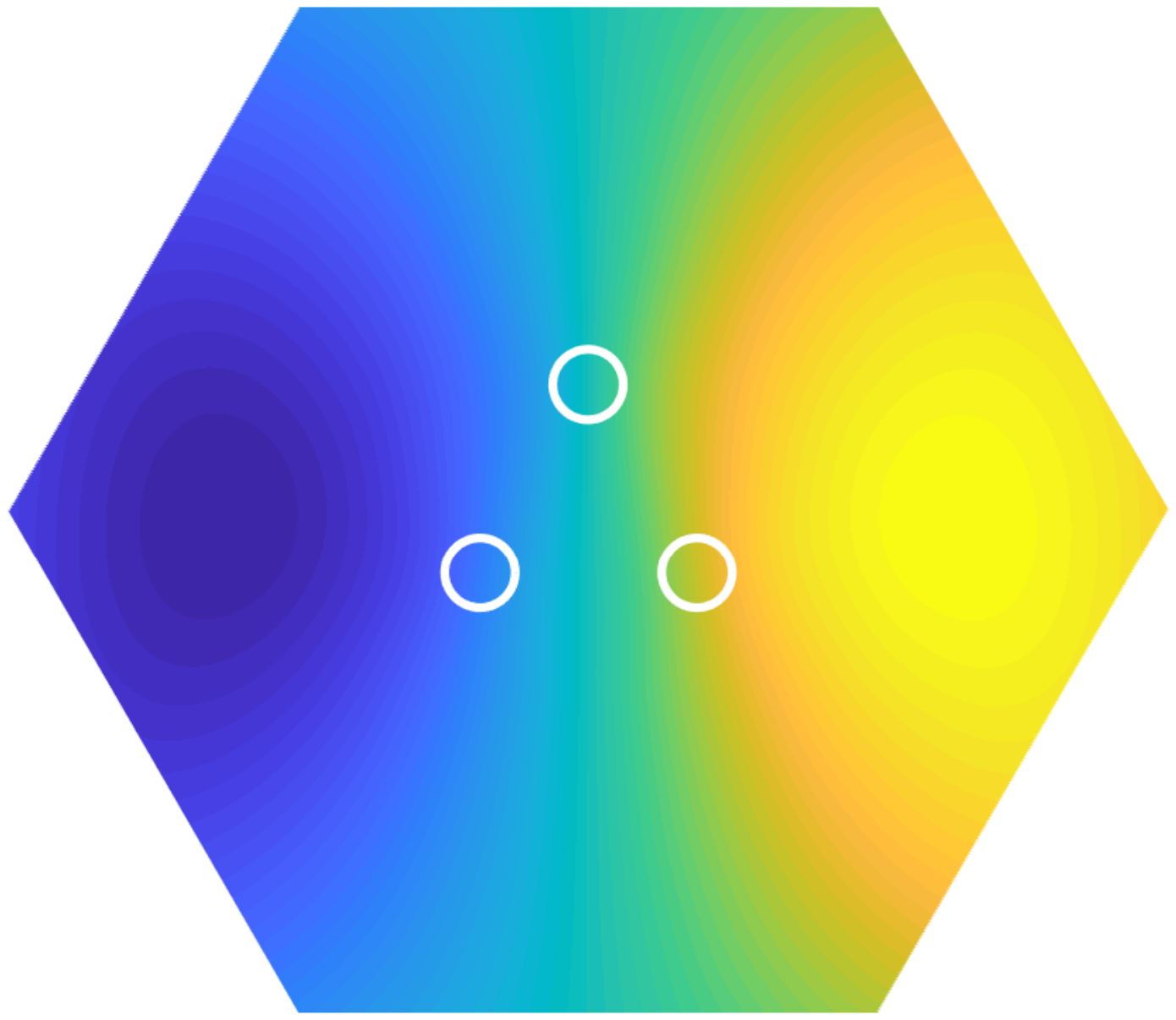}
               };
           }]{};
		\node[below, scale=4.4, black] at (-9.0,6.75) {$\displaystyle (f)$};           
\end{scope}  

\begin{scope}[xshift=23.9cm, yshift=52.025cm,scale=0.45]
		\node[regular polygon, regular polygon sides=6,draw, inner sep=6.25cm,rotate=0,line width=0.0mm, white,
           path picture={
               \node[rotate=0] at (-0.45,-0.35){
                   \includegraphics[scale=1.4]{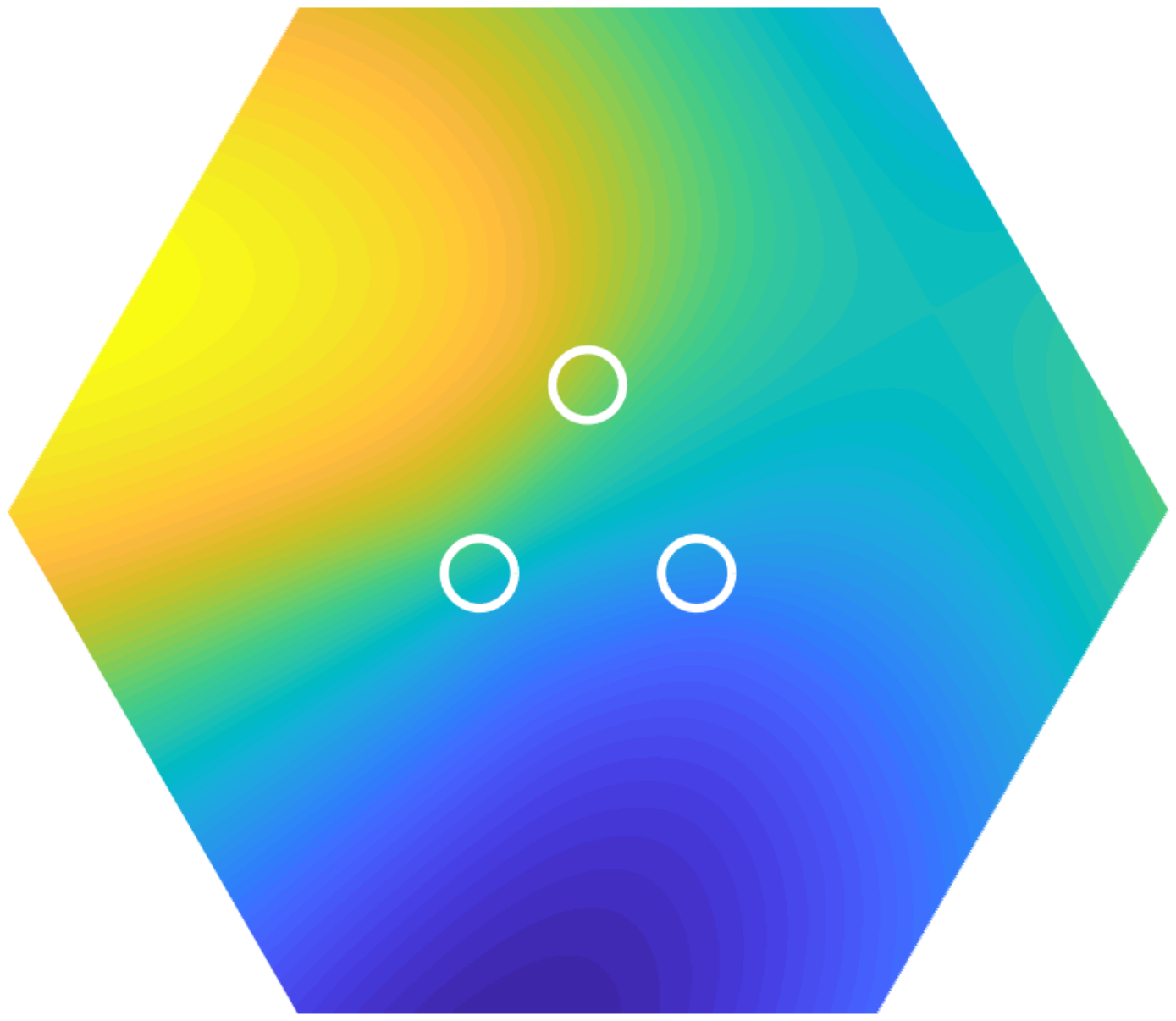}
               };
           }]{};
		\node[below, scale=4.4, black] at (9.0,6.75) {$\displaystyle (g)$};           
\end{scope}  

\begin{scope}[xshift=23.9cm, yshift=44.075cm,scale=0.45]
		\node[regular polygon, regular polygon sides=6,draw, inner sep=6.25cm,rotate=0,line width=0.0mm, white,
           path picture={
               \node[rotate=0] at (-0.45,-0.35){
                   \includegraphics[scale=1.4]{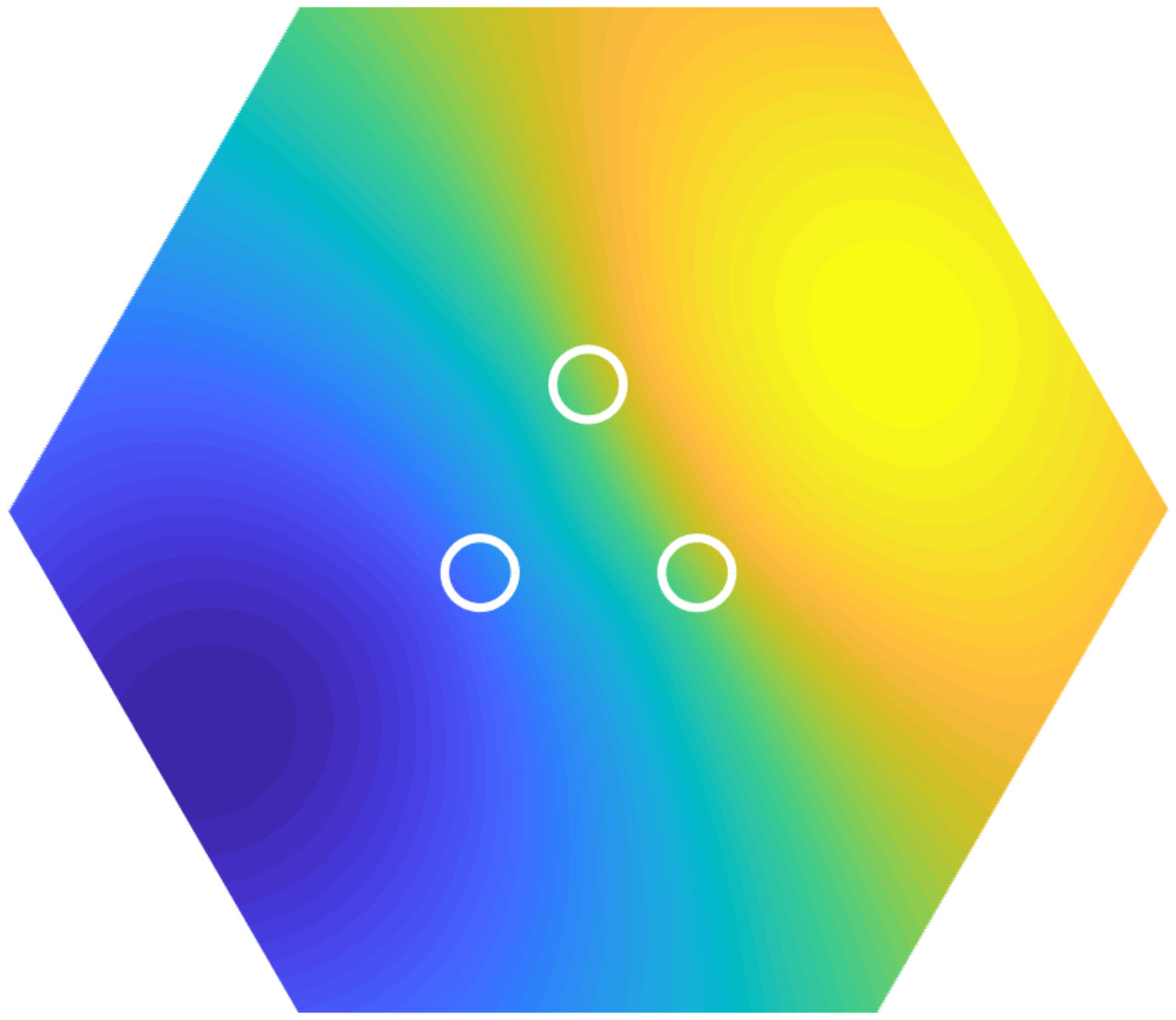}
               };
           }]{};
		\node[below, scale=4.4, black] at (9.0,6.75) {$\displaystyle (h)$};           
\end{scope}

\begin{scope}[xshift=-9.5cm, yshift=45.0cm,scale=1.5]
	\node[regular polygon, regular polygon sides=6, draw, inner sep=0.5*6.28*10.0 pt,rotate=90] at (0 pt,-6.28*10.0*1.5 pt) {};
	\draw[line width=0.5mm,gray,-] (0pt,-6.28*10.0*1.5 pt) -- (0+ 0.5*6.28*10*1.41 pt,-6.28*10.0*1.5 pt);
	\draw[line width=0.5mm,gray,-] (0+ 0.5*6.28*10*1.41 pt,-6.28*10.0*1.5 pt) -- (0+ 0.5*6.28*10*1.41  pt, 0.577*0.5*6.28*10*1.41 -6.28*10.0*1.5 pt);
	\draw[line width=0.5mm,gray,-] (0+ 0.5*6.28*10*1.41  pt, 0.577*0.5*6.28*10*1.41 -6.28*10.0*1.5 pt) -- (0pt,-6.28*10.0*1.5 pt);
	\node[below,left,scale=1.75] at (0pt,-6.28*10.0*1.5 pt) {$\displaystyle  \Gamma$}; 
	\node[below,right,scale=1.75] at (0+ 0.5*6.28*10*1.41 pt,-6.28*10.0*1.5 pt) {$\displaystyle  M$};
	\node[above,right,scale=1.75] at (0+ 0.5*6.28*10*1.41  pt, 0.577*0.5*6.28*10*1.41 -6.28*10.0*1.5 pt) {$\displaystyle  K$};
		\node[below,left,scale=1.75] at (0- 0.5*6.28*10*1.41  pt, -0.577*0.5*6.28*10*1.41 -6.28*10.0*1.5 pt) {$\displaystyle  K'$};
	\node[below,scale=1.333] at (0pt,-0.5*6.28*10*1.41 -6.28*10.0*1.5 -10pt) {$\displaystyle  (b)$}; 
\end{scope}  

\begin{scope}[xshift=-0.5cm, yshift=39.5cm,scale=0.5]
		\node[regular polygon, regular polygon sides=4,draw, inner sep=4.5cm,rotate=0,line width=0.0mm, white,
           path picture={
               \node[rotate=0] at (-0.5,0){
                   \includegraphics[scale=1.25]{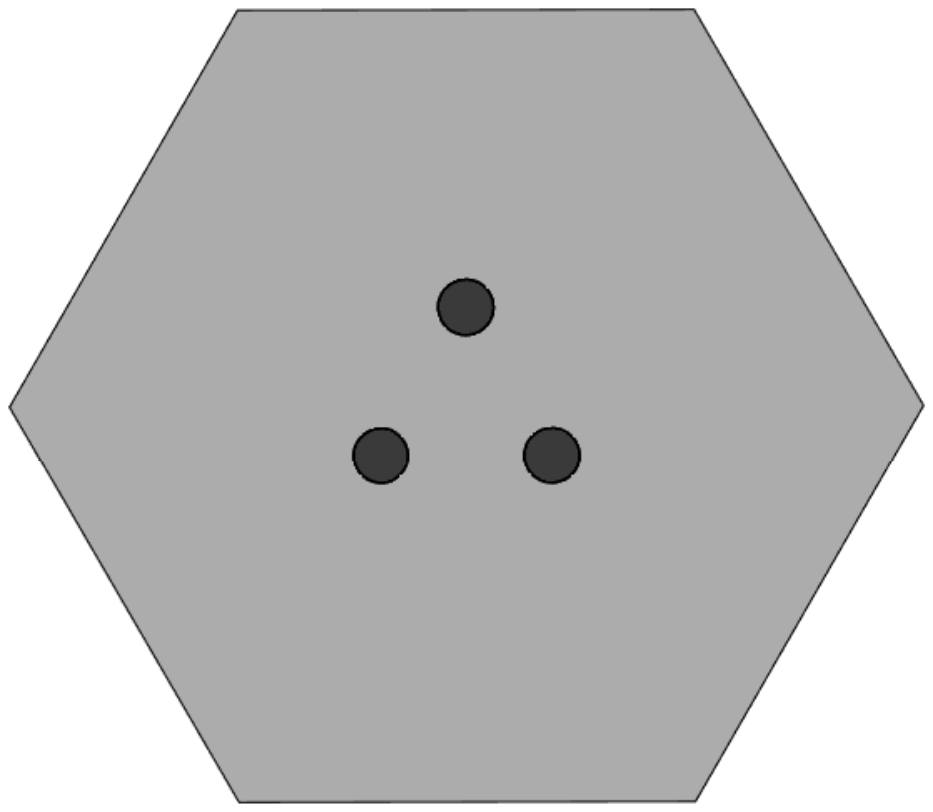}
               };
           }]{};
		\node[below, scale=4, black] at (-0.5,-4.95) {$\displaystyle (c)$};         
\end{scope}

\end{tikzpicture}
\caption{The same as fig. \ref{fig:SquareTopoArrangeDirac}, but with $\boldsymbol{\alpha}_{1} = \cos(\frac{\pi}{6}) \textbf{e}_{x} + \sin(\frac{\pi}{6}) \textbf{e}_{y}$ and $\hat{\ell}_{11} = \hat{\ell}_{12} = \hat{\ell}_{13}  = 11.25$. Here $(a)$ we plot the dimensionless dispersion curves $\Omega = \Omega (\boldsymbol{\kappa})$ from our eigenvalue problem \eqref{EVPactualalgebraicDispBloch} throughout the irreducible Brillouin zone $(b)$, appropriate for the fundamental cell given in $(c)$ (top view) and $(d)$ (side view). Panels $(e)$, $(f)$, $(g)$ and $(h)$ show the eigenmodes from $\boldsymbol{\bigcirc}$, $	\boldsymbol{\square}$, $\color{myRed} \boldsymbol{\square}$ and $\color{myRed} \boldsymbol{\bigcirc}$ in $(a)$ respectively, where the real part of the out-of-plane displacement is plotted in $(e)-(h)$. Here the dimensional frequency of the Dirac point is $0.38398$ Hz, and we expect a similar relative error as in fig. \ref{fig:SquareTopoArrangeDirac}. } 
\label{fig:HexTopoArrangeDirac}
\end{figure}

\begin{figure}[h!]
\centering
\hspace*{1.5cm} 
\begin{tikzpicture}[scale=0.3, transform shape]

\begin{scope}[xshift=14.5cm, yshift=22cm,scale=1.4]
		\node[regular polygon, regular polygon sides=4,draw, inner sep=7cm,rotate=0,line width=0.0mm, white,
           path picture={
               \node[rotate=0] at (1,0){
                   \includegraphics[scale=1.25]{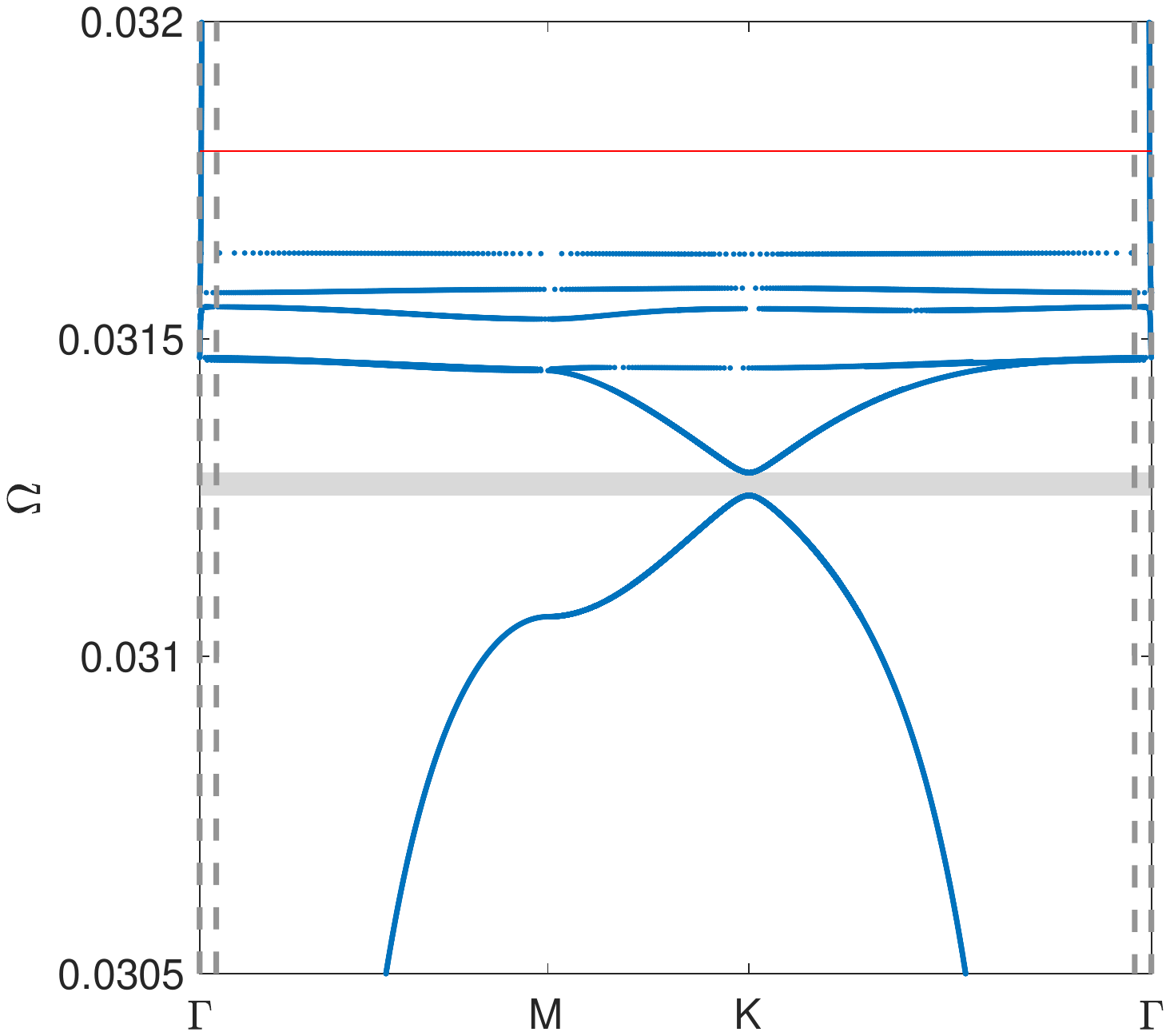}
               };
           }]{};
		\node[below, scale=2.86, black] at (-10,8.75) {$\displaystyle (a)$};           
\end{scope}  

\begin{scope}[xshift=35cm, yshift=27.5cm,scale=0.8]
		\node[regular polygon, regular polygon sides=4,draw, inner sep=6.5cm,rotate=0,line width=0.0mm, white,
           path picture={
               \node[rotate=0] at (2,0){
                   \includegraphics[scale=1.25]{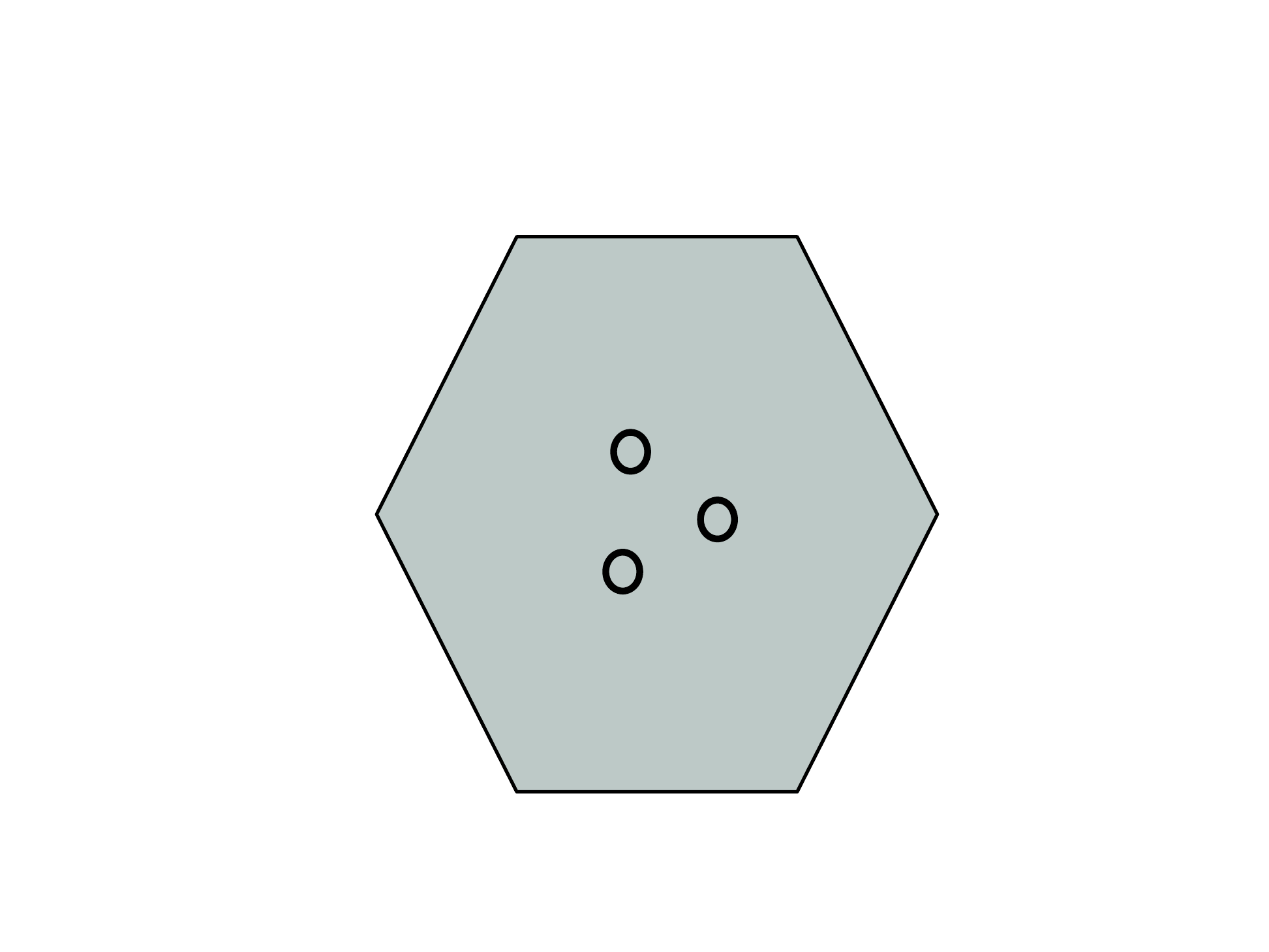}
               };
           }]{};
		\node[below, scale=5, black] at (-6,5.75) {$\displaystyle (b)$};           
\end{scope}

\begin{scope}[xshift=35cm, yshift=14.5cm,scale=0.8]
		\node[regular polygon, regular polygon sides=4,draw, inner sep=6.5cm,rotate=0,line width=0.0mm, white,
           path picture={
               \node[rotate=0] at (2,4){
                   \includegraphics[scale=1.25]{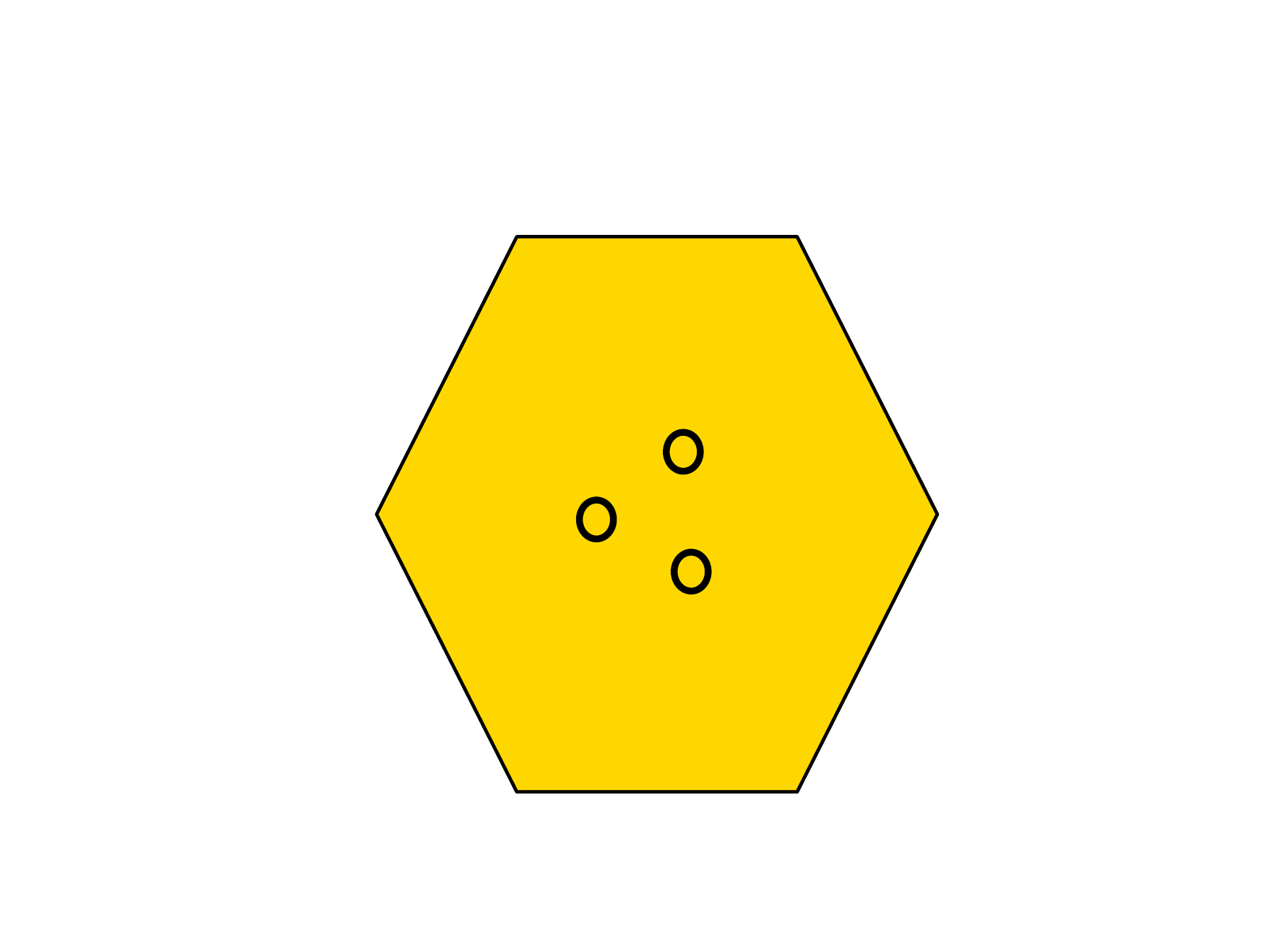}
               };
           }]{};
		\node[below, scale=5, black] at (-6,9.75) {$\displaystyle (c)$};           
\end{scope}  

\begin{scope}[xshift=14.5cm, yshift=-4cm,scale=1.4]
		\node[regular polygon, regular polygon sides=4,draw, inner sep=7cm,rotate=0,line width=0.0mm, white,
           path picture={
               \node[rotate=0] at (-9,0){
                   \includegraphics[scale=1.25]{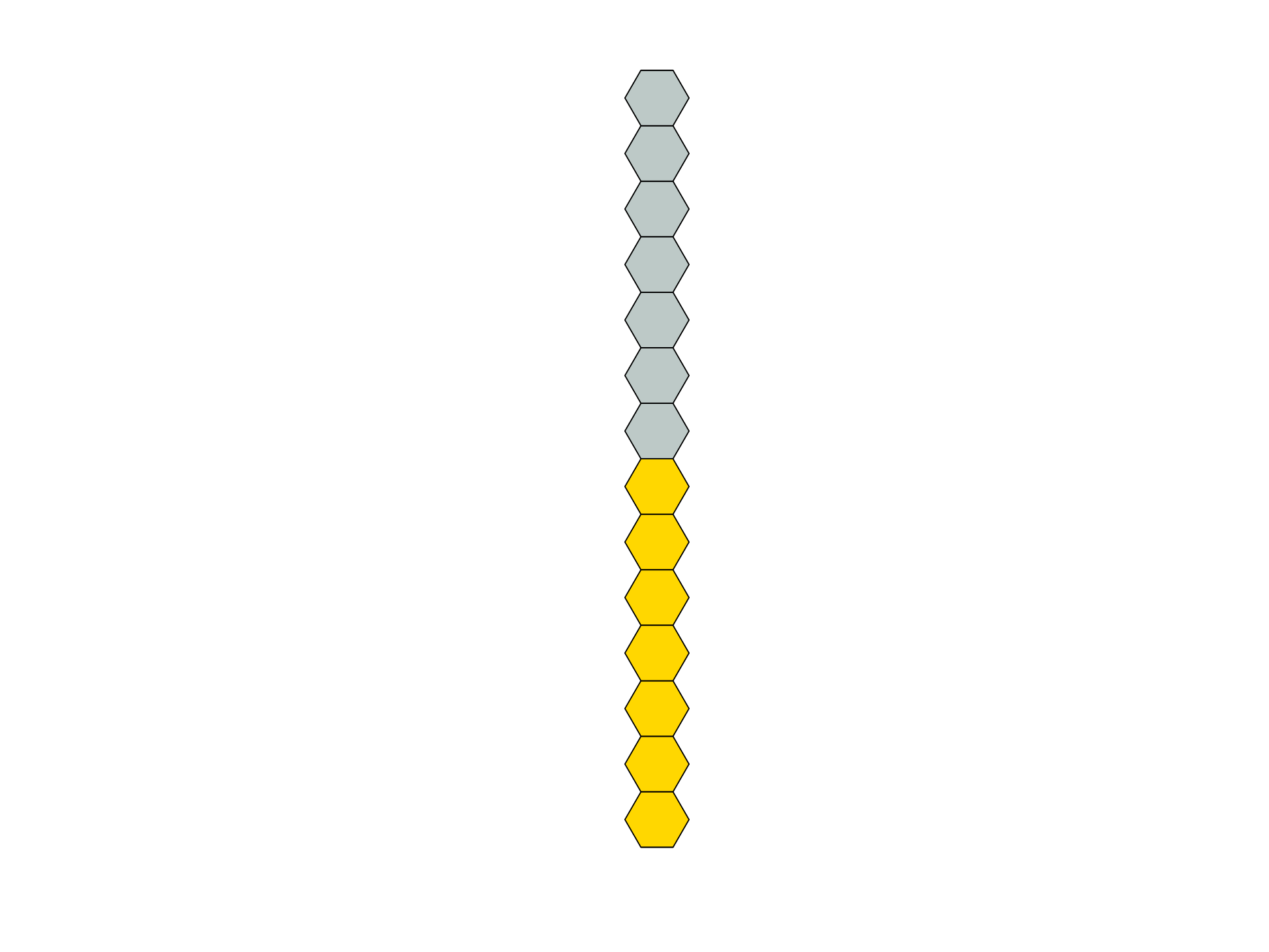}
               };
           }]{};
\end{scope}  

\begin{scope}[xshift=17.5cm, yshift=-4cm,scale=1.4]
		\node[regular polygon, regular polygon sides=4,draw, inner sep=7cm,rotate=0,line width=0.0mm, white,
           path picture={
               \node[rotate=0] at (-9,0){
                   \includegraphics[scale=1.25]{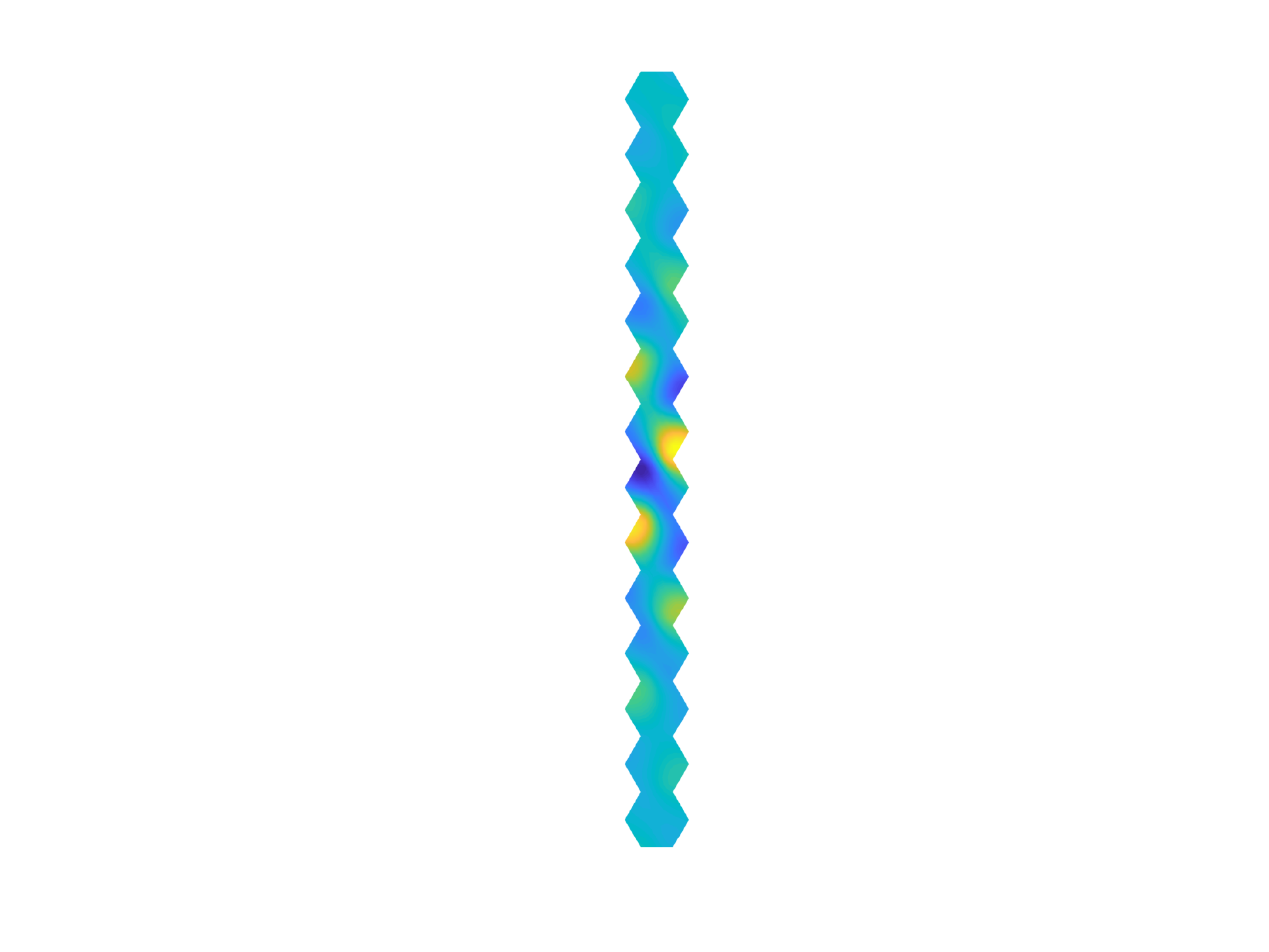}
               };
           }]{};
				\node[below, scale=2.86, black] at (-9.75,-7.5) {$\displaystyle (d)$};           
\end{scope}  

\begin{scope}[xshift=20.5cm, yshift=-4cm,scale=1.4]
		\node[regular polygon, regular polygon sides=4,draw, inner sep=7cm,rotate=0,line width=0.0mm, white,
           path picture={
               \node[rotate=0] at (1.5,0){
                   \includegraphics[scale=1.25]{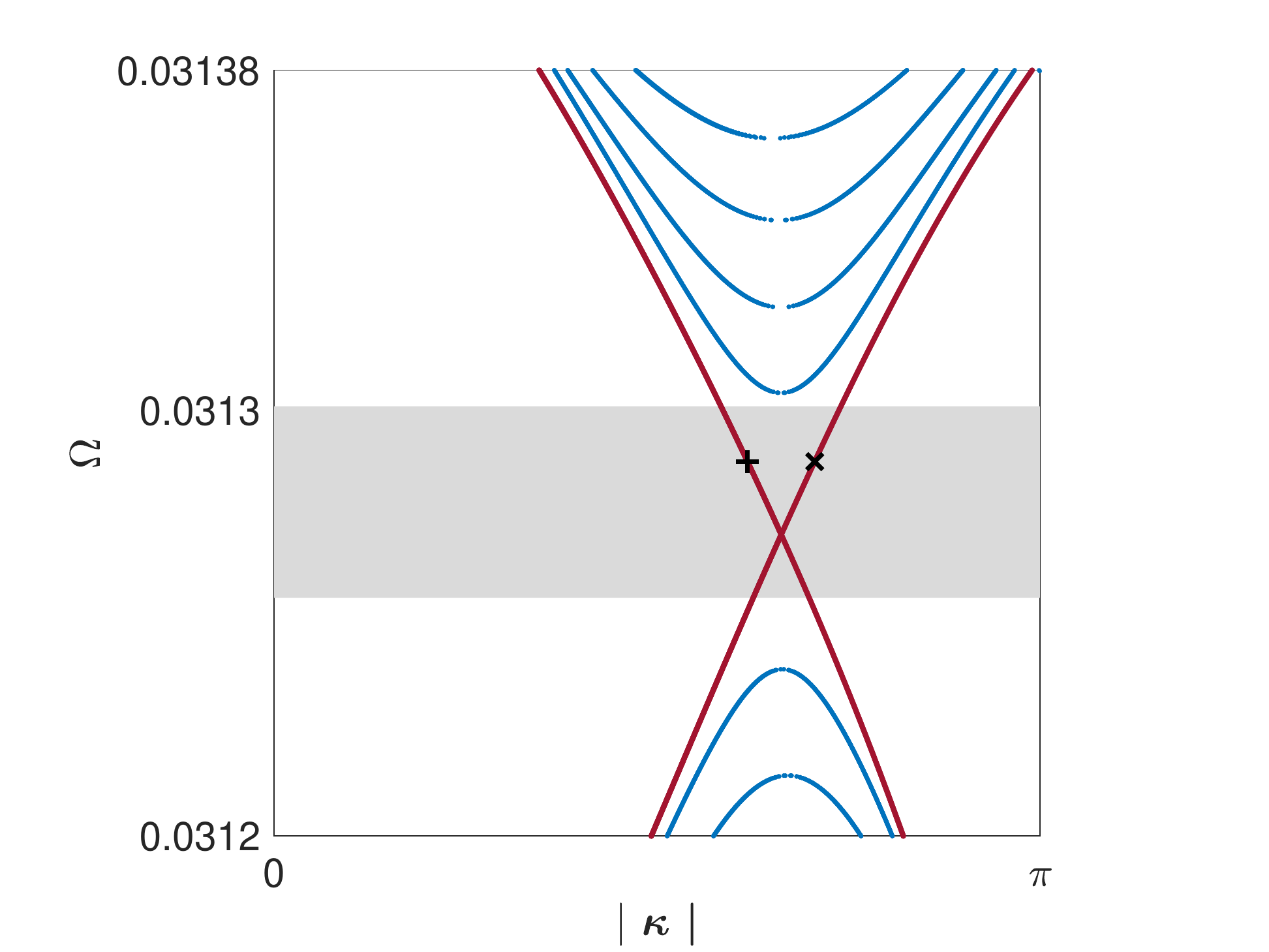}
               };
           }]{};
		\node[below, scale=2.86, black] at (-7,6.75) {$\displaystyle (e)$};           
\end{scope}  

\begin{scope}[xshift=48.5cm, yshift=-4cm,scale=1.4]
		\node[regular polygon, regular polygon sides=4,draw, inner sep=7cm,rotate=0,line width=0.0mm, white,
           path picture={
               \node[rotate=0] at (-9,0){
                   \includegraphics[scale=1.25]{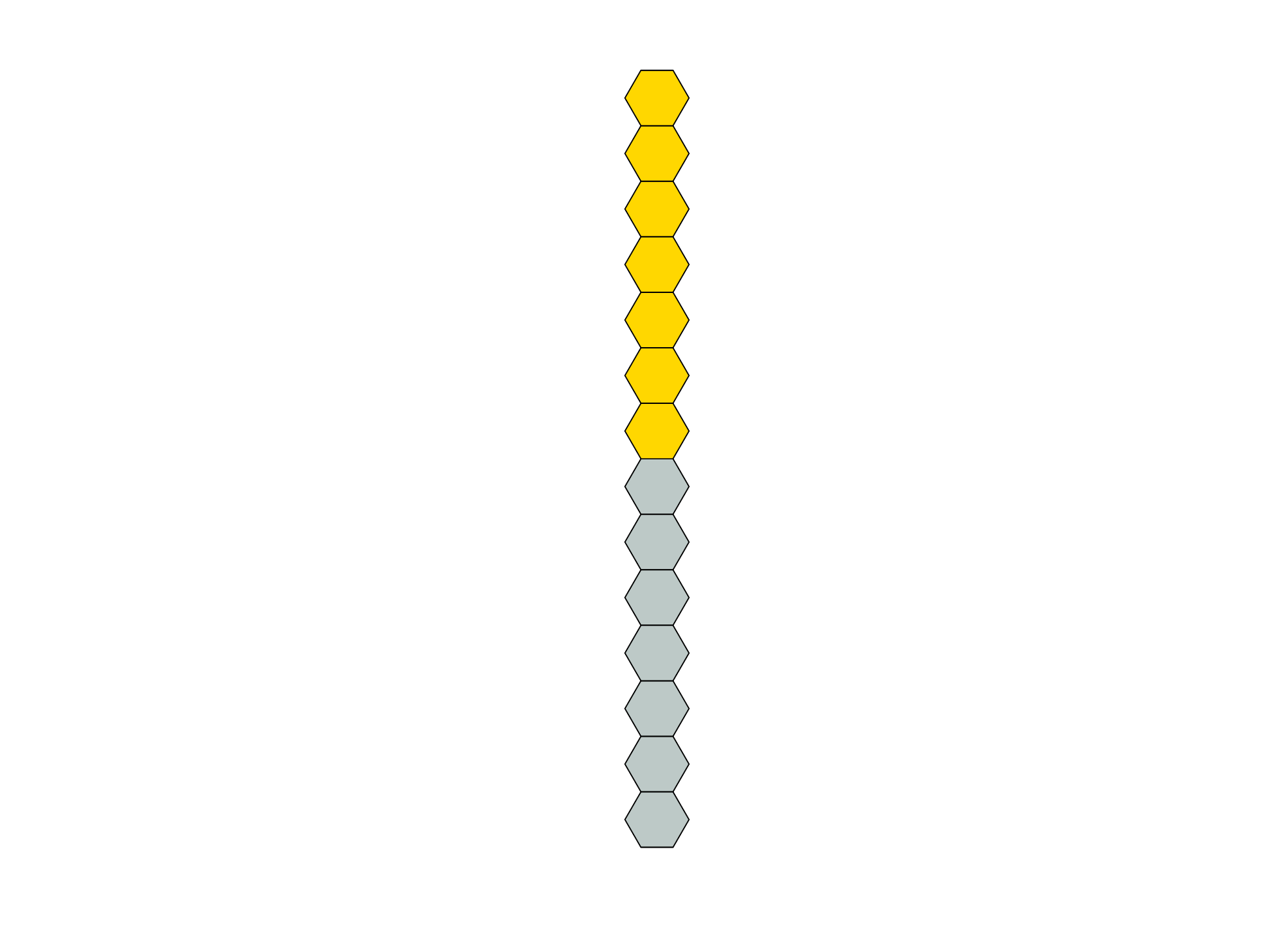}
               };
           }]{};
\end{scope}

\begin{scope}[xshift=51.5cm, yshift=-4cm,scale=1.4]
		\node[regular polygon, regular polygon sides=4,draw, inner sep=7cm,rotate=0,line width=0.0mm, white,
           path picture={
               \node[rotate=0] at (-9,0){
                   \includegraphics[scale=1.25]{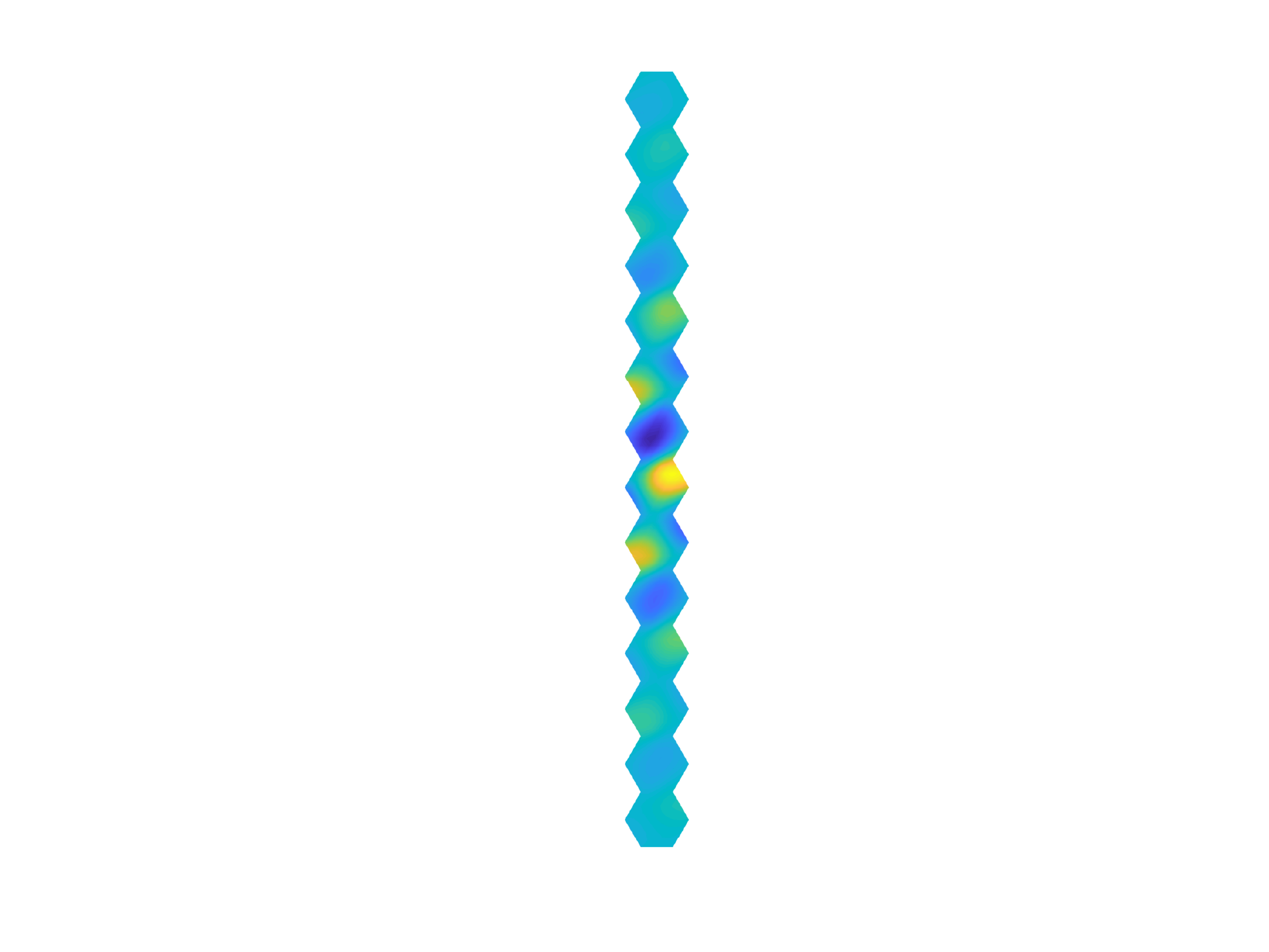}
               };
           }]{};
						\node[below, scale=2.86, black] at (-9.75,-7.5) {$\displaystyle (f)$}; 
\end{scope}  

\begin{scope}[xshift=54.5cm, yshift=-4cm,scale=1.4]
		\node[regular polygon, regular polygon sides=4,draw, inner sep=7cm,rotate=0,line width=0.0mm, white,
           path picture={
               \node[rotate=0] at (-10.25,0){
                   \includegraphics[scale=1.25]{Figs/colourBar.pdf}
               };
           }]{};
\end{scope}

\end{tikzpicture}
\caption{The Floquet-Bloch dispersion curves $(a)$ from the primitive cell $(b)$. Cells $(b)$ \& $(c)$ are obtained by perturbing the arrangement shown in fig. \ref{fig:HexTopoArrangeDirac} $(c)$ via rotations - setting $\textbf{X}_{1J} = 0.125\left[ \cos(\frac{2\pi (J-1)}{3} + \frac{\pi}{2} + \theta') \textbf{e}_{x} + \sin(\frac{2\pi (J-1)}{3} + \frac{\pi}{2} + \theta')  \textbf{e}_{y}\right]$. The grey $(b)$ and yellow $(c)$ cells respectively correspond to positive and negative rotations by setting $\theta' = \pm \frac{\pi}{7}$. The topological band gap is shaded in grey in $(a)$. Panels $(d)$ and $(f)$ show ribbon media, formed by stacking the cells in $(b)$ and $(c)$ as shown. The dispersion relations of these ribbon media, from our Floquet-Bloch analysis \eqref{EVPactualalgebraicDispBloch}, are plotted in $(e)$ where the band gap from panel $(a)$ is similarly shaded in $(e)$. The interfacial modes of interest are plotted in red in $(e)$, the even $(d)$ and odd $(f)$ ZLMs are plotted next to the ribbon media in which they persist;  the eigenstates in $(d)$ and $(f)$ correspond to the  points $\boldsymbol{+}$ and  $\boldsymbol{\times}$ in panel $(e)$ respectively.} 
\label{fig:PertHexArrange}
\end{figure}

\begin{table}[H]
\centering
\begin{tabular}{c|c c c|c}
\cline{2-4} & \multicolumn{3}{ c| }{Classes} \\ 
	\hline
\multicolumn{1}{ ||c|  }{IRs} & $E$ & $2C_{3}$ & $3\sigma_{v}$ & \multicolumn{1}{ |c||  }{Basis} \\
	\hline\hline
	\multicolumn{1}{ ||c|  }{$E$} & $+2$	 & $-1$ & $0$ &	\multicolumn{1}{ |c||  }{$\left\lbrace x , y \right\rbrace$} \\
	\hline	 
\end{tabular}
\caption{Excerpt of the $C_{3v}$ character table, only the linear basis functions spanning the irreducible representations (IRs) are required.} 
\label{table:C3VCharacter}
\end{table}

The hexagonal primitive cell in fig. \ref{fig:HexTopoArrangeDirac} $(c)$ has $3 \sigma_{v}$ spatial symmetries which induce protected degeneracies at the $KK'$ locations of the first Brillouin zone, provided $\lbrace G_{\Gamma}, G_{K, K'} \rbrace = \lbrace C_{3v}, C_{3v} \rbrace$ \cite{makwana2018geometrically,makwana2020hybrid}. Here $C_{3v} = C_{3} + 3 \sigma_{V}$ and $G_{K, K'}$ represents $G_{\boldsymbol{\kappa}}$ at the $K$ and $K'$ high symmetry points. Figs \ref{fig:HexTopoArrangeDirac} $(e)$ \& $(f)$ respectively correspond to the eigenstates at points $\boldsymbol{\bigcirc}$ \& $\boldsymbol{\square}$ in fig. \ref{fig:HexTopoArrangeDirac} $(a)$, observe the eigenmodes $\boldsymbol{\square}$ \& $\boldsymbol{\bigcirc}$ respectively match the $E$  basis functions $x$ and $y$ from the $C_{3v}$ character table \ref{table:C3VCharacter}. The states at fig. \ref{fig:HexTopoArrangeDirac} $(a)$ $\boldsymbol{\square}$ \& $\boldsymbol{\bigcirc}$ exist on the dispersionless sections approaching the crossing at $K$ and hence are compatible with $G_{K}$, and by time-reversal-symmetry $G_{K'}$ - therefore $G_{K, K'} = C_{3v}$. It follows $G_{\Gamma} = C_{3v}$, indeed since $G_{K} \le G_{\Gamma}$.

Furthermore, applying the compatibility relations within Sakoda \cite{sakoda2004optical}, we confirm the band following fig. \ref{fig:HexTopoArrangeDirac} $(a)$ points $\boldsymbol{\bigcirc}$ \& $\color{myRed} \boldsymbol{\bigcirc}$ also has irreducible representations corresponding to the $E$ basis and is even with respect to the symmetry line following the Bloch momentum wave vector, $\boldsymbol{\kappa}$, defining each state. Similarly, the states along the band following $\boldsymbol{\square}$ \& $\color{myRed} \boldsymbol{\square}$ are odd and match the $E$ basis. Therefore, the well ordered and opposite parity eigenmodes, with the correct symmetry along branches, allows the crossing to exist at $K$ and confirms its symmetry protected nature.
 
Similarly to section \ref{TopoSquareLattice}, we rotate the arrangement of beams in fig. \ref{fig:HexTopoArrangeDirac} $(c)$ to reduce the symmetry set: from $\lbrace G_{\Gamma}, G_{K,K'} \rbrace = \lbrace C_{3v}, C_{3v} \rbrace $ to $\lbrace G_{\Gamma}, G_{K,K'} \rbrace = \left\lbrace  C_{3}, C_{3} \right\rbrace $ in fig. \ref{fig:PertHexArrange} $(b)$ \& $(c)$ by $\sigma_{v}$ symmetry breaking. Again, the perturbation affects the well ordered parities of eigenstates along the branches near the crossing, causing band repulsion and gapping the $K K'$ Dirac points, as seen in fig. \ref{fig:PertHexArrange} $(a)$. Observe the creation of ZLMs within figs \ref{fig:PertHexArrange} $(d)$ and $(f)$, as described in fig. \ref{fig:BerryCurvature} $(d)$. Furthermore, when breaking $3 \sigma_{v}$ symmetries within a hexagonal lattice we expect \cite{makwana19a,makwana2020hybrid} even and odd ZLMs existing over two distinct interfaces, as observed in figs \ref{fig:PertHexArrange} $(d)$ and $(f)$ - now following the $\Gamma K$ direction of $\boldsymbol{\kappa}$. 

\begin{figure}[H]
\centering
\hspace*{-1.5cm} 
\begin{tikzpicture}[scale=0.29, transform shape]

\begin{scope}[xshift=14.5cm, yshift=22cm,scale=1.5]
		\node[regular polygon, regular polygon sides=4,draw, inner sep=6.5cm,rotate=0,line width=0.0mm, white,
           path picture={
              \node[rotate=0] at (-2,0.5){
                   \includegraphics[scale=1.25]{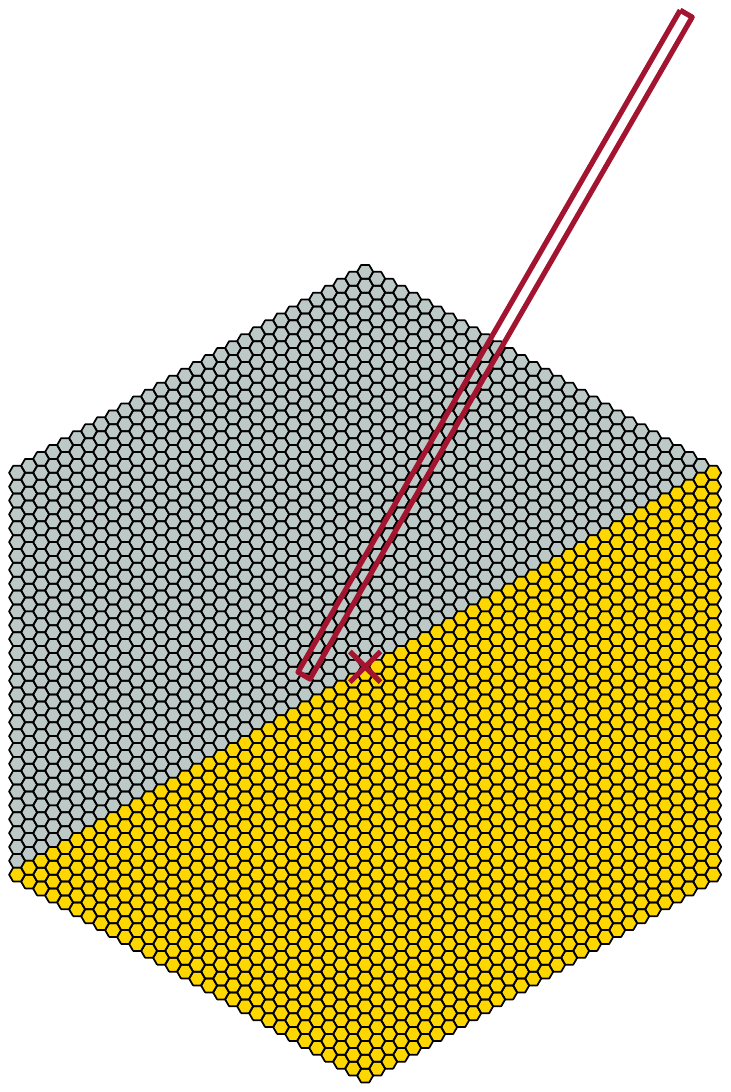}
               };    
           }]{};
		\node[below, scale=2.67, black] at (-8.5,8.75) {$\displaystyle (a)$};           
\end{scope}

\begin{scope}[xshift=40cm, yshift=22cm,scale=1.4]
		\node[regular polygon, regular polygon sides=4,draw, inner sep=5.5cm,rotate=0,line width=0.0mm, white,
           path picture={
               \node[rotate=0] at (-0.5,-0.25){
                   \includegraphics[scale=1.25]{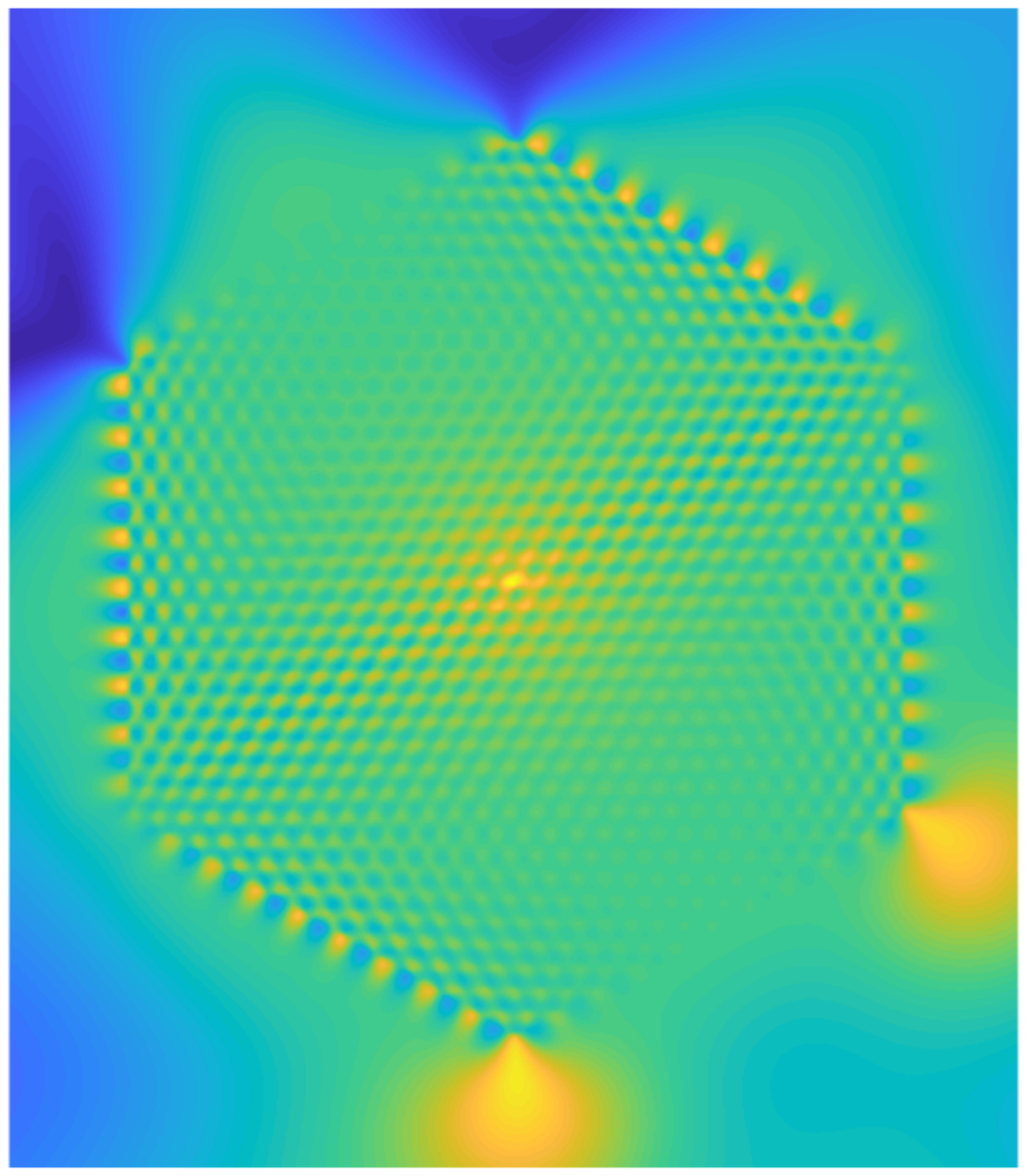}
               };
           }]{};
		\node[below, scale=2.86, black] at (-8.5,8.75) {$\displaystyle (b)$};           
\end{scope}

\begin{scope}[xshift=14.5cm, yshift=-2cm,scale=1.4]
               \node[rotate=0] at (-2,-0.25){
                   \includegraphics[scale=1.25]{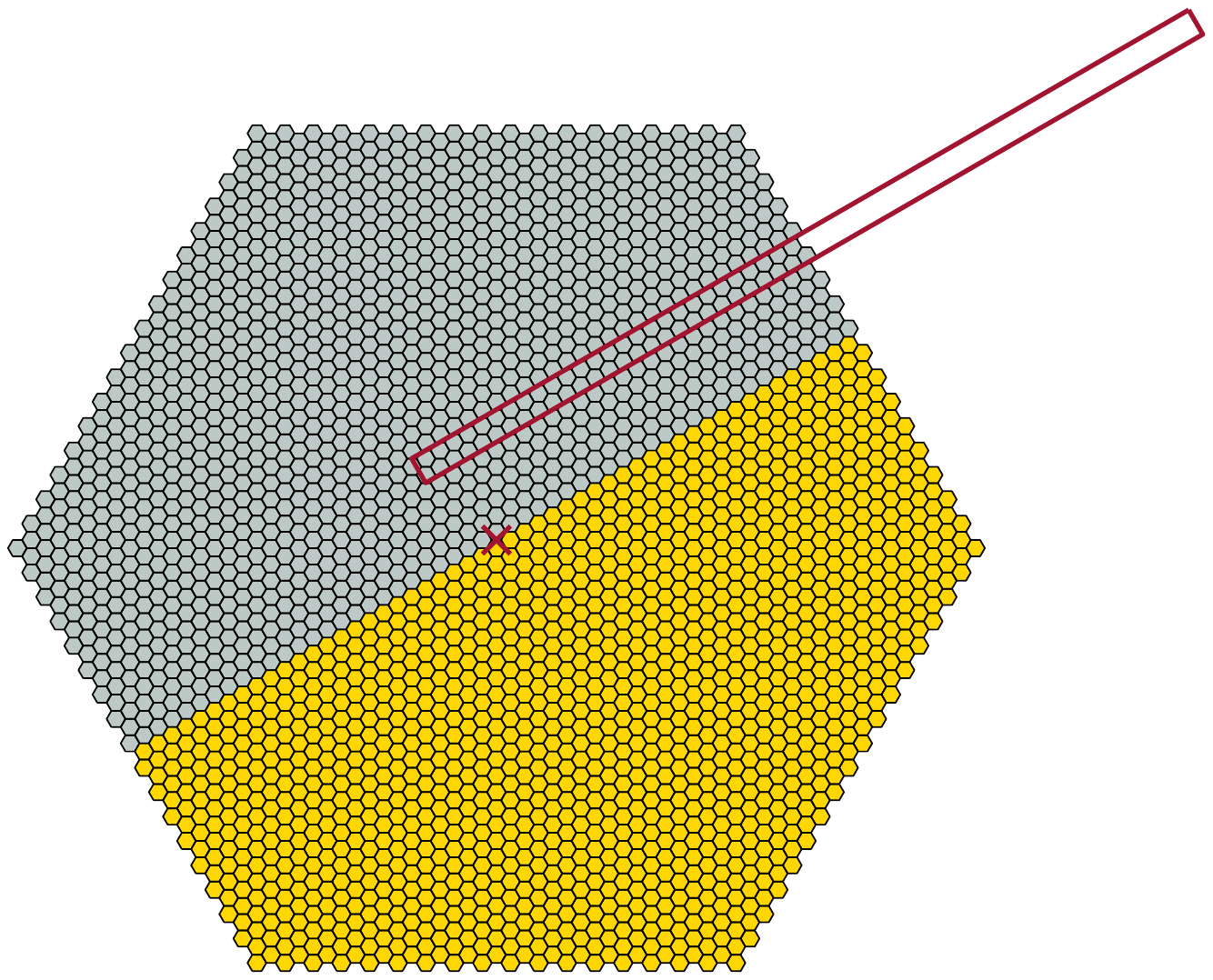}
               };
		\node[below, scale=2.86, black] at (-8.5,8.75) {$\displaystyle (c)$};           
\end{scope}

\begin{scope}[xshift=40cm, yshift=-2cm,scale=1.4]
		\node[regular polygon, regular polygon sides=4,draw, inner sep=5.5cm,rotate=0,line width=0.0mm, white,
           path picture={
               \node[rotate=0] at (-0.5,-0.25){
                   \includegraphics[scale=1.25]{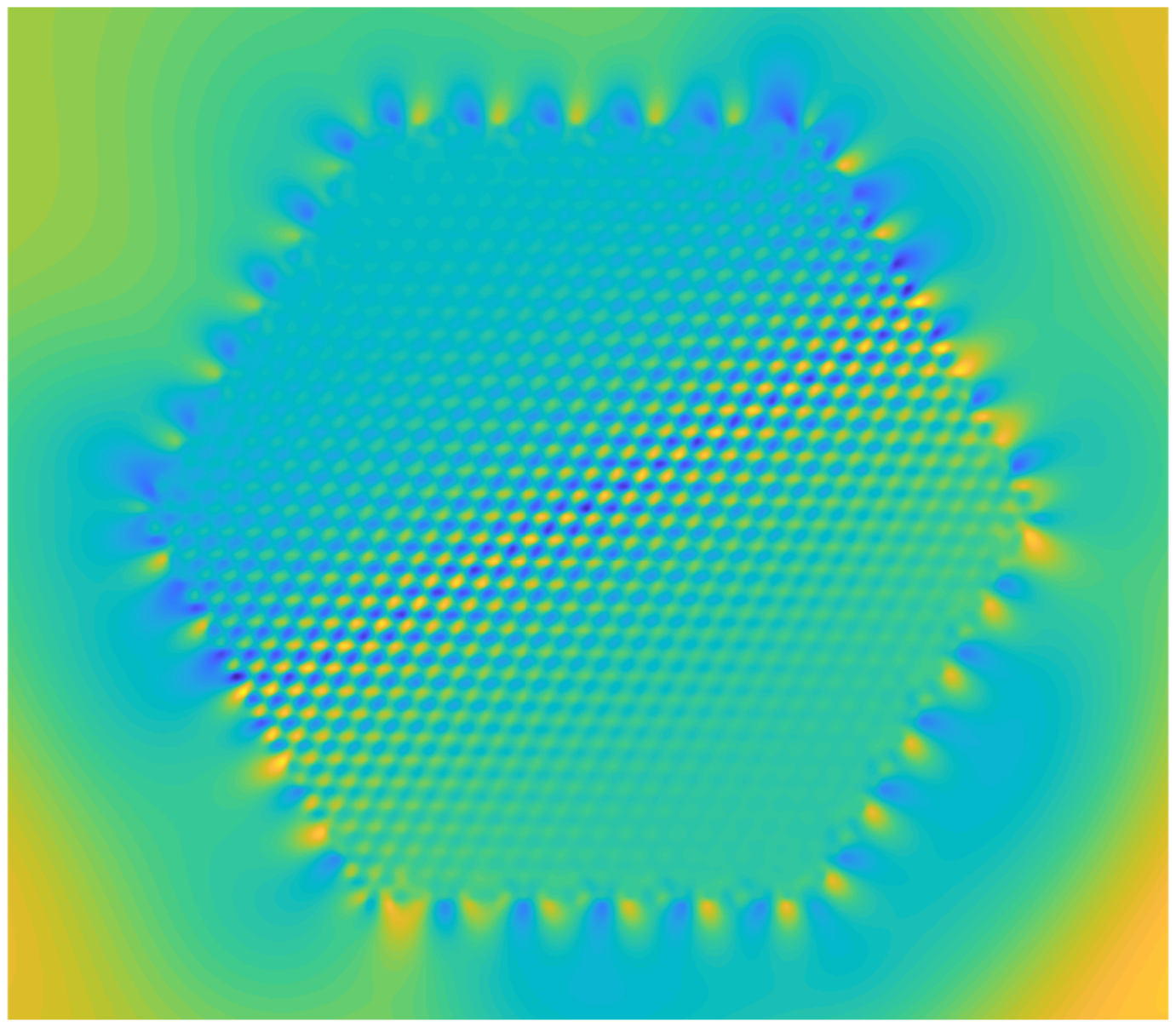}
               };
           }]{};
		\node[below, scale=2.86, black] at (-8.5,8.75) {$\displaystyle (d)$};           
\end{scope}  

\begin{scope}[xshift=65cm, yshift=9cm,scale=1.4]
		\node[regular polygon, regular polygon sides=4,draw, inner sep=7cm,rotate=0,line width=0.0mm, white,
           path picture={
               \node[rotate=0] at (-17.0,0){
                   \includegraphics[scale=1.25]{Figs/colourBarScattRe.pdf}
               };
           }]{};
\end{scope}

\begin{scope}[xshift=10.75cm, yshift=-26cm,scale=1.2]
		\node[regular polygon, regular polygon sides=4,draw, inner sep=7cm,rotate=0,line width=0.0mm, white,
           path picture={
               \node[rotate=0] at (-9,0){
                   \includegraphics[scale=1.25]{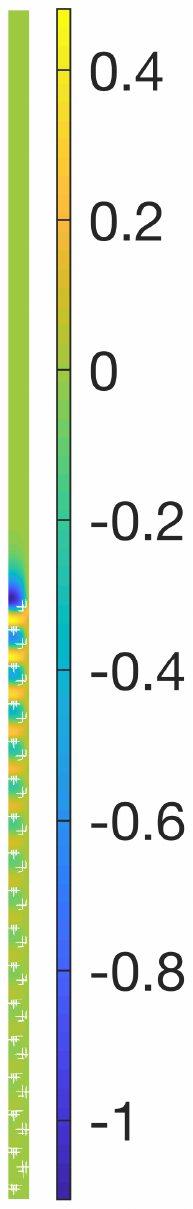}
               };
           }]{};
				\node[below, scale=2.45, black] at (-9.75,-7.5) {$\displaystyle (e)$};           
\end{scope}  

\begin{scope}[xshift=13.75cm, yshift=-26.5cm,scale=1.2]
		\node[regular polygon, regular polygon sides=4,draw, inner sep=7cm,rotate=0,line width=0.0mm, white,
           path picture={
               \node[rotate=0] at (1.5,0){
                   \includegraphics[scale=1.25]{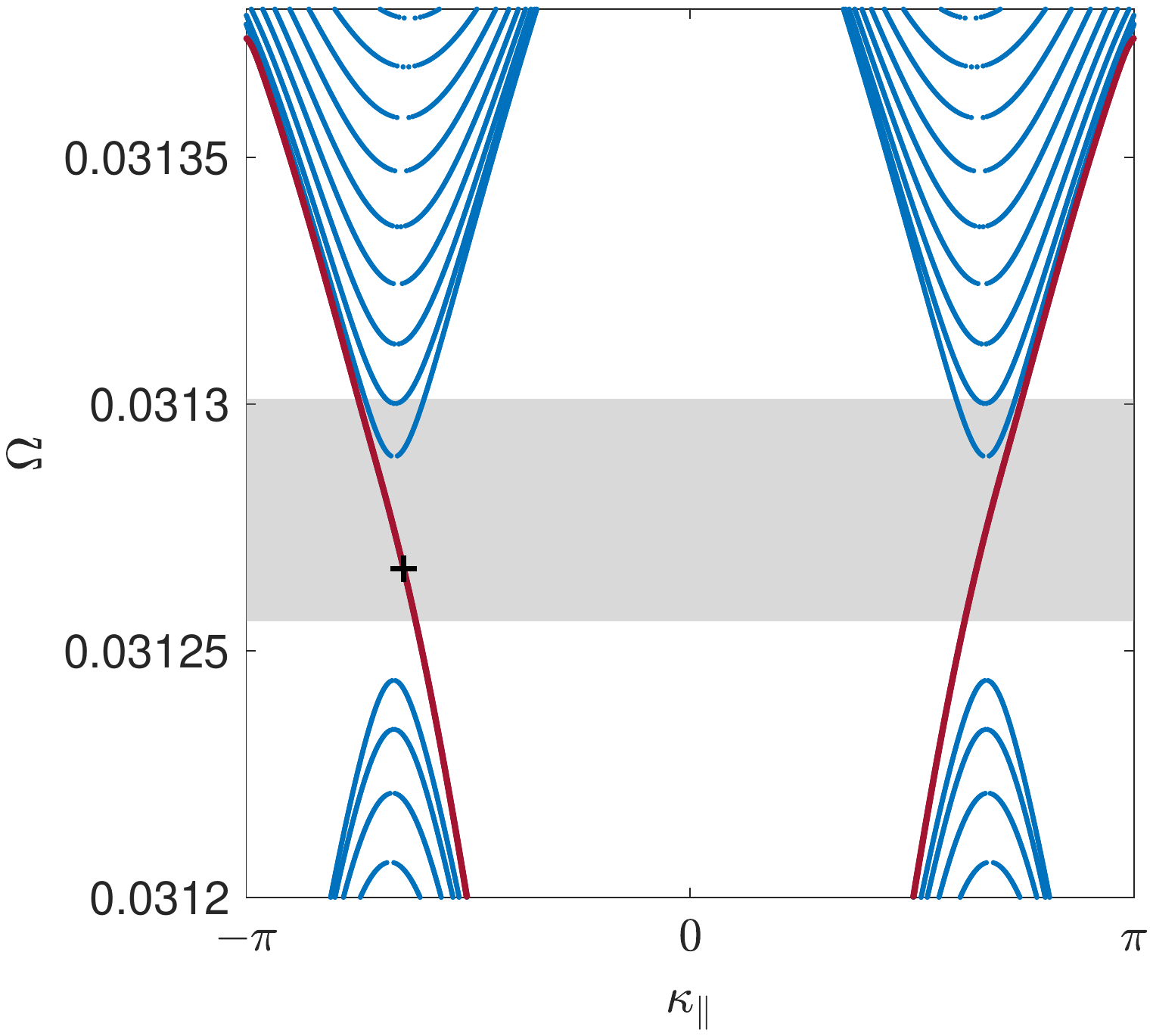}
               };
           }]{};
		\node[below, scale=2.45, black] at (-6.5,8.75) {$\displaystyle (f)$};           
\end{scope}  

\begin{scope}[xshift=38.0cm, yshift=-26cm,scale=1.2]
		\node[regular polygon, regular polygon sides=4,draw, inner sep=7cm,rotate=0,line width=0.0mm, white,
           path picture={
               \node[rotate=0] at (-9,0){
                   \includegraphics[scale=1.25]{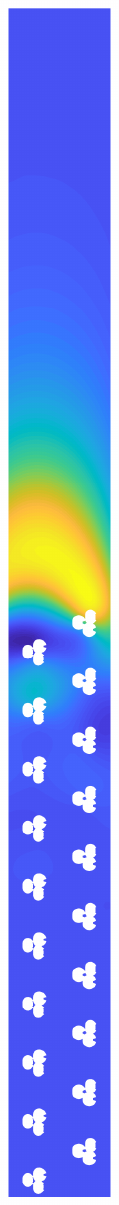}
               };
           }]{};
				\node[below, scale=2.45, black] at (-8.75,-7.5) {$\displaystyle (g)$};           
\end{scope} 

\begin{scope}[xshift=40.5cm, yshift=-26.5cm,scale=1.2]
		\node[regular polygon, regular polygon sides=4,draw, inner sep=7cm,rotate=0,line width=0.0mm, white,
           path picture={
               \node[rotate=0] at (1.5,0){
                   \includegraphics[scale=1.25]{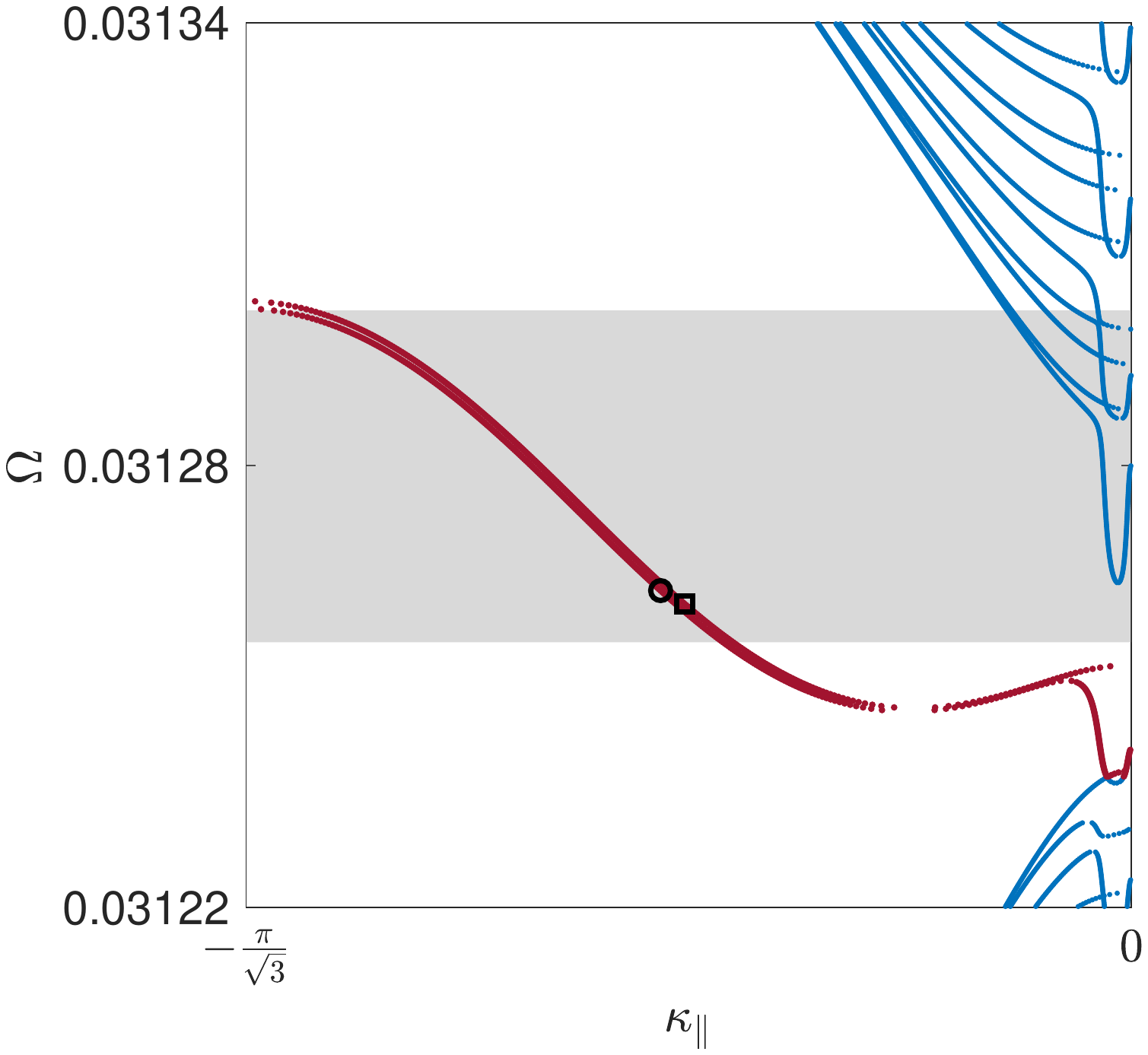}
               };
           }]{};
		\node[below, scale=2.45, black] at (-7,6.75) {$\displaystyle (h)$};           
\end{scope}  

\begin{scope}[xshift=63.5cm, yshift=-26.5cm,scale=1.2]
		\node[regular polygon, regular polygon sides=4,draw, inner sep=6.5cm,rotate=0,line width=0.0mm, white,
           path picture={
               \node[rotate=0] at (-9.25,0){
                   \includegraphics[scale=1.25]{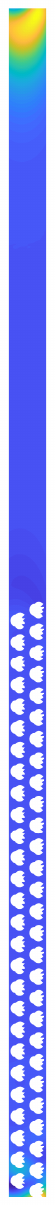}
               };
           }]{};
\end{scope}  

\begin{scope}[xshift=64.5cm, yshift=-26.5cm,scale=1.2]
		\node[regular polygon, regular polygon sides=4,draw, inner sep=6.5cm,rotate=0,line width=0.0mm, white,
           path picture={
               \node[rotate=0] at (-9,0){
                   \includegraphics[scale=1.25]{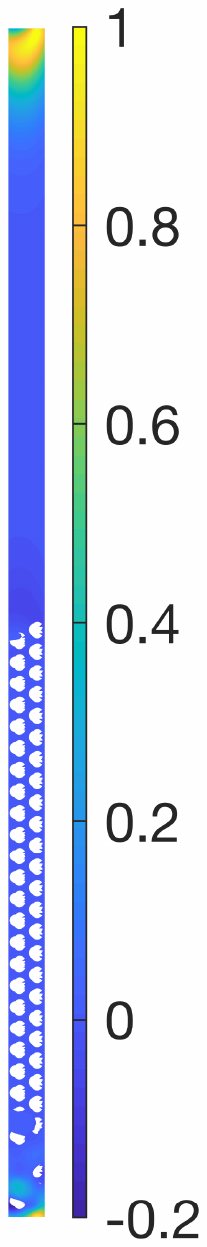}
               };
           }]{};
          \node[below, scale=2.45, black] at (-9.5,-7.5) {$\displaystyle (i)$};           
\end{scope}

\end{tikzpicture}
\caption{Generalised Foldy simulations $(b)$ and $(d)$ for the phononic crystal designs respectively shown in the schematics of panels $(a)$ and $(c)$; here $(a)$ and $(c)$ are respectively formed from $2611$ and $2653$ cells, with (nearly) half grey and half yellow cells from fig. \ref{fig:PertHexArrange} $(b)$ \& $(c)$ - for a total of $7833$ and $7959$ beams atop an elastic plate of infinite expanse. The incident sources, in $(b)$ and $(d)$, consider monopoles with $\varpi_{\mathrm{inc}}=1$ from \eqref{ForcePOPinc} and $\textbf{X}_{\mathrm{inc}}$ is marked by $\color{myRed} \boldsymbol{\times}$ in $(a)$ and $(c)$. The frequencies of the point-sources both lie within the band gaps from fig. \ref{fig:PertHexArrange} $(a)$,  and are set to $\Omega = 0.0312587$ for $(b)$ and $\Omega=0.0312852$ for $(d)$. The red rectangular strips in schematics $(a)$ and $(c)$ represent ribbon strips, half over the bare plate and half over the grey medium; panels $(f)$ and $(h)$ show the dispersion curves taken by considering Floquet-Bloch boundary conditions (from \eqref{EVPactualalgebraicDispBloch}) on the edges of these strips, the edge modes of interest are plotted in red and the bulk band gap for the constituent cells is again highlighted in grey from fig. \ref{fig:PertHexArrange} (a). Panel $(e)$ corresponds to the eigenmode at $\boldsymbol{+}$ within panel $(f)$, similarly $(g)$ and $(i)$ respectively correspond to eigenmodes from points $\boldsymbol{\bigcirc}$ (on left-most branch) and $\boldsymbol{\square}$ (on right-most branch) in panel $(h)$, the locations of the beams are shown by white circles in panels $(e)$, $(g)$ and $(i)$. The colour bars refers to the normalised displacement fields next to the computations they exist - where the real part of the out-of-plane displacement field has been plotted. Panels $(b)$, $(d)$ share the same colour bar. Also $(g)$ and $(i)$ share the same colour bar. } 
\label{fig:ScattPertHexTopoArrange}
\end{figure}

There are two types of interfaces which can be constructed by tessellations of hexagonal cells; zigzag or armchair interfaces, where zigzag interfaces are preferred for the construction of ZLMs since they offer greater topological protection \cite{makwana2018designing}. Figs \ref{fig:ScattPertHexTopoArrange} $(a)$ \& $(c)$ show two phononic crystals, both with zigzag interfaces supporting ZLMs,  with zigzag $(a)$ or armchair $(c)$ edges. In contrast to square lattice, fig. \ref{fig:ScattSquareTopoArrange}, the topologically non-trivial bulk derived from hexagonal primitive cells  meeting the free-space always produces edges with non-zero berry curvature - as in fig. \ref{fig:BerryCurvature} $(c)$. Therefore, by the bulk-edge correspondence we expect topologically protected edgestates to exist for both of the designs within fig. \ref{fig:ScattPertHexTopoArrange} $(a)$ \& $(c)$. 

The eigenmodal analysis in figs. \ref{fig:ScattPertHexTopoArrange} $(e)$-$(i)$ confirms the existence of these topologically non-trivial edgestates, where we see modes which decay rapidly into the topological bulk - localized by the existence of chiral flux by the QVHE. Again we operate far beneath the free-space `light' lines and observe modes confined to the edges of the crystal. In contrast to the ZLMs, the armchair edgestates (in figs. \ref{fig:ScattPertHexTopoArrange} $(g)$-$(i)$) offer a greater topological protection than the zigzag edgestates (figs. \ref{fig:ScattPertHexTopoArrange} $(e)$ \& $(f)$); the armchair edge produces two modes existing along two distinct edges within the ribbon, with much flatter bands and eigenmodes with superior localization of energy than the sole mode existing for the zigzag edge.

The topological protection of the ZLMs and edgestates is confirmed by the scattering simulations in figs. \ref{fig:ScattPertHexTopoArrange} $(b)$ \& $(d)$. Both of which show the ZLMs corresponding to the interfacial states of fig. \ref{fig:PertHexArrange} $(d)$, which efficiently allow navigation of energy at the junction between the interface and edges via modal conversion of states; the topological protection of which accounts for the efficiency of the conversion, around sharp corners, with negligible backscatter and strong localisation along the topological interfaces and edges within the phononic crystal. The two distinct edgemodes for a sole chirality in figs \ref{fig:ScattPertHexTopoArrange} $(g)$ \& $(i)$ readily couple into one another - as a consequence of their minimal Fourier separation \cite{makwana2018designing} in fig. \ref{fig:ScattPertHexTopoArrange} $(h)$ - therefore these modes are expected to efficiently navigate around the corners of the perimeter of the phononic crystal in fig. \ref{fig:ScattPertHexTopoArrange} $(c)$. In fig. \ref{fig:ScattPertHexTopoArrange} $(d)$ we observe conversion between the ZLMs and edgestates, which efficiently propagate around the entire perimeter and hence fig. \ref{fig:ScattPertHexTopoArrange} $(c)$ is truly a sub-wavelength phononic circuit whose efficiency and robustness is guaranteed by topological protection. 

The same cannot be said of the design in fig. \ref{fig:ScattPertHexTopoArrange} $(a)$. The ZLMs and edgestates within the design are topologically protected, and energy similarly navigates the corners by converting between interfacial and edgestates, where backscatter is still protected against. However, since only one topological edgestate exists the energy cannot propagate around the entire perimeter of the phononic crystal and must be shed at its vertices. This is observed in fig. \ref{fig:ScattPertHexTopoArrange} $(b)$ where corner states shed energy into the free-space.

\begin{figure}[H]
\centering
\hspace*{0.25cm} 
\begin{tikzpicture}[scale=0.29, transform shape]
\begin{scope}[xshift=14.5cm, yshift=-18cm,scale=1.4]
	\node[regular polygon, regular polygon sides=4,draw, inner sep=7.5cm,rotate=0,line width=0.0mm, black, opacity=0.0,
           path picture={
               \node[rotate=0,opacity=1.0] at (-0.5,-2.5){
                   \includegraphics[scale=1.25]{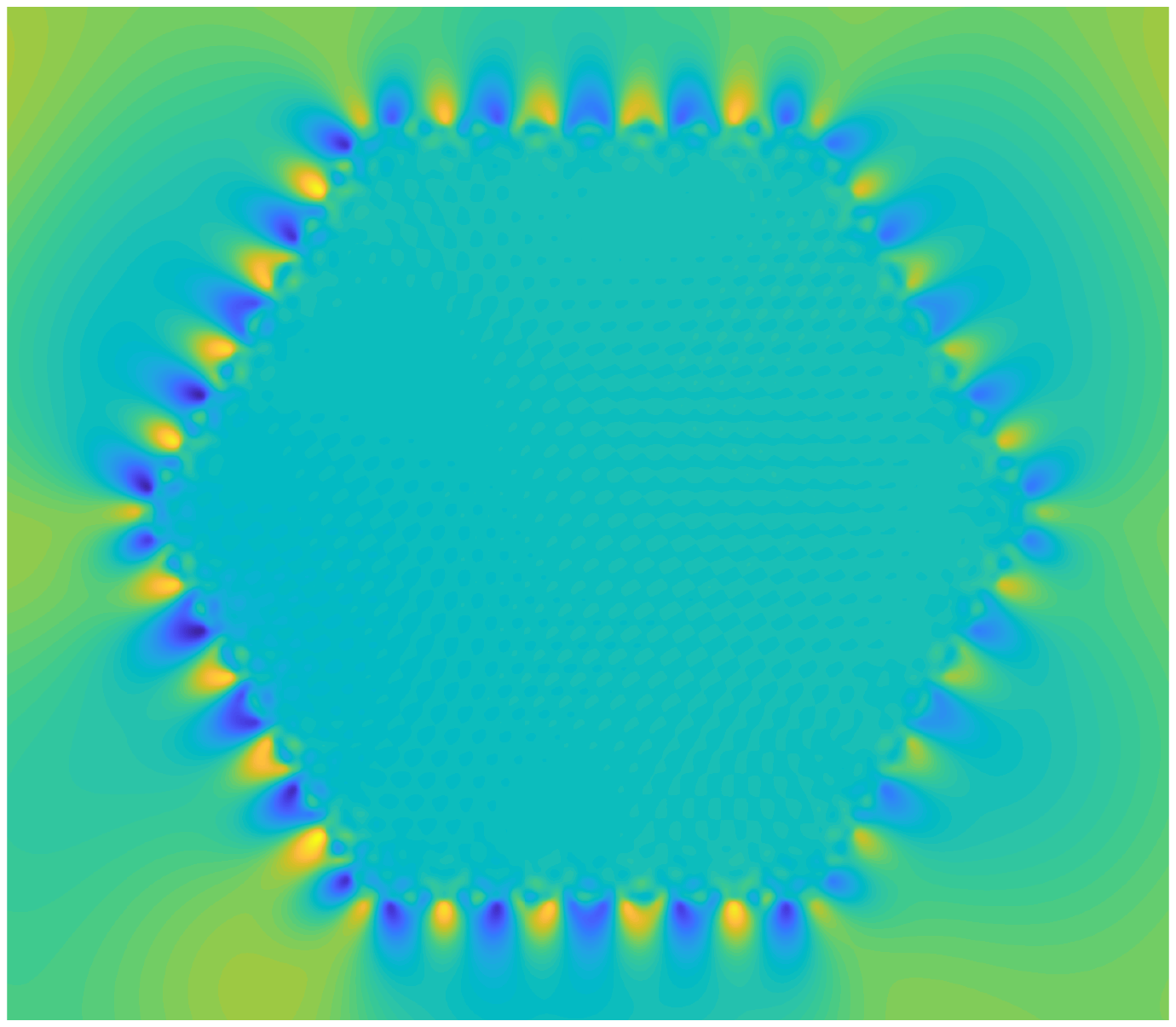}
               };
    	}] at (0,2.5){};           
		\node[below, scale=2.86, black] at (-8.5+1.0,8.75-3.5) {$\displaystyle (c)$};           
\end{scope}

\begin{scope}[xshift=14.5cm, yshift=22cm,scale=1.4]
	\node[regular polygon, regular polygon sides=6,draw, inner sep=4.5cm,rotate=0,line width=0.0mm, black,  opacity=0.0,
           	path picture={
               \node[rotate=0,opacity=1.0] at (-0.5,-0.5){
                   \includegraphics[scale=1.25]{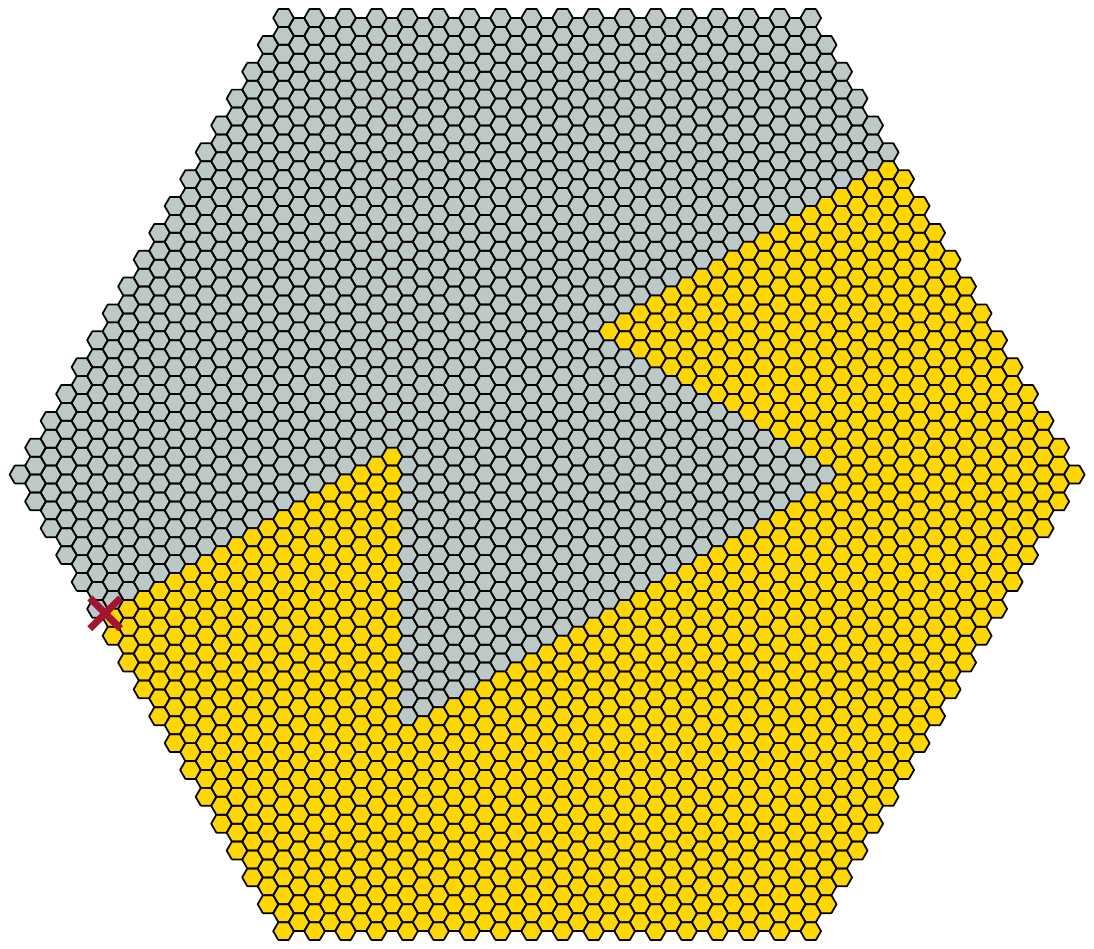}
               };
      }]{};          
		\node[below, scale=2.86, black] at (-8.5+1.0,8.75-3.5) {$\displaystyle (a)$};           
\end{scope}

\begin{scope}[xshift=14.4cm, yshift=3.25cm,scale=1.45]
           		\node[regular polygon, regular polygon sides=6,draw, inner sep=4.0cm,rotate=0,line width=0.0mm, black, opacity=0.0,
           			path picture={
              			 \node[rotate=0,opacity=1.0] at (-0.5,-0.25){
              		    	 \includegraphics[scale=1.25]{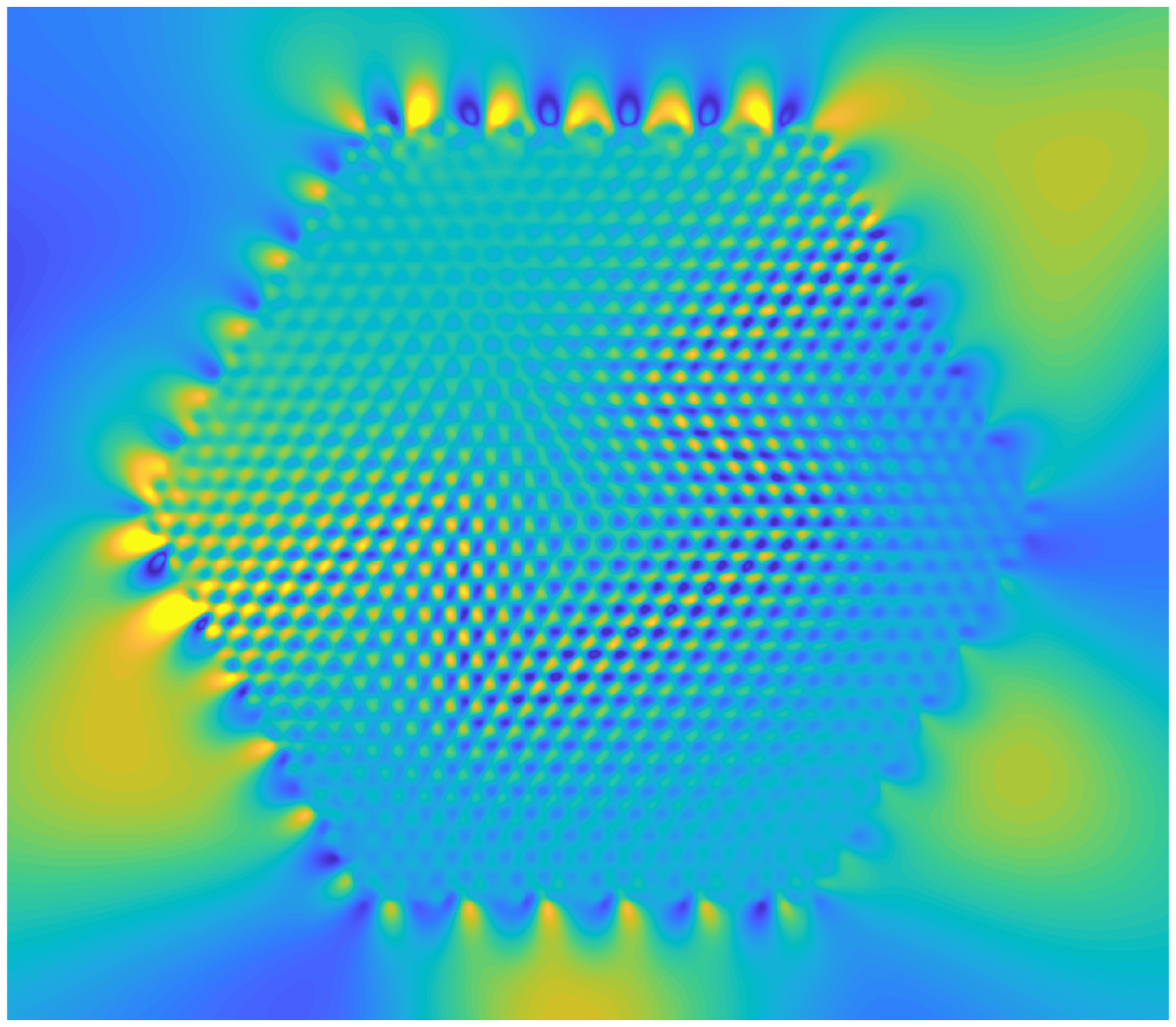}
             		  		};
             		}]{};	  		
           \node[below, scale=2.86, black] at (-8.5+1.0,8.75-3.5) {$\displaystyle (b)$};           
\end{scope}

\begin{scope}[xshift=40.0cm, yshift=-18cm,scale=1.4]
	\node[regular polygon, regular polygon sides=4,draw, inner sep=7.5cm,rotate=0,line width=0.0mm, black, opacity=0.0,
           path picture={
               \node[rotate=0,opacity=1.0] at (-0.5,-2.5){
                   \includegraphics[scale=1.25]{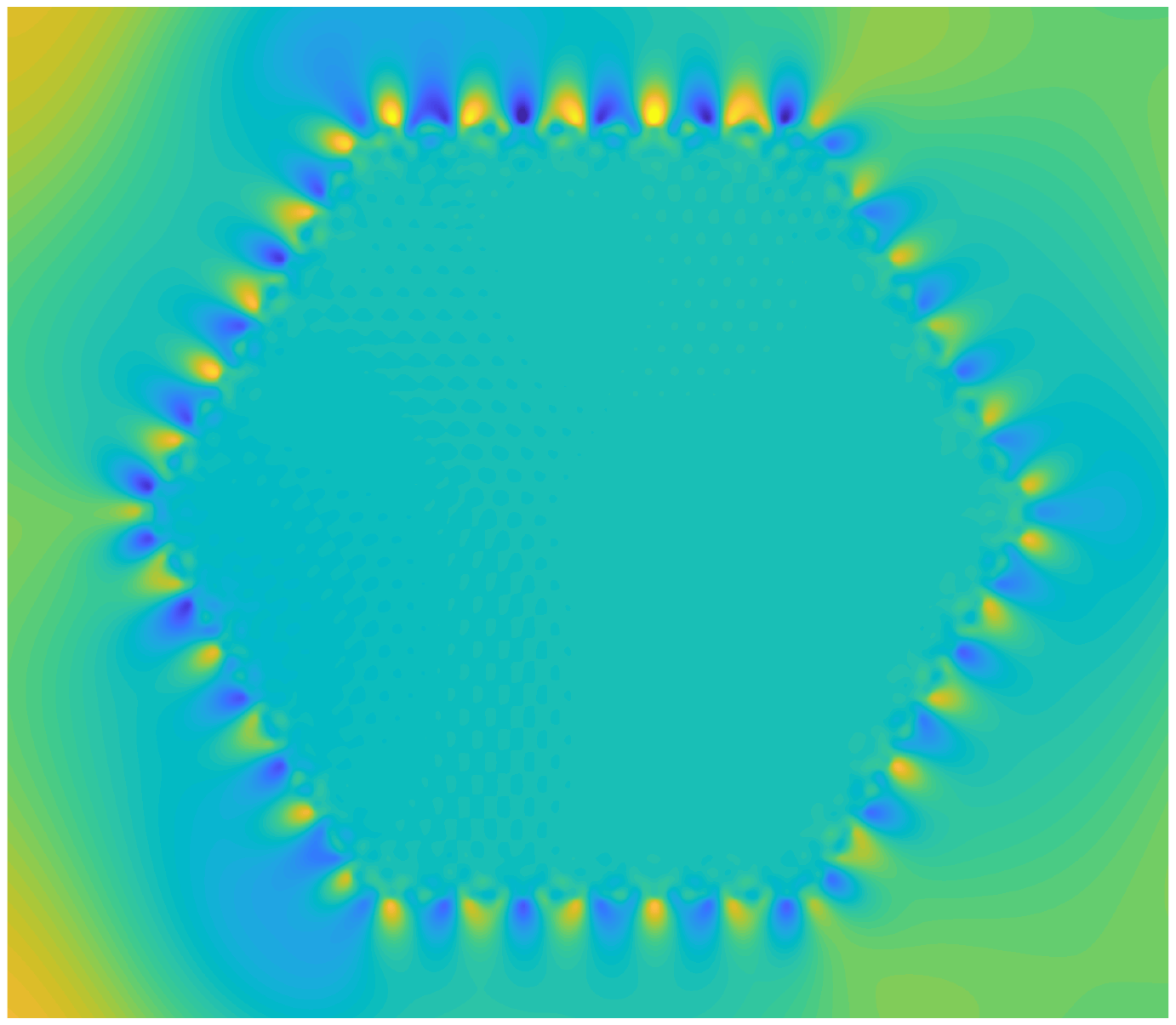}
               };
    	}] at (0,2.5){};           
		\node[below, scale=2.86, black] at (-8.5+1.0,8.75-3.5) {$\displaystyle (f)$};           
\end{scope}

\begin{scope}[xshift=40.0cm, yshift=22cm,scale=1.4]
	\node[regular polygon, regular polygon sides=6,draw, inner sep=4.5cm,rotate=0,line width=0.0mm, black,  opacity=0.0,
           	path picture={
               \node[rotate=0,opacity=1.0] at (-0.5,-0.5){
                   \includegraphics[scale=1.25]{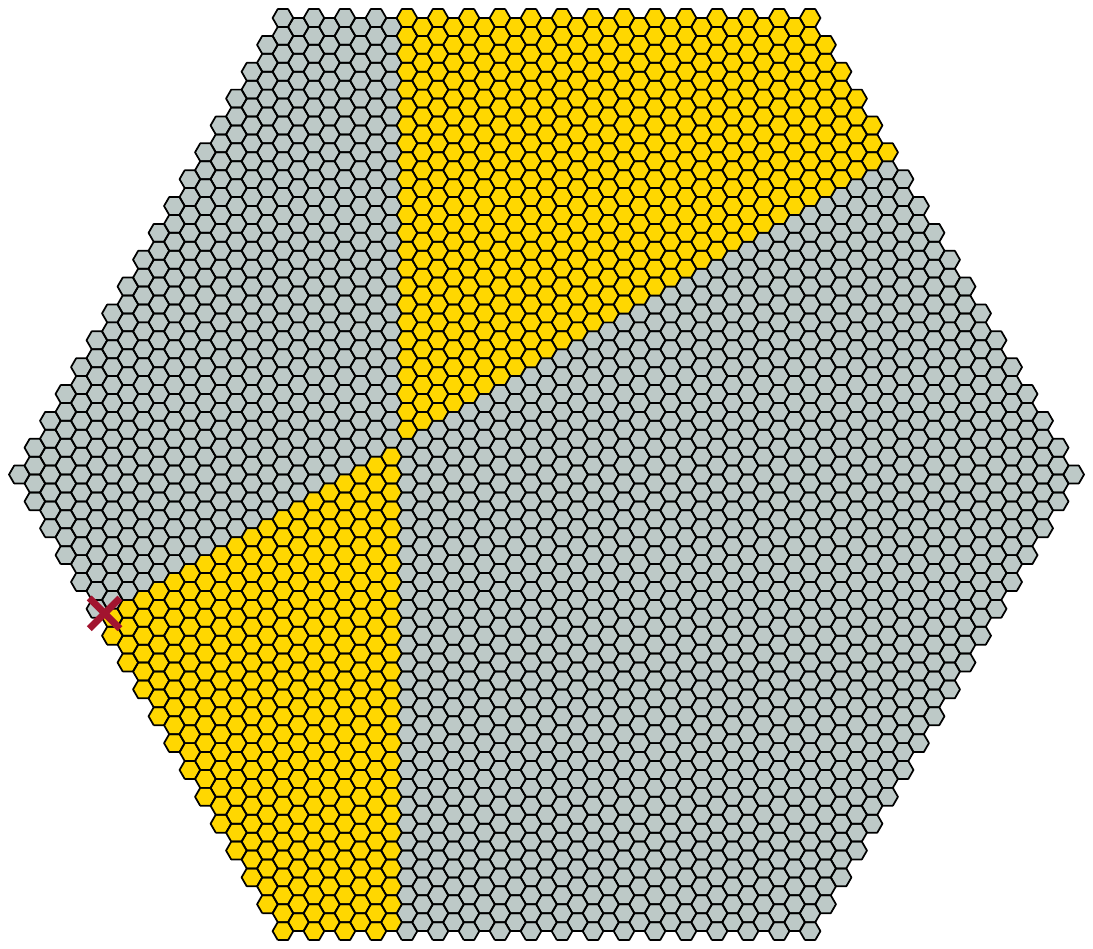}
               };
      }]{};          
		\node[below, scale=2.86, black] at (-8.5+1.0,8.75-3.5) {$\displaystyle (d)$};           
\end{scope}

\begin{scope}[xshift=40.2cm, yshift=3.25cm,scale=1.45]
           		\node[regular polygon, regular polygon sides=6,draw, inner sep=4.0cm,rotate=0,line width=0.0mm, black, opacity=0.0,
           			path picture={
              			 \node[rotate=0,opacity=1.0] at (-0.5,-0.25){
              		    	 \includegraphics[scale=1.4]{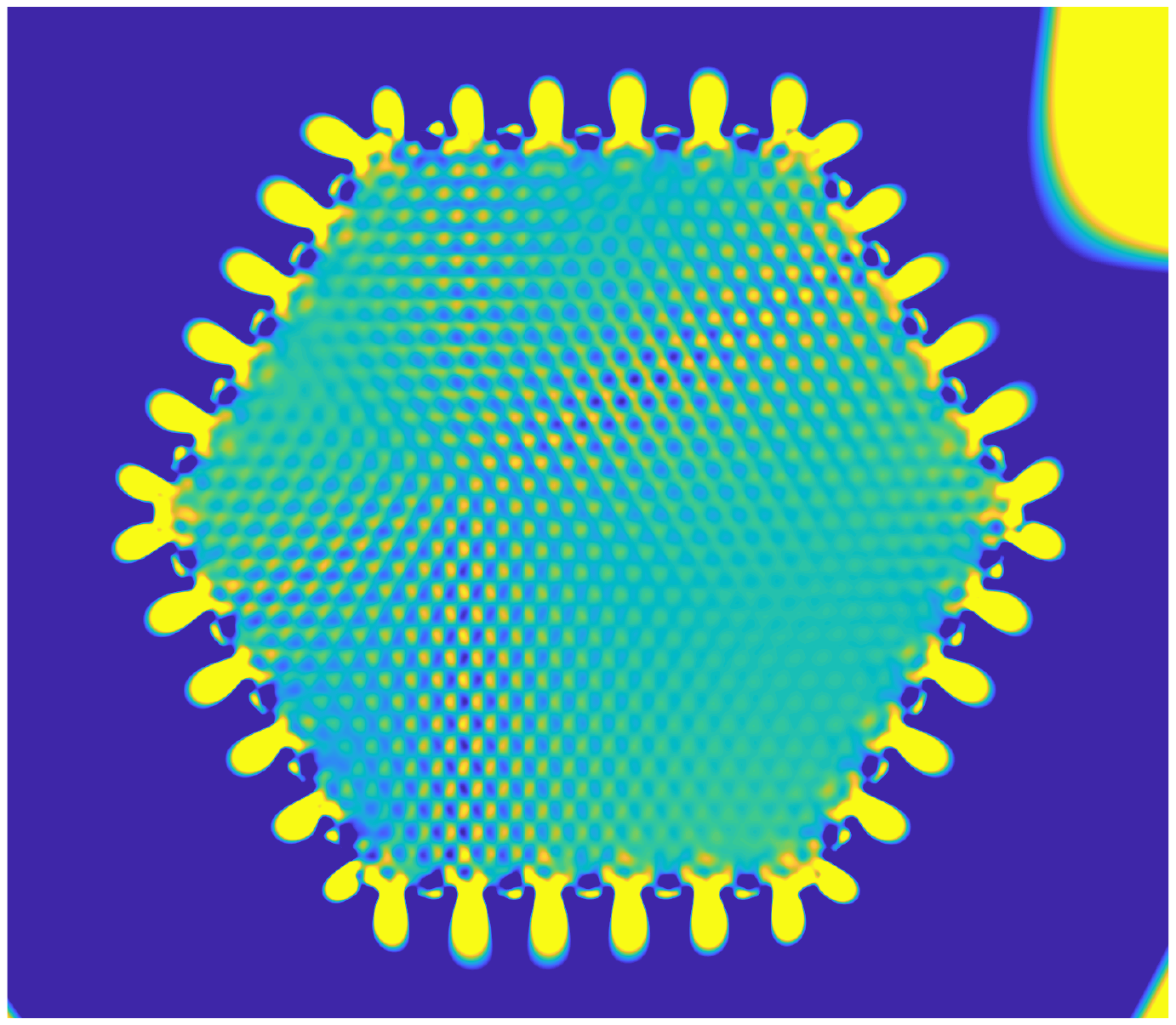}
             		  		};
             		}]{};	  		
           \node[below, scale=2.86, black] at (-8.5+1.0,8.75-3.5) {$\displaystyle (e)$};           
\end{scope}

\begin{scope}[xshift=65cm, yshift=9cm,scale=1.4]
		\node[regular polygon, regular polygon sides=4,draw, inner sep=7cm,rotate=0,line width=0.0mm, white,
           path picture={
               \node[rotate=0] at (-17.0,0){
                   \includegraphics[scale=1.25]{Figs/colourBarScattRe.pdf}
               };
           }]{};
\end{scope}  
\end{tikzpicture}
\caption{Phononic crystal designs demonstrating ZLM steering and energy splitting in panels $(a)$ and $(d)$. Here we modify the design of fig. \ref{fig:ScattPertHexTopoArrange} $(c)$ to test the robustness of the ZLMs. Panels $(a)$ and $(d)$ are both formed from $2653$ cells with $1320$ grey in $(a)$ and $1826$ grey in $(d)$, for designs with a total of $7959$ beams. The generalised Foldy simulations all consider monopole incident sources with $\varpi_{\mathrm{inc}}=1$ from \eqref{ForcePOPinc} and $\textbf{X}_{\mathrm{inc}}$ is marked by $\color{myRed} \boldsymbol{\times}$ in $(a)$ and $(d)$; the frequencies of the sources all lie within the deep-subwavelength topological band gaps from fig. \ref{fig:PertHexArrange} $(a)$,  and are set to $\Omega = 0.0312816$ for $(b)$, $\Omega=0.0312692$ for $(c)$, $\Omega=0.0312826$ for $(e)$ and $\Omega=0.0312760$ for $(f)$. The colour bar refers to the real part of the normalised displacement fields in all simulations.} 
\label{fig:ScattPertHexTopoRobustArrange}
\end{figure}

When joining any edges of grey and yellow hexagonal cells together, it is always the case that locations of opposite Berry curvature are projected onto one another and ZLMs are generated - refer to fig.\ref{fig:BerryCurvature} $(c)$. Therefore, we may modify the design of fig.\ref{fig:ScattPertHexTopoArrange} $(c)$ to introduce a variety of acute and obtuse bends along the interfaces, to demonstrate the robustness of ZLMs by achieving topological mode steering (fig.\ref{fig:ScattPertHexTopoRobustArrange} $(a)$) or energy splitting (fig.\ref{fig:ScattPertHexTopoRobustArrange} $(d)$). Acute ($\frac{\pi}{3}$) and obtuse ($\frac{2 \pi}{3}$) bends are respectively associated with model preservation and conversion. We now excite modes by placing our source at the edge of the phononic crystals. Consider fig.\ref{fig:ScattPertHexTopoRobustArrange} $(b)$, observe the ZLM of fig.\ref{fig:PertHexArrange} $(d)$ is conserved and able to navigate the acute bends within fig.\ref{fig:ScattPertHexTopoRobustArrange} $(a)$. Within fig.\ref{fig:ScattPertHexTopoRobustArrange} $(e)$, observe the ZLM of fig.\ref{fig:PertHexArrange} $(d)$ is excited, at $\color{myRed} \boldsymbol{\times}$ in fig.\ref{fig:ScattPertHexTopoRobustArrange} $(d)$, and is readily split 3-ways; model conversion to the ZLM of fig.\ref{fig:PertHexArrange} $(f)$ occurs at the $\frac{2 \pi}{3}$ and $\pi$ bends, and the ZLM of fig.\ref{fig:PertHexArrange} $(d)$ is preserved at the $\frac{\pi}{3}$ bend. Note ZLMs are not always excited when operating within the topological band gap, the cases of figs \ref{fig:ScattPertHexTopoRobustArrange} $(c)$ \& $(f)$ show this, observe the edgemodes in figs \ref{fig:ScattPertHexTopoArrange} $(g)$-$(i)$ are excited in figs \ref{fig:ScattPertHexTopoRobustArrange} $(a)$ \& $(d)$ at $\color{myRed} \boldsymbol{\times}$, and propagate around the entire perimeter of the phononic crystals. The topological mode steering and energy splitting occurs at the deep-subwavelength scale and in absence of backscatter, therefore demonstrating the robustness of modes.

\section{Conclusion}
Herein, we present our analytical solutions approximating dispersion and scattering from structured media built from arrays of elastic beams atop an elastic plate, appropriate for designing phononic crystals. The solutions are expressed through singular Green's functions, via evaluating Fourier series (for Floquet-Bloch bands) or Fourier transforms (scattering with finite arrays) in the limits where they diverge. The problem is considerably simplified by the hypothesis of Euler-Bernoulli beam theory \cite{graff1975wave}, which allows us to determine the forces and moments which couple into the plate from the constituent beams, and allows us to approximate the junction between the beams and plates as rigid disks. In sections \ref{sec:modelling} \& \ref{sec:modellingBCs} we show how to replace each beam with point monopoles and dipoles, whose coefficients determine forces and moments in terms of quantities defining arbitrary finite displacement and rotation - which encapsulates the assumed rigid portion of the plates motion. 

In section \ref{sec:EigProb}, we show how to relate the singular Green's functions for the Fourier series and Fourier transform representations of solutions in the limit as they diverge, and how this allows one to cancel singularities from either side of the relation; allowing construction of an eigenvalue problem determining the Floquet-Bloch dispersion branches from eigenvalues and eigenmodes from eigenvectors. In section \ref{sec:Foldy}, we apply Foldy's method to simulate scattering by a finite collection of beams under incidence. These solutions are tested against FEM computations in section \ref{sec::Testing}, where they are found to be accurate, and reveal two regimes for motion of the beams; either flexing like clamped-free or pinned-free beams depending on the ratio of the radius of beams to the thickness of the plate.

Finally in section \ref{Sec:Topo}, we apply our analytical solutions to consider topological arrangements of beams, again in good agreement with FEM computations, to form degeneracies with non-trivial symmetry protection. These degeneracies coincide with the first flexural resonance of the arrangements of the beams and hence can be tuned into a deep-subwavelength regime; additionally, they are simple to gap and produce the phononic analogue of the QVHE, where we demonstrate the existence of protected interfacial and edge states. These states coexist within the band gap of the bulk media, and propagate along the interfaces between two adjoining topological chiral mirrored bulk media and the edges where a topological bulk of one chirality meets the free space. We build on this knowledge to design phononic circuits, where interfacial and edge states efficiently convert between one another and navigate corners, all in the presence of deep subwavelength topological protection; which we further demonstrate by designing crystals which permit topological ZLM steering and energy splitting, where we show our ZLMs are able to navigate corners and split energy through modal conversion and preservation, all in absence of backscatter.

These demonstrations elucidate the creation of topological waveguides and energy splitters, at very low frequencies in a phononic setting, which could aid the creation of future seismic protection or energy harvesting devices to collect/mitigate vibrational energy. Similar examples have been discussed in the context of phononics and photonics, where the idea of melding topology and resonance is quite general and promising in controlling waves in many physical settings, from Bragg to deep-subwavelength scales.

\section*{Acknowledgements}
R. W. is thankful to Mehul P. Makwana for interesting discussions regarding topology and the QVHE. R. W. would also like to thank Daniel J. Colquitt for the intuition gained from discussing related problems \cite{colquitt2017seismic,carta2020chiral}.
J.M.D.P. and R.V.C acknowledge the financial support from the H2020 FET-proactive project MetaVEH under grant agreement No. 952039.

R. W. appreciates funding from the EPSRC Centre for Doctoral Training in Fluid Dynamics across Scales, UK, reference EP/L016230/1. R. W. and R.V.C. also acknowledge funding from the, EU, H2020 project MetaVEH grant agreement number 952039.

\appendix

\section{Modelling and continuity conditions for elastic beams attached to the surface of thin plates} \label{AppModelling}
Consider fig. \ref{fig:PeriodicBeamsSkem}, showing a plate whose displacement field is denoted $\boldsymbol{\mathfrak{u}}$. We consider a global Cartesian system such that the $x-y$ plane coincides with the neutral plane of the undeformed plate and consider beams attached to the surface of the plate, the neutral axis of the undeformed beam is aligned with the $z$ axis. We assume a time harmonic displacement field of the form
\begin{equation}
\boldsymbol{\mathfrak{u}}(\textbf{x},z,t) = \boldsymbol{\mathfrak{u}}(\textbf{x},z) \exp( -i \omega t),
\end{equation}
where $\textbf{x}$ denotes some in-plane position vector $\textbf{x} = x \textbf{e}_{x} + y \textbf{e}_{y}$ and $z$ characterises some out-of-plane position vector $z \textbf{e}_{z}$. We define regular and hatted quantities to denote the displacement fields of the plate or beams respectively. We make the classic assumptions of thin plates and beams, see for example Landau and Lifshitz \cite{d1989theory} or Graff \cite{graff1975wave}, the deformation in the plate is assumed to act under plane stress hence 
\begin{equation}
    \boldsymbol{\mathfrak{u}} = \Big( u(\textbf{x}) - z \frac{\partial w}{\partial x} \Big) \textbf{e}_{x} +  
    \Big( v(\textbf{x}) - z \frac{\partial w}{\partial y} \Big) \textbf{e}_{y} + w(\textbf{x}) \textbf{e}_{z}. \label{PlateDispFe}
\end{equation}
Here the total displacement field is a superposition of $w(\textbf{x})$ (pure bending deformations), and $\textbf{u}(\textbf{x}) = u(\textbf{x}) \textbf{e}_{x} + v(\textbf{x}) \textbf{e}_{y}$, (purely longitudinal deformations) in the plate. For the beams we need only consider the time harmonic displacement field along the neutral axis, hence
\begin{equation}
\hat{\boldsymbol{\mathfrak{u}}} = \hat{\boldsymbol{\mathfrak{u}}}(z).
\end{equation}
We non-dimensionalise quantities by introducing some length scale $L$ as follows
\begin{equation}
\textbf{x} = L \widetilde{\textbf{x}}, \quad z = L \widetilde{z}, \quad \boldsymbol{\mathfrak{u}} = L \widetilde{\boldsymbol{\mathfrak{u}}},
\end{equation} 
similarly for hatted quantities. Dropping tildes, the dimensionless system of equations governing longitudinal and flexural deformations within the beams and plate are given by \cite{d1989theory, graff1975wave} 
\begin{align}
\left[ \nabla^{4} - \Omega^{2} \right] \textbf{w} = \frac{L^{3}}{D} P_{3}(\textbf{x}) \textbf{e}_{z},  \label{KL}\\
\nabla (\nabla \cdot \textbf{u} ) - \frac{\beta^{2}}{\alpha^{2}} \nabla \times \nabla \times \textbf{u} + \frac{\Omega^{2} h^{2}}{12 L^{2}} \textbf{u} + \frac{L}{\rho h \alpha^{2}} \textbf{P} (\textbf{x}) = \textbf{0}, \label{2DE}\\
\left\lbrace \frac{d^{2}}{d z^{2}} + \hat{\alpha}^{2} \Omega^{2} \right\rbrace \hat{\textbf{w}}= \textbf{0}, \label{BeamComp} \\
\left\lbrace\frac{d^{4}}{d z^{4}} - \hat{\beta}^{4} \Omega^{2}  \right\rbrace \hat{\textbf{u}} = \textbf{0} \label{BeamFlex},
\end{align}
$\textbf{P}$ and $P_{3}(\textbf{x})$ respectively denote any in-plane or out-of-plane forces per unit area acting on the plate. Torsional beam motion is assumed to be negligible and hence ignored. It is convenient to express $\textbf{u}$ by means of the dilational and shear potentials
\begin{equation*}
\textbf{u} = \nabla \phi + \nabla \times \boldsymbol{\psi},
\end{equation*}
where $\boldsymbol{\psi} = \psi \textbf{e}_{z}$. 

Consider a beam, of length $\hat{\ell}$, whose base is centred and attached to the surface of the plate at $z=\frac{h}{2L}$ and $\textbf{x} = \textbf{X}$ (as in fig. \ref{fig:PeriodicBeamsSkem}). At $z = \hat{\ell} + \frac{h}{2L}$ we apply free end conditions
\begin{equation}
\frac{\partial \hat{\textbf{w}}}{\partial z} \Big|_{z = \hat{\ell} + \frac{h}{2L}} = \textbf{0}, \quad \quad \frac{\partial^{2} \hat{\textbf{u}}}{\partial z^{2}} \Big|_{z = \hat{\ell} + \frac{h}{2L}}  = \frac{\partial^{3} \hat{\textbf{u}}}{\partial z^{3}} \Big|_{z = \hat{\ell} + \frac{h}{2L}}  = \textbf{0}. \label{FreeEnd}
\end{equation}
To apply continuity of displacement and rotation conditions, where the beam and plate intersect, we introduce unknown plate displacements $\boldsymbol{\mathfrak{u}}(\textbf{X},\frac{h}{2L})$ at the center of the beam. We require 
\begin{equation}
    \hat{\boldsymbol{\mathfrak{u}}} \Big|_{z = \frac{h}{2L}} = \boldsymbol{\mathfrak{u}}(\textbf{X},\frac{h}{2L}), \quad \mbox{and} \quad \Big[ (\nabla + \textbf{e}_{z} \frac{\partial}{\partial z}) \times \hat{\boldsymbol{\mathfrak{u}}} \Big] \Big|_{z = \frac{h}{2L}} = \Big[ (\nabla + \textbf{e}_{z} \frac{\partial}{\partial z}) \times \boldsymbol{\mathfrak{u}} \Big] \Big|_{\substack{ \textbf{x} = \textbf{X} \\ \, z = \frac{h}{2L}}}
\end{equation}
Hence the following conditions must be satisfied
\begin{align}
\hat{\textbf{w}} \Big|_{z = \frac{h}{2L}} = \textbf{w}(\textbf{X}), \quad \quad  \hat{\textbf{u}} \Big|_{z = \frac{h}{2L}} = \textbf{u}(\textbf{X}) - \frac{h}{2L} \Big[ \nabla w \Big] \Big|_{\textbf{x} = \textbf{X}}, & \quad \quad \frac{\partial \hat{\textbf{u}}}{\partial z} \Big|_{z = \frac{h}{2L}} = - 2\Big[ \nabla w \Big] \Big|_{\textbf{x} = \textbf{X}}.  \label{NotFreeEnd} \\
\Big[ \nabla \times \boldsymbol{u} \Big] \Big|_{\textbf{x} = \textbf{X}} & = \textbf{0}. \label{plateCondRot}
\end{align}

Equations \eqref{BeamComp} and \eqref{BeamFlex} subject to \eqref{FreeEnd} and \eqref{NotFreeEnd} may be solved in terms of the unknowns $\textbf{w}(\textbf{X})$, $\textbf{u}(\textbf{X})$ and $\Big[ \nabla \textbf{w} \Big] \Big|_{\textbf{x} = \textbf{X}}$. For continuity of moments we require $\textbf{u}(\textbf{X}) = \textbf{0}$, otherwise one finds the bending moment coupling into plate does not arise from pure bending, and hence invalidates the Kirchhoff–Love theory of thin elastic plates; therefore, the gradients associated with any potentials are not required. 

\section{Complementary Coefficients} \label{AppCompCoeffs}
The coefficients present in \eqref{Wc} - \eqref{psiC} are simple to determine, by satisfying equations \eqref{DispCond1} - \eqref{PotentialConds} we find 
\begin{equation}
\begin{split}
    \begin{pmatrix}
A_{0} \\
B_{0}
\end{pmatrix} = \frac{1}{\mathcal{W}[J_{0}(r \sqrt{\Omega}), J_{0}(i r \sqrt{\Omega})]} \Bigg\lbrace  w(\textbf{X})  \begin{pmatrix}
\mathcal{W}[1, J_{0}(i r \sqrt{\Omega})] \\
\mathcal{W}[J_{0}(r \sqrt{\Omega}), 1 ]
\end{pmatrix} 
-  \\
- \frac{i L \textbf{F} \cdot \textbf{e}_{z}}{D 8 \Omega} \begin{pmatrix}
\mathcal{W}[H_{0}(r \sqrt{\Omega}) - H_{0}(i r \sqrt{\Omega}), J_{0}(i r \sqrt{\Omega})] \\
\mathcal{W}[J_{0}(r \sqrt{\Omega}), H_{0}(r \sqrt{\Omega}) -  H_{0}(i r \sqrt{\Omega}) ]
\end{pmatrix} 
 \Bigg\rbrace \Bigg|_{r=\epsilon}, 
\end{split} \label{EqnAB0comp}
\end{equation}
\begin{equation}
\begin{split}
    \begin{pmatrix}
A_{i} \\
B_{i}
\end{pmatrix} = - \frac{1}{\mathcal{W}[J_{1}(r \sqrt{\Omega}), J_{1}(i r \sqrt{\Omega})]} \Bigg\lbrace \frac{M_{i}}{D 8 \sqrt{\Omega}} \begin{pmatrix}
\mathcal{W}[\frac{H_{1}(r \sqrt{\Omega})}{i} - H_{1}(i r \sqrt{\Omega}), J_{1}(i r \sqrt{\Omega})] \\
\mathcal{W}[J_{1}(r \sqrt{\Omega}), \frac{H_{1}(r \sqrt{\Omega})}{i} -  H_{1}(i r \sqrt{\Omega}) ]
\end{pmatrix} + \\
+ [\nabla \times \textbf{w}] \Big|_{\textbf{x} = \textbf{X}} \cdot \textbf{e}_{i} \begin{pmatrix}
\mathcal{W}[r, J_{1}(i r \sqrt{\Omega})] \\
\mathcal{W}[J_{1}(r \sqrt{\Omega}), r ]
\end{pmatrix} \Bigg\rbrace \Bigg|_{r=\epsilon}  \quad \mbox{for subscript $i = 1,2$}. \label{EqnAB12comp}
\end{split} 
\end{equation}
\begin{equation}
   C = \phi(\textbf{X}),
\end{equation}
\begin{equation}
   D = \psi(\textbf{X}).
\end{equation}
The coefficients for $A_{0}, B_{0}, A_{i}, B_{i}$ are found in a similar fashion to O'Neill \textit{et al}. \cite{ONeill2015active}, by utilizing the Wronskian operator $\mathcal{W}[ \quad ]$. Note $\textbf{C}$ and $\textbf{D}$ within \eqref{phiC} and \eqref{psiC} are found in a similar way to \eqref{EqnAB0comp} and \eqref{EqnAB12comp}, however the contribution arising from the $J_{1}$ terms within the potentials can be disregarded at leading order - such a statement would not be true if $\nabla \phi \Big|_{\textbf{x} = \textbf{X}}$ or $\nabla \times \boldsymbol{\psi}\Big|_{\textbf{x} = \textbf{X}}$ were required. All of the Wronskian terms in  \eqref{EqnAB0comp} - \eqref{EqnAB12comp} are evaluated at $r = \epsilon$ and asymptotically expanded treating $\epsilon$ as a small parameter, and hence assuming functions have small arguments.

\section{Singular asymptotics of conditionally convergent series} \label{ResidualTermsApp}

\begin{figure}[H]
\centering
\begin{tikzpicture}
\begin{scope}[shift = {(6,0)}]
 \draw[->] (-1,0) -- (3.5,0) coordinate[label=below: $\displaystyle G_{1}$] (x);
\draw[->] (0,-1) -- (0,3.5) node[left] {$\displaystyle G_{2}$} ; 
\draw[thick,->] (0,0) -- (0.5,1.5);
\draw[thick,->] (0,0) -- (-0.75,0.75);
\draw[thick,->] (0,0) -- (1.5,-0.75);
\draw[thick,->] (0,0) -- (2.0,0.55);
\draw[thick,->] (0,0) -- (2.5, 2.75 );
\draw[dashed] (2.0,0.55) -- (2.75, 0.55);
\node[above right] at (2.5,2.75) {$\displaystyle \textbf{x}$}; 
\node[below] at (2.0,0.55) {$\displaystyle \quad \textbf{X}_{11}$}; 
\node[left] at (2.25,1.6) {$\displaystyle \quad \quad r$}; 
\node[right] at (1.5,-0.75) {$\displaystyle \boldsymbol{\kappa}$}; 
\draw[-] (2.0,0.55) -- (2.5, 2.75 );

  \draw
  	(2.5,0) coordinate (a) node[above right] {}
     (0,0) coordinate (b) node[above right] {}
     (1.5,-0.75) coordinate (c) node[above left] {}
    pic["$\quad \quad \quad \, \, \displaystyle (2 \pi - \theta_{\kappa})$", draw=black, <->, angle eccentricity=1.2, angle radius=1.0cm]
    {angle=c--b--a};

 \draw
  	(2.5,0) coordinate (a) node[above right] {}
     (0,0) coordinate (b) node[above right] {}
     (-0.75,0.75) coordinate (c) node[above left] {$\displaystyle \textbf{N}$}
    pic["$\displaystyle (2 \pi - \theta_{N}) \quad \quad \quad \quad \, \,$", draw=black, <->, angle eccentricity=1.2, angle radius=0.5cm]
    {angle=c--b--a};   
    
\draw
  	(2.5,0) coordinate (a) node[above right] {}
     (0,0) coordinate (b) node[above right] {}
     (0.5,1.5) coordinate (c) node[above] {$\displaystyle \, \, \, \textbf{G}$}
    pic["$\displaystyle  \, \, \, \, \theta_{G}$", draw=black, <->, angle eccentricity=1.2, angle radius=0.750cm]
    {angle=a--b--c};    
    
\draw
  	(2.75,0.5) coordinate (a) node[above right] {}
     (2.0,0.55) coordinate (b) node[above right] {}
     (2.5,2.75) coordinate (c) node[above right] {}
    pic["$\quad \displaystyle \theta_{r}$", draw=black, <->, angle eccentricity=1.2, angle radius=0.50cm]
    {angle=a--b--c}; 
\end{scope}          
\end{tikzpicture}
\caption{The required vector quantities in Fourier space to derive the  double sum asymptotics of our Fourier series solutions beyond truncation. Here $\textbf{N}$ refers to either $\textbf{M}$ or $\textbf{V}$ as appropriate, to be regarded as a constant for some $\boldsymbol{\kappa} = \boldsymbol{\kappa}(\Omega)$ wave of fixed phase.
}
\label{GandN}
\end{figure}
We require the singular asymptotics of several series which diverge in the limit as $r \to 0$. These terms are calculated analytically and expressed through Van der Pol's Bessel-integral function of zero order, $Ji_{0}(x)$, as defined in Humbert \cite{humbert1933bessel} where the following behaviour is of importance
\begin{equation}
Ji_{0}(x) = Ci(x) - \log(2) = \log(\frac{x}{2}) + \gamma_{E} - \frac{x^{2}}{4} + \mathcal{O}(x^{4}) \quad \mbox{as $x \to 0$}, \label{BesselIntBehave}
\end{equation}
here $\gamma_{E}$ is the Euler–Mascheroni constant. Utilizing the Euler-Maclaurin formula \cite{olver1997asymptotics} it is shown, using the notation as in fig. \ref{GandN}
\begin{equation}
\lim_{\substack{r \to 0 \\ R \to \infty }} \nabla w_{\mathrm{res}} \sim \frac{M_{p} \exp( i \boldsymbol{\kappa} \cdot \textbf{x})}{4 \pi^{2} D} \int_{R}^{\infty } \int_{0}^{2 \pi} \frac{\cos \theta_{G} \textbf{e}_{x} + \sin \theta_{G} \textbf{e}_{y}}{G} \sin(\theta_{G} - \theta_{M} ) \exp(i \textbf{G} \cdot r) \, d \theta_{G} \, dG + o(1). \label{ResidualGradW}
\end{equation} 
Refer to the appendices of \cite{schnitzer2017bloch, wiltshaw2020asymptotic} for more detail. The integral over $\theta_{G}$ is performed by means of the integral representations of the Bessel functions of integer order \cite{abramowitz1964handbook}, finally integrating over $G$ it can be shown, neglecting $o(1)$ terms
\begin{align}
\lim_{\substack{r \to 0 \\ R \to \infty }} \textbf{e}_{x} \cdot \nabla w_{\mathrm{res}} = - \frac{M_{p}}{4 \pi D} \Big\lbrace \sin(2 \theta_{r} - \theta_{M_{p}}) \frac{J_{1} (Rr)}{Rr} - \sin \theta_{m} Ji_{0}(Rr) \Big\rbrace \exp( i \boldsymbol{\kappa} \cdot \textbf{x}), \\
\lim_{\substack{r \to 0 \\ R \to \infty }} \textbf{e}_{y} \cdot \nabla w_{\mathrm{res}} = - \frac{M_{p}}{4 \pi D} \Big\lbrace \cos \theta_{m} Ji_{0}(Rr) -  \cos(2 \theta_{r} - \theta_{M_{p}}) \frac{J_{1} (Rr)}{Rr}  \Big\rbrace \exp( i \boldsymbol{\kappa} \cdot \textbf{x}).
\end{align}
Additionally, setting $a$ and $\textbf{b}$ as appropriate from Appendix A of \cite{wiltshaw2020asymptotic}, we find
\begin{equation}
\begin{split}
\lim_{\substack{r \to 0 \\ R \to \infty }} i \alpha^{2} \phi_{\mathrm{res}} \exp(-i \boldsymbol{\kappa} \cdot \textbf{X}) =\frac{3 i L }{\pi \rho \Omega^{2} h^{3}}  \exp(i \boldsymbol{\kappa} \cdot \textbf{r} )  \Big\lbrace \frac{2 \textbf{e}_{r} \cdot \textbf{V}_{p} }{r} J_{0}(Rr) -  \textbf{e}_{r} \cdot \textbf{V}_{p} \frac{\Omega^{2} h^{2}}{12 L^{2}} r  Ji_{0}(Rr) - \quad \quad     \\
- \frac{J_{1}(Rr)}{Rr} [ 2i V_{p} \kappa \cos(2 \theta_{r} - \theta_{\kappa} - \theta_{V_{p}}) - r \frac{\Omega^{2} h^{2}}{12 L^{2}} \textbf{e}_{r} \cdot \textbf{V}_{p} ] - 2 V_{p} \kappa^{2} \cos(\theta_{V_{p}} + 2 \theta_{\kappa} - 3 \theta_{r}) \frac{J_{2}(Rr)}{R^{2} r} \Big\rbrace, \label{APP:Exp1}
\end{split}
\end{equation}

\begin{equation}
\begin{split}
\lim_{\substack{r \to 0 \\ R \to \infty }} \frac{\alpha^{2}}{i} \psi_{\mathrm{res}} \exp(-i \boldsymbol{\kappa} \cdot \textbf{X}) = \frac{3 i L }{\pi \rho \Omega^{2} h^{3}}  \exp(i \boldsymbol{\kappa} \cdot \textbf{r} )   \Big\lbrace \frac{2 \textbf{e}_{\theta} \cdot \textbf{V}_{p} }{r} J_{0}(Rr) -  \textbf{e}_{\theta} \cdot \textbf{V}_{p}  \frac{\Omega^{2} h^{2}}{12 L^{2}} \frac{\alpha^{2}}{\beta^{2}} r  Ji_{0}(Rr) + \quad \quad  \\
+ \frac{J_{1}(Rr)}{Rr} [ 2i V_{p} \kappa \sin(2 \theta_{r} - \theta_{\kappa} - \theta_{V_{p}}) + r \frac{\Omega^{2} h^{2}}{12 L^{2}} \frac{\alpha^{2}}{\beta^{2}} \textbf{e}_{\theta} \cdot \textbf{V}_{p}  ] - 2 V_{p} \kappa^{2} \sin(\theta_{V_{p}} + 2 \theta_{\kappa} - 3 \theta_{r}) \frac{J_{2}(Rr)}{R^{2} r} \Big\rbrace, \label{APP:Exp2}
\end{split}
\end{equation}

Expressions \eqref{APP:Exp1} \& \eqref{APP:Exp2} are differentiated if associated gradients of potentials are required.

\section{The computationally least expensive eigenvalue problem determining dispersion} \label{CompLE}
The polynomial eigenvalue problem in \eqref{EVPdispBloch} can be expressed as a generalized eigenvalue problem by use of the linear companion matrix method \cite{bridges1984differential}, which further increase the dimension of the system but considers a computationally less expensive eigenvalue problem. The linear companion matrix method can be made even more efficient, provided the matrix multiplying the highest degree term in the polynomial eigenvalue problem is non-singular \cite{bridges1984differential}. However, our system only contains $\eta^{4}$ terms from \eqref{WFourCoeffs}, hence most rows of this matrix will be zero. The inverse of $\mathcal{A}$ is not available and \eqref{EVPdispBloch} cannot be simplified further without another change of variables. By denoting $\gamma = \frac{1}{\eta}$ one can consider 
\begin{equation}
\Big[ \gamma^{4} \mathcal{E}(\Omega) + \gamma^{3} \mathcal{D}(\Omega) + \gamma^{2} \mathcal{C}(\Omega) + \gamma \mathcal{B}(\Omega) +  \mathcal{A}(\Omega) \Big] \boldsymbol{ \Theta } = \textbf{0}. \label{EVPdispBlochCOMPANION}
\end{equation}
Since equations \eqref{WFourCoeffs} - \eqref{PsiFourCoeffs}, \eqref{Disp_rTo0}, \eqref{GradDisp_rTo0} - \eqref{Psi_rTo0} all have $\boldsymbol{\kappa}^{0}$ terms, $\mathcal{E}(\Omega)$ is naturally readily invertible, the computational eigenvalue problem is further simplified to 
\begin{equation}
\left( \widetilde{\mathcal{A}}- \gamma \mathcal{I} \right) \textbf{Y} = \textbf{0}, \label{EVPactualalgebraicDispBloch}
\end{equation}
where 
\begin{equation}
\widetilde{\mathcal{A}} = - \begin{bmatrix}
\mathcal{E}^{-1} \mathcal{D}  & \mathcal{E}^{-1} \mathcal{C}  & \mathcal{E}^{-1} \mathcal{B}  & \mathcal{E}^{-1} \mathcal{A} \\
-\mathcal{I} & \mathcal{Z} & \mathcal{Z} &  \mathcal{Z} \\
 \mathcal{Z} & -\mathcal{I} & \mathcal{Z} &  \mathcal{Z} \\
 \mathcal{Z} & \mathcal{Z} & -\mathcal{I} &  \mathcal{Z} 
\end{bmatrix}, \quad \quad \textbf{Y} = \begin{bmatrix}
\gamma^{3} \boldsymbol{\Theta} \\
\gamma^{2} \boldsymbol{\Theta} \\
\gamma^{1} \boldsymbol{\Theta} \\
\boldsymbol{\Theta}
\end{bmatrix}. \label{EVPalgebraicDispBloch}
\end{equation}
Here $\mathcal{I}$ and $\mathcal{Z}$ respectively denote the identity matrix and a matrix of zeros, both of dimension $(3 M + 5P) \times (3 M + 5P)$ . Finally one should note $\mathcal{E}$ contains terms from $\textbf{F}_{p}$, $\textbf{M}_{p}$ and $\textbf{V}_{p}$ which will be singular at resonance, therefore approaching resonance $\mathcal{E}$ will be poorly conditioned hence $\mathcal{E}^{-1} \mathcal{D}$ should be computed with care (similarly for other terms within $\widetilde{\mathcal{A}}$) -  these products were computed using the minimum norm least-squares solution (MATLAB's lsqminnorm function), which was found to be sufficiently numerically stable for our needs.

\section{The external fields} \label{AppExtFie}
In order to evaluate the unknowns within \eqref{GreensCrystalW}-\eqref{GreensCrystalpsi}, one needs to consider the following approaching the $n$th scatterer
\begin{equation}
\begin{split}
\lim_{\textbf{x} \to \textbf{X}_{n}} w(\textbf{x}) = w(\textbf{X}_{n}) = w_{\mathrm{inc}} (\textbf{X}_{n}) +  \lim_{\textbf{x} \to \textbf{X}_{n}} \sum_{j=1}^{m} \Big\lbrace \frac{i L}{8D} F_{j} w(\textbf{X}_{j}) \Big[ H_{0} (r_{j} \sqrt{\Omega} ) - H_{0} ( i r_{j} \sqrt{\Omega} ) \Big]  - \\ 
- \textbf{e}_{\theta \, j} \cdot (\nabla \times \textbf{w})\Big|_{\textbf{x} = \textbf{X}_{j}} \frac{M_{j}}{8D}  \Big[ i H_{1} (r_{j} \sqrt{\Omega} ) + H_{1} ( i r_{j} \sqrt{\Omega} ) \Big]  \Big\rbrace ,
\end{split} \label{GreensCrystalWXn}
\end{equation}

\begin{equation}
\begin{split}
\lim_{\textbf{x} \to \textbf{X}_{n}} \nabla w(\textbf{x}) = \nabla w |_{\textbf{x} = \textbf{X}_{n}} = \nabla w_{\mathrm{inc}} (\textbf{X}_{n})  + \lim_{\textbf{x} \to \textbf{X}_{n}}  \nabla \sum_{j=1}^{m} \Big\lbrace \frac{i L}{8D} F_{j} w(\textbf{X}_{j}) \Big[ H_{0} (r_{j} \sqrt{\Omega} ) - H_{0} ( i r_{j} \sqrt{\Omega} ) \Big]  - \\ 
- \textbf{e}_{\theta \, j} \cdot (\nabla \times \textbf{w})\Big|_{\textbf{x} = \textbf{X}_{j}} \frac{M_{j}}{8D}  \Big[ i H_{1} (r_{j} \sqrt{\Omega} ) + H_{1} ( i r_{j} \sqrt{\Omega} ) \Big]  \Big\rbrace ,
\end{split} \label{GreensCrystalGradWXn}
\end{equation}

\begin{equation}
\lim_{\textbf{x} \to \textbf{X}_{n}}  \phi(\textbf{x}) = \phi(\textbf{X}_{n}) = \phi_{\mathrm{inc}} (\textbf{X}_{n}) + \lim_{\textbf{x} \to \textbf{X}_{n}}  \sum_{j=1}^{m} \frac{i V_{j} \sqrt{3}}{2 \rho  h^{2} \alpha^{2}} \textbf{e}_{r \, j}  \cdot \nabla w\Big|_{\textbf{x} = \textbf{X}_{j}}  H_{1} (\frac{\Omega h}{2 L \sqrt{3}} r_{j} ), \label{GreensCrystalphiXn}
\end{equation}

\begin{equation}
\lim_{\textbf{x} \to \textbf{X}_{n}}  \psi(\textbf{x}) = \psi(\textbf{X}_{n})  = \psi_{\mathrm{inc}} (\textbf{X}_{n}) + \lim_{\textbf{x} \to \textbf{X}_{n}}  \sum_{j=1}^{m}  \frac{V_{j} \sqrt{3}}{2 i \rho h^{2} \alpha \beta} \textbf{e}_{\theta \, j} \cdot \nabla w\Big|_{\textbf{x} = \textbf{X}_{j}}  H_{1} (\frac{\alpha}{\beta} \frac{\Omega h}{2 L \sqrt{3}} r_{j} ), \label{GreensCrystalpsiXn}
\end{equation} where we denote the $\lim_{\textbf{x} \to \textbf{X}_{n}} r_{j} = |\textbf{X}_{n} - \textbf{X}_{j}| = r_{nj}$. The $j=n$th contribution from the above sums, as $r_{nn} \to 0$ for arbitrary angle, introduces logarithmic singularities within \eqref{GreensCrystalGradWXn} and $\frac{1}{r_{nn}}$ singularities occurring in \eqref{GreensCrystalphiXn} \& \eqref{GreensCrystalpsiXn}. These singularities are removed by considering the generalised Foldy method, as outlined by Martin \cite{martin2006multiple, martin2015scattering}, by defining the external displacement and potential fields about the $n$th scatterer as follows
\begin{equation}
\begin{split}
w_{n}(\textbf{x}) = w(\textbf{x}) - \Big\lbrace \frac{i L}{8D} F_{n} w(\textbf{X}_{n}) \Big[ H_{0} (r_{n}\sqrt{\Omega} ) - H_{0} ( i r_{n}\sqrt{\Omega} ) \Big] -   \quad \quad \quad \quad\\ 
\quad \quad \quad \quad - \textbf{e}_{\theta \, n} \cdot (\nabla \times \textbf{w})\Big|_{\textbf{x} = \textbf{X}_{n}} \frac{M_{n}}{8D}  \Big[ i H_{1} (r_{n}\sqrt{\Omega} ) + H_{1} ( i r_{n}\sqrt{\Omega} ) \Big]  \Big\rbrace ,
\end{split} \label{wExternal}
\end{equation}
\begin{equation}
\phi_{n}(\textbf{x}) = \phi(\textbf{x}) - \Big\lbrace \frac{i V_{n} \sqrt{3}}{2 \rho  h^{2} \alpha^{2}} \textbf{e}_{r \, n}  \cdot \nabla w\Big|_{\textbf{x} = \textbf{X}_{n}}  H_{1} (\frac{\Omega h}{2 L \sqrt{3}} r_{n} )  \Big\rbrace, \label{phiExternal}
\end{equation}
\begin{equation}
\psi_{n}(\textbf{x}) = \psi(\textbf{x}) - \Big\lbrace \frac{V_{n} \sqrt{3}}{2 i \rho h^{2} \alpha \beta} \textbf{e}_{\theta \, n} \cdot \nabla w\Big|_{\textbf{x} = \textbf{X}_{n}}  H_{1} (\frac{\alpha}{\beta} \frac{\Omega h}{2 L \sqrt{3}} r_{n} )  \Big\rbrace. \label{psiExternal}
\end{equation}

\printbibliography

 \end{document}